\documentclass[a4paper,10pt]{article}
\usepackage[english]{babel}
\usepackage{jheppub}
\usepackage{subfig}
\usepackage[T1]{fontenc} 
\usepackage{graphicx,calc}
\usepackage{epsfig}
\usepackage{amsmath}
\usepackage{amssymb}
\usepackage{float}
\usepackage{placeins}
\usepackage{braket}
\usepackage{slashed}
\usepackage{mathdots}
\usepackage{lipsum}
\usepackage{feynmf}
\usepackage{bbm}
\usepackage{color}
\usepackage{caption}
\usepackage{array}
\usepackage[export]{adjustbox}

\allowdisplaybreaks[4]

\title{Two-loop master integrals for a planar topology contributing to $pp \rightarrow t\bar{t}j$}

\author[a]{Simon Badger}
\author[a]{Matteo Becchetti}
\author[a,b]{Ekta Chaubey}
\author[c,d]{Robin Marzucca}
\affiliation[a]{Physics Department, Torino University and INFN Torino, Via Pietro Giuria 1, I-10125 Torino, Italy}
\affiliation[b]{Laboratoire de Physique Théorique et Hautes Energies (LPTHE), UMR 7589, Sorbonne Université
et CNRS, 4 place Jussieu, 75252 Paris Cedex 05, France}
\affiliation[c]{Niels Bohr Institute, Copenhagen University, Blegdamsvej 17, 2100 Copenhagen \O , Denmark}
\affiliation[d]{Physik-Institut, Universität Zürich, Winterthurerstrasse 190, 8057 Zürich, Switzerland}
\emailAdd{simondavid.badger@unito.it}
\emailAdd{matteo.becchetti@unito.it}
\emailAdd{ekta@lpthe.jussieu.fr}
\emailAdd{robin.marzucca@uzh.ch}

\abstract{
We consider the case of a two-loop five-point pentagon-box integral
configuration with one internal massive propagator that contributes to
top-quark pair production in association with a jet at hadron colliders. We
construct the system of differential equations for all the master integrals in
a canonical form where the analytic form is reconstructed from numerical
evaluations over finite fields. We find that the system can be represented as a
sum of d-logarithmic forms using an alphabet of 71 letters. Using high
precision boundary values obtained via the auxiliary mass flow method, a
numerical solution to the master integrals is provided using generalised power
series expansions.
}

\newcommand{\beq}{\begin{equation}}

\newcommand{\eeq}{\end{equation}}
\newcommand{\nn}{\nonumber}
\newcommand{\bea}{\begin{eqnarray}}
\newcommand{\eea}{\end{eqnarray}}
\newcommand{\bfig}{\begin{figure}}
\newcommand{\efig}{\end{figure}}
\newcommand{\bc}{\begin{center}}
\newcommand{\ec}{\end{center}}

\newcommand{\eps}{{\varepsilon}}
\newcommand{\tb}{{\bar{t}}}

\newcommand{\cI}{{\mathcal{I}}}

\definecolor{mypink}{RGB}{219, 48, 122}
\definecolor{mygreen}{rgb}{0,0.7,0}
\definecolor{raspberry}{rgb}{0.53,0.15,0.34}

\date{}
\begin{document}
\maketitle
\flushbottom

\section{Introduction}

As the heaviest particle in the Standard Model (SM) of particle physics, the
top quark has many important implications for the nature of the fundamental
forces. The stability of the SM vacuum is highly sensitive to the value of the
top mass whose precision measurement is a high priority at the Large Hadron
Collider (LHC). Top quark pair production at hadron colliders is known
extremely precisely both theoretically and experimentally and can be used to
constrain SM parameters and parton distribution
functions~\cite{Czakon:2019yrx,Cooper-Sarkar:2020twv}. It has been argued that
top-quark pair production in association with a jet is even more sensitive to
the value of the top quark
mass~\cite{Alioli:2013mxa,Bevilacqua:2017ipv,Alioli:2022lqo}, yet the
theoretical predictions for this process are not currently at the same level of
precision as the experimental measurements. Current theoretical predictions are
represented by the next-to-leading order (NLO) QCD
corrections~\cite{Dittmaier:2007wz,Dittmaier:2008uj} with state-of-the-art
predictions including complete decay information and interfaces with a parton
shower~\cite{Melnikov:2010iu,Alioli:2011as,Czakon:2015cla,Bevilacqua:2015qha,Bevilacqua:2016jfk}.
Mixed QCD and EW corrections are now also available~\cite{Gutschow:2018tuk}.
In order to match the experimental precision, see for
example~\cite{ATLAS:2019guf,CMS:2020grm}, next-to-next-to-leading order (NNLO)
corrections are required. Indeed, fully differential cross-section predictions
at NNLO in the strong coupling would open up opportunities for the most precise
determination of the top-quark mass, yet substantial computational bottlenecks
remain.

The two-loop scattering amplitudes that form part of the NNLO correction are
currently unknown. In general, amplitudes with massive internal propagators
represent a considerable increase in complexity compared to the massless
internal propagators that have been considered so far for five particle
processes. In addition to the growth in algebraic complexity that comes from
the increased number of scales, the analytic complexity contained in the
Feynman integrals that appear can lead to difficulties in identifying a
numerically well-defined function space. In some cases, of which $pp\to t\tb$
is one, analytic evaluation of the integrals leads to elliptic integrals that
still require a better mathematical understanding. While in the case of leading
colour $pp\to t\tb j$ elliptic functions should not appear~\footnote{We
refrain from making a stronger statement though the pattern established in
$pp\to t \tb$ would mean elliptic curves (and more complicated geometries) would only appear in closed
heavy fermion loops or sub-leading colour, non-planar topologies.}, the
evaluation of the master integrals is still a substantial challenge.

A lot of experience in these type of problems has been gained from the study of
massless propagator five-point integrals which form a good starting point for
the integrals we study in this article. The kinematic case of five massless
external particles has now been fully classified into a basis numerically
well-defined pentagon
functions~\cite{Gehrmann:2015bfy,Papadopoulos:2015jft,Gehrmann:2018yef,Abreu:2018aqd,Chicherin:2018old,Chicherin:2020oor}.
For the case of one off-shell external leg and four massless legs the situation
is also almost complete with the planar
\cite{Abreu:2020jxa,Canko:2020ylt,Chicherin:2021dyp} and the non-planar
hexa-box \cite{Abreu:2021smk} now known. This progress has allowed the
calculation of several five-point two-loop scattering amplitudes
\cite{Gehrmann:2015bfy,Badger:2018enw,Abreu:2018aqd,Chicherin:2018yne,Abreu:2018zmy,Abreu:2019odu,Abreu:2018jgq,Badger:2019djh,Abreu:2020cwb,Abreu:2021oya,Chawdhry:2020for,Hartanto:2019uvl,Agarwal:2021grm,Chawdhry:2021mkw,Agarwal:2021vdh,Badger:2021imn,Badger:2021ega,Badger:2022ncb}
and led to the first NNLO theoretical predictions for $2 \rightarrow 3$
processes
\cite{Chawdhry:2019bji,Kallweit:2020gcp,Chawdhry:2021hkp,Badger:2021ohm,Hartanto:2022qhh}.

In this article we make a small step towards the two-loop amplitudes for $pp\to
t\tb j$ by considering the computation of the master integrals associated to a
five-point pentagon-box configuration with one internal massive propagator (see
figure \ref{fig:t431def}). This builds upon previous work considering the
one-loop helicity amplitudes expanded up to $\mathcal{O}(\eps^2)$ in the
dimensional regulator. Our methodology to determine a set of master integrals
follows by the means of the differential equation method
\cite{Kotikov:1990kg,Remiddi:1997ny}. In particular, we write the system of differential equations
in a canonical form \cite{Henn:2013pwa}, where the dependence on the
dimensional regulator factorises. The canonical form requires the
identification of a uniform transcendental weight (UT) basis of master
integrals and the solution to a large system of Integration-by-Parts (IBP)
relations \cite{Tkachov:1981wb,Chetyrkin:1981qh}. For the later we employ the
Laporta algorithm~\cite{Laporta:2001dd} which can be implemented within a
numerical framework using finite field
arithmetic~\cite{vonManteuffel:2014ixa,Peraro:2016wsq,Peraro:2019svx}. The
derivation of the differential equation system can be implemented entirely
within the dataflow graphs provided by the \textsc{FiniteFlow}
library~\cite{Peraro:2019svx} allowing us to sidestep traditional limitations
due to huge intermediate expressions. The determination of a UT basis also
presents a significant challenge and has a significant effect on the simplicity
of the differential equation system. While considerable effort has been spent
to determine automated, or semi-automated techniques for the determination of
UT bases yet they are still difficult to apply to situations with a large
number of kinematic scales. In this work we will describe how the UT system
can instead be inferred by observing patterns in known examples to provide a
suitable ansatz.

Once the differential equation system has been determined we employ the
semi-analytic approach to provide the solution of the master integrals.  The
generalised power series method
\cite{Lee:2017qql,Mandal:2018cdj,Francesco:2019yqt} provides a practical way to
evaluate the integrals at given numerical values through contour integration
from a boundary point. In this work we use the implementation of the method
discussed in the Ref.~\cite{Francesco:2019yqt} into the \textsc{Mathematica}
package \textsc{DiffExp}~\cite{Hidding:2020ytt}. For a successful implementation
the boundary value must be given with a sufficiently high numerical precision.
The development of the auxiliary mass flow method
\cite{Liu:2017jxz,Liu:2021wks,Liu:2022tji} and in particular the \textsc{Mathematica}
package \textsc{AMFlow}~\cite{Liu:2022chg} offers a simple and practical
solution to this task. We are therefore able to offer a solution for the master integrals which has the potential for phenomenological applications, as has been done for other processes
\cite{Bonciani:2016qxi,Bonciani:2019jyb,Frellesvig:2019byn,Abreu:2020jxa,Becchetti:2020wof,Abreu:2021smk,Armadillo:2022bgm,Bonciani:2021zzf,Badger:2022mrb}.

Beyond our semi-analytic solution for the master integrals, we also derive the
analytic representation for the system of differential equations in terms of logarithmic one-forms. The
alphabet for this system is written in a compact form and it shows the same
analytic structure as in the five-point massless \cite{Gehrmann:2015bfy} and in
the one-mass \cite{Abreu:2020jxa} cases. As a consequence, this paper lays the
groundwork for a fully analytic solution, in terms of an extension of the
pentagon functions \cite{Gehrmann:2018yef,Chicherin:2020oor,Chicherin:2021dyp}, to the case of
top-pair plus jet production.

The paper is structured as follows. In section \ref{sec:setup}
we define the topology that is under study and we discuss the computational
framework. In section \ref{sec:deqs} we describe our approach to construct the
canonical differential equations and the UT basis of master integrals. In section \ref{sec:dlog} we present the
logarithmic one-forms representation of the differential equations and the analytic form of the
alphabet, while in section \ref{sec:num} we discuss the numerical evaluation of the
master integrals. Finally in \ref{sec:conclusion} we give our conclusions and we analyse
future developments.

\section{Notation and definitions} \label{sec:setup}

\begin{figure}[t!]
\begin{center}
\includegraphics[width = 0.4\linewidth]{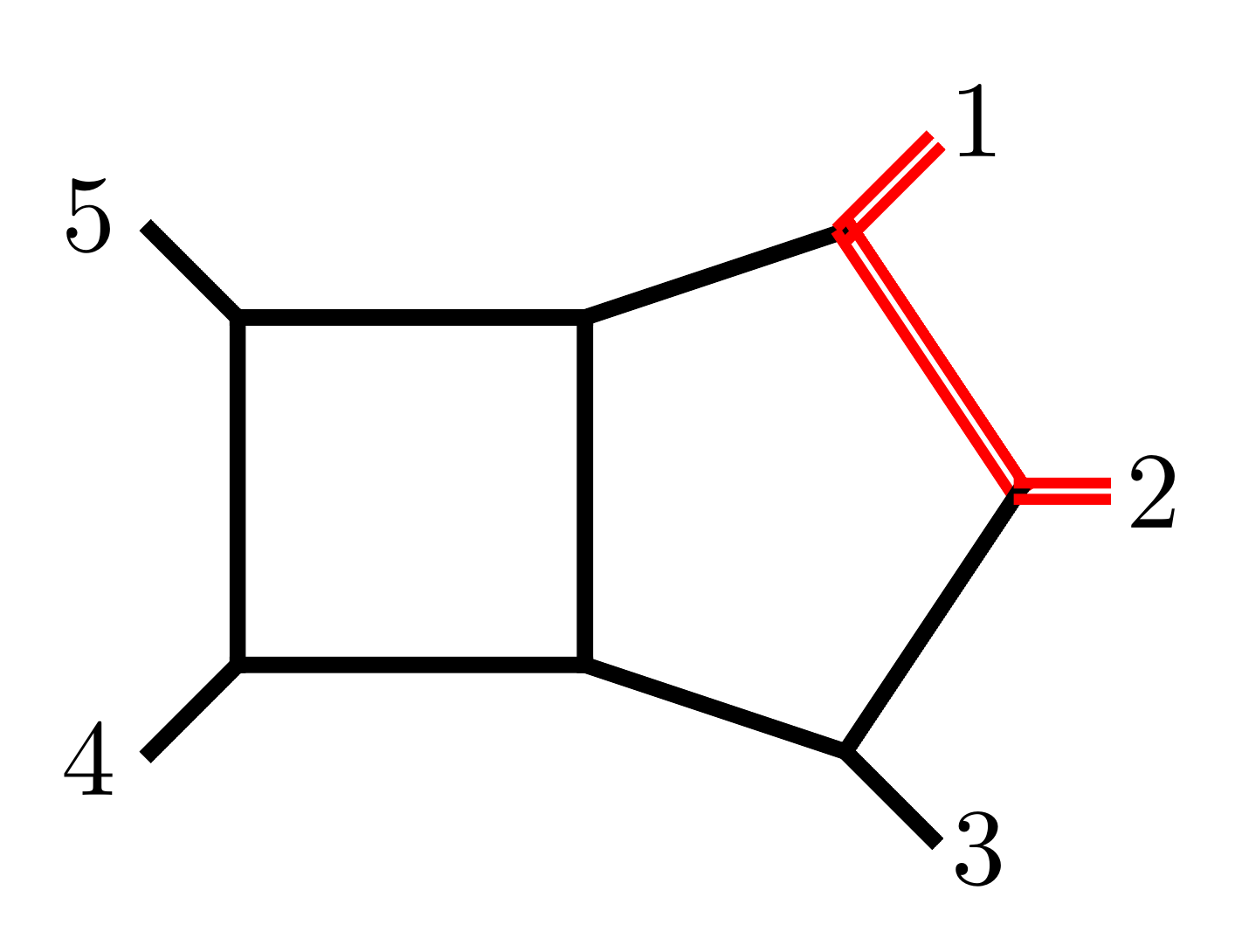}
\end{center}
\caption{The pentagon-box topology contributing to  $pp \rightarrow t\bar{t}j$. Black lines denote massless particles and red double-lines denote massive particles.}
\label{fig:t431def}
\end{figure}

\begin{figure}[h]
\captionsetup[subfigure]{labelformat=empty}
\centering
\subfloat[$\cI_{1},\cI_{2},\cI_{3}$]{\includegraphics[width = 2.7 cm]{figs/diag1.pdf}} \quad
\subfloat[$\cI_{4},\cI_{5},\cI_{6},\cI_{7}$]{\includegraphics[width = 2.7 cm]{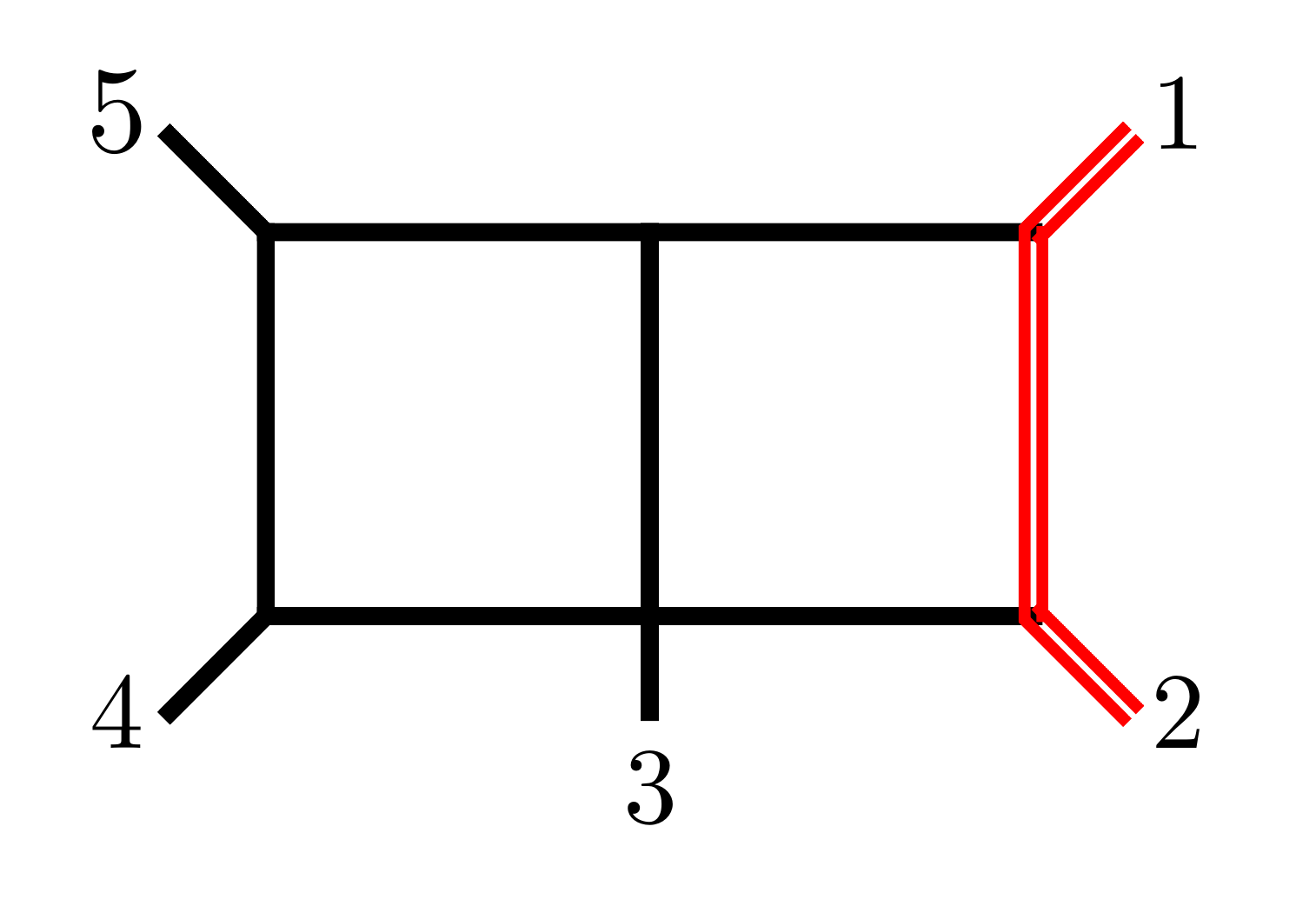}} \quad
\subfloat[$\cI_{8},\cI_{9},\cI_{10}$]{\includegraphics[width = 2.7 cm]{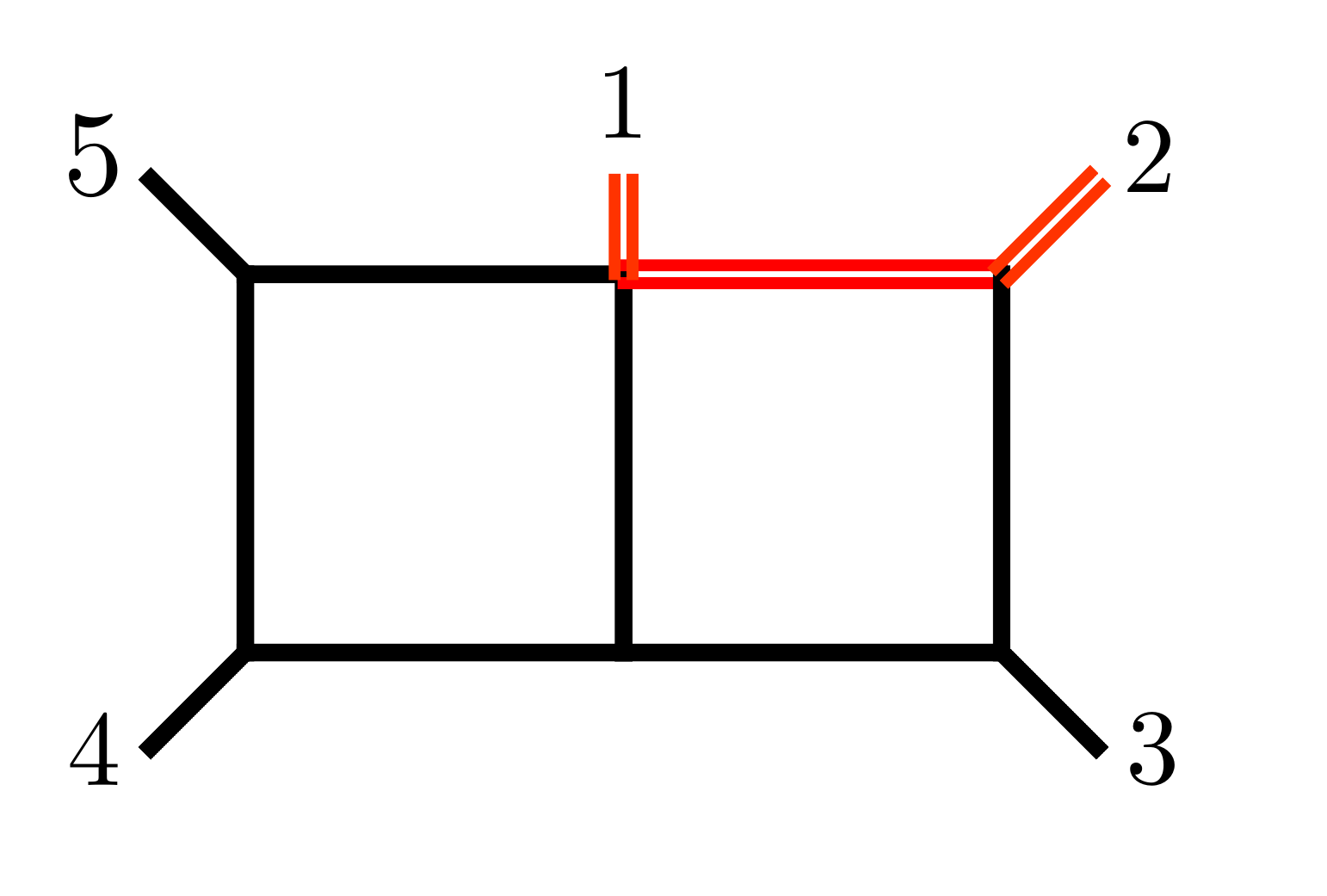}} \quad
\subfloat[$\cI_{11},\cI_{12}$]{\includegraphics[width = 2.7 cm]{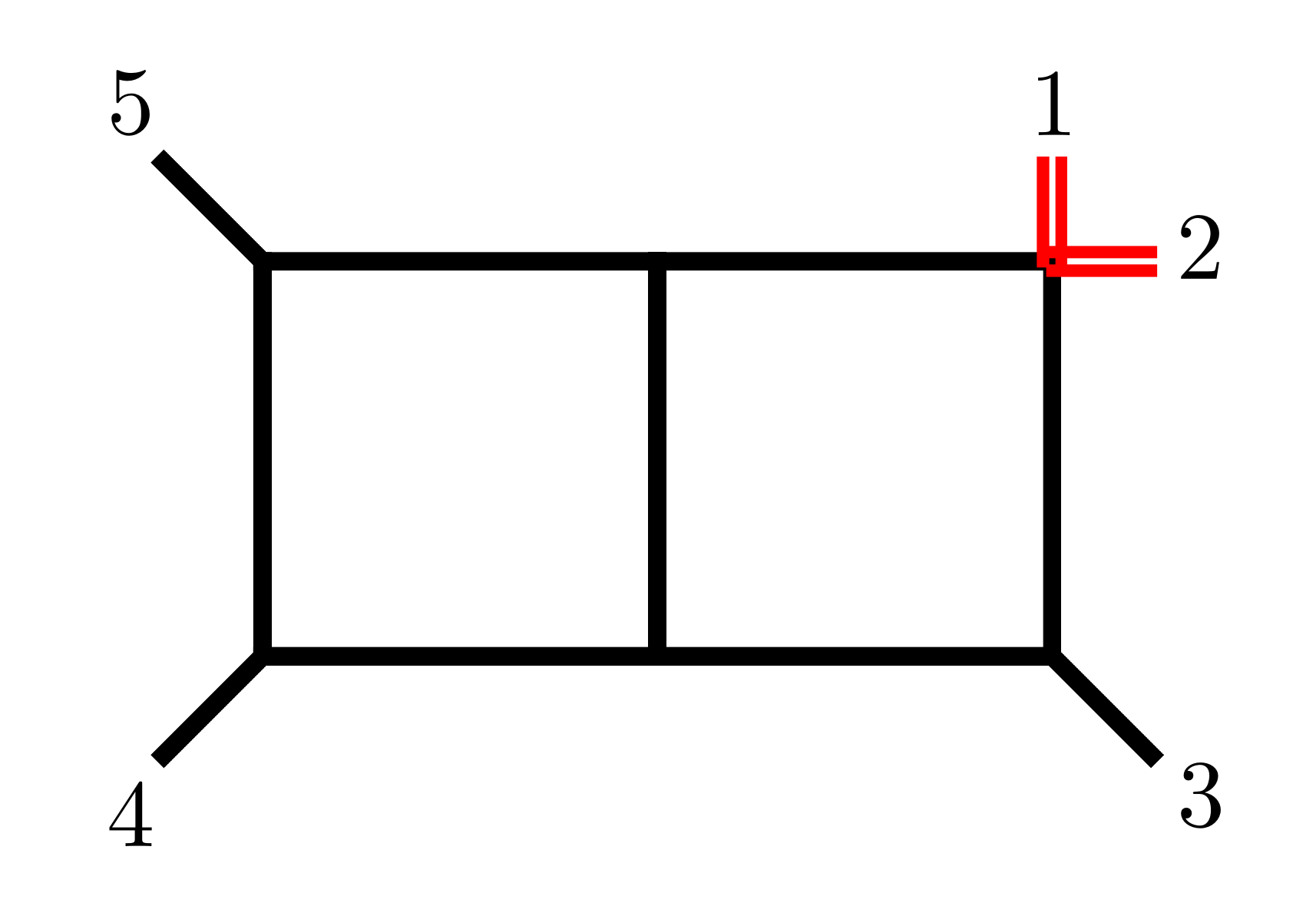}} \quad
\subfloat[$\cI_{13},\cI_{14},\cI_{15}$]{\includegraphics[width = 2.7 cm]{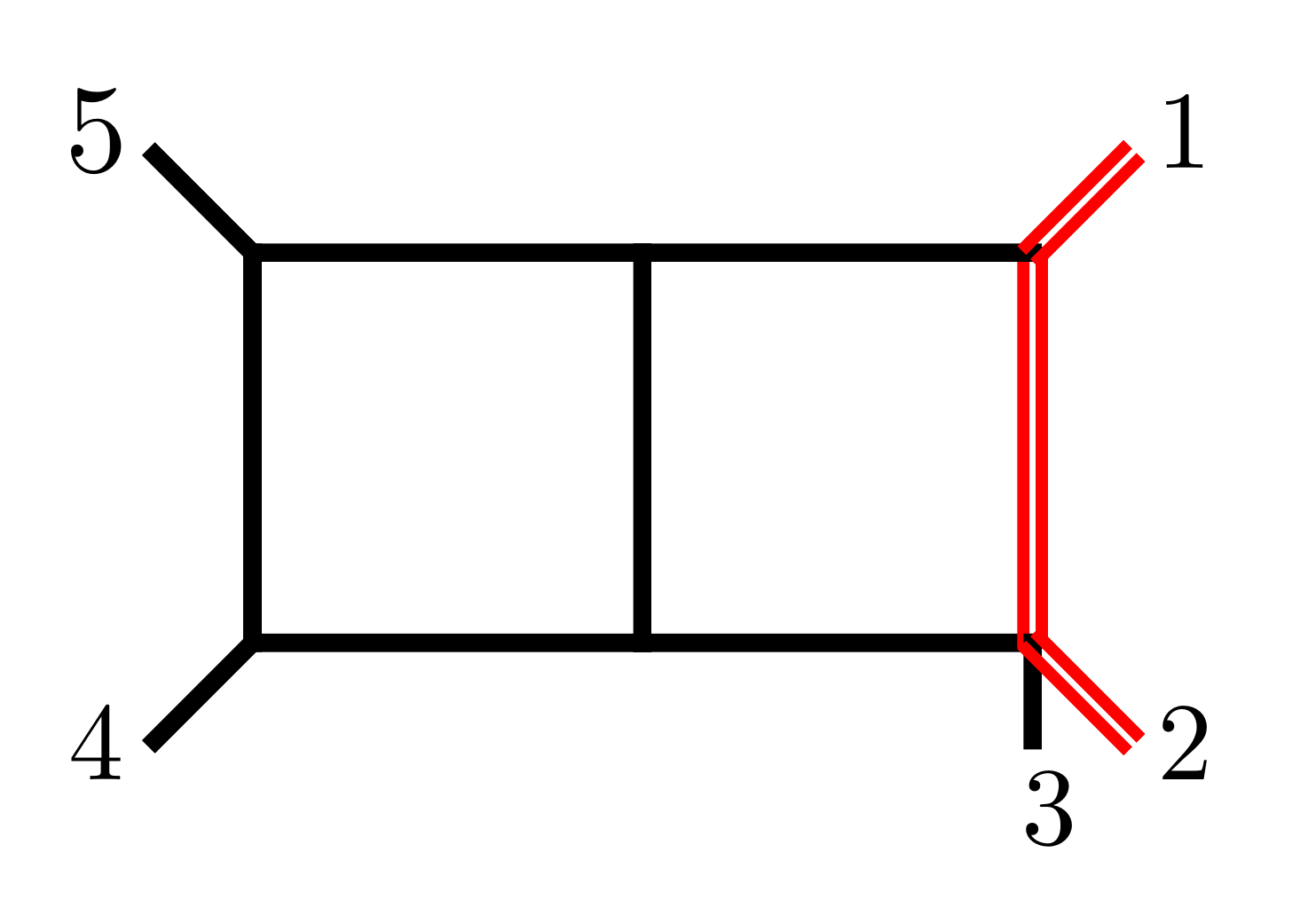}} \quad
\subfloat[$\cI_{16},\cI_{17}$]{\includegraphics[width = 2.5 cm]{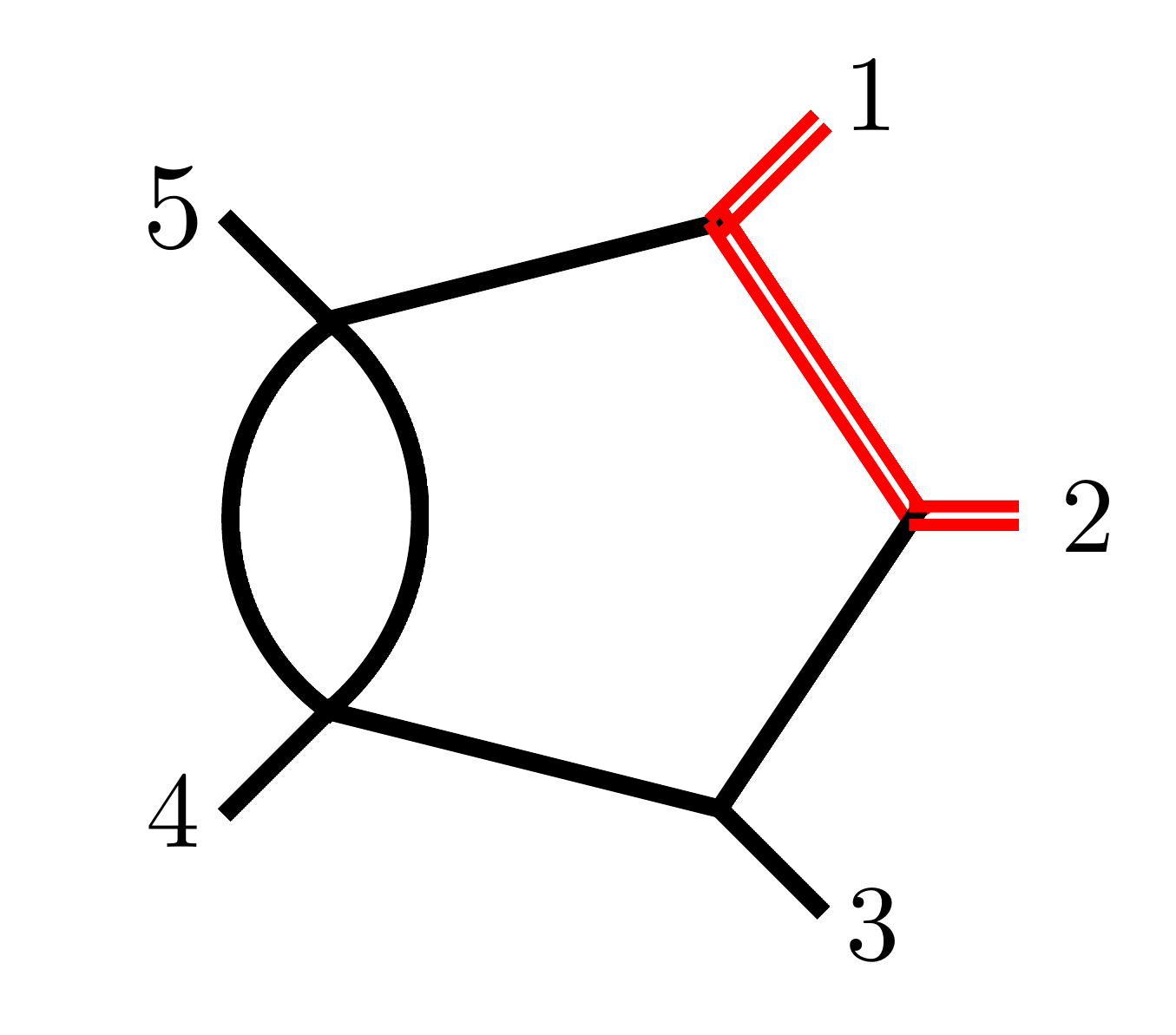}} \quad
\subfloat[$\cI_{18},\cI_{19}$]{\includegraphics[width = 2.5 cm]{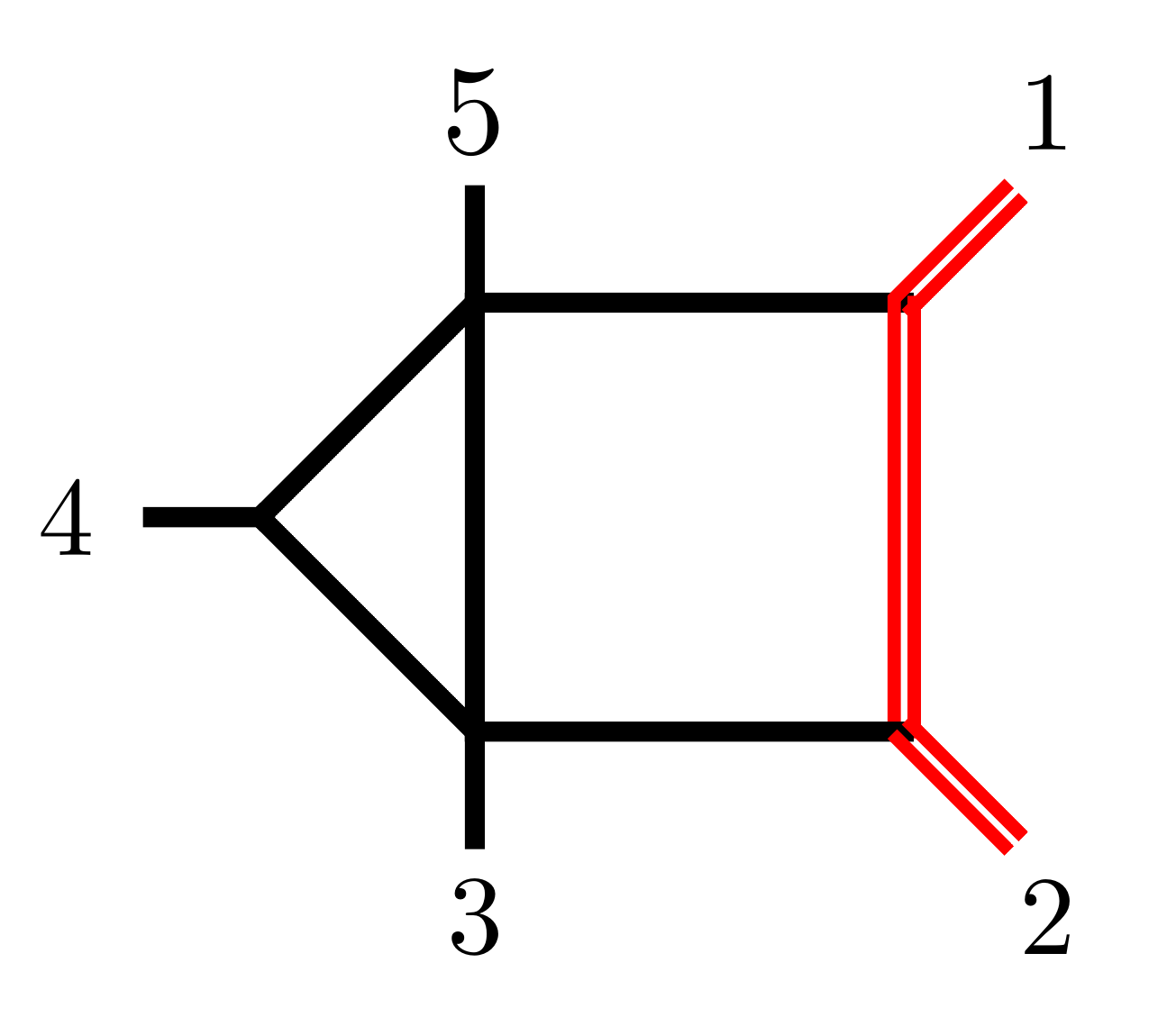}} \quad
\subfloat[$\cI_{20},\cI_{21}$]{\includegraphics[width = 2.6 cm]{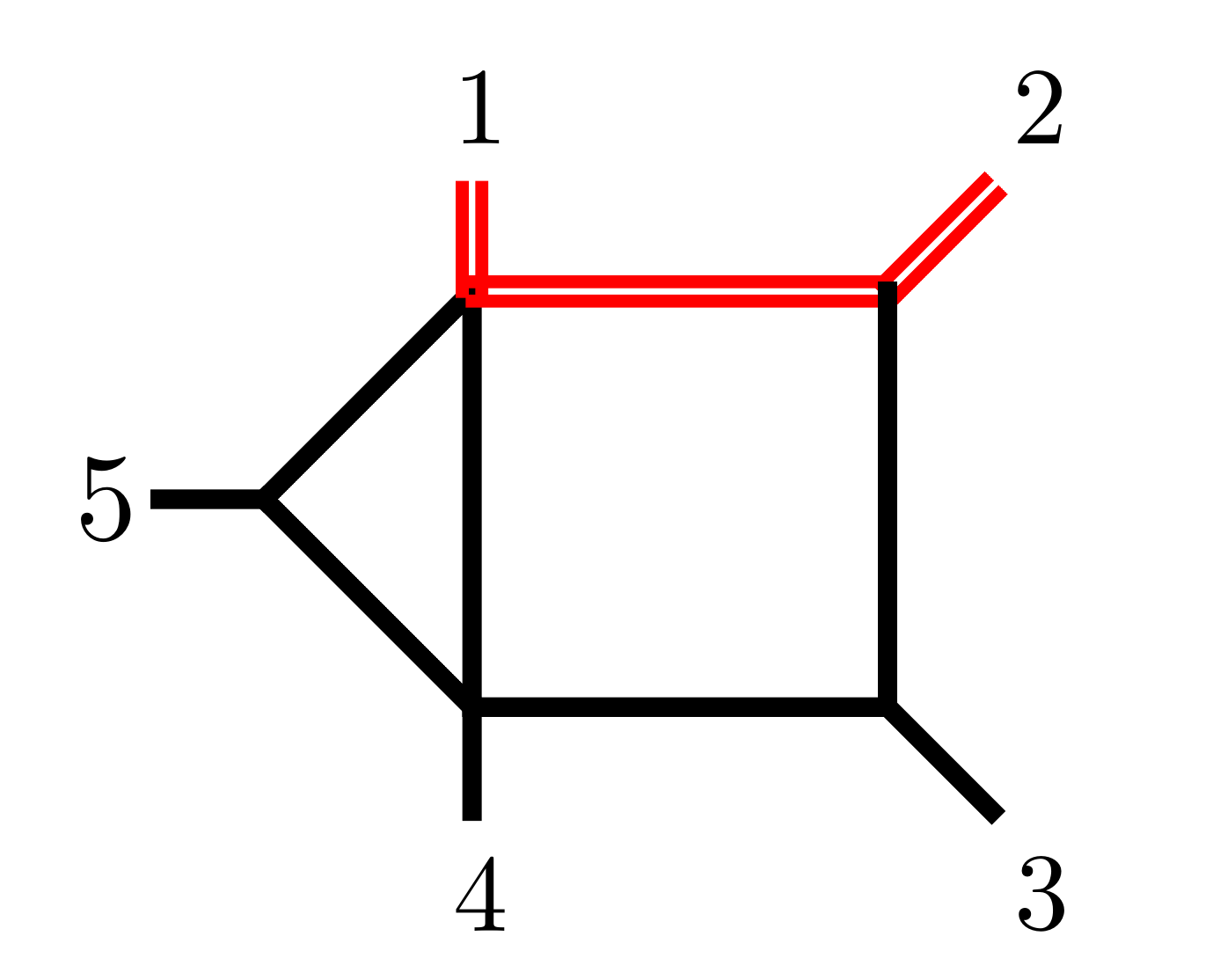}} \quad
\subfloat[$\cI_{22},\cI_{23},\cI_{24},\cI_{25},\cI_{26}$]{\includegraphics[width = 2.5 cm]{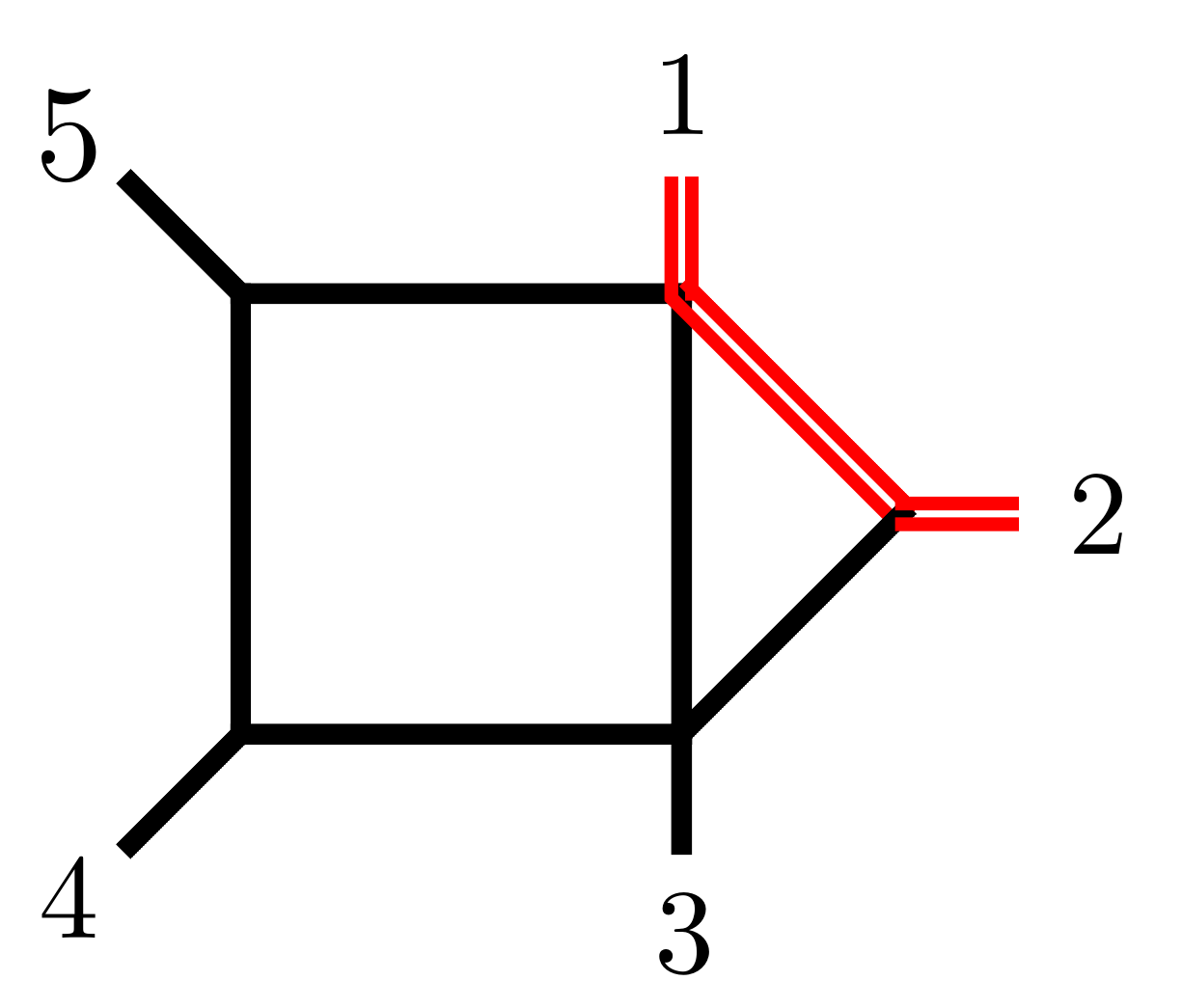}} \quad
\subfloat[$\cI_{27}$]{\includegraphics[width = 2.5 cm]{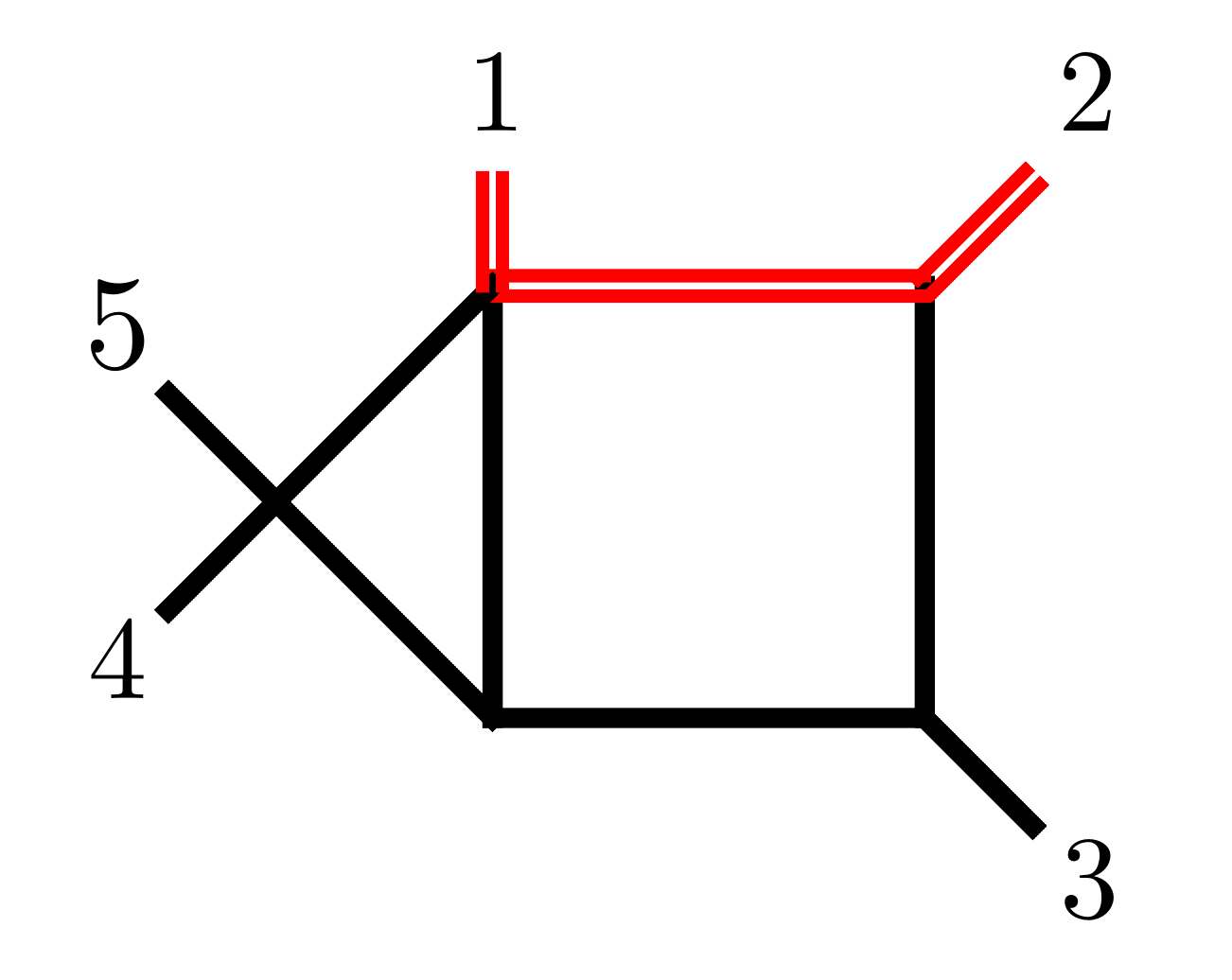}} \quad
\subfloat[$\cI_{28}$]{\includegraphics[width = 2.5 cm]{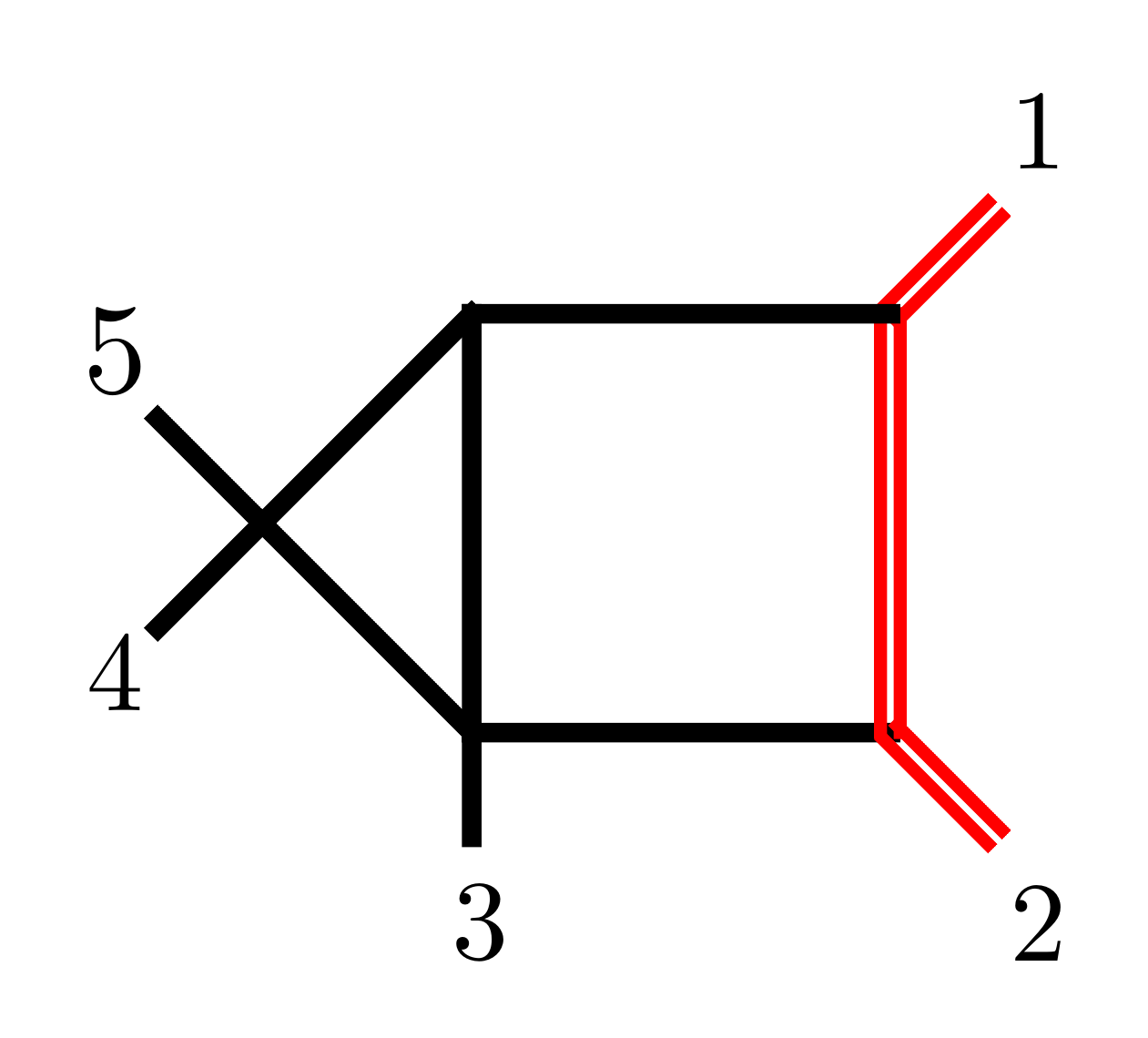}} \quad
\subfloat[$\cI_{29}$]{\includegraphics[width = 2.5 cm]{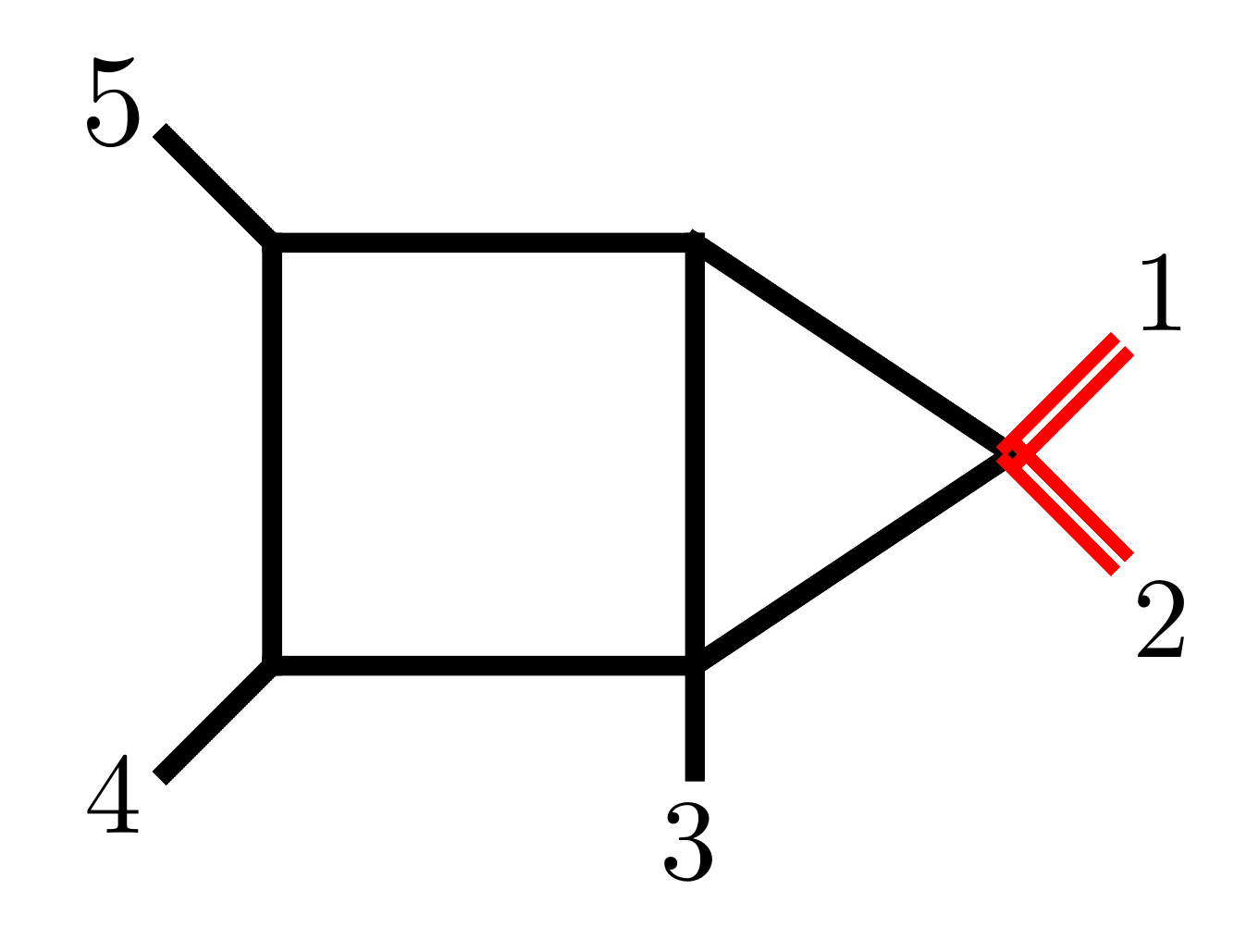}} \quad
\subfloat[$\cI_{30}$]{\includegraphics[width = 2.6 cm]{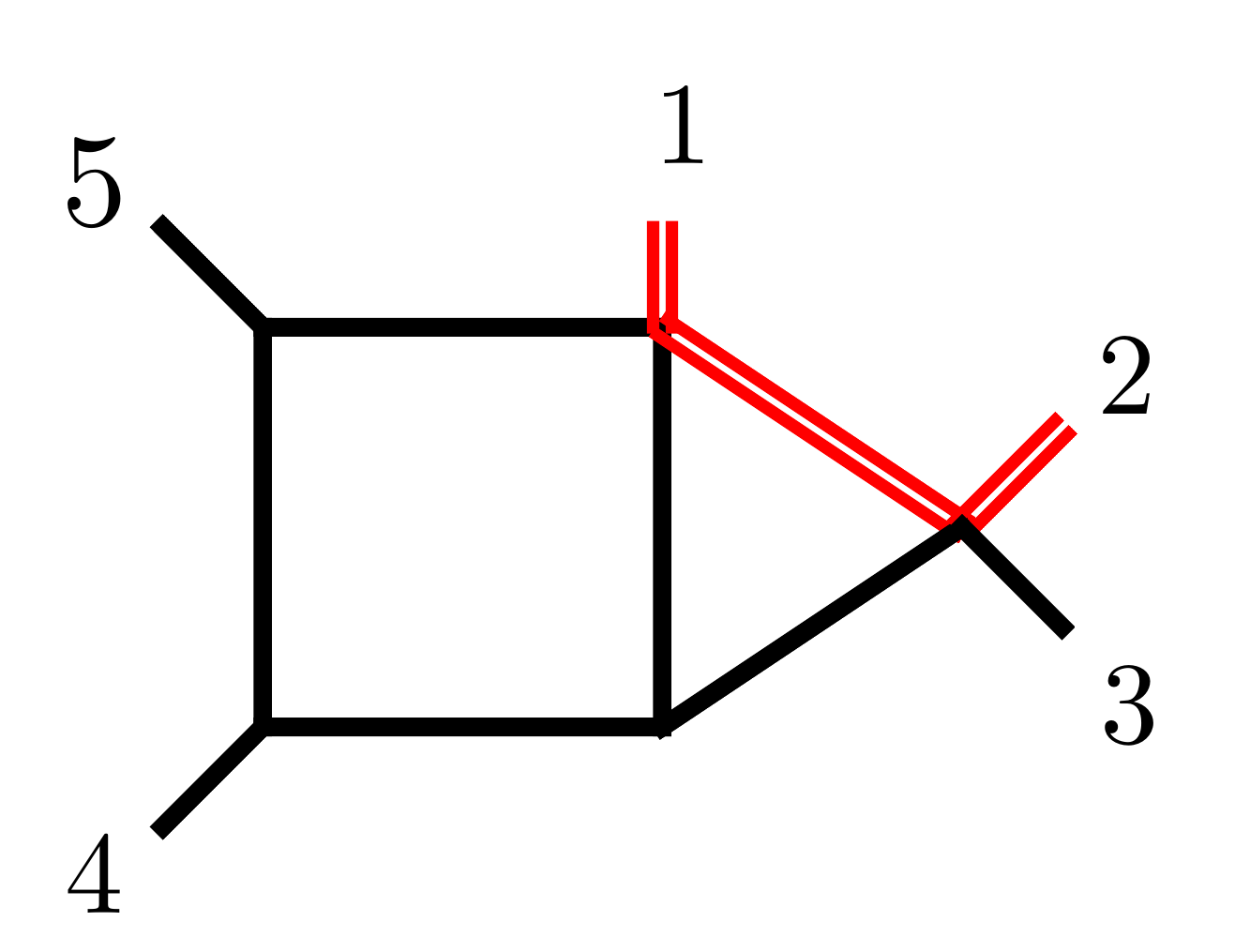}} \quad
\subfloat[$\cI_{31},\cI_{32},\cI_{33},\cI_{34},\cI_{35}$]{\includegraphics[width = 2.5 cm]{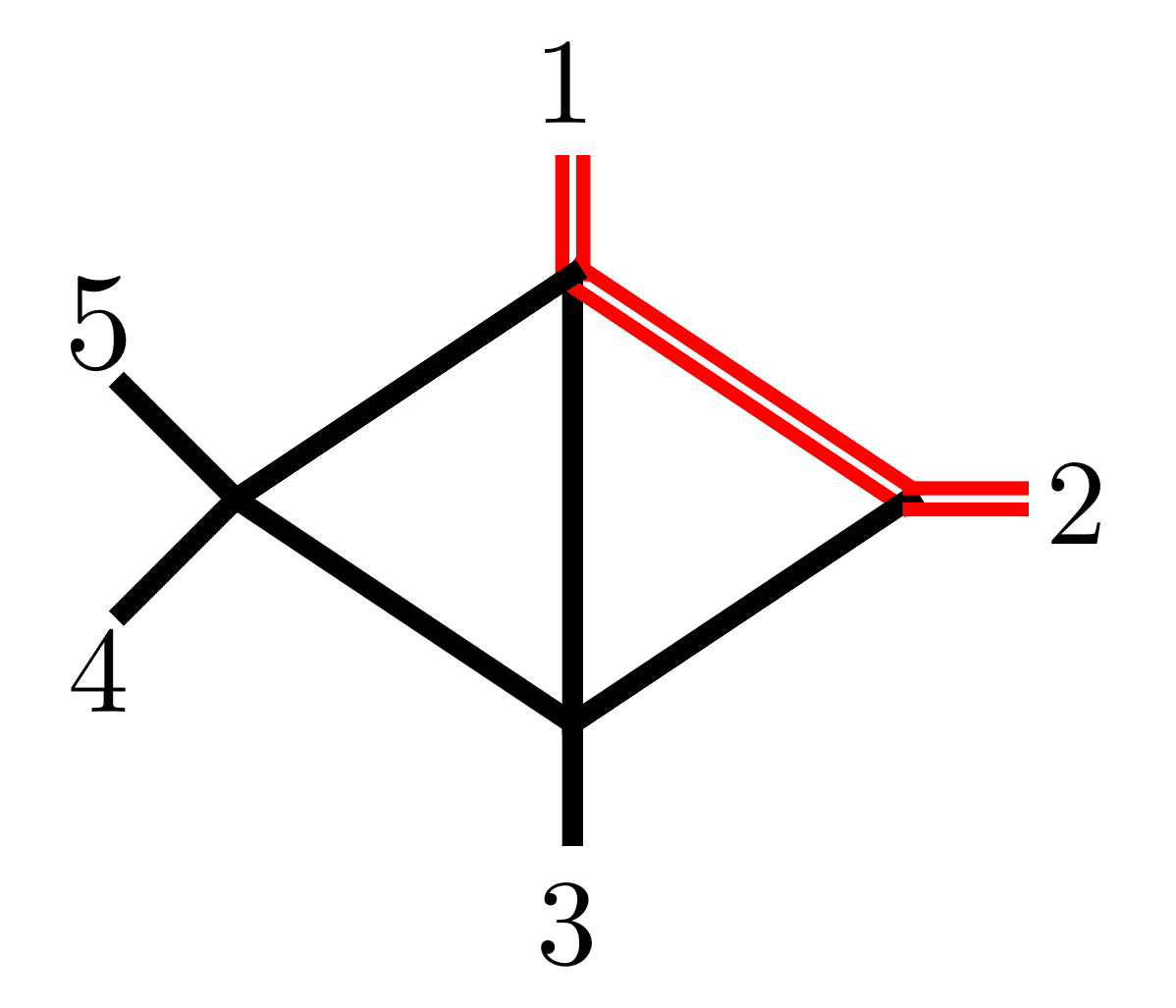}} \quad
\subfloat[$\cI_{36},\cI_{37}$]{\includegraphics[width = 2.5 cm]{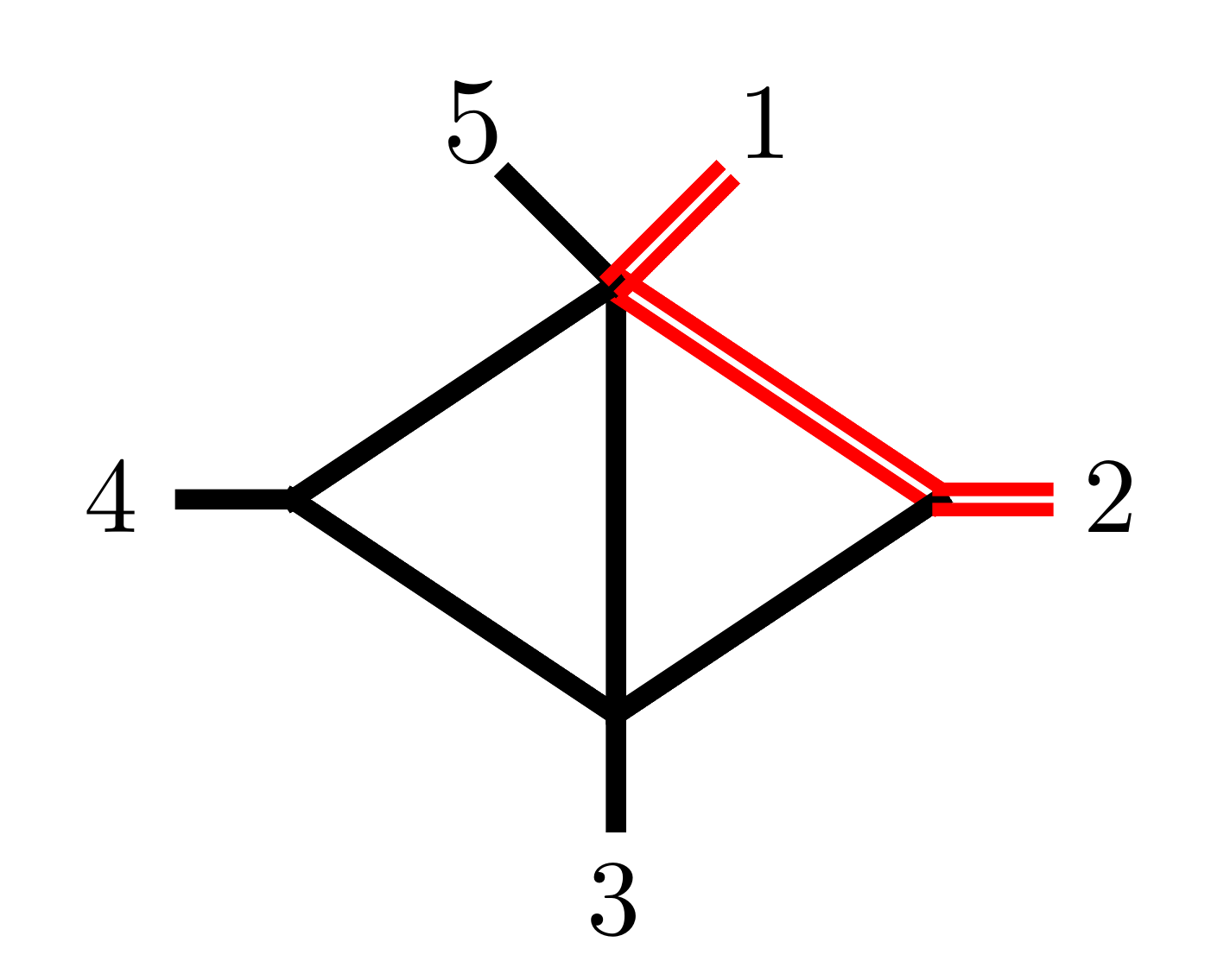}} \quad
\subfloat[$\cI_{38},\cI_{39}$]{\includegraphics[width = 2.5 cm]{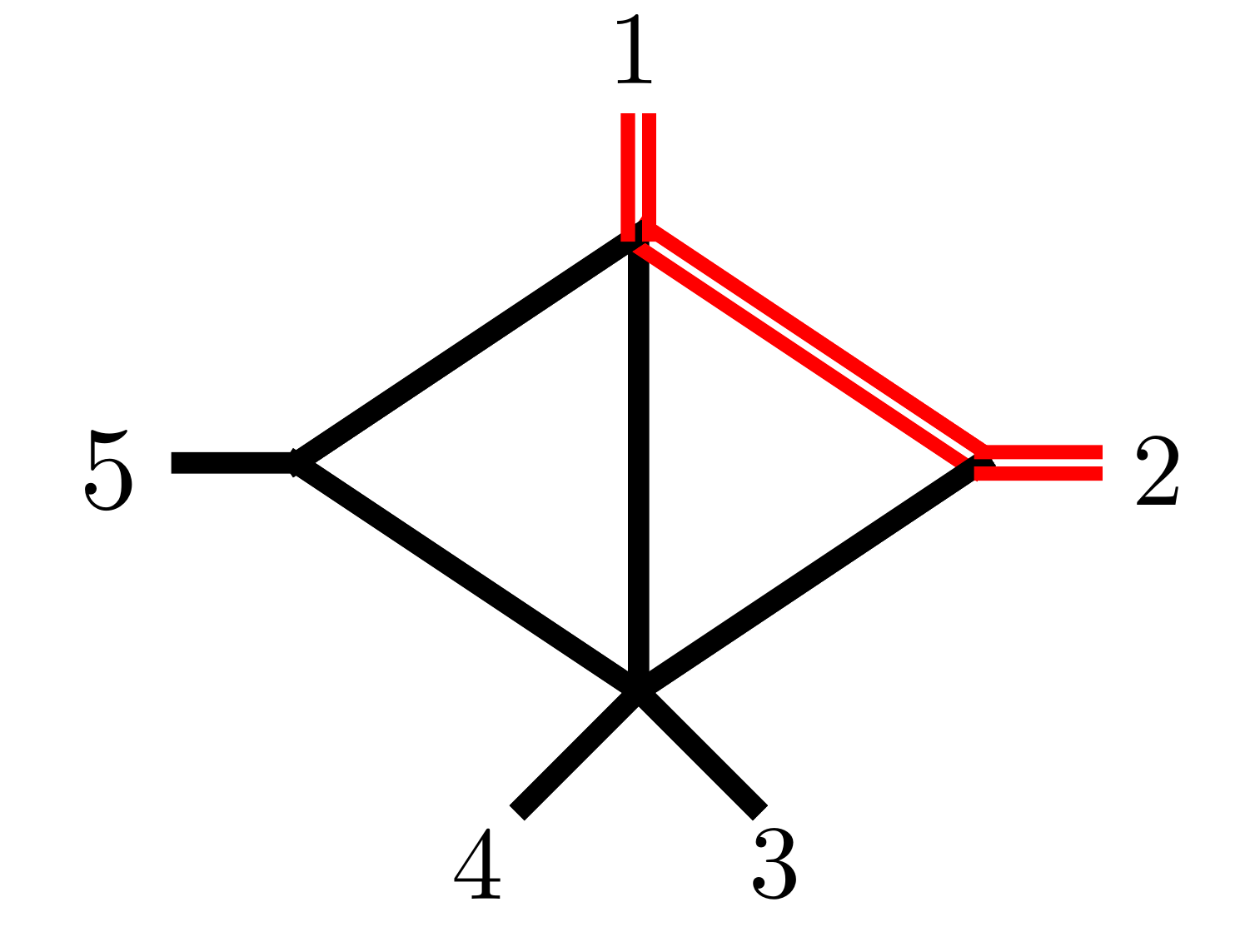}} \quad
\raisebox{-0.1cm}{\subfloat[$\cI_{40},\cI_{41}$]{\includegraphics[width = 2.5 cm]{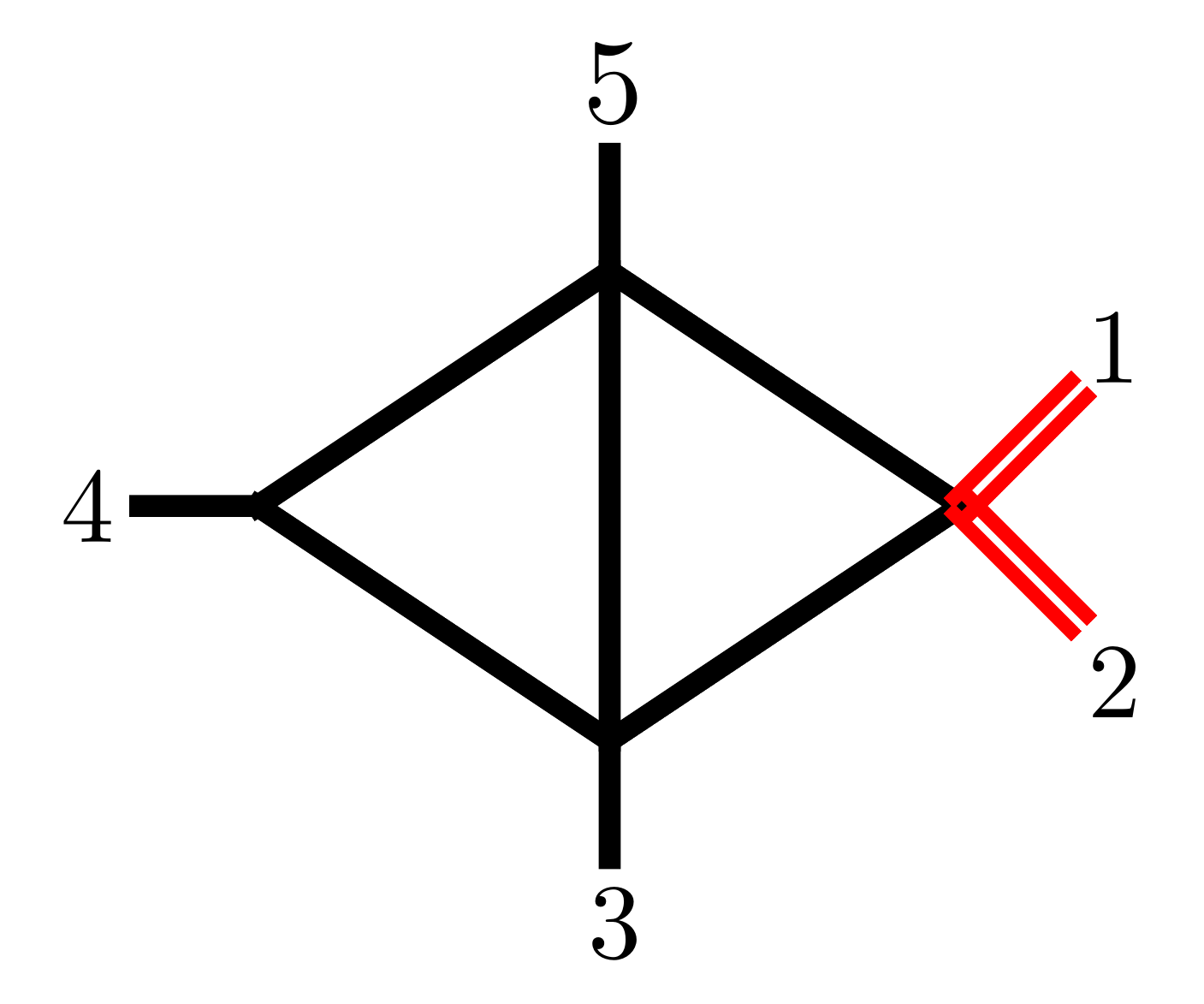}}} \quad
\subfloat[$\cI_{42}$]{\includegraphics[width = 2.5 cm]{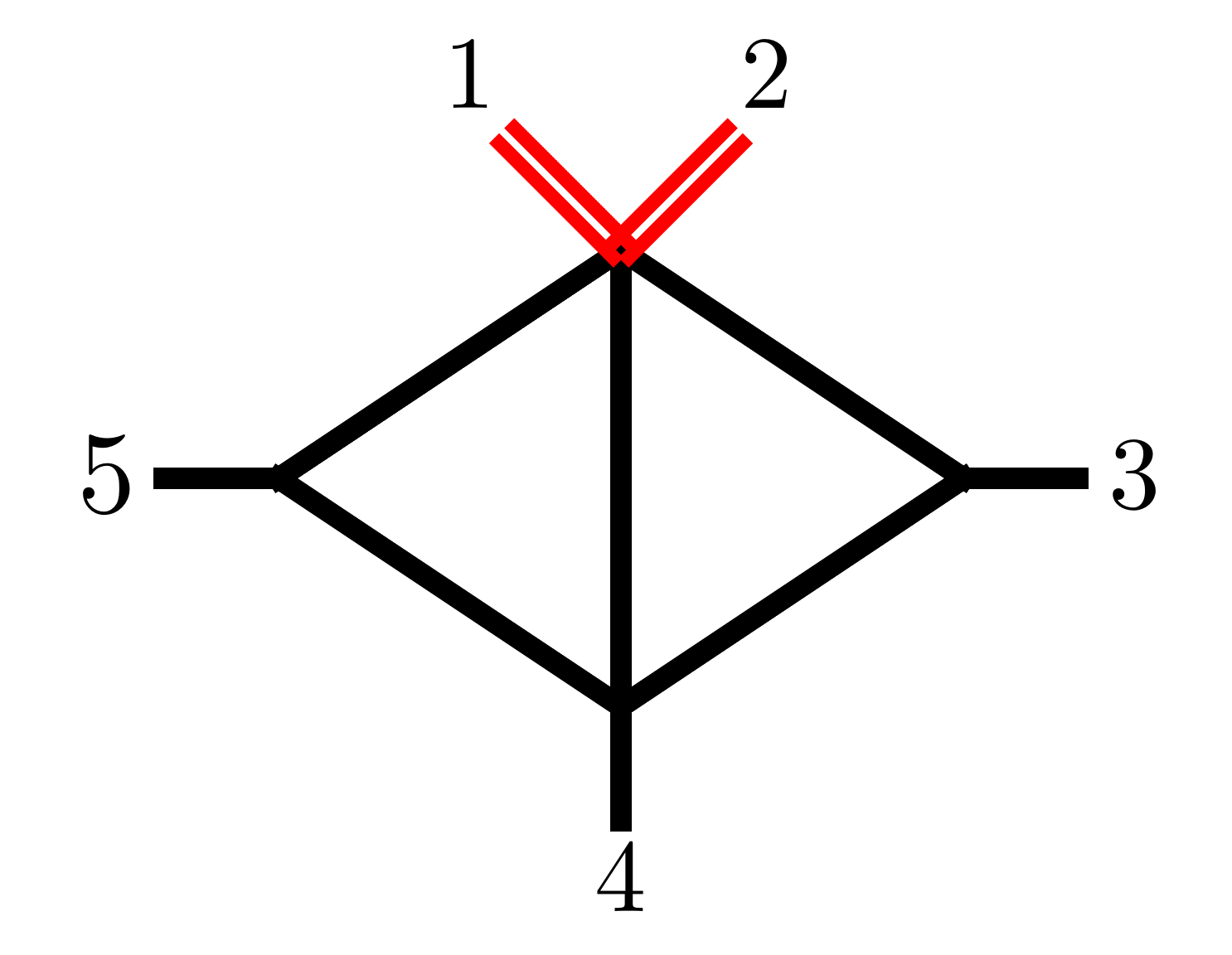}} \quad
\subfloat[$\cI_{43}$]{\includegraphics[width = 2.5 cm]{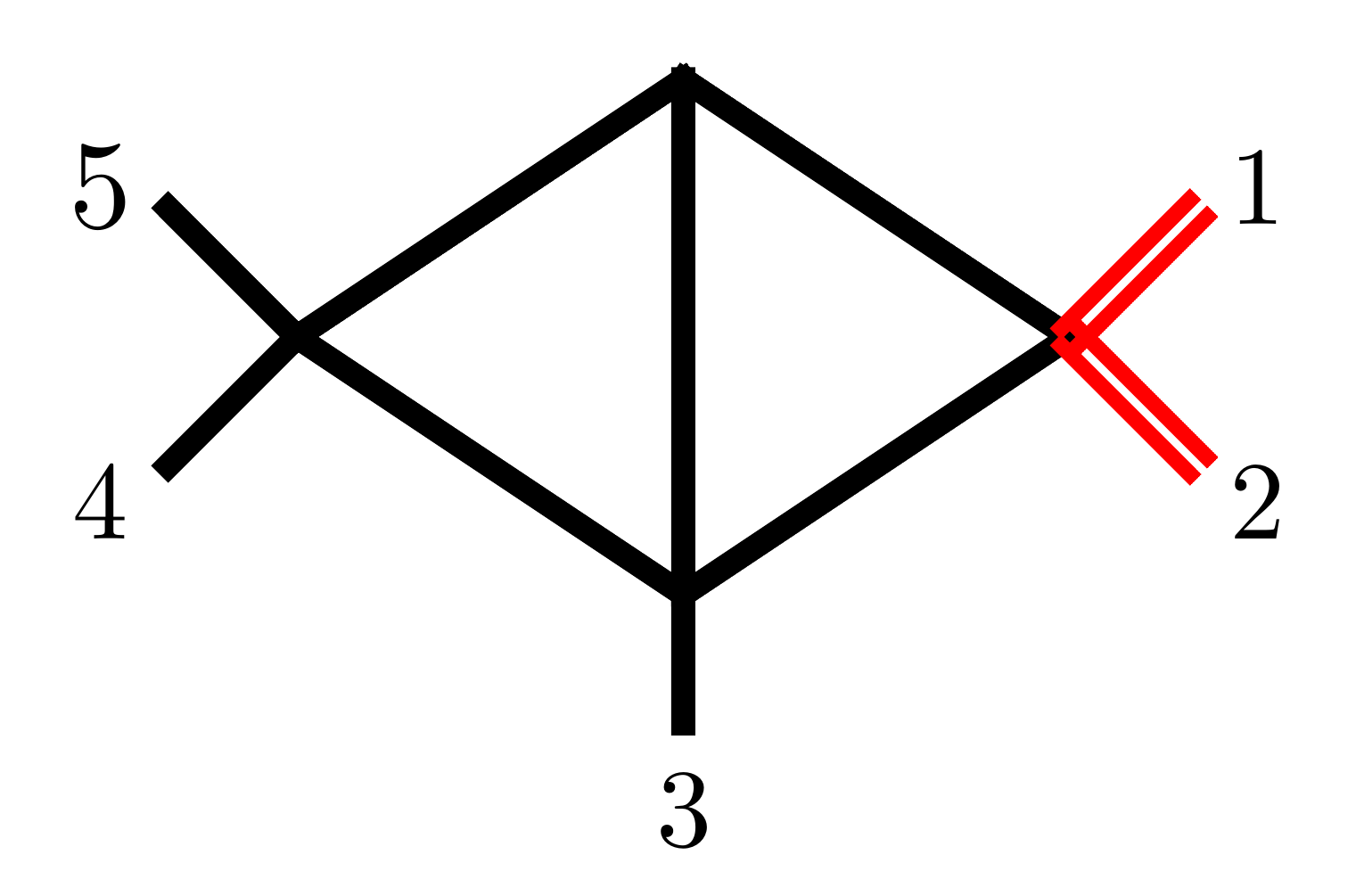}} \quad
\subfloat[$\cI_{44},\cI_{45}$]{\includegraphics[width = 2.5 cm]{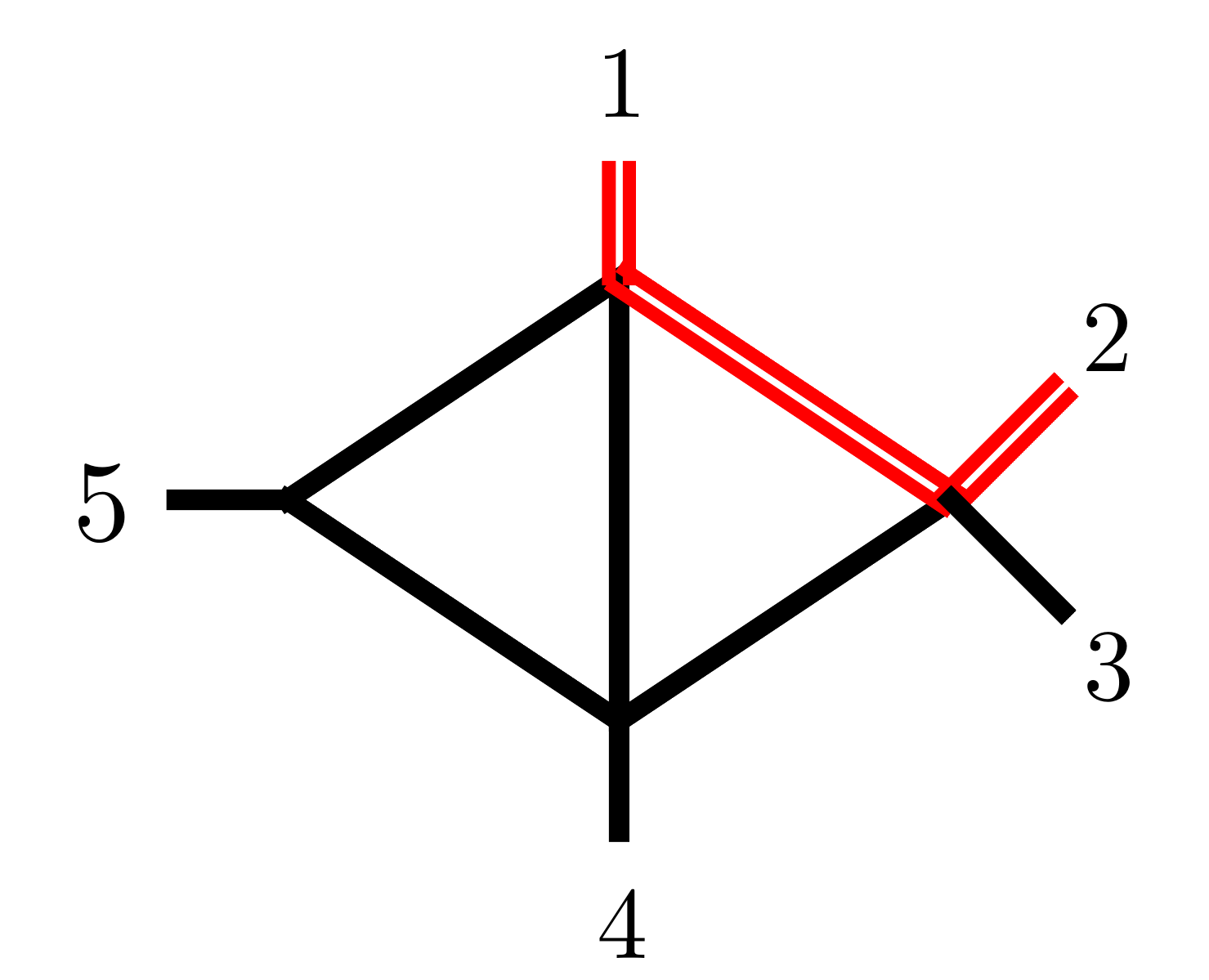}} \quad
\subfloat[$\cI_{46},\cI_{47}$]{\includegraphics[width = 2.5 cm]{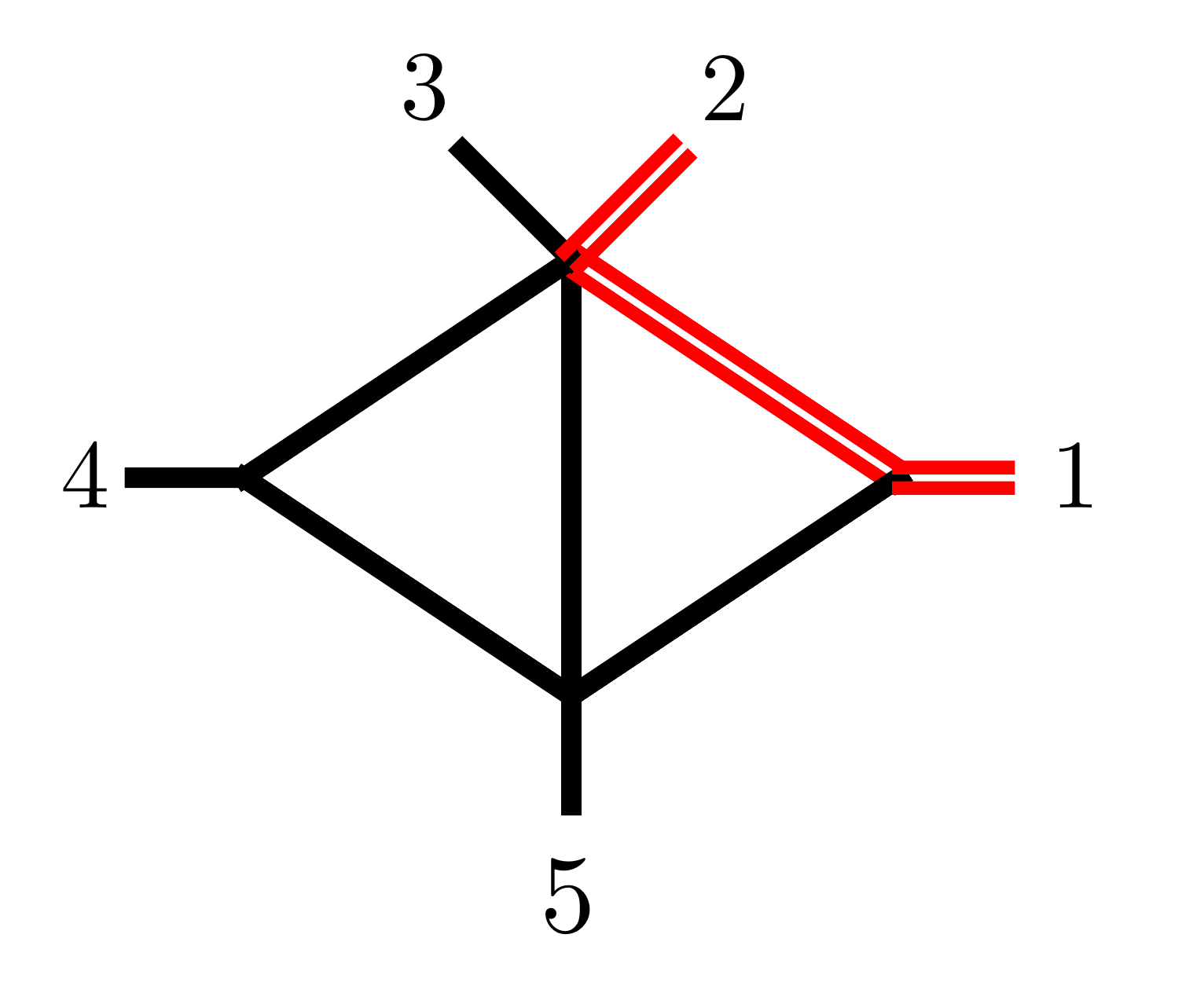}} \quad
\subfloat[$\cI_{48}$]{\includegraphics[width = 2.4 cm]{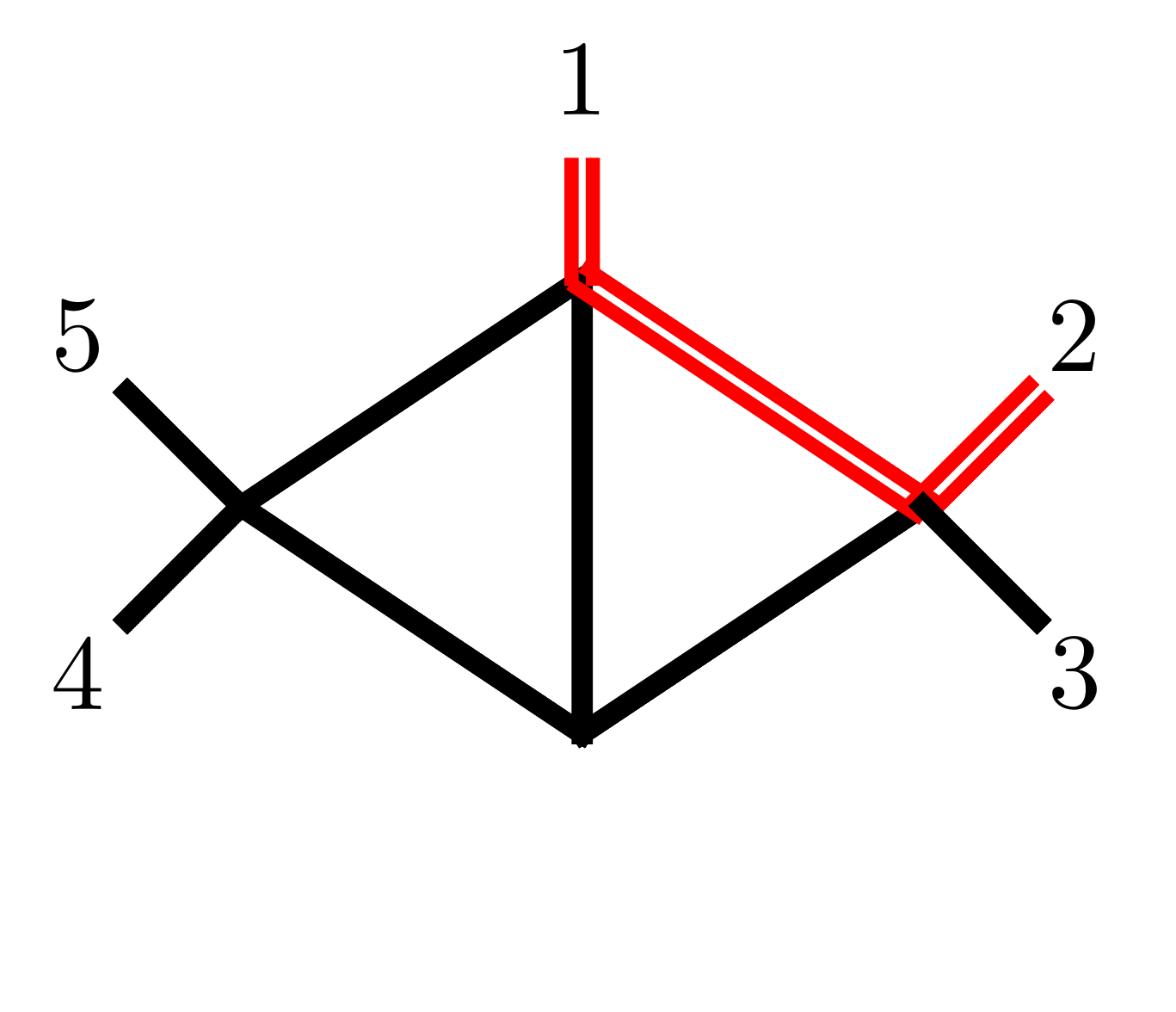}} \quad
\subfloat[$\cI_{49},\cI_{50}$]{\includegraphics[width = 2.1 cm]{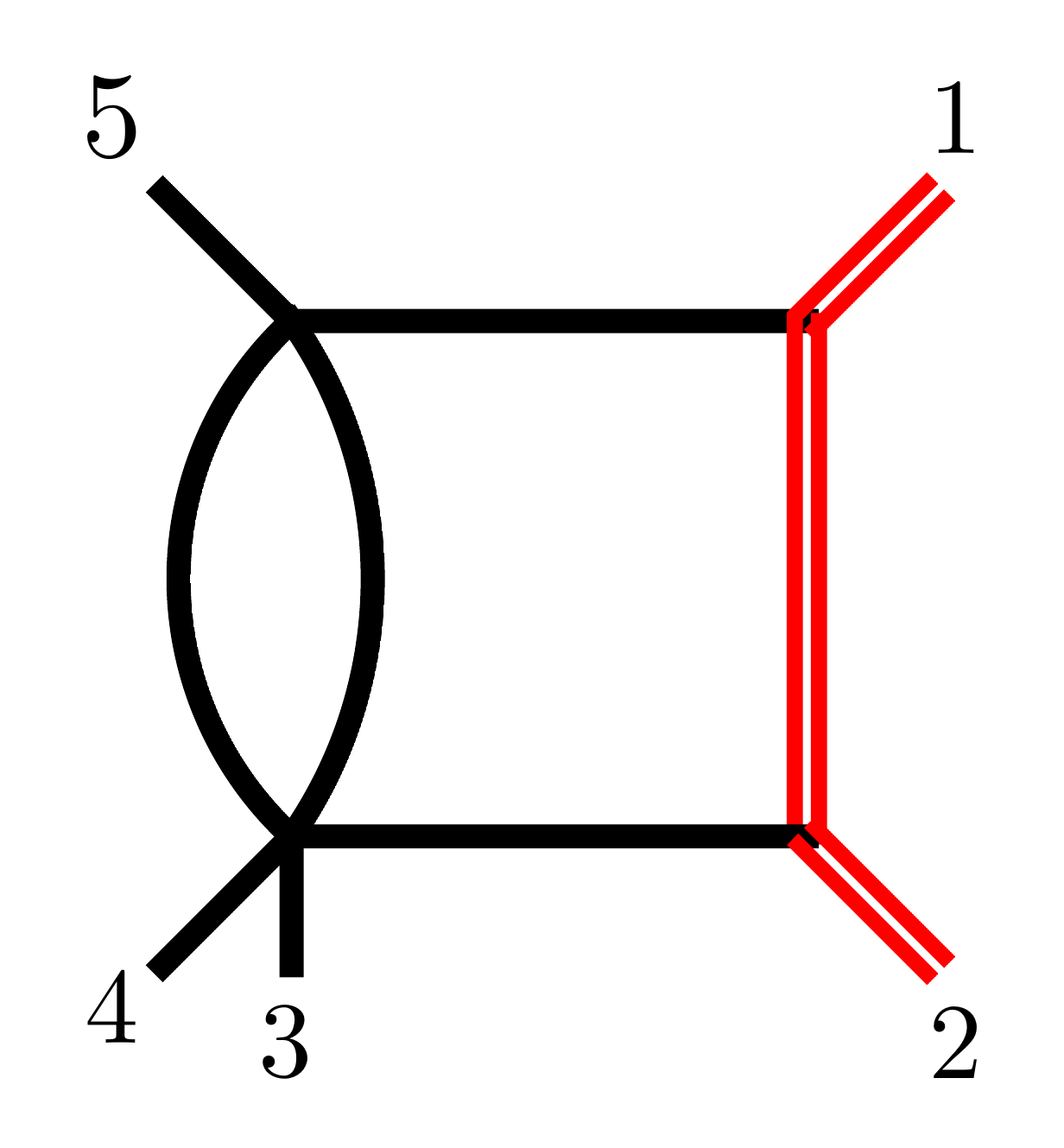}} \quad
\subfloat[$\cI_{53},\cI_{54}$]{\includegraphics[width = 2.1 cm]{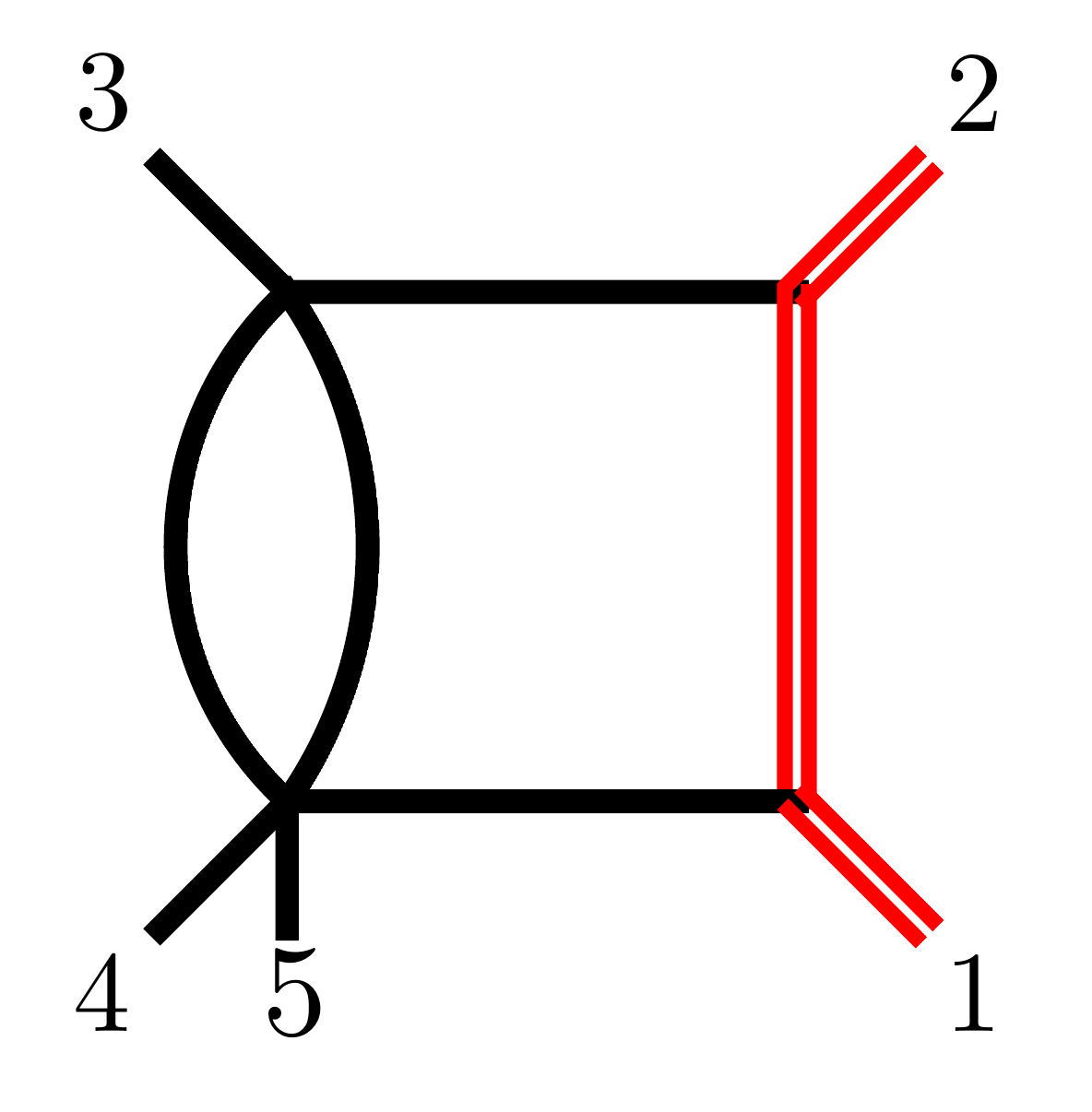}} \quad
\subfloat[$\cI_{51}$]{\includegraphics[width = 2.1 cm]{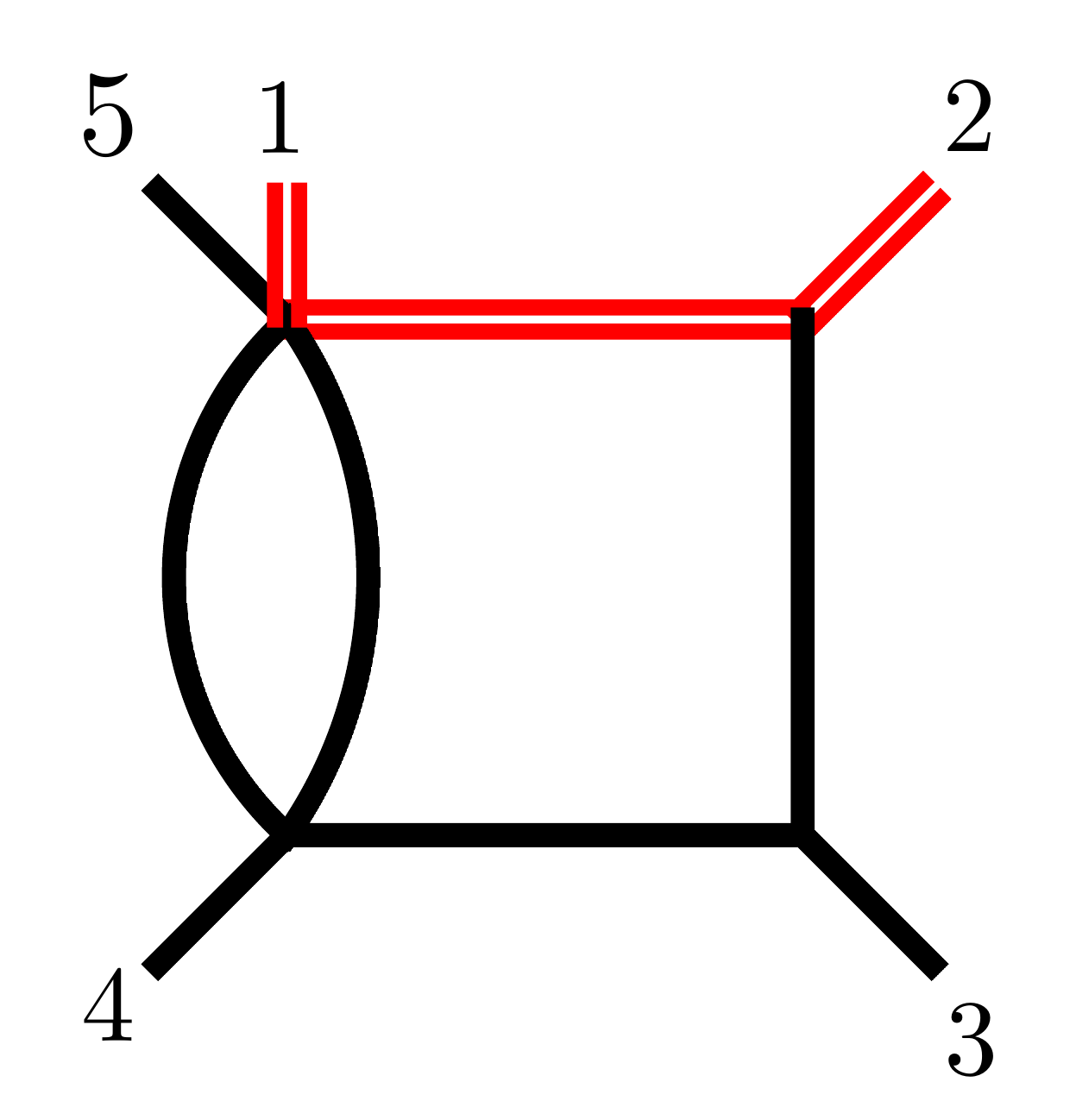}} \quad
\subfloat[$\cI_{52}$]{\includegraphics[width = 2.1 cm]{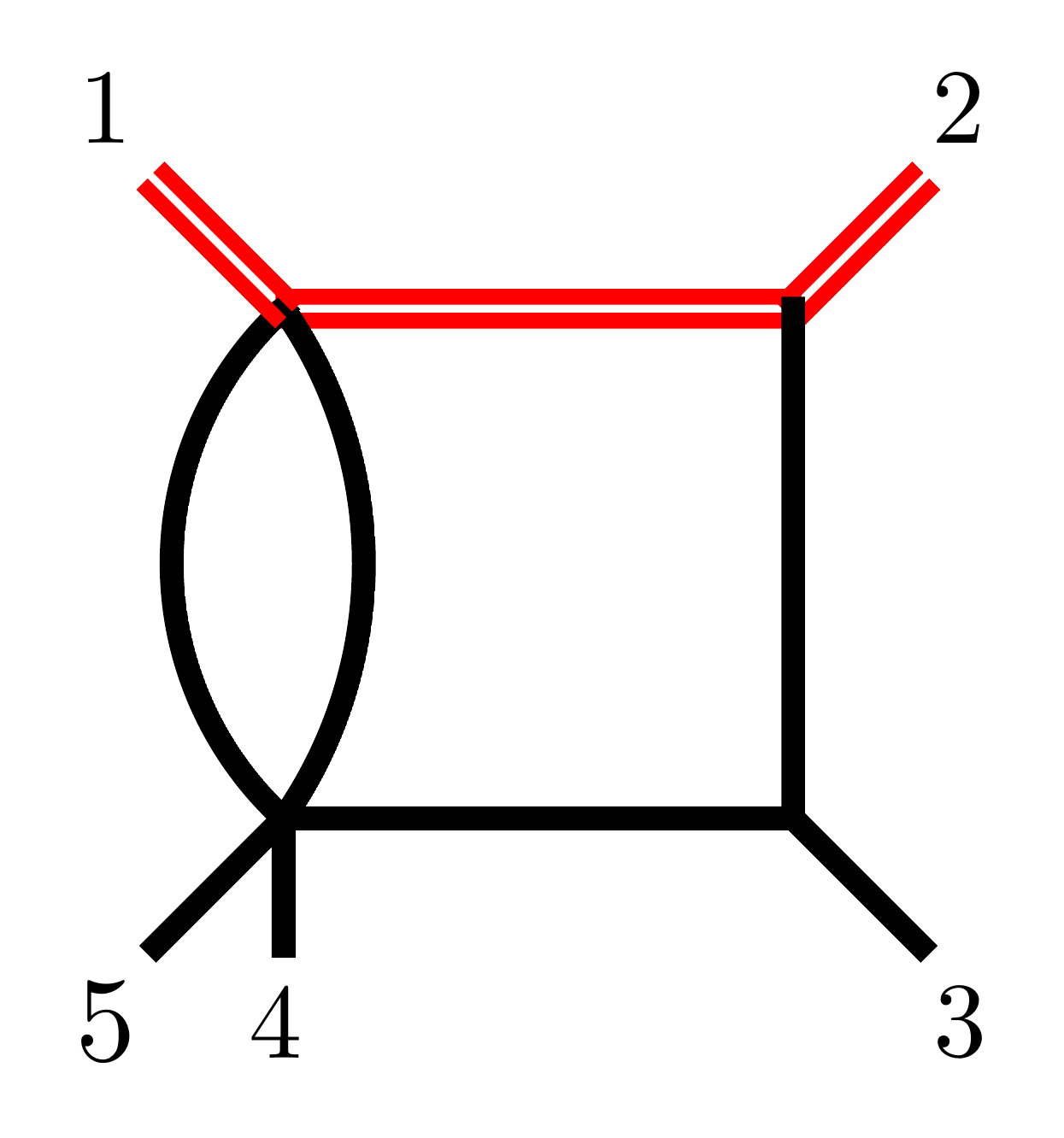}} \quad
\subfloat[$\cI_{55}$]{\includegraphics[width = 2.1 cm]{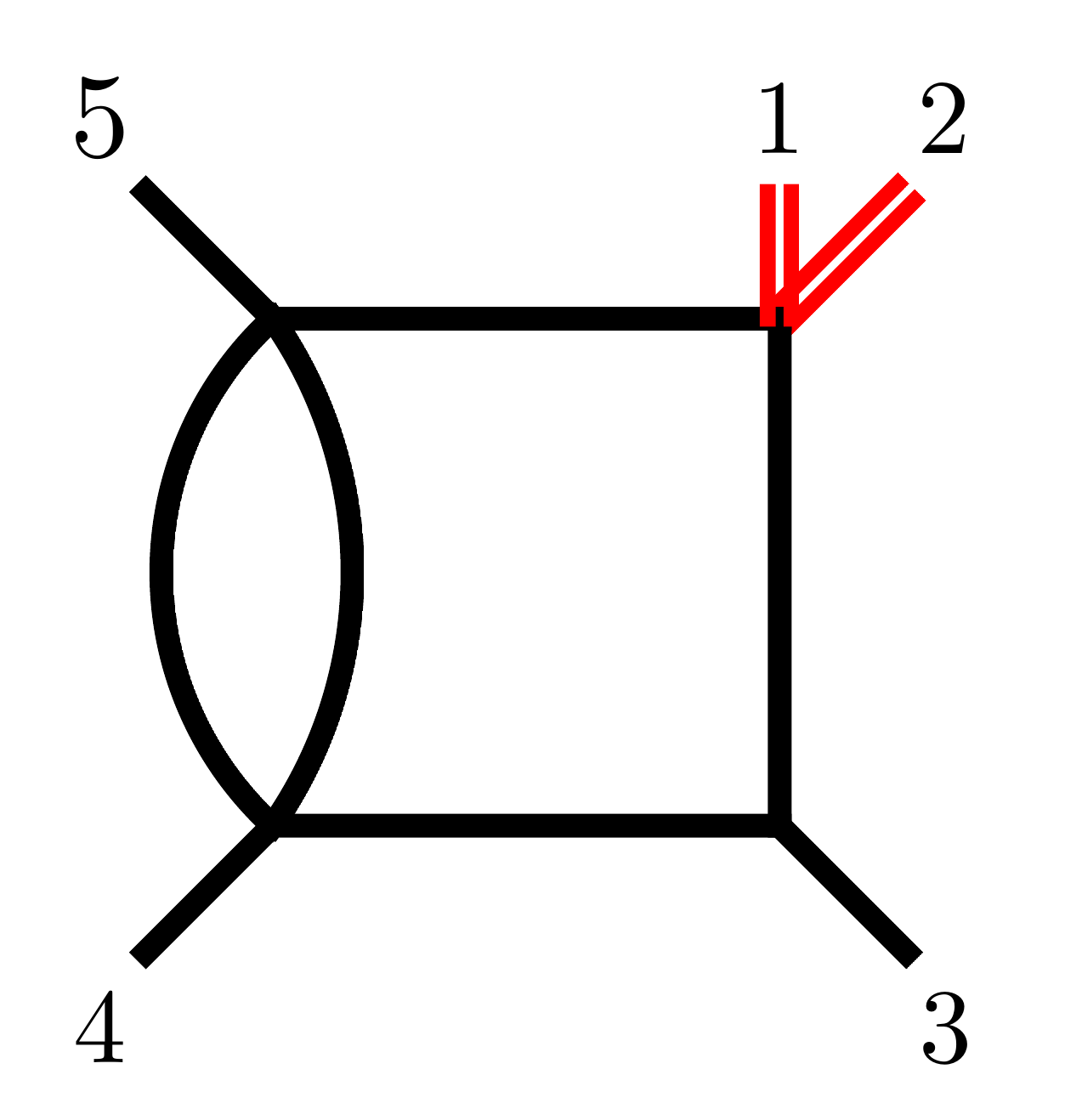}} \quad
\subfloat[$\cI_{56},\cI_{57}$]{\includegraphics[width = 2.1 cm]{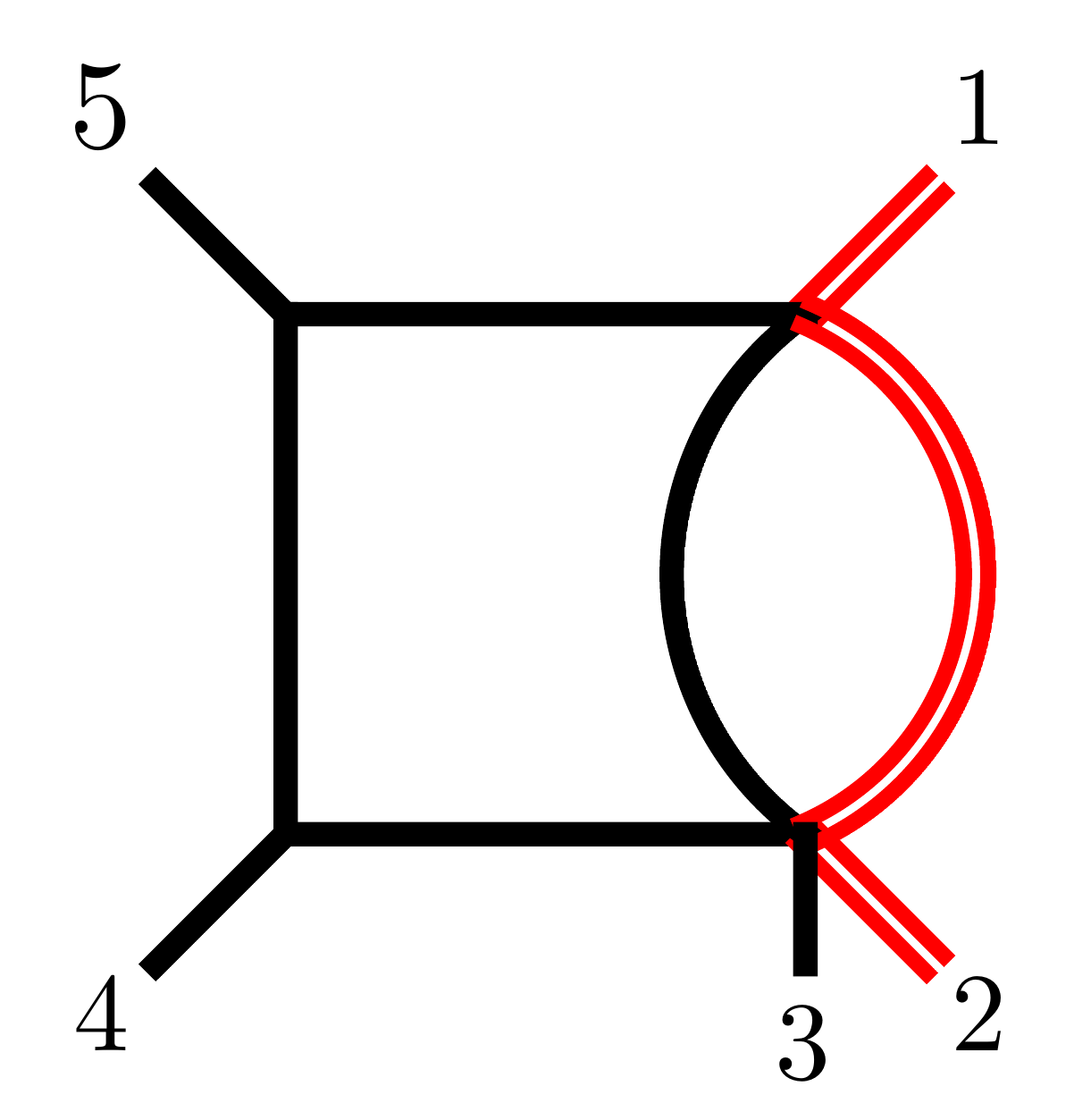}} \quad
\subfloat[$\cI_{58},\cI_{59}$]{\includegraphics[width = 2.25 cm]{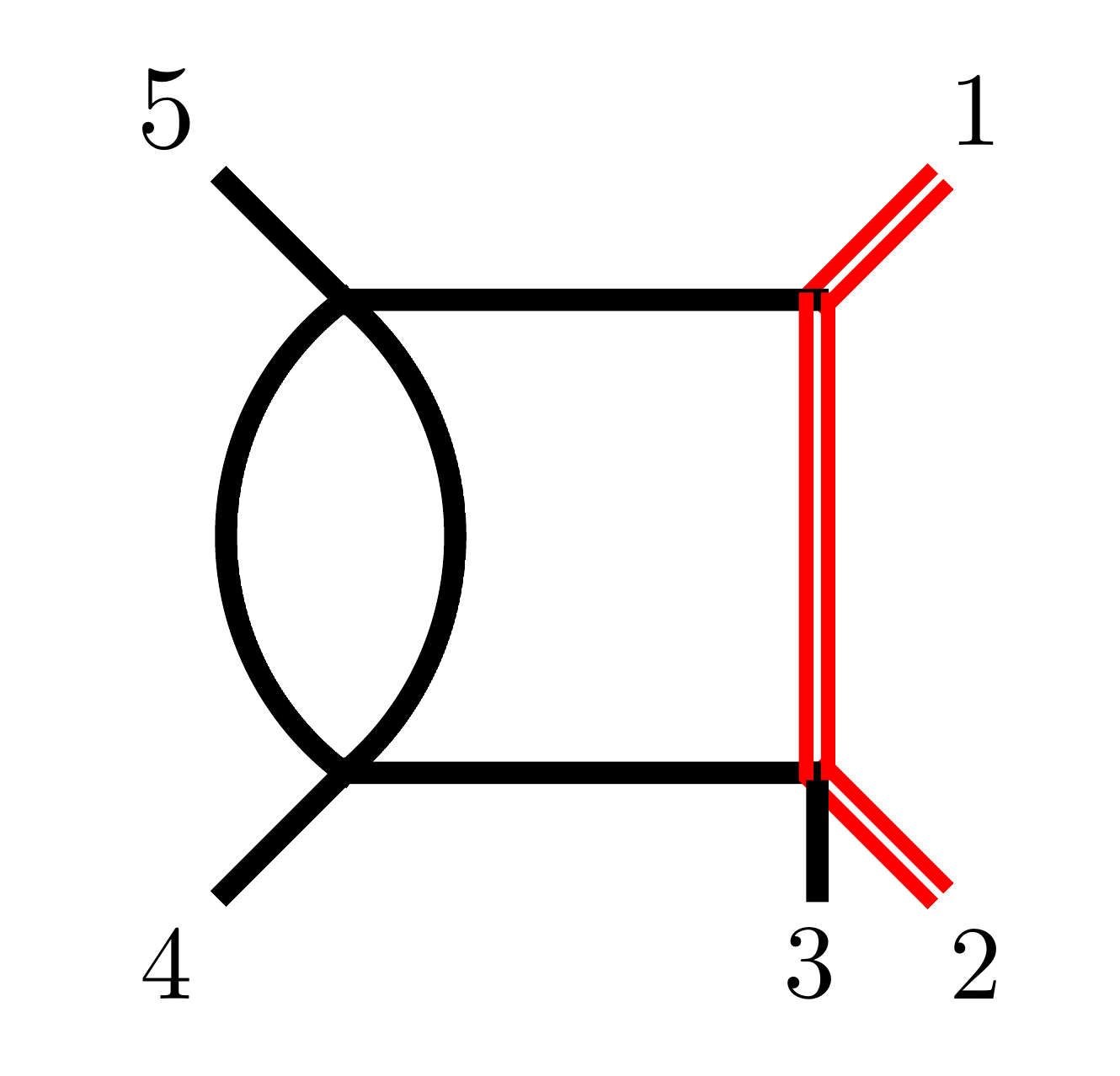}} \quad
\subfloat[$\cI_{60}$]{\includegraphics[width = 2.1 cm]{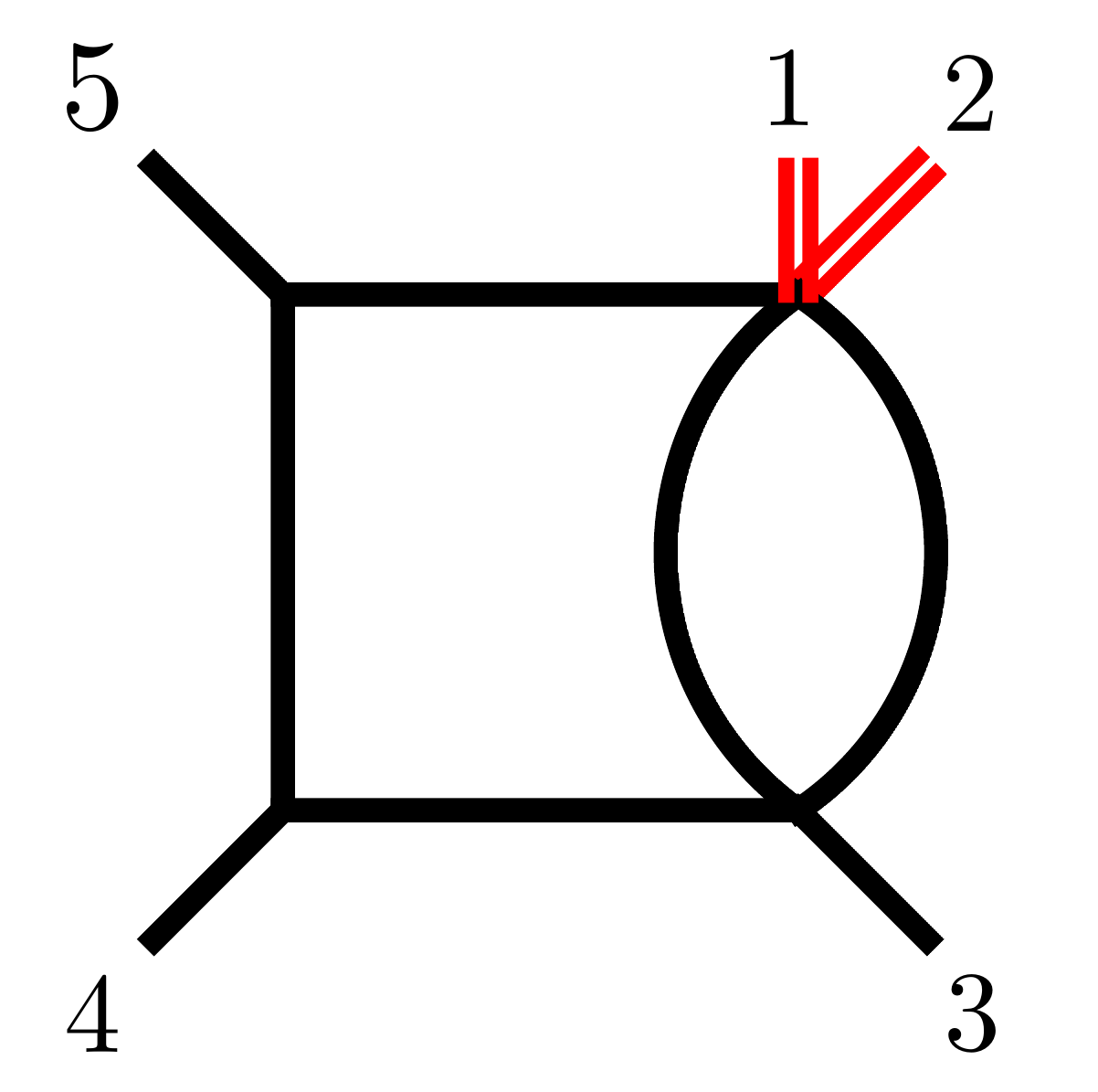}} \quad
\caption{The first 30 diagram topologies describing 60 out of 88 master integrals. The label of the individual sub-figures lists the master integrals belonging to the corresponding topology. Massive propagators and massive external momenta are indicated by red double-lines.}
\label{fig:graph_topos1}
\end{figure}

\begin{figure}[h]
\captionsetup[subfigure]{labelformat=empty}
\centering
\subfloat[$\cI_{61}$]{\includegraphics[width = 2.7 cm]{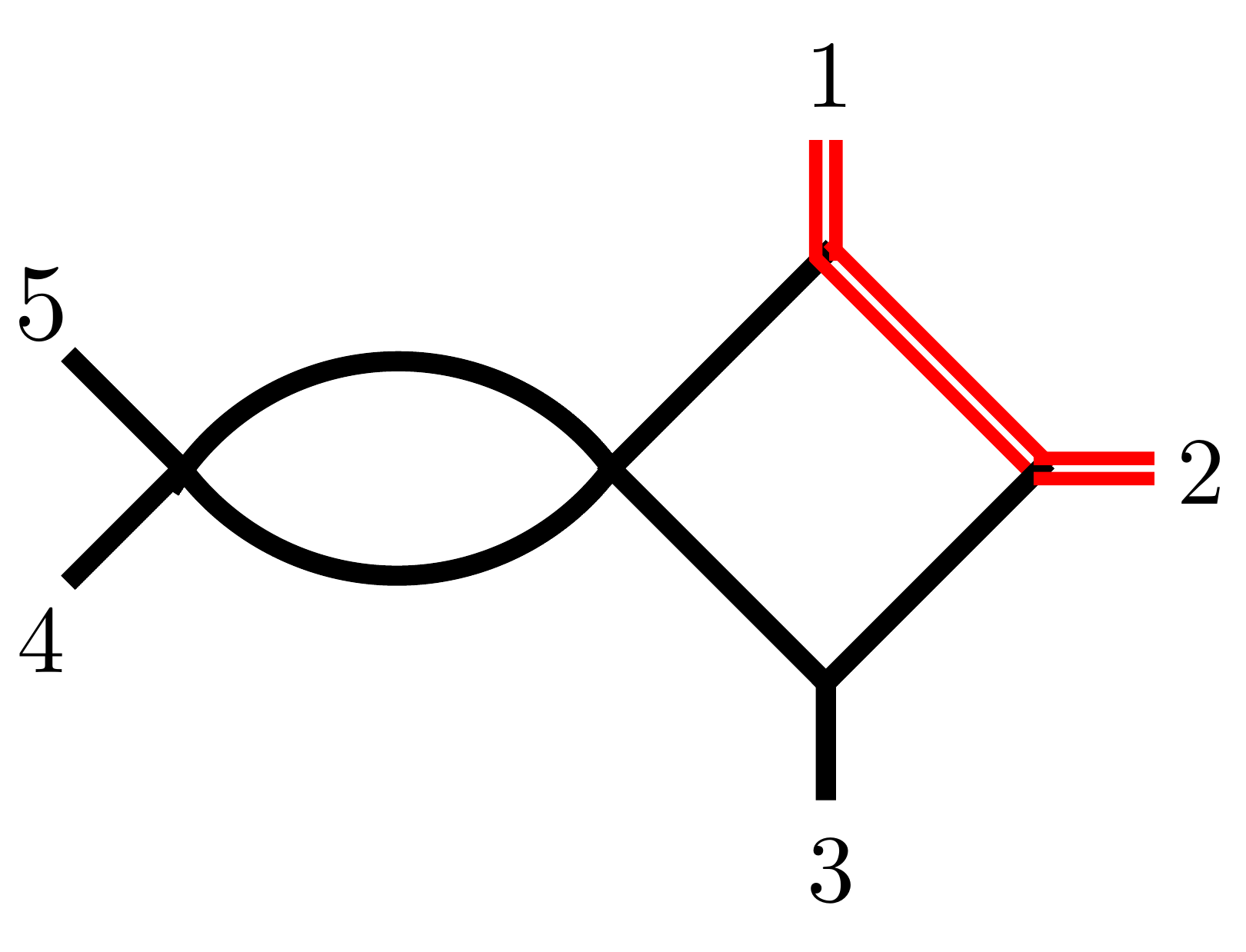}} \quad
\subfloat[$\cI_{62}$]{\includegraphics[width = 2.7 cm]{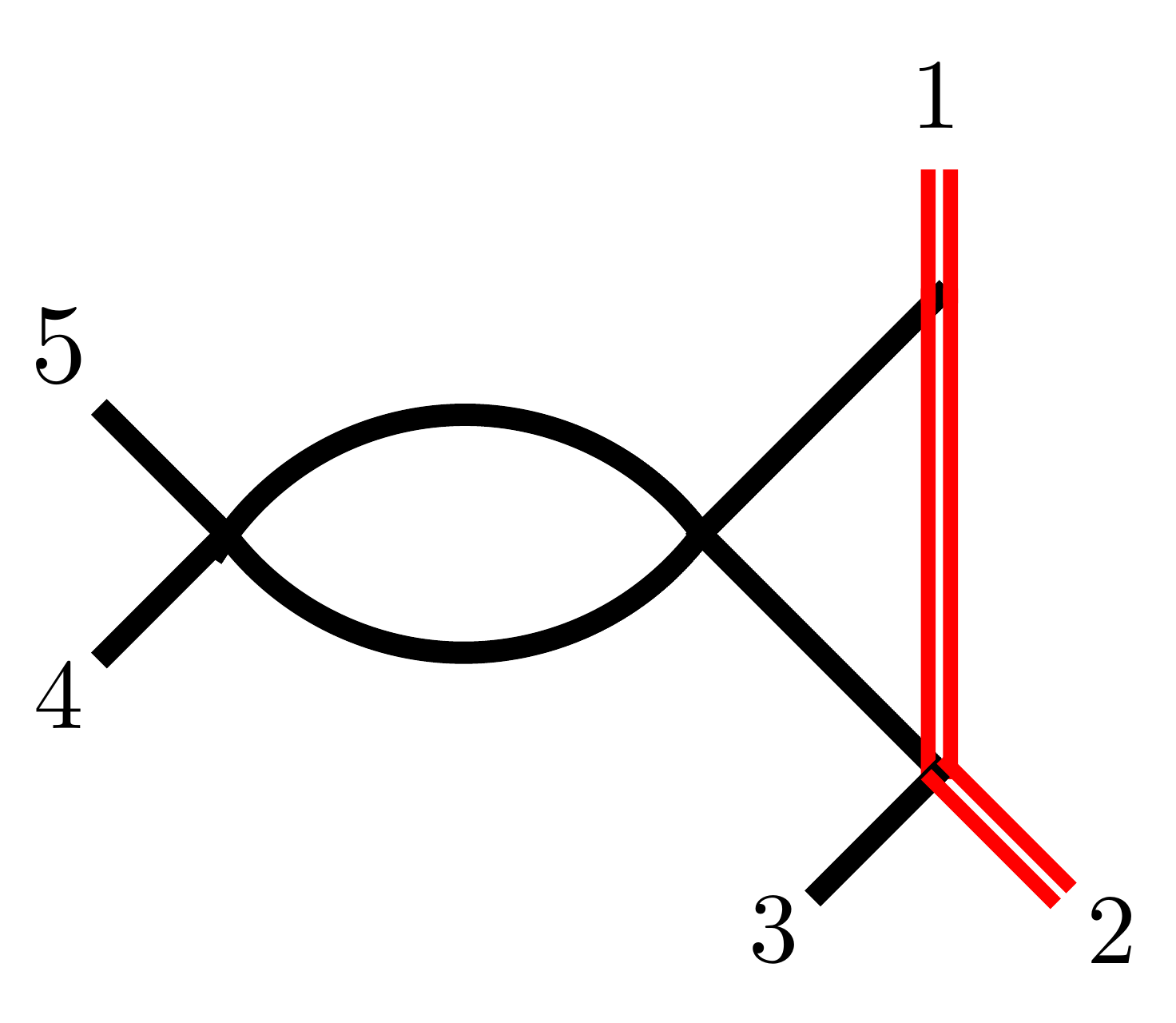}} \quad
\subfloat[$\cI_{63}$]{\includegraphics[width = 2.1 cm]{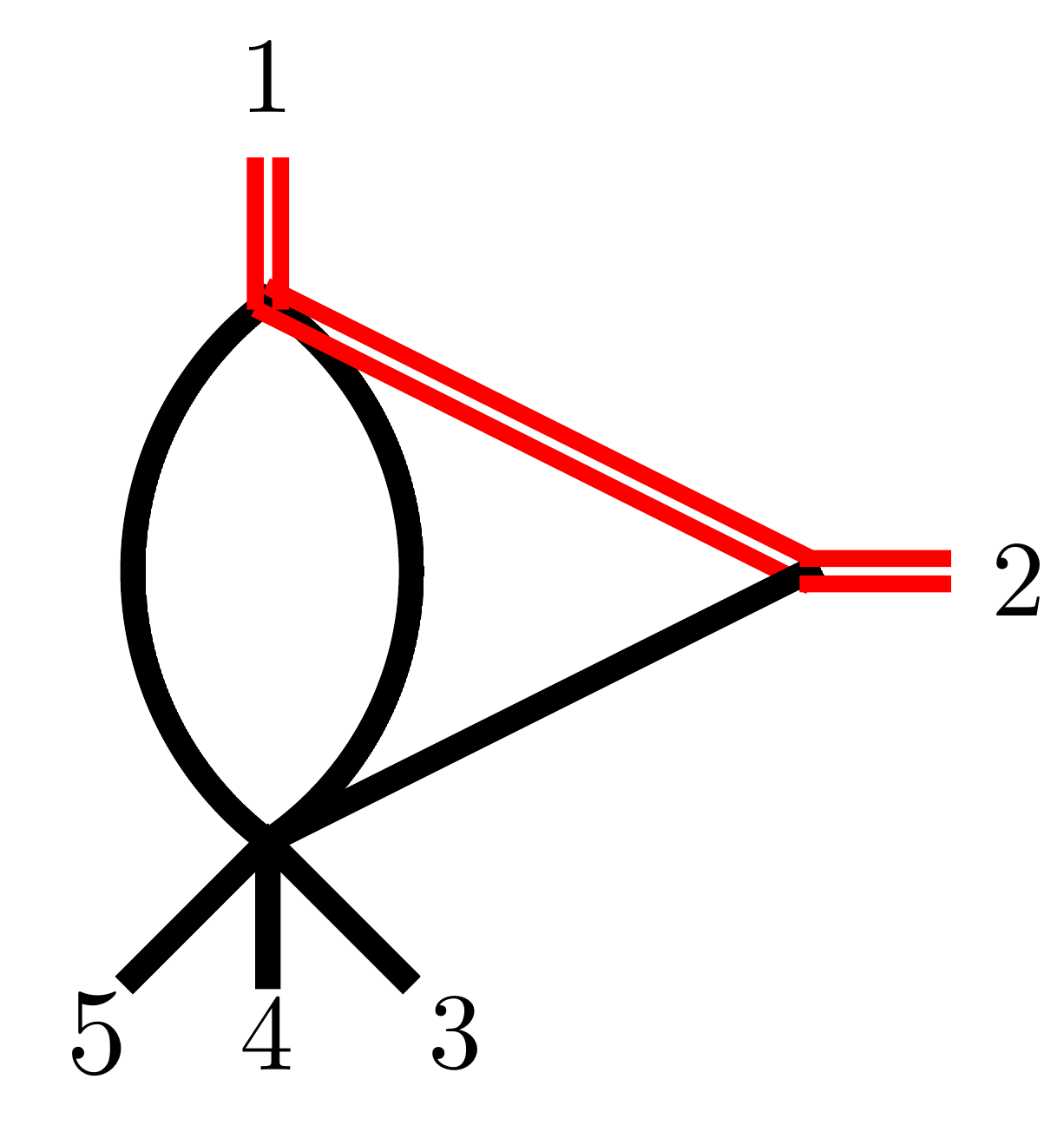}} \quad
\subfloat[$\cI_{64}$]{\includegraphics[width = 2.1 cm]{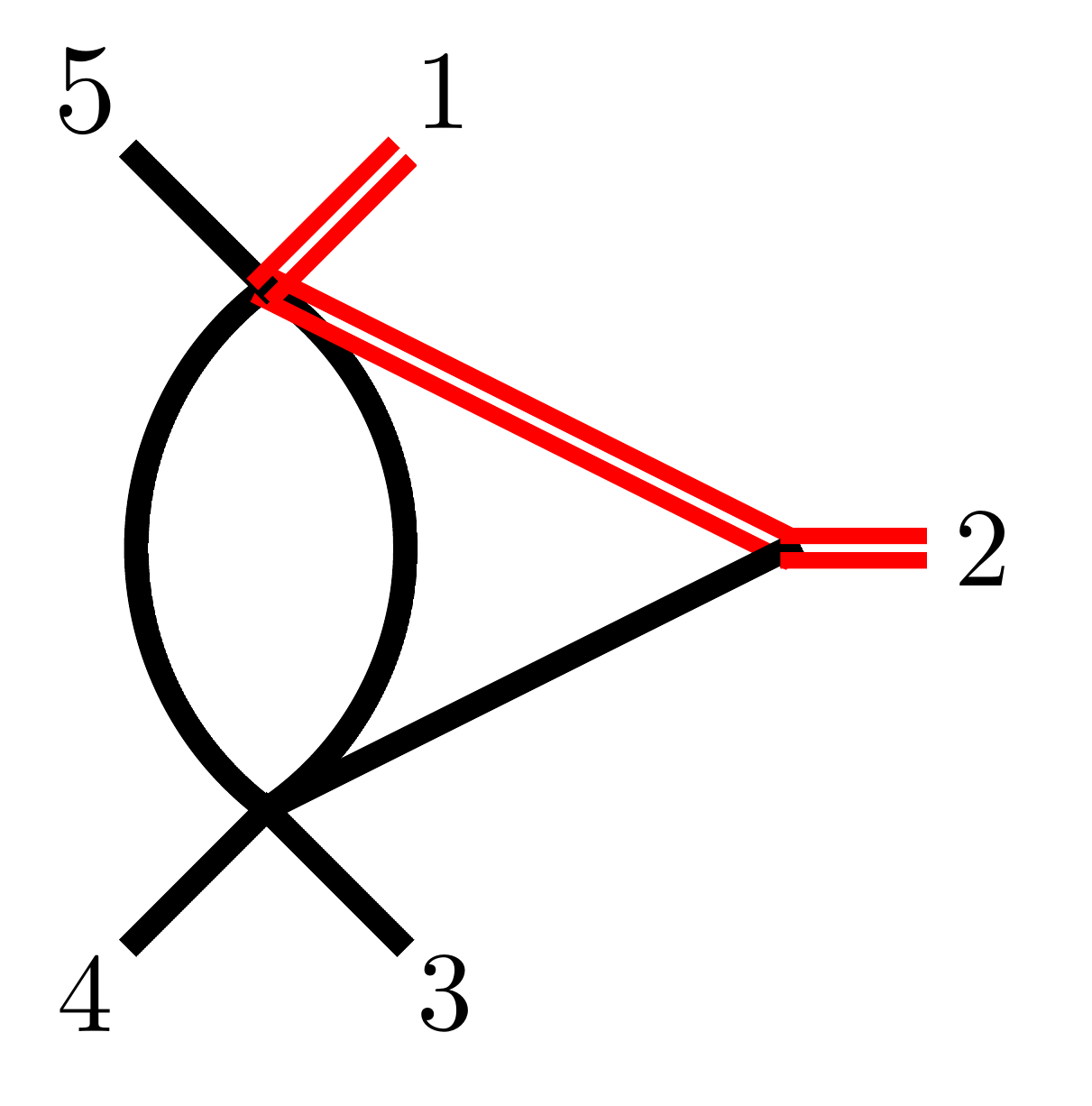}} \quad
\subfloat[$\cI_{65}$]{\includegraphics[width = 2.1 cm]{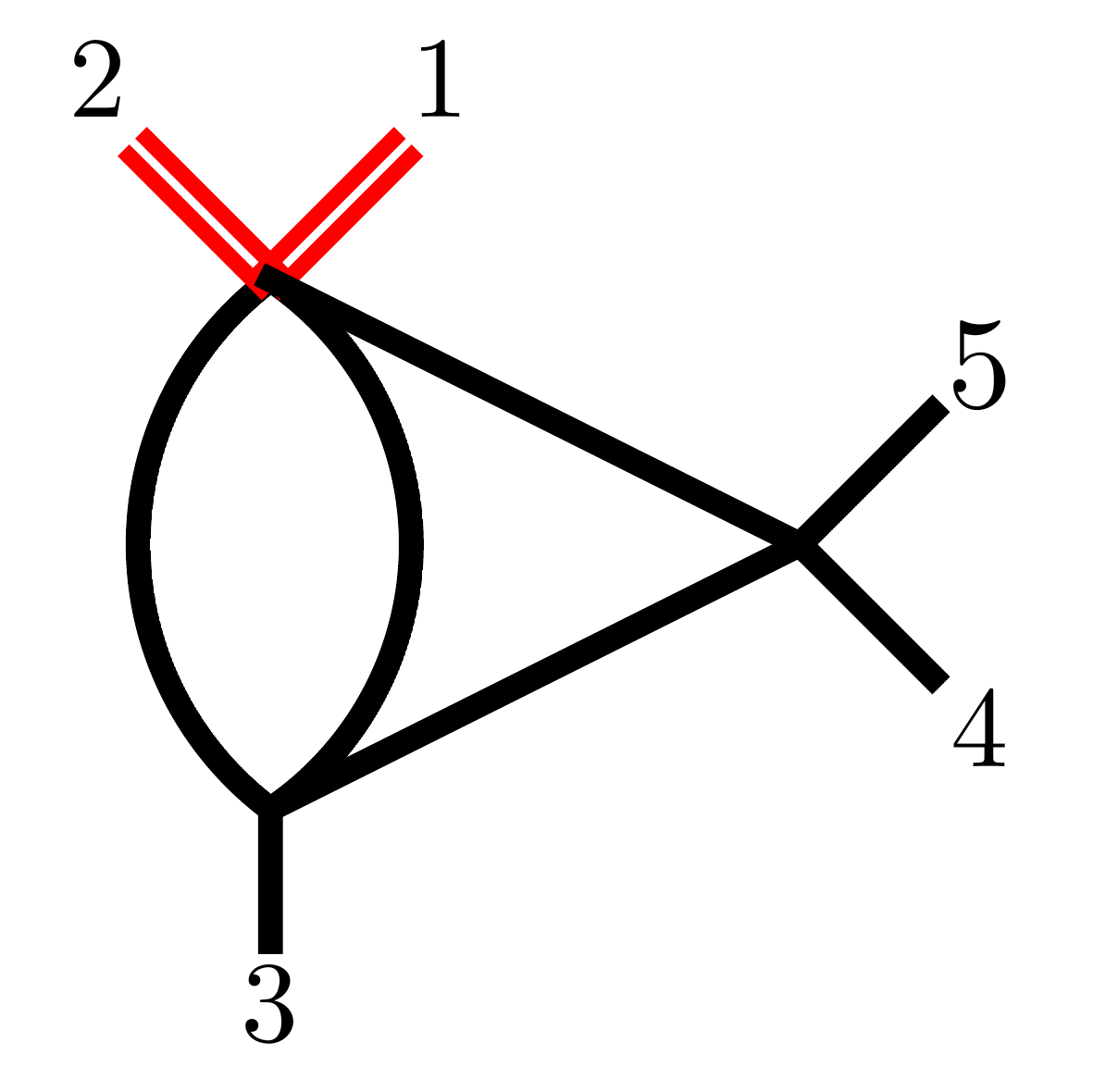}} \quad
\subfloat[$\cI_{66}$]{\includegraphics[width = 2.1 cm]{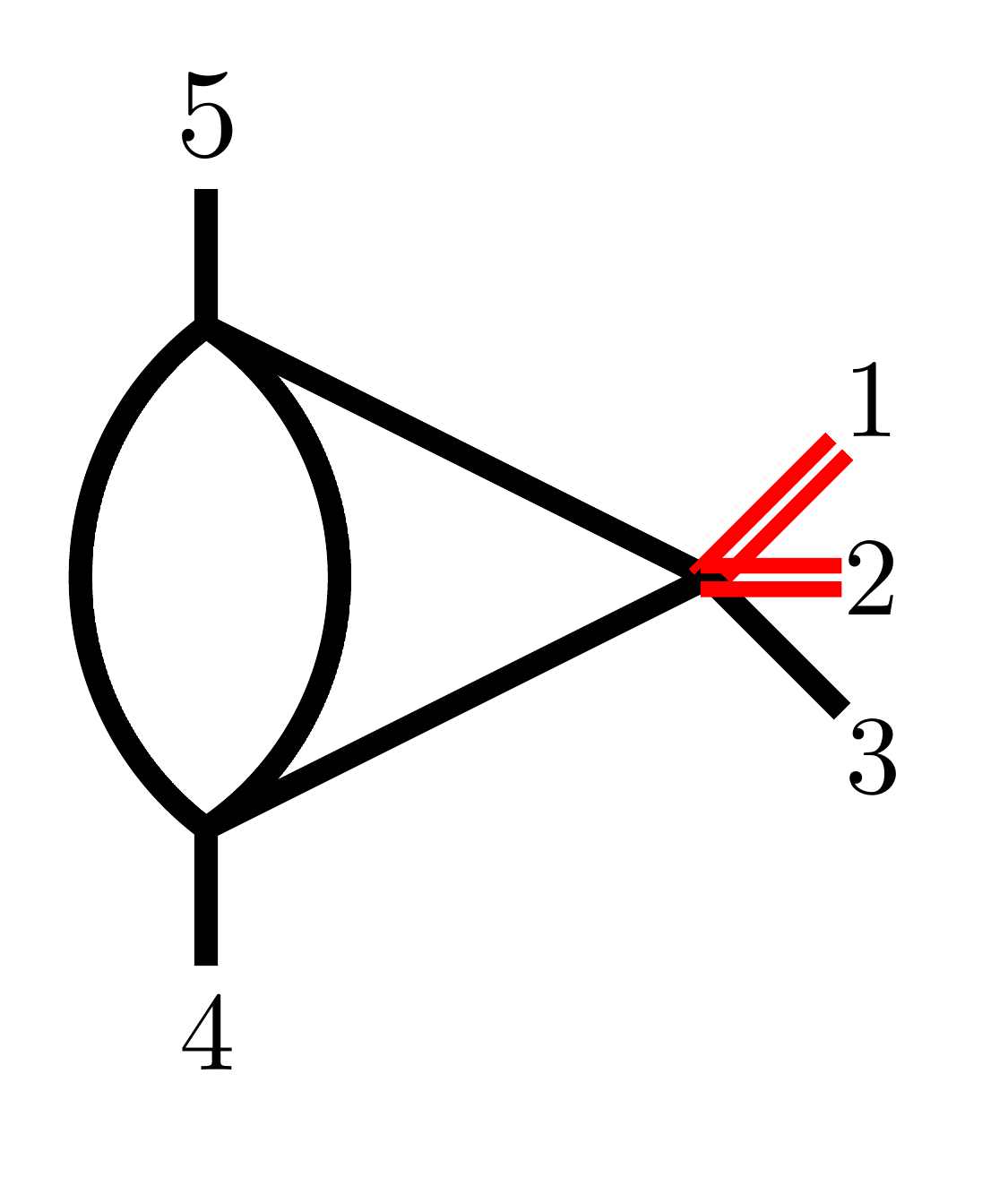}} \quad
\subfloat[$\cI_{67}$]{\includegraphics[width = 2.4 cm]{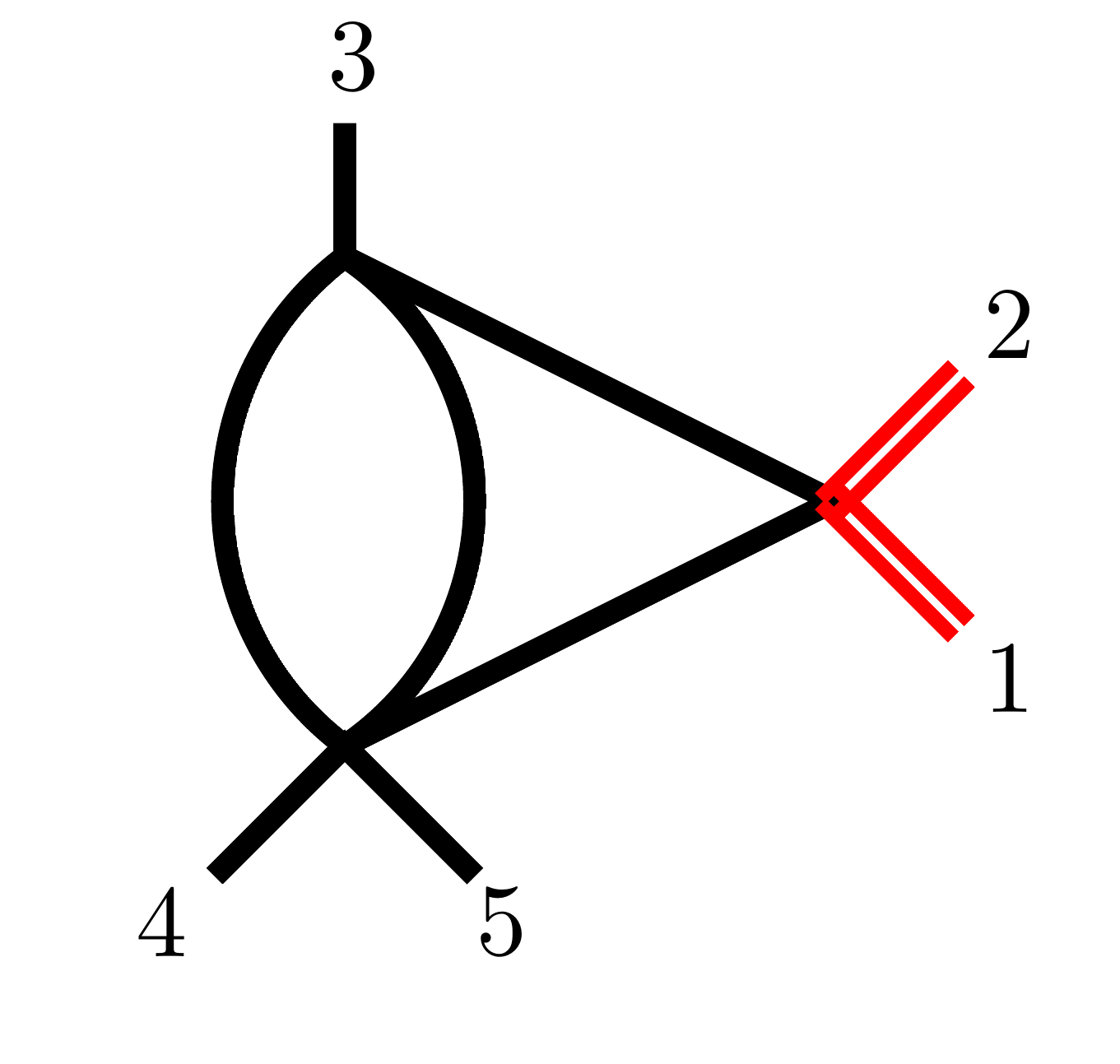}} \quad
\subfloat[$\cI_{68}$]{\includegraphics[width = 2.1 cm]{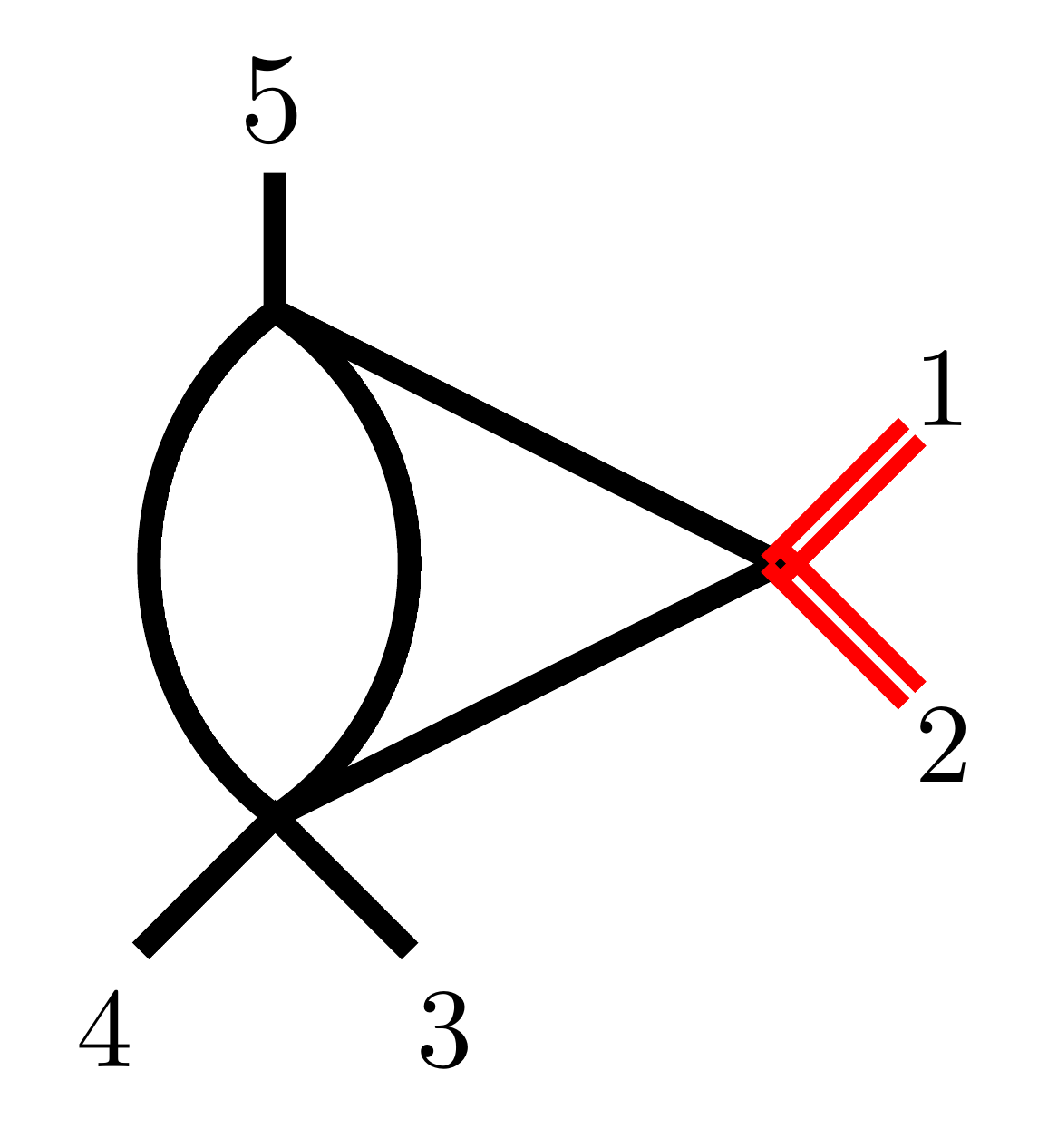}} \quad
\subfloat[$\cI_{69},\cI_{70},\cI_{71}$]{\includegraphics[width = 2.2 cm]{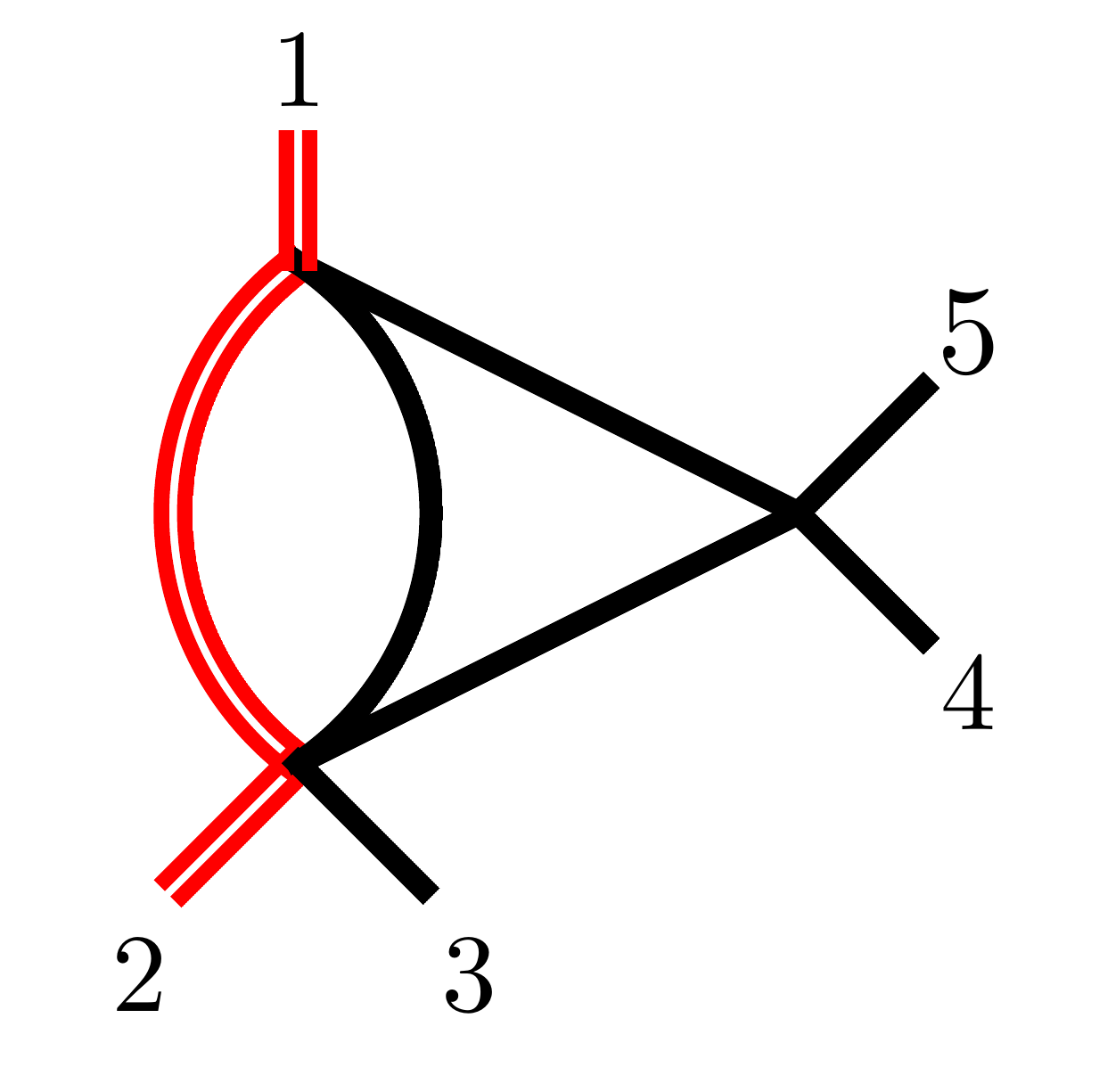}} \quad
\subfloat[$\cI_{72},\cI_{73}$]{\includegraphics[width = 2.3 cm]{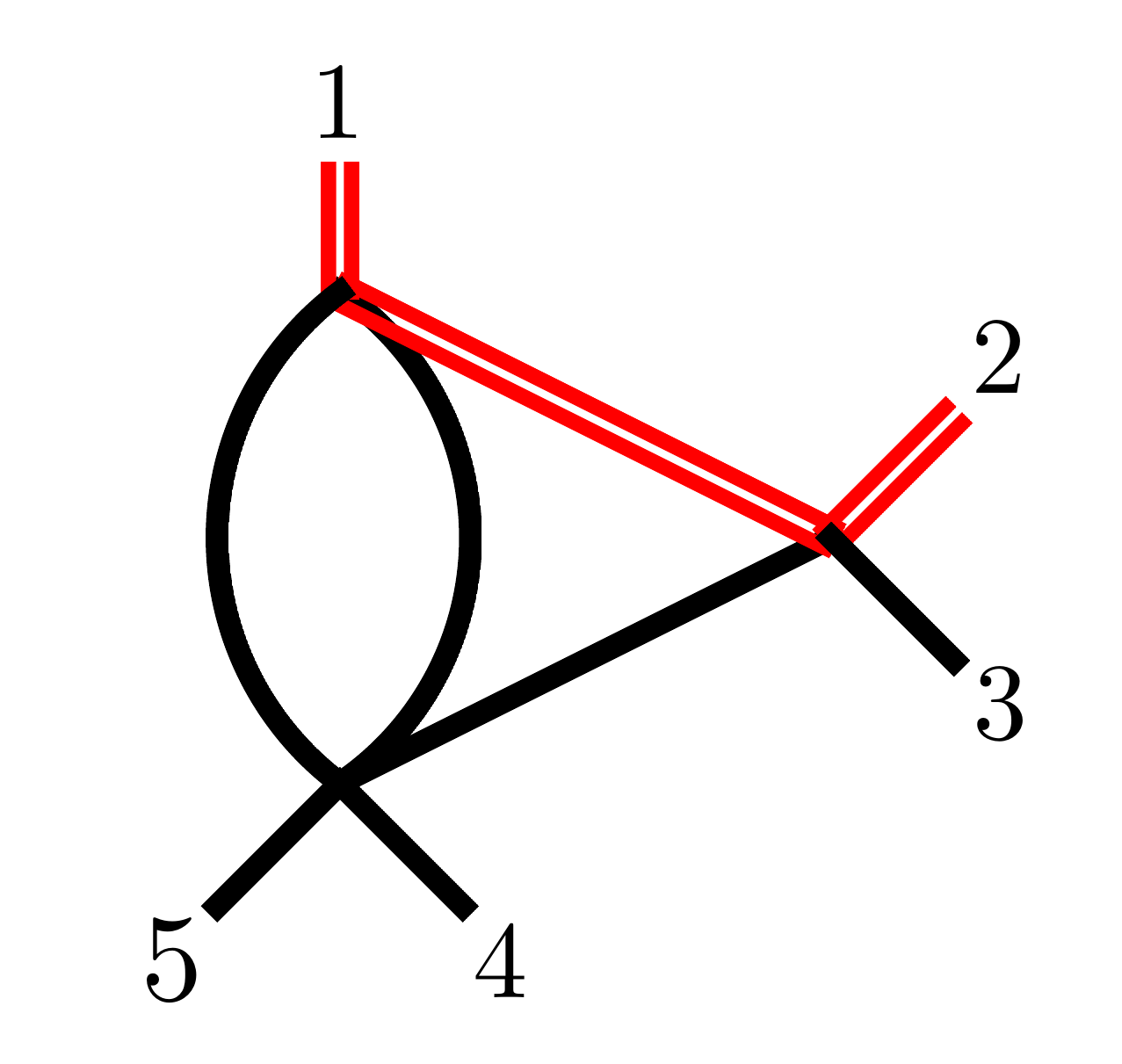}} \quad 
\subfloat[$\cI_{74}$]{\includegraphics[width = 2.1 cm]{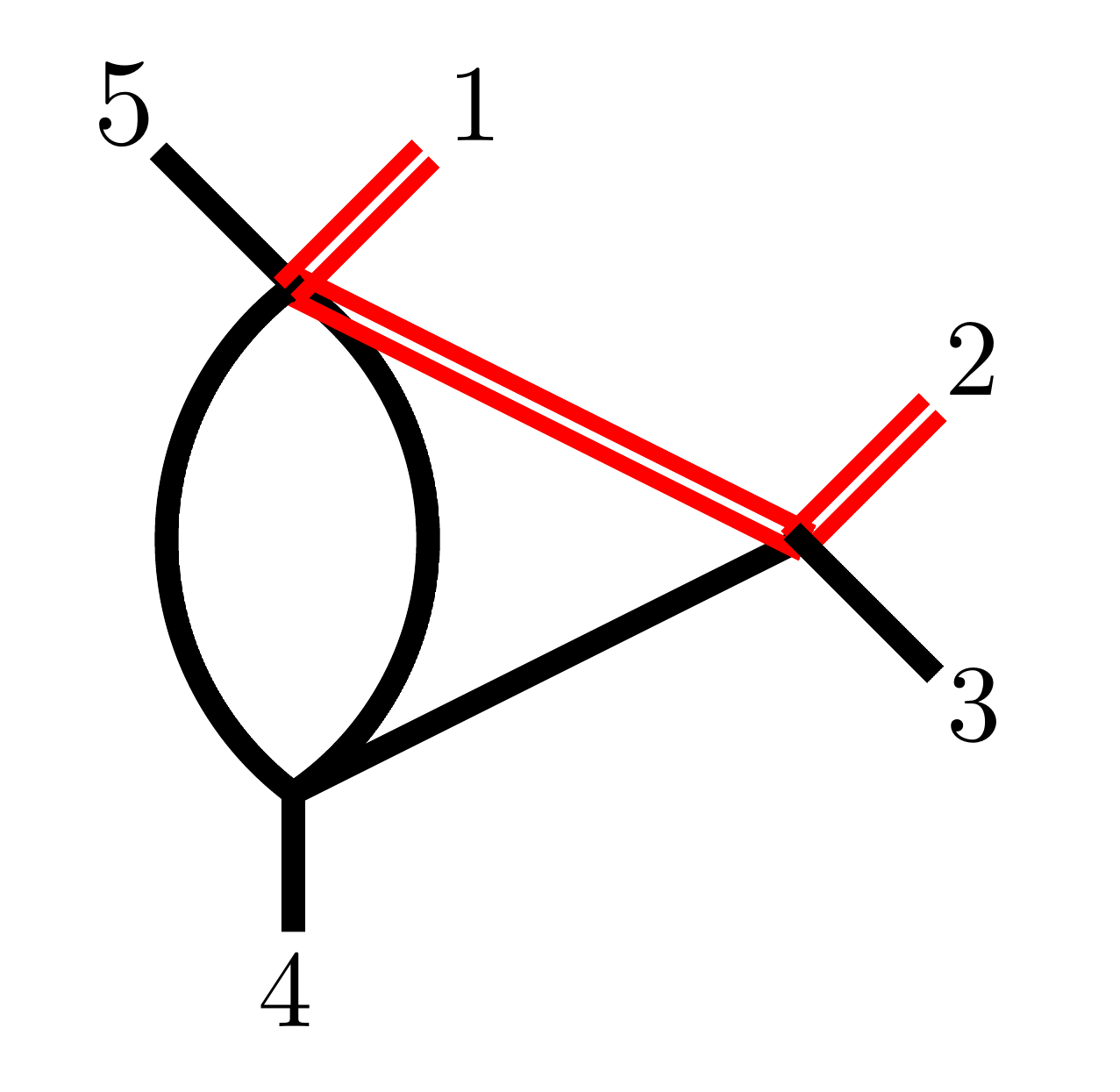}} \quad 
\subfloat[$\cI_{75}$]{\includegraphics[width = 2.2 cm]{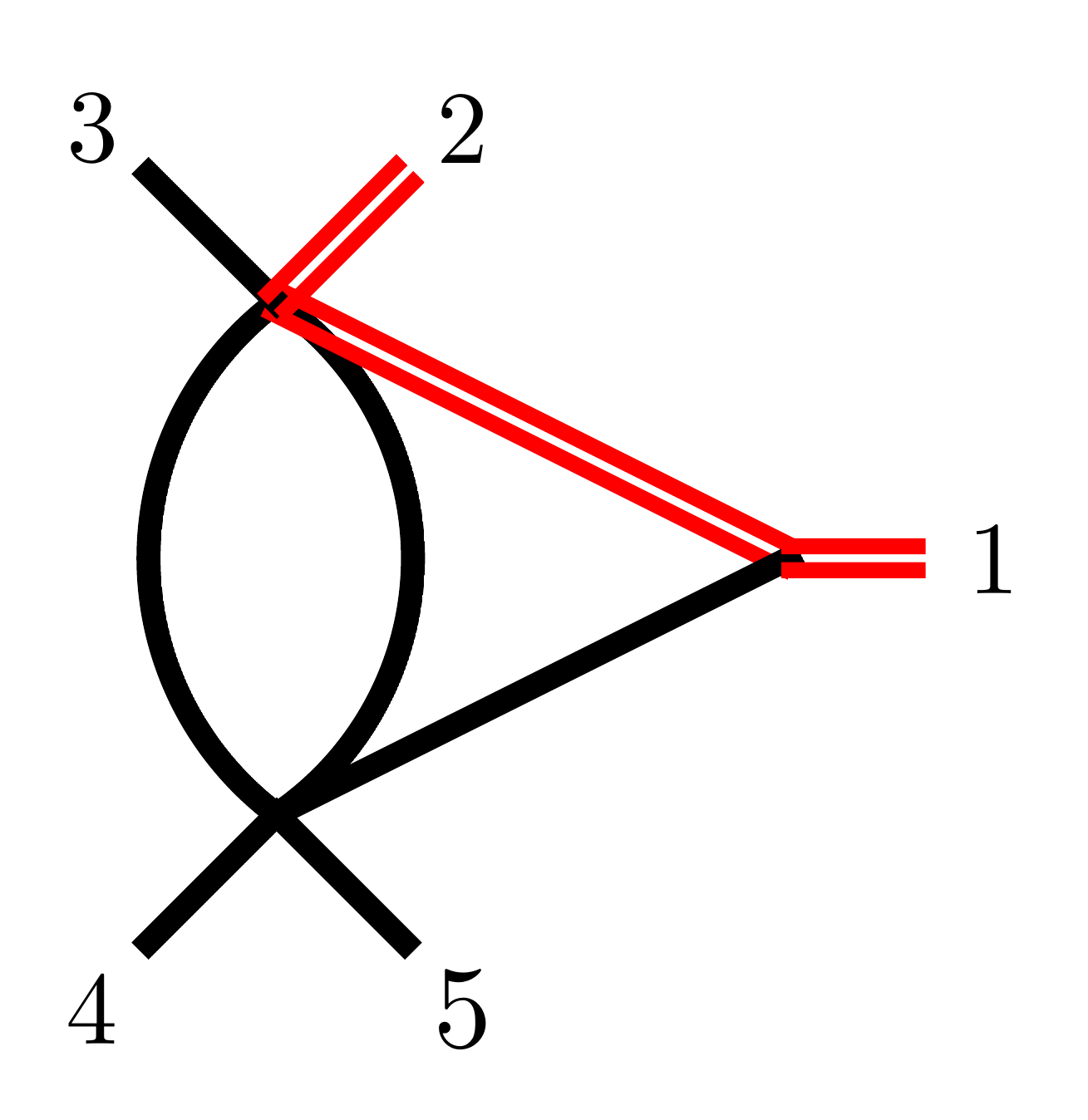}} \quad
\subfloat[$\cI_{76}$]{\includegraphics[width = 2.6 cm]{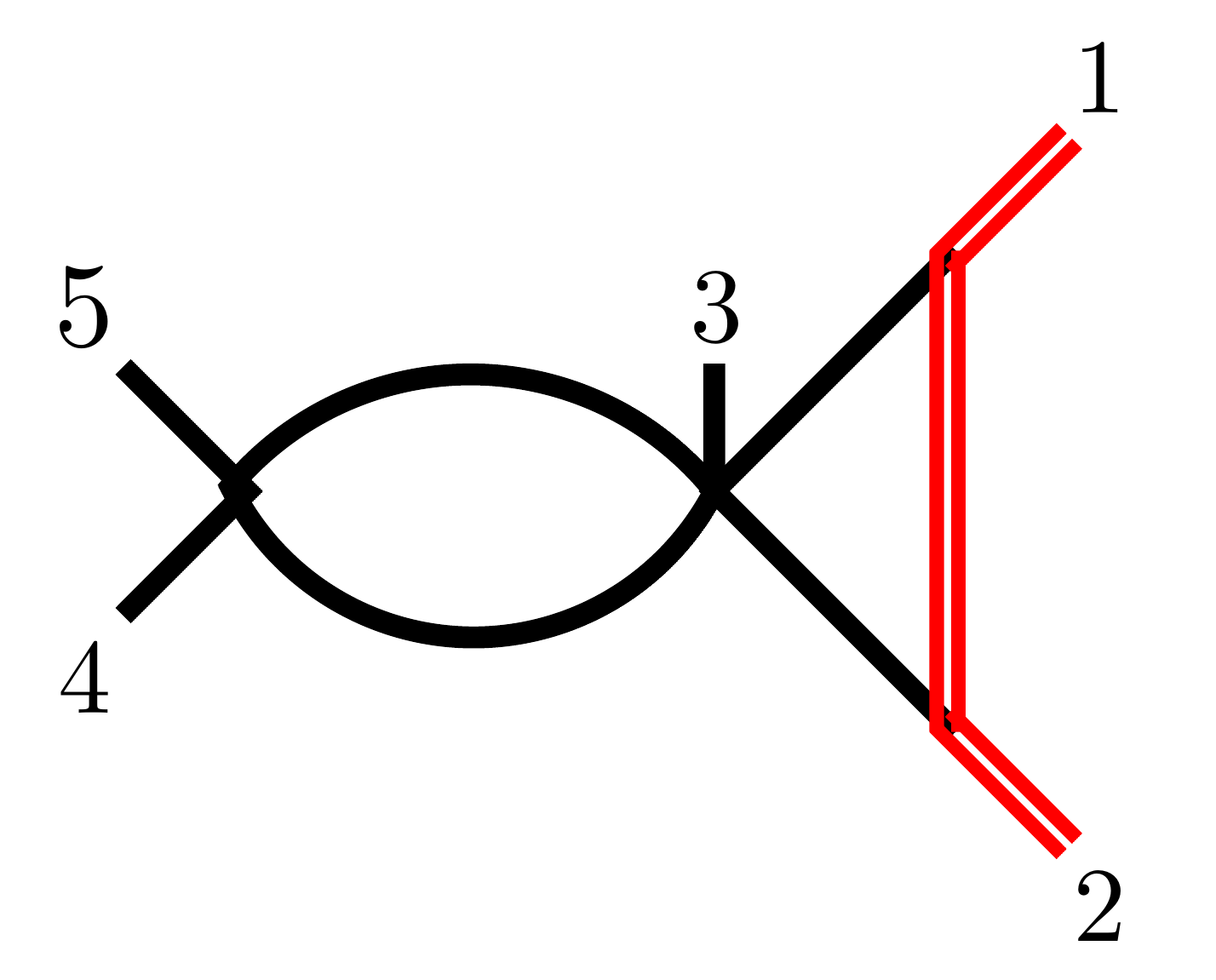}} \quad
\subfloat[$\cI_{77}$]{\includegraphics[width = 3.2 cm]{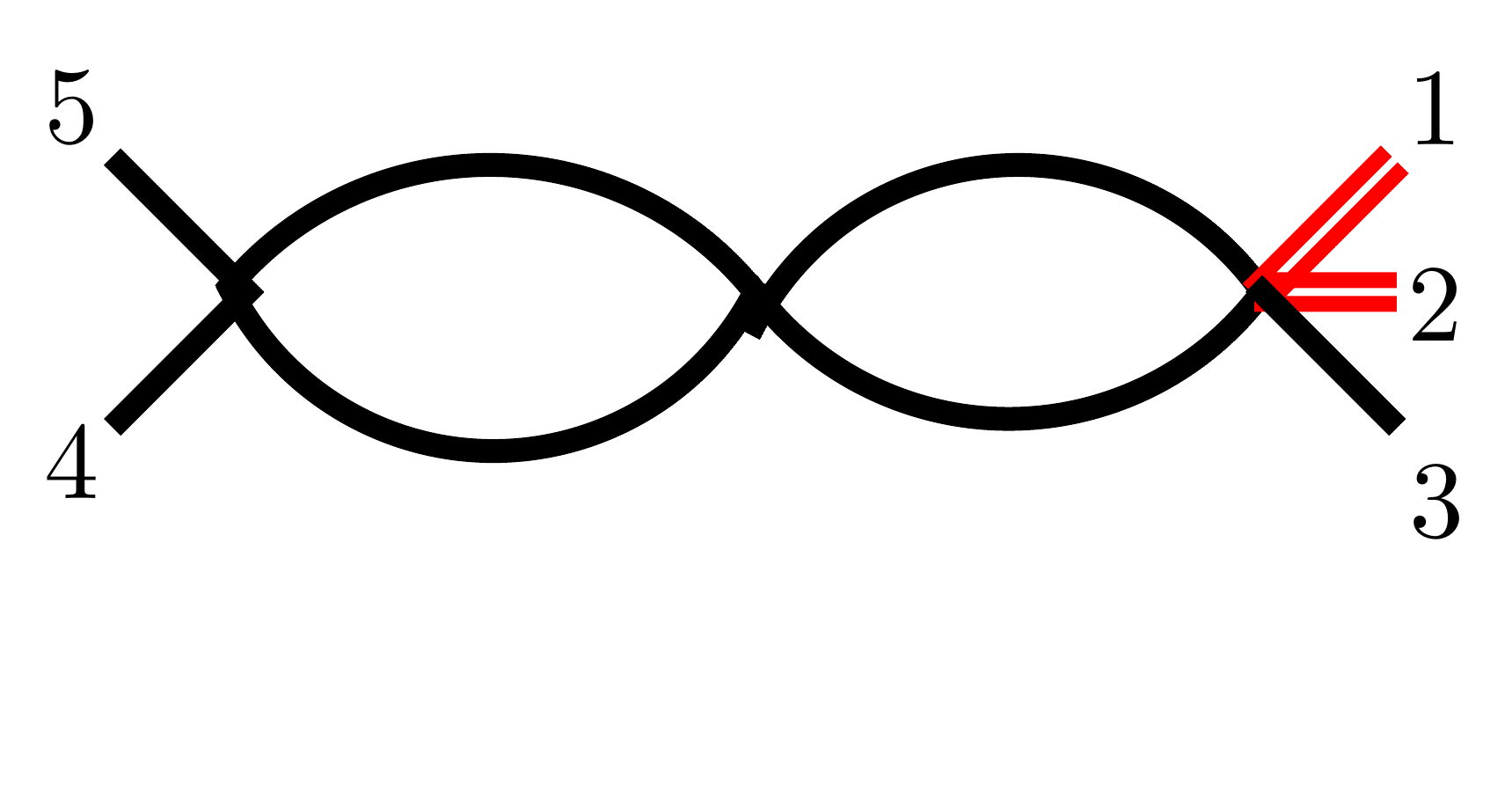}} \quad
\subfloat[$\cI_{78}$]{\includegraphics[width = 3.2 cm]{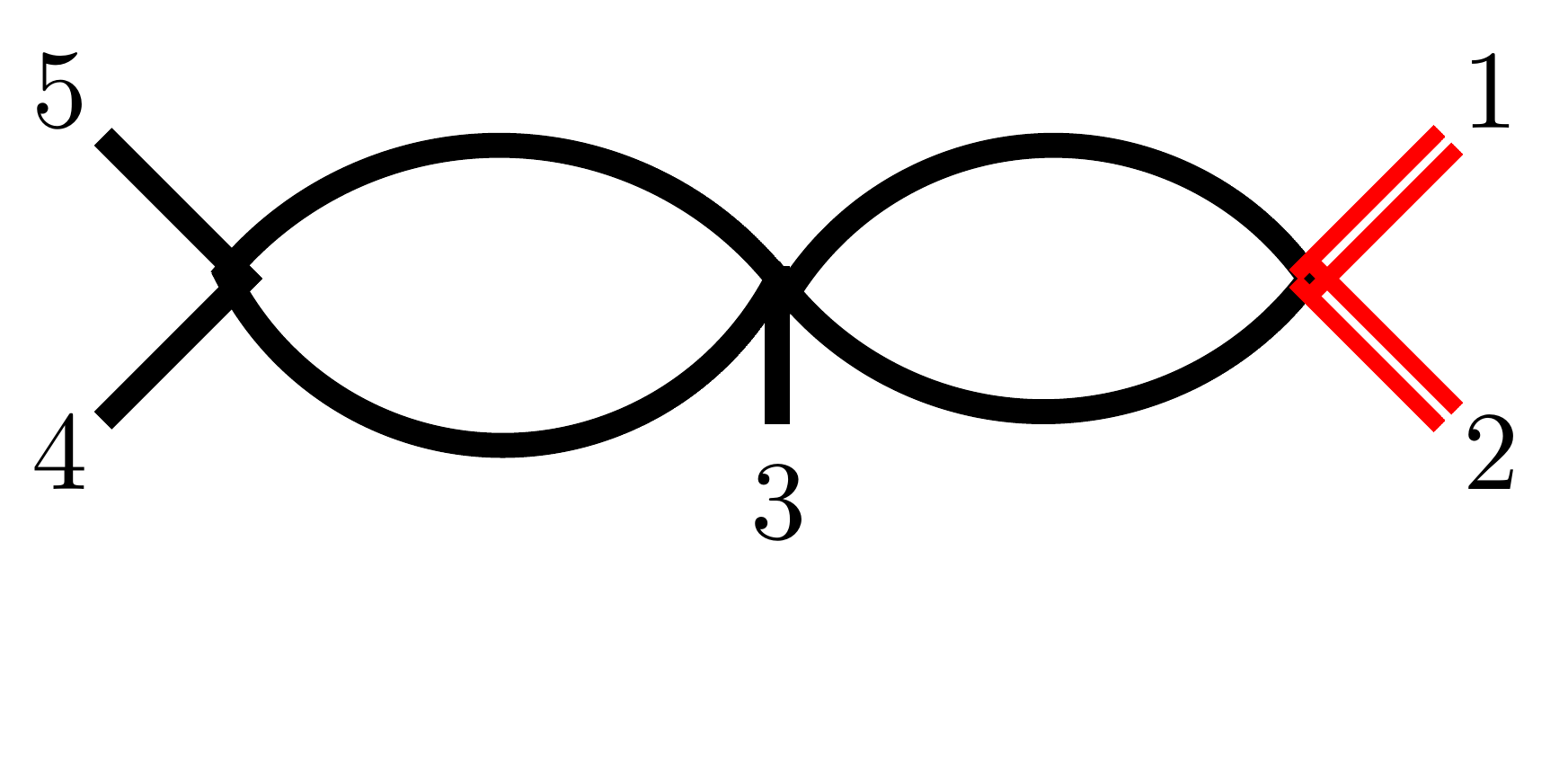}} \quad
\subfloat[$\cI_{79}$]{\includegraphics[width = 3.2 cm]{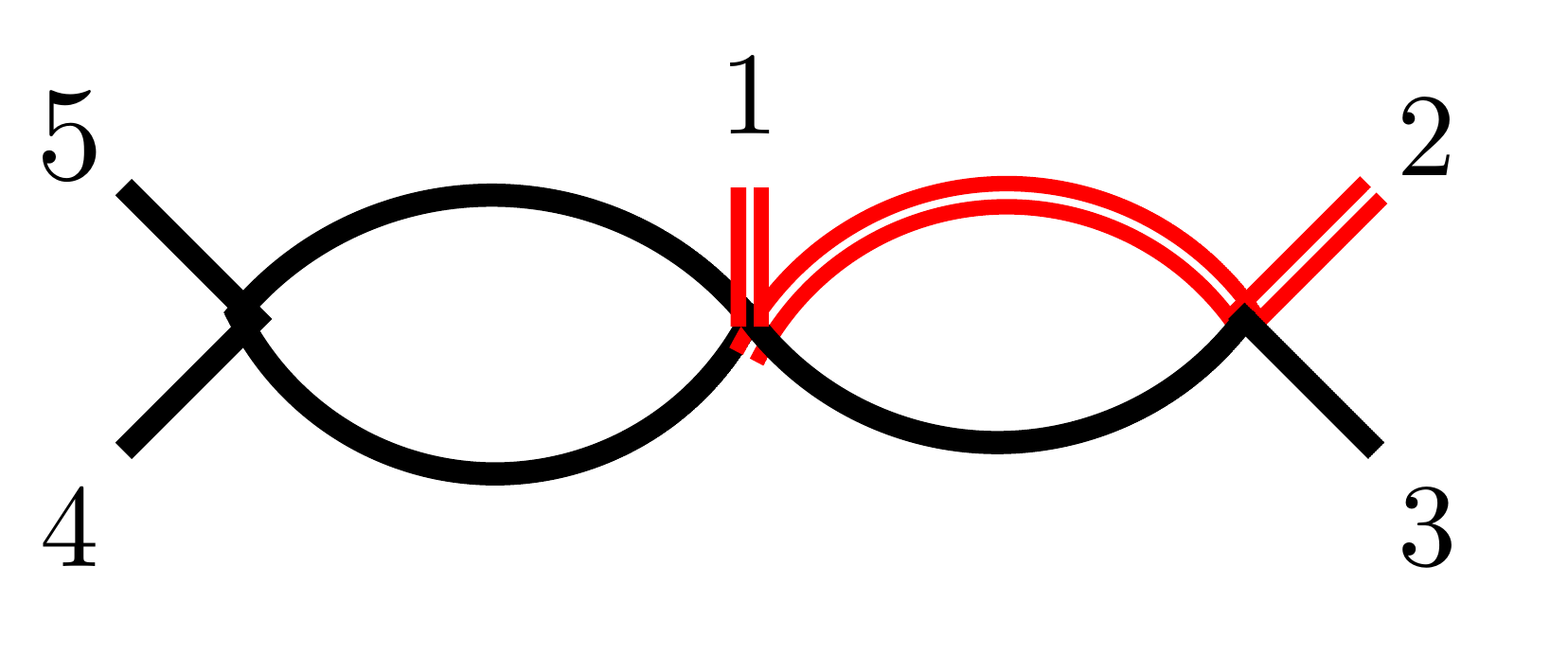}} \quad
\subfloat[$\cI_{80}$]{\includegraphics[width = 2.4 cm]{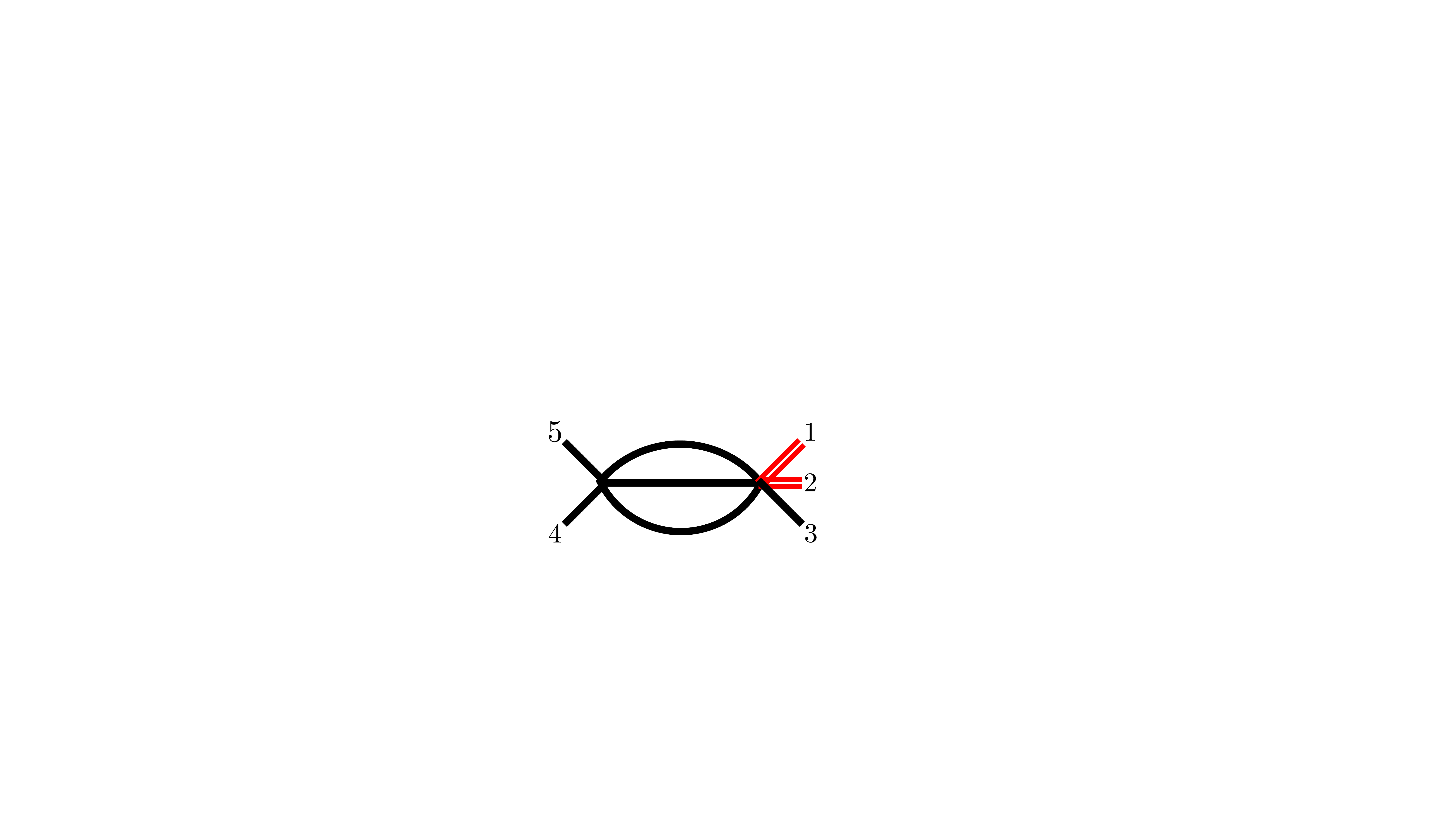}} \quad
\subfloat[$\cI_{81}$]{\includegraphics[width = 2.4 cm]{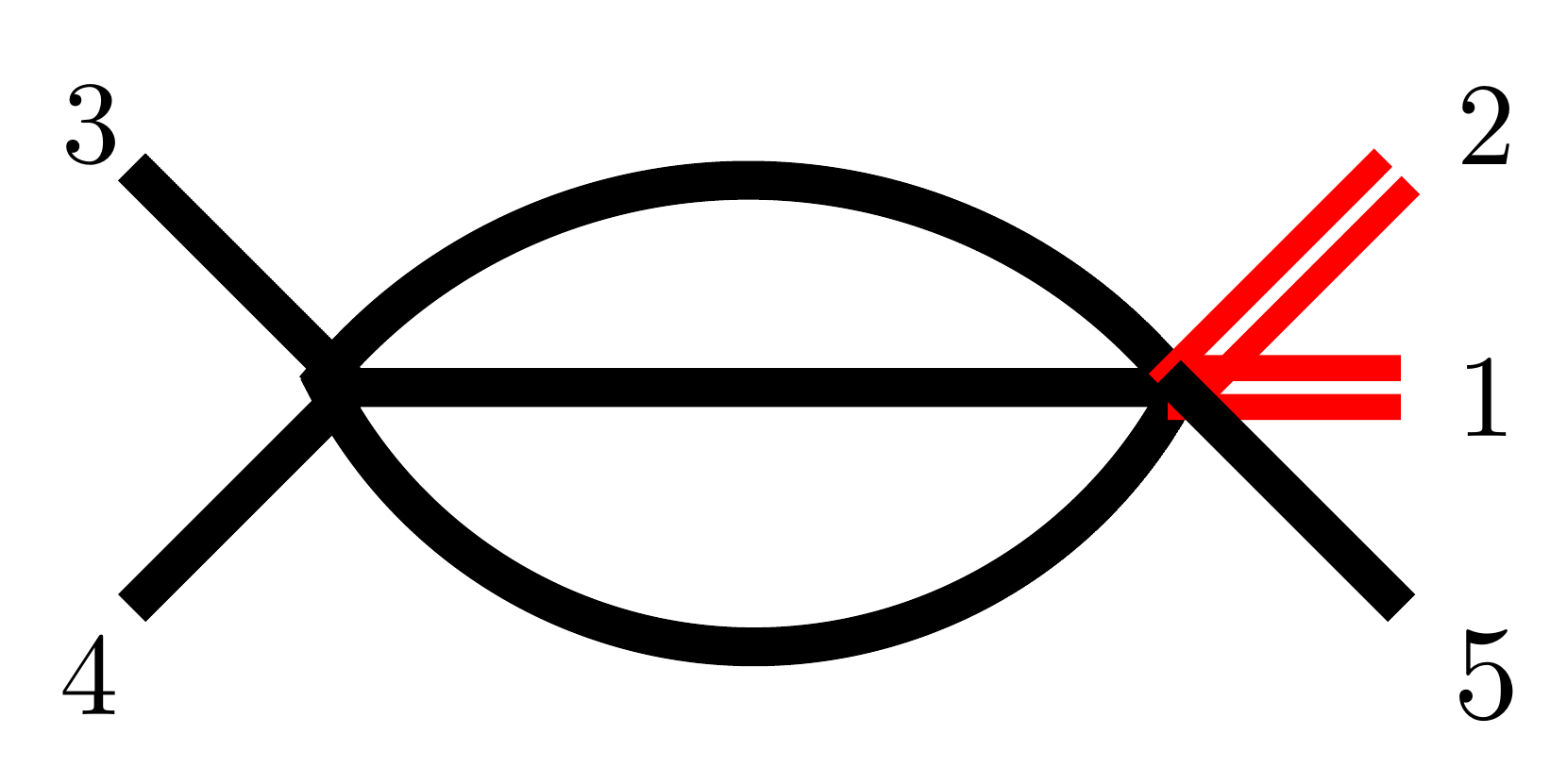}} \quad
\subfloat[$\cI_{82}$]{\includegraphics[width = 2.4 cm]{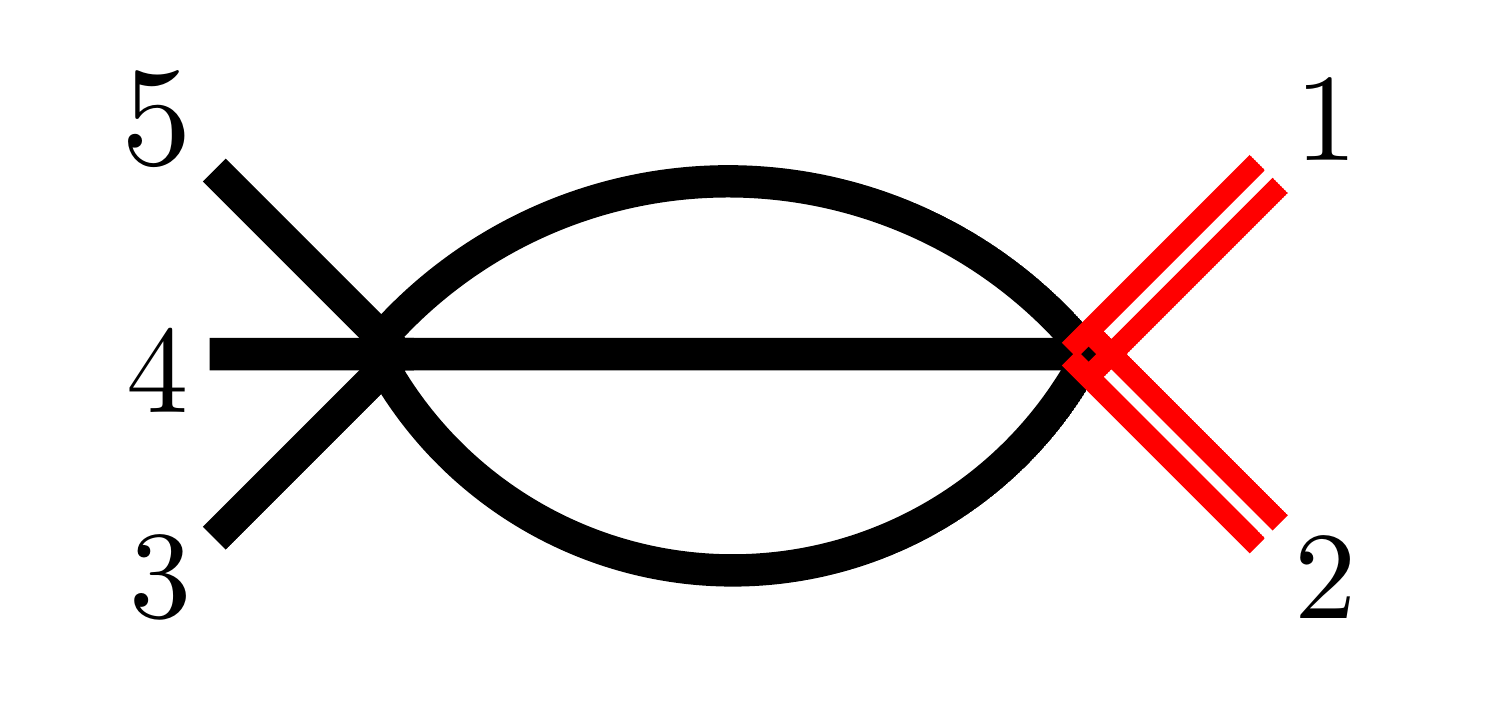}} \quad
\subfloat[$\cI_{83},\cI_{84}$]{\includegraphics[width = 2.4 cm]{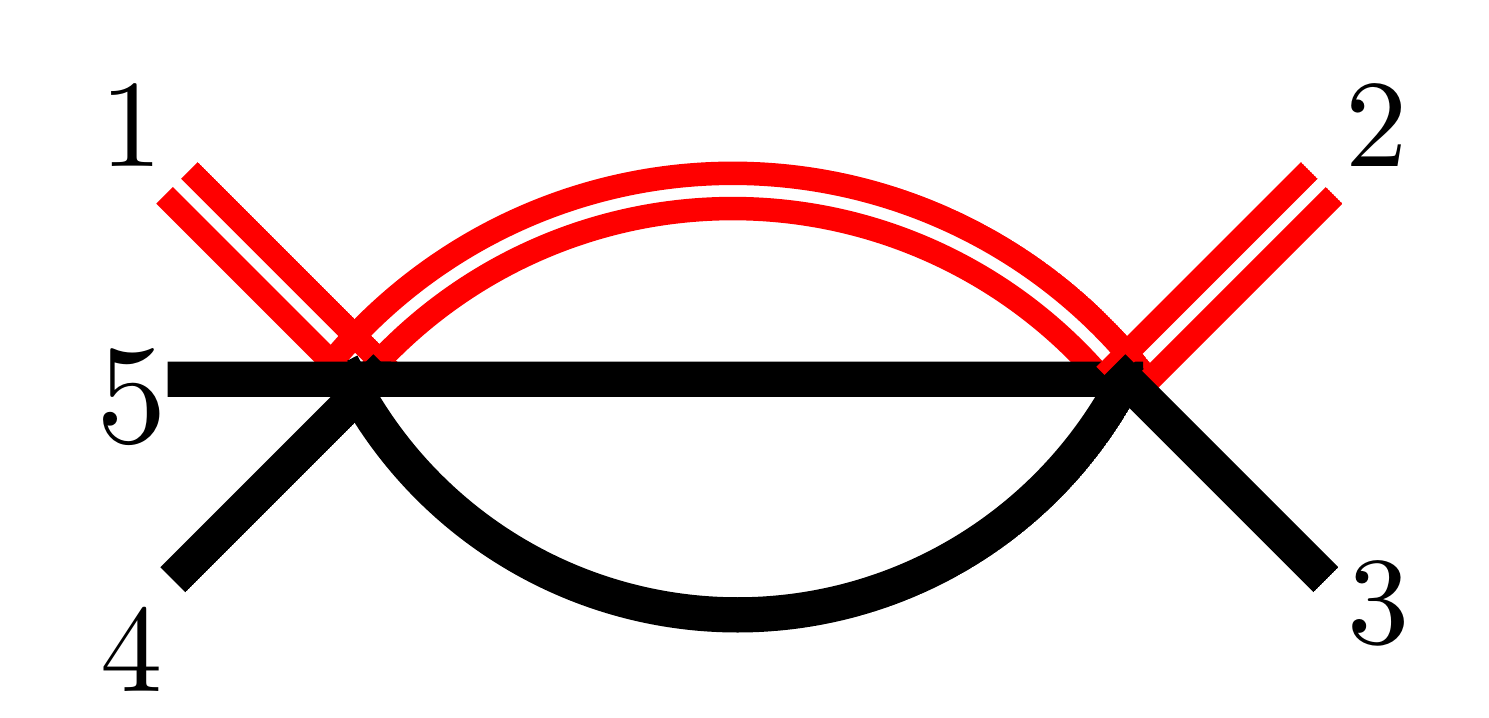}} \quad
\subfloat[$\cI_{85},\cI_{86}$]{\includegraphics[width = 2.4 cm]{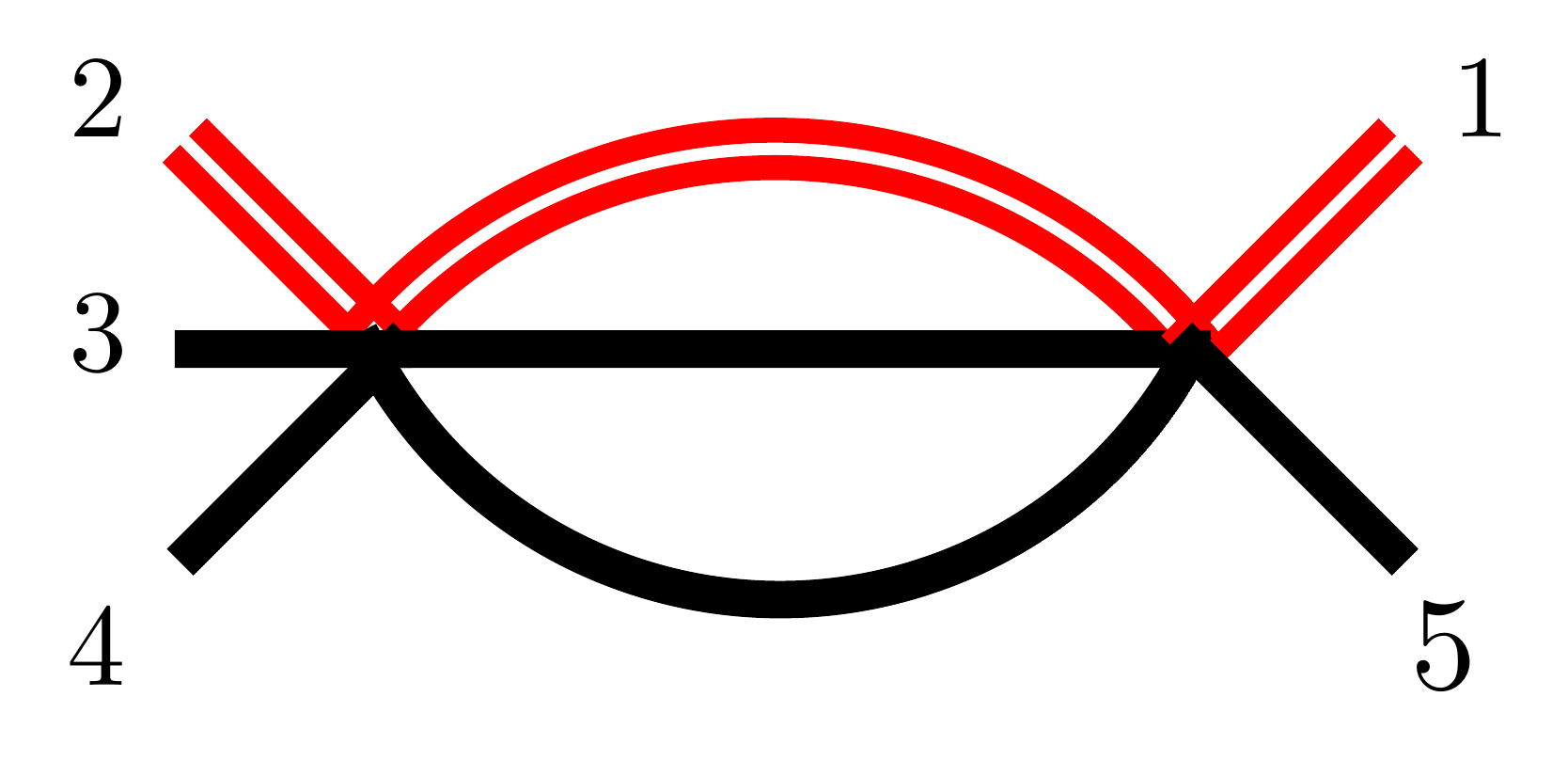}} \quad
\subfloat[$\cI_{87}$]{\includegraphics[width = 2.4 cm]{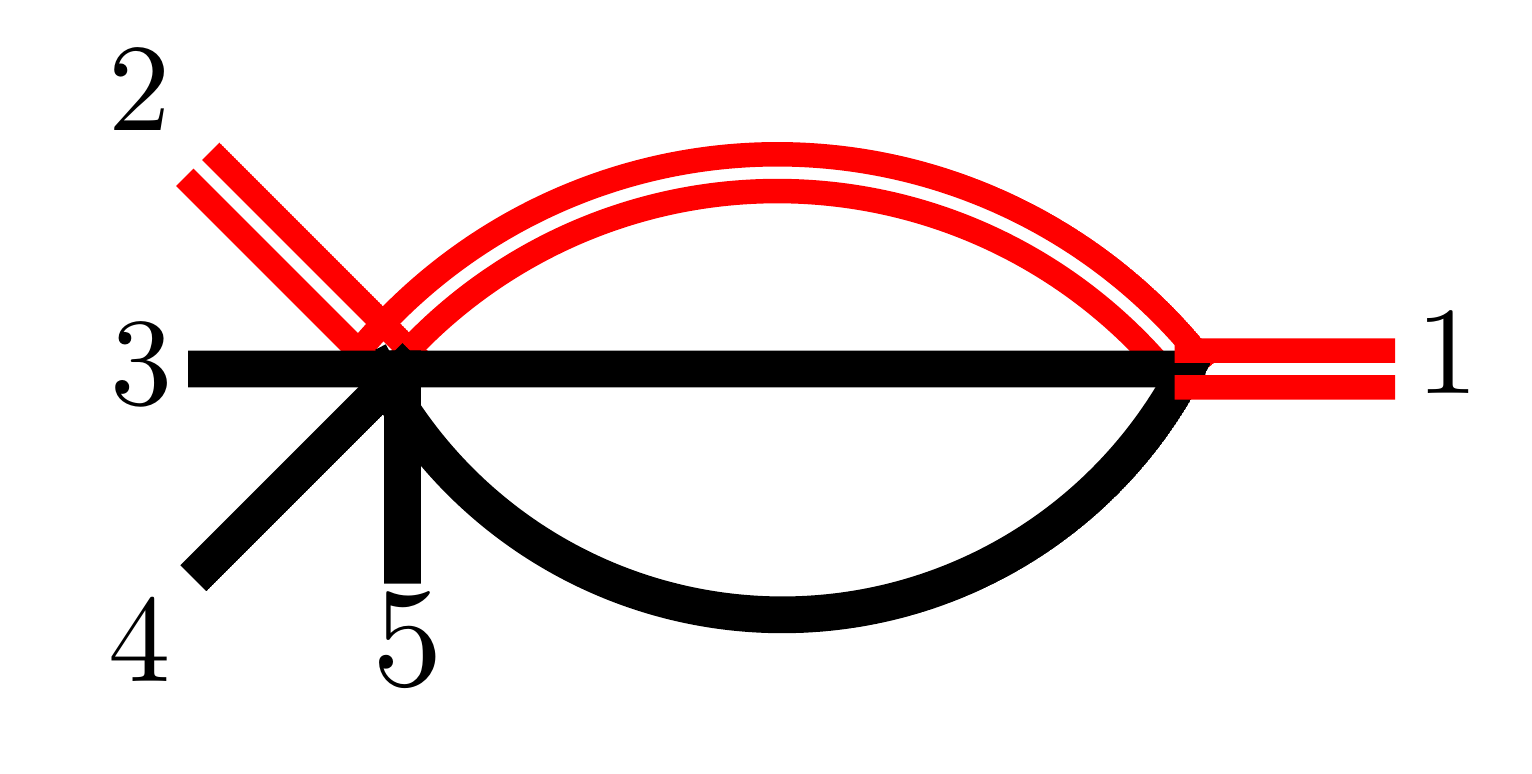}} \quad
\subfloat[$\cI_{88}$]{\includegraphics[width = 2.4 cm]{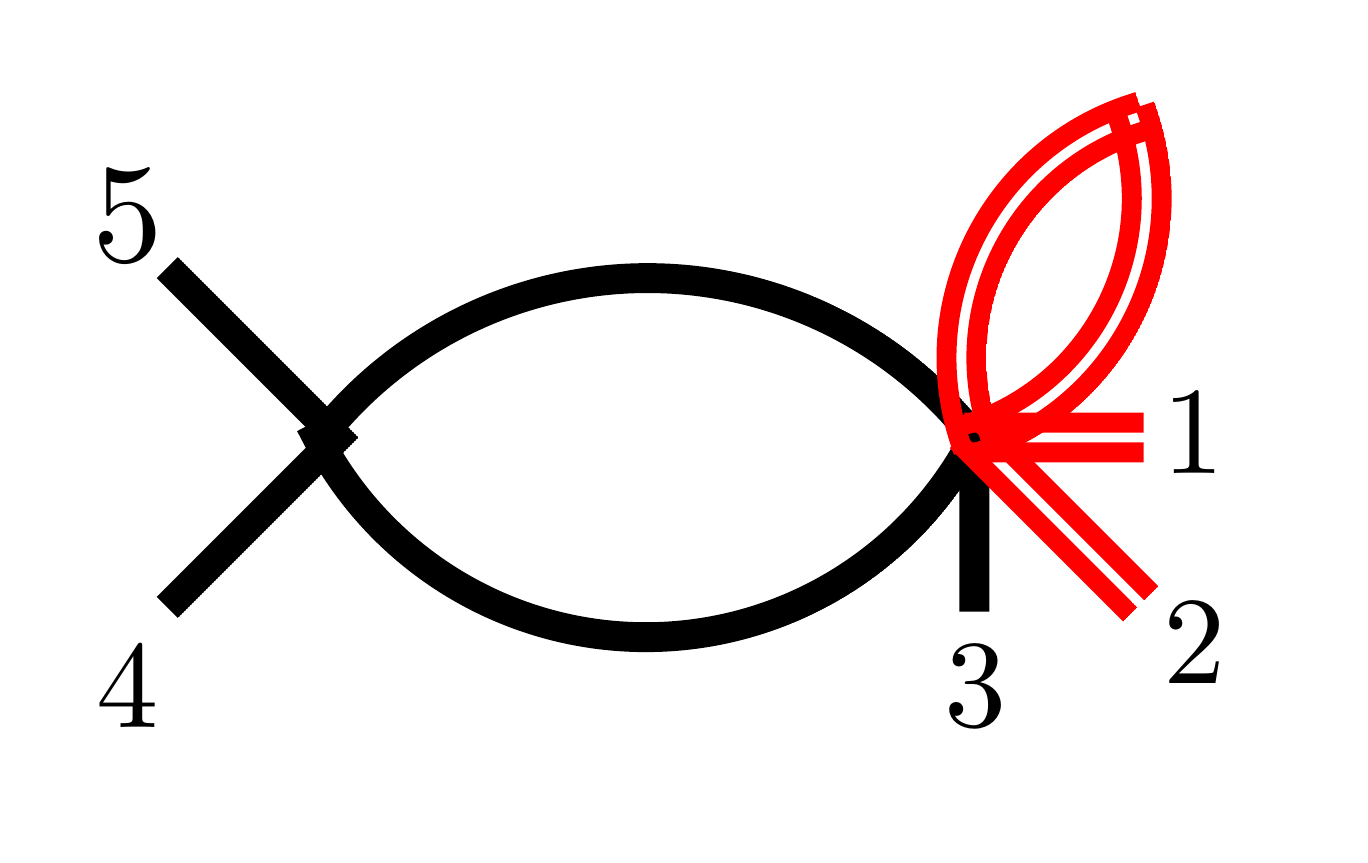}} \quad
\caption{The remaining 23 diagram topologies describing 28 out of 88 master integrals. The label of the individual sub-figures lists the master integrals belonging to the corresponding topology. Massive propagators and massive external momenta are indicated by red double-lines.}
\label{fig:graph_topos2}
\end{figure}

We consider the Feynman integral topology in $d=4-2\eps$ dimensions with eight propagators as shown in figure \ref{fig:t431def}. This can be written as,
\begin{equation}
  I_{a_1,a_2,a_3,a_4,a_5,a_6,a_7,a_8}^{a_9,a_{10},a_{11}} = \int \mathcal{D}^{4-2\epsilon} k_1 \mathcal{D}^{4-2\epsilon} k_2 
  \frac{D_{9}^{a_9} D_{10}^{a_{10}} D_{11}^{a_{11}}}{D_{1}^{a_1}\cdots D_{8}^{a_8}}\,
  \label{eq:t431def}
\end{equation}
where $a_1,\cdots,a_{11} \geq 0$. The propagators, and numerators, are defined as
\begin{align}
  D_1 & =  k_1^2, & D_2 &= (k_1 - p_1)^2 - m_t^2, & D_3 &= (k_1 - p_1 - p_2)^2,  \nonumber \\ 
  D_4 &= (k_1 - p_1 - p_2 - p_3)^2, & D_5 &= k_2^2, & D_6  & =  (k_2 - p_5)^2, \nonumber \\
  D_7 &= (k_2 - p_4 - p_5)^2, & D_8 &= (k_1 +k_2)^2,    &
  D_9 &= (k_1 + p_5)^2, \nonumber \\ \label{eq:props}
  D_{10} &  =   (k_2 + p_1)^2 -m_t^2, &  D_{11} &= (k_2 + p_1 + p_2)^2,& & 
\end{align}
and the integration measure is:
\begin{equation}
  \mathcal{D}^d k_i = \dfrac{d^d k_i}{i \pi^{\frac{d}{2}}} e^{\eps \gamma_E}  \,.
\end{equation}
Momenta are considered outgoing from the graphs and all the particles are on-shell, i.e. $p_1^2 = p_2^2 = m_t^2$ while $p_3^2 = p_4^2 =
p_5^2=0$. The kinematics of the integrals can be described in terms of six
independent invariants. Here we choose the top-quark mass $m_t$ and the five dot products, $\vec{x}=\{d_{12}, d_{23}, d_{34}, d_{45}, d_{15}, m_t^2\}$, where
\begin{equation}
  d_{ij} = p_i \cdot p_j.
  \label{eq:dijdef}
\end{equation}
The minimal set of master integrals (MIs) is obtained by IBP reduction \cite{Chetyrkin:1981qh,Chetyrkin:1979bj}, as implemented in the software \textsc{LiteRed} \cite{Lee:2012cn,Lee:2013mka}
and \textsc{FiniteFlow} \citep{Peraro:2019svx}. We found a total number of 88 MIs which are shown in Fig. \ref{fig:graph_topos1} and \ref{fig:graph_topos2}.

We wish to find a basis of MIs, $\vec{\mathcal{I}}$, which satisfies a system of differential equations in canonical form \cite{Henn:2013pwa}:
\begin{equation} \label{eq:deqsCan}
 d\, \vec{\mathcal{I}}(\vec{x},\eps) = \eps \, d A(\vec{x}) \, \vec{\mathcal{I}}(\vec{x},\eps),
\end{equation}   
where $d$ is the total differential with respect to the kinematic invariants, and the matrix $A(\vec{x})$ is a linear combination of logarithms:
\begin{equation}\label{eq:dematrix}
 A(\vec{x}) = \sum c_i \log (w_i (\vec{x})).
\end{equation}
The $c_i$ are matrices of rational numbers, and the \emph{alphabet} $\left\{w_i (\vec{x})\right\}$ consists of
algebraic functions of the kinematic invariants $\vec{x}$. We discuss the details of the canonical basis of MIs and the alphabet structure
in Sec. \ref{sec:deqs}.

The system of differential equations depends on a set of square roots which we define here for later convenience:
\begin{align}
  \beta & = \sqrt{1-\frac{4 m_t^2}{s_{12}}},  \nn \\
  \Delta_1 &= \sqrt{\operatorname{det}G(p_{23},p_1)},  && \Delta_2 = \sqrt{\operatorname{det}G(p_{15},p_2)}, \nn \\
  \Delta_3 & = \sqrt{1 - \frac{4 s_{45} m_t^2}{(s_{12}+s_{23}-m_t^2)^2}},  && \Delta_4 = \sqrt{1 + \frac{4 s_{34} s_{45} m_t^2}{s_{12} (s_{15}-s_{23})^2}}, \nn \\
  \Delta_5 & = \sqrt{1-\frac{s_{45} m_t^2}{4 d_{15} d_{23}}},  && \Delta_6 = \sqrt{1-\frac{s_{34} s_{45} m_t^2}{4 d_{15} d_{23} s_{12}}}, \nn \\
  \operatorname{tr}_5 & = 4 \sqrt{\operatorname{det}G(p_3,p_4,p_5,p_1)} = {\rm tr}(\gamma_5 \slashed{p}_3 \slashed{p}_4 \slashed{p}_5 \slashed{p}_1), 
  \label{eq:sqrt}
\end{align}
where $G_{ij}(\vec{v}) = v_i\cdot v_j $ is the Gram matrix and $s_{ij} = (p_i+p_j)^2$. The square roots $\Delta_5$ and $\Delta_6$ appear in some intermediate steps of the differential equations reconstruction but they are not related to the normalisation of any master integral. We nevertheless list them here, as some letters of the alphabet can be written in terms of their squared expression and therefore they can be used to match factors appearing in the denominator of the differential equation system.

In order to be able to build a canonical system of differential equations in a rather compact form, our basis of MIs contains integrals with insertions of local numerators \cite{Arkani-Hamed:2010zjl,Arkani-Hamed:2010pyv,Gehrmann:2015bfy,Badger:2016ozq,Abreu:2020jxa}. We will therefore need to extend the notation introduced in Eq.~\eqref{eq:t431def} to allow for insertions of these local numerators into the integrand. For the scope of this paper it will suffice to extend the notation to the local numerators $\mu_{ij}$, which are defined after splitting the loop momenta into four dimensional and $(-2\eps)$ dimensional components,
\begin{align}
  k_i &= k_i^{[4]} + k_i^{[-2\eps]}, & \mu_{ij} &= -k_i^{[-2\epsilon]}\cdot k_j^{[-2\epsilon]}.
\end{align}
Hence, we introduce the minimal extensions
\begin{align}
  I_{a_1,a_2,a_3,a_4,a_5,a_6,a_7,a_8}^{[ij],a_9,a_{10},a_{11}} & = \int \mathcal{D}^{4-2\epsilon} k_1 \,  \mathcal{D}^{4-2\epsilon} k_2 \, \mu_{ij} \,
  \frac{D_{9}^{a_9} D_{10}^{a_{10}} D_{11}^{a_{11}}}{D_{1}^{a_1}\cdots D_{8}^{a_8}}\, ,\nn \\
  I_{a_1,a_2,a_3,a_4,a_5,a_6,a_7,a_8}^{[ij,kl],a_9,a_{10},a_{11}} & = \int \mathcal{D}^{4-2\epsilon} k_1 \,  \mathcal{D}^{4-2\epsilon} k_2 \, \mu_{ij} \, \mu_{kl} \,
  \frac{D_{9}^{a_9} D_{10}^{a_{10}} D_{11}^{a_{11}}}{D_{1}^{a_1}\cdots D_{8}^{a_8}}\,.
\end{align}

\section{Canonical form differential equations and a basis of uniform transcendental weight master integrals} \label{sec:deqs}

In this section we describe the structure of the canonical basis of UT master integrals. The canonical basis approach
\cite{Henn:2013pwa} for systems of differential equations greatly improved the
effectiveness of this method for computing Feynman integrals. As a consequence,
a great effort has been put into developing techniques aimed at identifying a
basis of MIs which satisfy canonical differential equations
\cite{Henn:2013pwa,Gehrmann:2014bfa,Argeri:2014qva,Lee:2014ioa,Lee:2020zfb,Gituliar:2017vzm,Prausa:2017ltv,Dlapa:2020cwj,Dlapa:2021qsl}.
Given the complexity of the kinematics, automated approaches are difficult to
apply in our case yet we find a relatively compact form that demonstrates an
emerging pattern in $2\to3$ scattering problems~\cite{Gehrmann:2015bfy,Papadopoulos:2015jft,Abreu:2018rcw,Chicherin:2018yne,Chicherin:2019xeg,Chicherin:2018old,Abreu:2018aqd,Abreu:2019rpt,Abreu:2020jxa,Abreu:2021smk,Canko:2020ylt}.

Our approach relies on our ability to perform IBP reduction and evaluate the
differential equation matrix over finite fields. This means it is relatively
easy to extract information about the $\eps$ structure of the differential
equations from a univariate slice. Combining this with cuts to identify the
homogeneous parts of each sector means that it is very quick to check whether
particular choices of MIs are suitable. The second important part
of our approach is the availability of a sufficiently good set of potential
choices. Even though we do not attempt to
provide any algorithmic way to generate such a set there is an
increasingly large set of known UT bases for $2\to 3$ scattering problems and
many subtopologies that gives us an excellent starting point. In particular the
existence of known topologies for massless and one-mass five-point~\cite{Gehrmann:2018yef,Chicherin:2018old,Abreu:2020jxa,Abreu:2021smk} (for e.g.
$pp\to W+2j$ and $pp\to 3j$), two-mass four-point for $pp\to Wt$ scattering~\cite{Chen:2021gjv} provide a lot
of information about the subtopologies in our 88 integral system and so only 40
were completely unknown in UT form.

Owing to the large number of square roots appearing in the problem we do not
attempt to construct the canonical form of Eq.~\eqref{eq:deqsCan} directly but
instead search for a form linear in $\eps$ with purely rational matrices.
The square roots appearing in the UT basis can be arranged to be overall
normalisations of individual integrals and can thus be removed for the purposes of
simple finite field evaluations. This approach is explained in reference
\cite{Peraro:2019svx}. Specifically,
\begin{equation}
  d\, \vec{\mathcal{J}}(\vec{x},\eps) = \, d \, \left(\hat{A}^{(0)}(\vec{x}) + \eps \widehat{A}^{(1)}(\vec{x}) \right) \, \vec{\mathcal{J}}(\vec{x},\eps),
  \label{eq:deqsCan2}
\end{equation}   
where
\begin{equation}
  \mathcal{I}_i = N_{ij}(\vec{x}) \mathcal{J}_j
  \label{eq:MISnorm}
\end{equation}
and both of the $88\times 88$ matrices $\widehat{A}^{(0)}$ and $N$ are diagonal. The canonical form differential equation is then easy to obtain via,
\begin{equation}
  d\, \vec{\mathcal{I}}(\vec{x},\eps) = \eps d\left( N(\vec{x})\widehat{A}^{(1)}(\vec{x})N^{-1}(\vec{x}) \right) \, \vec{\mathcal{I}}(\vec{x},\eps)
\end{equation}
after fixing the normalisation through,
\begin{equation}
  \widehat{A}^{(0)} - \frac{1}{2} N^2  d N^{(-2)}  = 0.
\end{equation}
Since the matrix $N$ is diagonal the inverse and square operations are trivial. We write the latter relations using $N^2$ to demonstrate it contains only rational functions.

The set of 88 MIs shown in Fig. \ref{fig:graph_topos1} and
\ref{fig:graph_topos2} are split into genuine two-loop integrals and one-loop
factorisable (one-loop squared) integrals. These integrals are grouped into 52
different sectors of which 6 are of one-loop squared type. We can also
subdivide the two-loop topologies by the number of external legs and we will refer to
the topologies according to the shape of each loop:
\begin{itemize}
\item \textbf{Five-point integrals}: this class contains pentagon-box, pentagon-bubble, double-box and box-triangle topologies;
\item \textbf{Four-point integrals}: this class contains double-box, box-triangle, box-bubble and kite topologies;
\item \textbf{Three-point integrals}: this class contains kite-like and triangle-bubble topologies;
\item \textbf{Two-point integrals}: this class contains just the sunrise topology.
\end{itemize}

The guide for selecting candidate MIs then follows from patterns already observed in previously studied cases and can be justified by considering the leading singularities and local numerator insertions:
\begin{itemize}
  \item In the two-point and three-point class the canonical MI candidates can involve scalar integrals with dotted denominators;
  \item In the four-point class the canonical MI candidates can involve scalar integrals with dotted denominators or the numerators $D_9,D_{10},D_{11}$;
  \item In the five-point class the canonical MI candidates can involve scalar integrals with the numerators $D_9,D_{10},D_{11}$ and local integrand insertions $\mu_{ij}$. 
\end{itemize}
Another important feature in the selection of candidates is to ensure that the
maximum numerator rank and number of dotted propagators is minimised. Including
high rank numerators and large numbers of dotted propagators quickly causes the
number of required IBP relations to explode and requires excessive computational
resources. We therefore build up from a Laporta style minimisation of numerator
rank and dotted denominators and add dots and numerators until each sector has
a homogeneous differential equations (i.e. on the maximal cut of each sector)
of the form of Eq.~\eqref{eq:deqsCan2}. During this
process we can also use the univariate slice in $\eps$ to determine factorised
prefactors that would allow us to rotate the homogeneous differential equation
matrix into the desired form. As a result we can use integrals with fewer dots
and substitute with prefactors depending only on $\eps$.

After checking each homogeneous system, the remaining $\eps$ dependent factors
can be determined from a univariate slice of the full system. After this
procedure we find that some sectors require additional rotations in
sub-sectors. In our case this step was particularly simple and only involved the treatment of $2\times2$ systems, yet it would be
interesting to understand why this is necessary in some cases so a better
selection of candidates could be made. Interestingly, such problems did not
arise in any of the most complicated five-point topologies where the
(extra-dimensional) local numerator insertions worked well.

For the remainder of this section we present explicit forms for all integrals
in the five-point sectors. A complete list of the remaining UT integrals is
given in Appendix \ref{app:utbasis} as well as in computer readable form in the ancillary
files. 

\subsection{Pentagon-box sector}

\begin{figure}[H]
\begin{center}
\includegraphics[width = 0.3\linewidth]{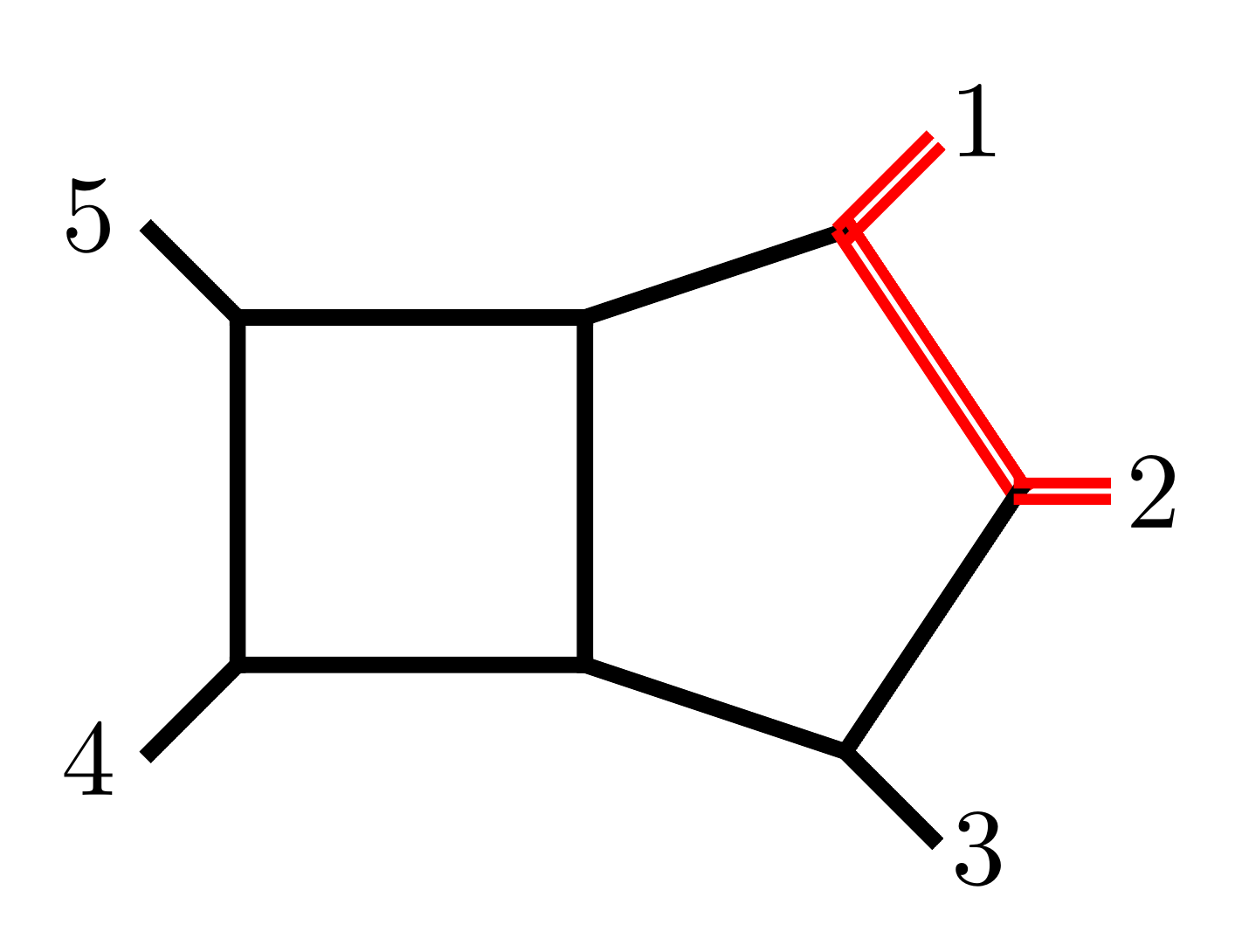}
\end{center}
\caption{The pentagon-box sector with the master integrals $\cI_{1} $, $\cI_{2}$ and $\cI_{3}$.}
\label{fig:pentabox}
\end{figure}

The eight propagator pentagon-box sector shown in figure
\ref{fig:pentabox} contains three MIs. As the topology
with the maximal number of propagators it is particularly important to find a
simple basis choice in order to avoid technical complications with the size of
the IBP system. In particular we find a convenient choice of UT integrals with
a lower tensor rank than in previous five-point bases which simplified the
analytic reconstruction. 

In these massless and one-mass five-point planar cases \cite{Gehrmann:2015bfy,Abreu:2020jxa} a basis of canonical MIs was obtained that involved the following integrals:
\begin{equation}
\left\{
I_{1,1,1,1,1,1,1,1}^{1,0,0},
  I_{1,1,1,1,1,1,1,1}^{[11,22],0,0,0}-I_{1,1,1,1,1,1,1,1}^{[12,12],0,0,0},
  I_{1,1,1,1,1,1,1,1}^{[12],0,0,0}
\right\}.
\end{equation}
The local numerator $\mu_{11}\mu_{22} - \mu_{12}^2$, requires the reduction of
rank 4 numerators which puts a considerable strain on the system of IBP
equations. We find that a different local numerator insertion of rank 2,
\begin{equation}
I_{1,1,1,1,1,1,1,1}^{[11],0,0,0},
\end{equation}
also leads to a UT basis which allows for a simple analytic reconstruction. We note that this choice is also UT for the other five-point configurations mentioned above.

We then find that a canonical basis of MIs for this sector is:
\begin{align}
  \cI_1 &= \epsilon^4 \, 8 \, d_{23} \, d_{45} \left(d_{12}+m_t^2\right) I_{1,1,1,1,1,1,1,1}^{1,0,0},\\   \nonumber 
\cI_2 &= \epsilon^4 \, \frac{d_{45}}{2} \operatorname{tr}_5  I_{1,1,1,1,1,1,1,1}^{[11],0,0,0},\\   \nonumber 
\cI_3 &= \epsilon^4 \, \frac{d_{45}}{2} \operatorname{tr}_5 I_{1,1,1,1,1,1,1,1}^{[12],0,0,0} \,.
\end{align}
One should be aware that this simplification in the rank of the IBP system is
only valid for the differential equation system. Rank five numerators cannot,
at least with the current technology, be avoided in the reduction of the amplitude.
However since the differential equation system requires the reduction of many
more dotted propagators than the amplitude, we may still avoid the need for a
system requiring simultaneous reduction of high ranks and multiple dots.

\subsection{Double-box sectors}

\begin{figure}[H]
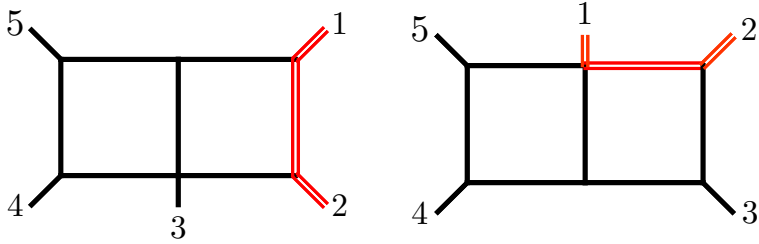

\centering
\subfloat{}{\includegraphics[width = 0.35\linewidth]{figs/diag2.pdf}}
\raisebox{0.1cm}{\subfloat{}{\includegraphics[width = 0.36\linewidth]{figs/diag3.pdf}}}
\caption{The two five-point double-box topologies, containing the canonical MIs $\cI_4, \, \cI_5, \, \cI_6$, $\cI_7$, and $\cI_8, \, \cI_9$, $\cI_{10}$ respectively.}
\label{fig:5ptDoubleBox}
\end{figure}

There are two sectors with a double-box topology, as shown in figure
\ref{fig:5ptDoubleBox}. As for the pentagon-box, a
compact form of the canonical basis for these two sectors can be constructed
using local numerators. Specifically, we choose as canonical MIs for the first sector
in figure \ref{fig:5ptDoubleBox} the set:
\begin{align}
  \cI_4&= \epsilon^4 \, 8\,  d_{15}\, d_{45} \left(d_{12}+m_t^2\right) I_{1,1,1,0,1,1,1,1}^{0,0,0},\\  \nonumber
  \cI_5&= \epsilon^4 \, 4\, \beta \, d_{45} \left(d_{12}+m_t^2\right) I_{1,1,1,0,1,1,1,1}^{1,0,0},\\  \nonumber
  \cI_6&= \epsilon^4 \, \frac{1}{4} \operatorname{tr}_5  I_{1,1,1,0,1,1,1,1}^{[12],0,0,0},\\  \nonumber
  \cI_7&= \epsilon^4 \, \left(d_{12}+m_t^2\right)\left(4 \left(d_{15}-d_{23}\right) I_{1,1,1,0,0,1,1,1}^{0,0,0}+4 d_{45}  I_{1,1,1,0,1,1,1,1}^{0,1,0}\right).
\end{align}
We note that in the massless limit there are only three master integrals in this sector. The fourth integral in this set was identified by a simple analysis on the maximal cut of the sector, and it required a rotation to remove contribution from a sub-sector.
For the second sector in figure  \ref{fig:5ptDoubleBox} we have the following set of canonical MIs:
\begin{align}
  \cI_8&= \epsilon^4 \, 4 \, d_{23}\, d_{34} \, d_{45} \, I_{0,1,1,1,1,1,1,1}^{0,0,0},\\   \nonumber 
  \cI_9&= \epsilon^4 \, 4 \, d_{23} \, d_{45} \, I_{0,1,1,1,1,1,1,1}^{1,0,0},\\   \nonumber 
  \cI_{10}&= \epsilon^4 \, \frac{1}{4} \operatorname{tr}_5  I_{0,1,1,1,1,1,1,1}^{[12],0,0,0}.
\end{align}
These integrals line up precisely with previously considered five-point kinematics.

\subsection{Pentagon-bubble sector}

For the pentagon-bubble sector, differently from the previous cases, we find a choice of canonical basis which involves also a dotted denominator.
\begin{figure}[H]
\begin{center}
\includegraphics[width = 0.3\linewidth]{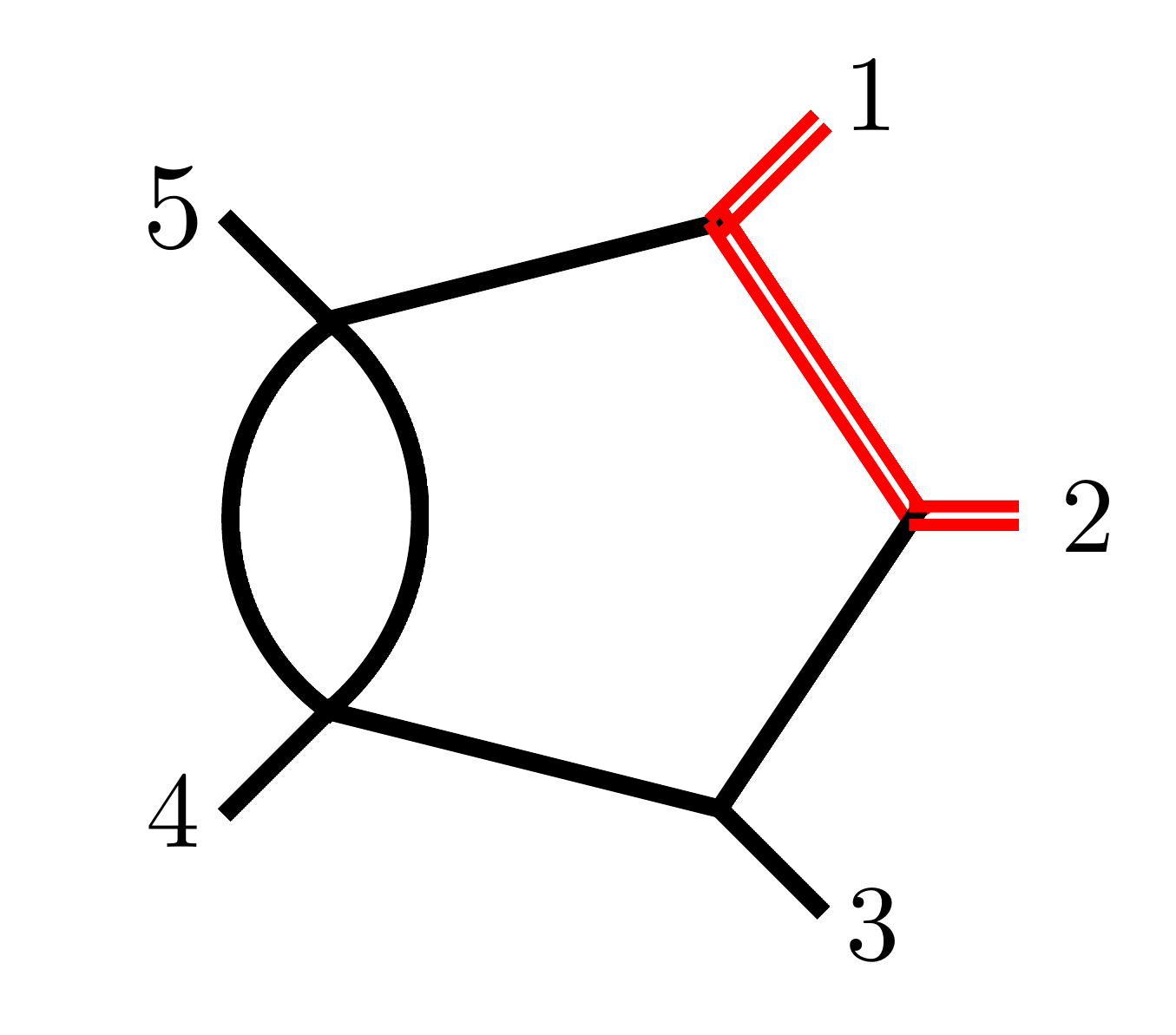}
\end{center}
\caption{The pentagon with a bubble insertion covers the master integrals $\cI_{16} $ and $\cI_{17}$.}
\label{fig:pentabub}
\end{figure}
The dotted denominator corresponds to one of the one-loop bubble propagators. Hence, we define the canonical basis for this sector as follows:
\begin{align}
  \cI_{16}&= \epsilon^3(1- 2 \epsilon)\,4 \, d_{23} \, \left(d_{12}+m_t^2\right) I_{1,1,1,1,0,1,0,1}^{0,0,0},\\   \nonumber 
\cI_{17}&= \epsilon^3\frac{1}{4} \operatorname{tr}_5 I_{1,1,1,1,0,1,0,2}^{[11],0,0,0}.
\end{align}

\subsection{Box-triangle sectors}

\begin{figure}[H]
\centering
\subfloat[]{\includegraphics[width = 0.25\linewidth]{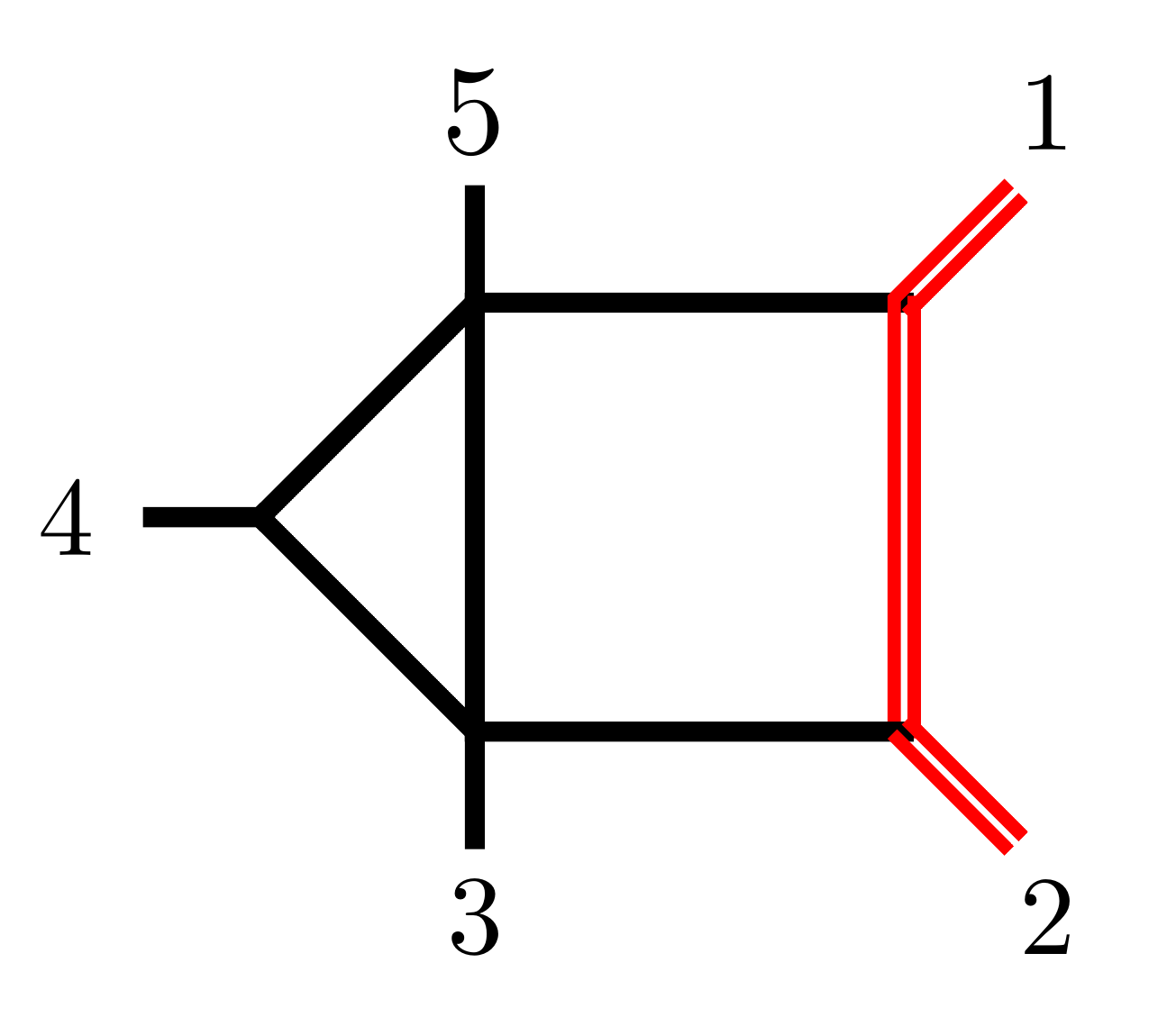}}
\subfloat[]{\includegraphics[width = 0.27\linewidth]{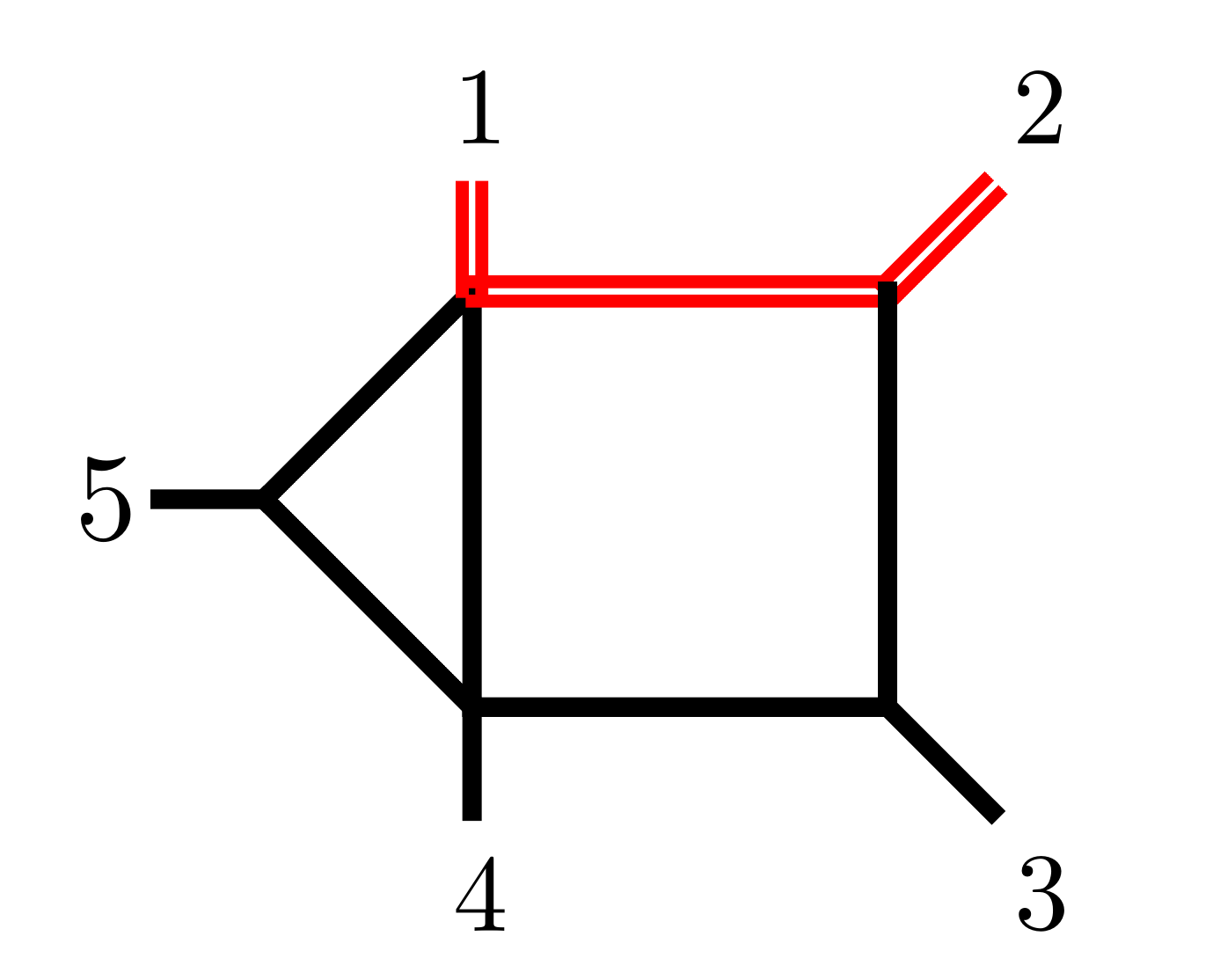}}
\subfloat[]{\includegraphics[width = 0.25\linewidth]{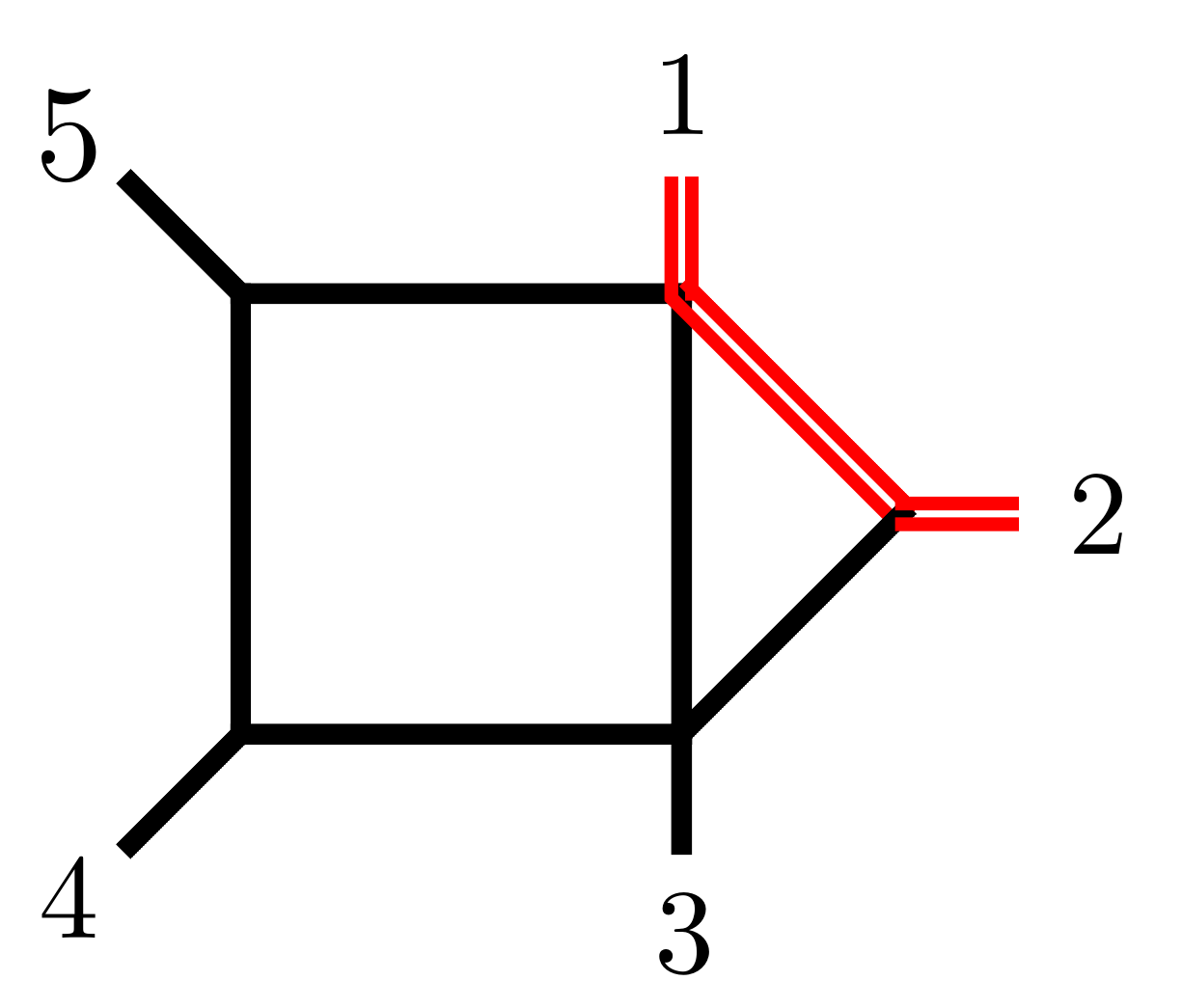}}
\caption{The three genuine five-point box-triangle topologies covering the master integrals $\cI_{18} $ and $\cI_{19}$ (a), $\cI_{20} $ and $\cI_{21}$ (b), and $\cI_{22} $ - $\cI_{26}$ (c), respectively.}
\label{fig:boxtris}
\end{figure}

There are three distinct box-triangle sectors with genuine five-point
kinematics displayed in figure~\ref{fig:boxtris}. Four of the nine master
integrals require the insertion of local numerators.
The explicit form of the canonical MIs in these topologies is given by
\begin{align}
  \cI_{18}&= \epsilon^4  \left(d_{15}-d_{23}\right) \left(d_{12}+m_t^2\right)\, \Delta_4 \, I_{1,1,1,0,0,1,1,1}^{0,0,0},\\   \nonumber 
  \cI_{19}&= \epsilon^3 \frac{1}{4} \operatorname{tr}_5 I_{1,1,1,0,0,1,1,2}^{[11],0,0,0},\\ \nonumber
  \cI_{20}&= \epsilon^4  d_{23} \left(d_{12}-d_{34}+m_t^2\right) I_{0,1,1,1,1,1,0,1}^{0,0,0},\\   \nonumber 
  \cI_{21}&= \epsilon^3 \frac{1}{4} \operatorname{tr}_5 I_{0,1,1,1,1,1,0,2}^{[11],0,0,0}, \\ \nonumber
  \cI_{22}&= \epsilon^4  d_{45}\, \Delta_2 \, I_{0,1,1,0,1,1,1,1}^{0,0,0},\\  \nonumber 
  \cI_{23}&= \epsilon^3 d_{34}\, d_{45} \, m_t^2 \, I_{0,2,1,0,1,1,1,1}^{0,0,0}-\epsilon^4\left(d_{15}-d_{34}\right) d_{45}\, I_{0,1,1,0,1,1,1,1}^{0,0,0},\\  \nonumber 
  \cI_{24}&= \epsilon^3 d_{45}\, m_t^2 \, I_{0,2,1,0,1,1,1,1}^{0,0,1}+ \epsilon^3 d_{34} \, d_{45} \, m_t^2 \, I_{0,2,1,0,1,1,1,1}^{0,0,0} \\ \nonumber 
& - 3\,\epsilon^4  \left(d_{15}-d_{34}\right) d_{45} \, I_{0,1,1,0,1,1,1,1}^{0,0,0},\\  \nonumber 
  \cI_{25}&= \epsilon^3 \frac{1}{4} \operatorname{tr}_5 I_{0,1,1,0,1,1,1,2}^{[12],0,0,0},\\  \nonumber 
  \cI_{26}&= \epsilon^3 \frac{1}{4} \operatorname{tr}_5 I_{0,1,1,0,1,1,1,2}^{[22],0,0,0}.
\end{align}

\subsection{Rational function reconstruction}

Having identified an integral basis in the form of Eq.~\eqref{eq:deqsCan2},
we find the maximal polynomial degree (numerators/denominators) in the
variables $\eps$ and $d_{ij}$ drop from $53/57$ to $15/15$. Since many denominators
align with the one-loop case considered recently \cite{Badger:2022mrb}, matching
factors on a univariate slice also simplifies the final analytic reconstruction which was eventually
achieved in just a couple of hours on a 32 (physical) core workstation.

\section{Analytic structure of the differential equations} \label{sec:dlog}

The reconstructed, $\eps$-factorised form of the DEQ system can be used directly in the generalised series expansion method. However, for a more detailed understanding and the first steps towards constructing a well defined special function basis, we demonstrate that the system can also be written compactly in terms of d-logarithmic forms using an alphabet which is made of 71 letters $w_i$:
\begin{equation}
 d\, \vec{\mathcal{I}}(\vec{x},\eps) = \eps \, d A(\vec{x}) \, \vec{\mathcal{I}}(\vec{x},\eps), \,\,\, A(\vec{x}) = \sum_{i = 1}^{71} c_i \log (w_i (\vec{x})).
\end{equation}
In situations such as these where there are many square roots it can be
difficult to identify the complete alphabet but we find the following a
strategy along the lines of those described in
Refs.~\cite{Heller:2019gkq,Zoia:2021zmb,Chaubey:2022hlr} is sufficient in this
case. We proceed in two steps, first we identify a set of rational letters
(i.e. without square roots). The remaining algebraic letters containing square
roots can then be constructed by examining the denominator structure of a
particular element of the total derivative matrix. It is useful to first
determine the linear relations in the total derivative matrix to minimise the
number of times the strategy must be followed. Given an independent entry of
the derivative matrix one looks for all square roots appearing in the
denominators. One can then construct an ansatz containing free polynomials in
the variables $d_{ij}$ which depends on the number of square roots. If there is
one square root we may try a letter of the form,
\begin{equation}
  \Omega(a,b) := \frac{a + \sqrt{b}}{a - \sqrt{b}},
  \label{eq:lettertemplate1}
\end{equation}
and in the case of two square roots,
\begin{equation}
  \tilde{\Omega}(a,b,c) := \frac{(a + \sqrt{b} + \sqrt{c})(a - \sqrt{b} - \sqrt{c})}{(a + \sqrt{b} - \sqrt{c} )(a -\sqrt{b} + \sqrt{c} )}.
  \label{eq:lettertemplate2}
\end{equation}
Such forms have appeared in numerous of previously studied examples including
five-particle kinematics~\cite{Chicherin:2017dob,Gehrmann:2018yef,Abreu:2020jxa,Abreu:2021smk}.
We note that one can expand the form of Eq. \eqref{eq:lettertemplate2} into one
similar to Eq. \eqref{eq:lettertemplate1} where the single square root is the
product $\sqrt{bc}$. The structure in Eq. \eqref{eq:lettertemplate2} is
preferable as the polynomial degree of the unknown element $a$ is lower as
noted in Ref.~\cite{Abreu:2020jxa}. Using an ansatz for $a$ up to a particular
order it is simple to compute the quantity $d(\log(\Omega))$ and check for a
solution in the unknown numerical coefficients in $a$. Taking the polynomial
factors inside the square roots and the dimensions into account allows a simple
template to be constructed where the polynomial order is kept as low as
possible. We note that if the square root appearing in the letter is ${\rm tr}_5$ we may find another compact representation of the form,
\begin{equation}
  \operatorname{tr}_{\pm}(ij\cdots k) = \frac{1}{2}\operatorname{tr}((1 \pm \gamma_5)\slashed{p}_i\slashed{p}_j \cdots \slashed{p}_k).
  \label{eq:lettertemplate3}
\end{equation}
As before, this follows the structure identified previously in the literature~\cite{Chicherin:2017dob,Abreu:2020jxa,Abreu:2021smk}.

Following this strategy we identify an alphabet for our case in which the rational and algebraic letters can be divided into subsets which we describe in turn. For the rational letters we define,
\begin{equation}
\mathbf{W}_{R} := \mathbf{W}_K \cup \mathbf{W}_T \cup \mathbf{W}_S := \left\{w_1, \cdots , w_{17} \right\}\cup \left\{w_{18},\cdots,w_{25}\right\} \cup \left\{w_{26}, \cdots, w_{33} \right\},
\end{equation}
and for the algebraic letters
\begin{equation}
\mathbf{W}_{A} := \mathbf{W}_{SR-1} \cup \mathbf{W}_{TR} \cup \mathbf{W}_{SR-2} := \left\{w_{34},\cdots,w_{51}\right\}\cup\left\{w_{52}, \cdots, w_{60} \right\}\cup\left\{w_{61}, \cdots, w_{71} \right\}.
\end{equation}
The rational set of letters $\mathbf{W}_R$ is made of linear combinations of the kinematic invariants. However, we can identify three different kind of subsets in $\mathbf{W}_R$. The subset $\mathbf{W}_K$ can be written in terms of the Mandelstam variables $s_{ij} = (p_i + p_j)^2$ and it is defined as:
\begin{align} \label{eq:alp_rat}
\mathbf{W}_K := & \left\{ m_t^2,\, s_{12},\, s_{23}, \, s_{34}, \, s_{45}, \, s_{15}, \, s_{35}, \,s_{23} -m_t^2 , \, s_{14} -m_t^2, \, s_{15} -m_t^2, \, s_{24} -m_t^2, \, s_{25} -m_t^2, \right. \nn \\
& \left. s_{12} -s_{34}, \, s_{12} - s_{45}, \, s_{12} - s_{35}, \, s_{23} - s_{15}, \, s_{23} - s_{14}\right\}.
\end{align}
The subset $\mathbf{W}_T$ consists of letter that can be written as traces over $\gamma$-matrices. Defining,
\begin{align}
  \operatorname{tr}(ij\cdots k) &= \operatorname{tr}(\slashed{p}_i\slashed{p}_j \cdots \slashed{p}_k),
\end{align}
we can then write the 8 letters as,
\begin{align}
\mathbf{W}_T := & \left\{
\operatorname{tr}(4151), \,
\operatorname{tr}(4232), \,
\operatorname{tr}(5242), \,
\operatorname{tr}(3252), \,
\operatorname{tr}(32[1+2]4[1+2]2), \,
\operatorname{tr}(312312)\right. \nn \\
& \left. \operatorname{tr}(412412), \,
\operatorname{tr}(512512) \right\}.
\end{align}
Finally, the rational letters that belong to the third subset, $\mathbf{W}_S$, can be related to the roots defined in Eq.~\eqref{eq:sqrt}:
\begin{align}
\mathbf{W}_S := & \left\{\beta^2, \, (\Delta_1)^2, \, (\Delta_2)^2, \,4 (d_{12}+d_{23}+m_t^2)^2 (\Delta_3)^2, \, (\Delta_5)^2, \, (\Delta_4)^2, \, (\Delta_6)^2, \, \operatorname{tr}_5^2\right\}.
\end{align}
We identity three different classes of algebraic letters which
involve square roots of the kinematic invariants.  The first class,
$\mathbf{W}_{SR-1}$, is made by letters in terms of $\Omega$ as defined above in Eq. \eqref{eq:lettertemplate1},
\begin{align}
\mathbf{W}_{SR-1} := & \left\{\Omega(1,\beta^2), \, \Omega\left(1 + \frac{m_t^2 (d_{12}-d_{34}+m_t^2)}{d_{15} (d_{12}+m_t^2)}, \, \beta^2 \right), \right. \nn \\
& \left. \Omega\left(\frac{d_{12} d_{23}+d_{12} m_t^2+d_{23} m_t^2-d_{45} m_t^2+m_t^4}{d_{23} (d_{12}+m_t^2)}, \, \beta^2 \right), \right. \nn \\
& \left. \Omega\left(\frac{d_{12} d_{15}-d_{12} d_{23}+d_{12} d_{45}+d_{15} m_t^2-d_{23} m_t^2-d_{34} m_t^2}{(d_{12}+m_t^2) (d_{15}-d_{23}+d_{45})}, \, \beta^2 \right), \, \Omega\left(d_{23} - d_{45}, \, (\Delta_1)^2\right), \right. \nn \\
& \left. \Omega\left(d_{23} - 2d_{15} -d_{45}, \, (\Delta_1)^2 \right), \, \Omega\left(\frac{d_{12} (2 d_{23}+m_t^2)+(d_{23}+m_t^2) (d_{23}-d_{45}+m_t^2)}{d_{23}}, \, (\Delta_1)^2 \right), \right. \nn \\
& \left. \Omega\left(\frac{d_{23}^2-d_{23} d_{45}-d_{45} m_t^2}{d_{23}}, \, (\Delta_1)^2\right), \, \Omega\left(d_{15}-d_{34}, \, (\Delta_2)^2\right), \, \Omega\left(d_{15}-2d_{23}-d_{34}, \, (\Delta_2)^2\right), \right. \nn \\
& \left. \Omega\left(\frac{d_{12} (2 d_{15}+m_t^2)+(d_{15}+m_t^2) (d_{15}-d_{34}+m_t^2)}{d_{15}}, (\Delta_2)^2 \right), \right. \nn \\
& \left. \Omega\left(\frac{d_{15}^2-d_{15} d_{34}-d_{34} m_t^2}{d_{15}}, \, (\Delta_2)^2 \right), \, \Omega\left(1, (\Delta_3)^2 \right), \, \Omega\left(1 + \frac{2 (d_{15}-d_{23}-d_{34})}{d_{12}+d_{23}+m_t^2}, \, (\Delta_3)^2 \right), \right. \nn \\
& \left. \Omega\left(\frac{d_{12} d_{23}+d_{12} m_t^2+d_{23}^2+d_{23} m_t^2-d_{45} m_t^2+m_t^4}{d_{23} (d_{12}+d_{23}+m_t^2)}, \, (\Delta_3)^2 \right), \, \Omega\left(1, \, (\Delta_4)^2 \right), \right. \nn \\
& \left. \Omega\left(\frac{d_{15}+d_{23}}{d_{15}-d_{23}}, \, (\Delta_4)^2 \right), \, \Omega\left(\frac{d_{15} d_{34}-d_{15} d_{45}-d_{23} d_{34}+d_{23} d_{45}+2 d_{34} d_{45}}{(d_{15}-d_{23}) (d_{34}+d_{45})}, \, (\Delta_4)^2 \right)\right\}.
\end{align}
The letters associated to the class $\mathbf{W}_{TR}$, contain dependence $\gamma_5$ and are of the form defined above in Eq. \eqref{eq:lettertemplate3},
\begin{align}
\mathbf{W}_{TR} := & \left\{\frac{\operatorname{tr}_{+}(5241)}{\operatorname{tr}_{-}(5241)}, \, \frac{\operatorname{tr}_{+}(35[1+2]2)}{\operatorname{tr}_{-}(35[1+2]2)}, \, \frac{\operatorname{tr}_{+}(34[1+2]2)}{\operatorname{tr}_{-}(34[1+2]2)}, \, \frac{\operatorname{tr}_{-}(341542)}{\operatorname{tr}_{+}(341542)}, \right. \nn \\
& \left. \frac{\operatorname{tr}_{+}(5142[1+2]4)}{\operatorname{tr}_{-}(5142[1+2]4)}, \, \frac{\operatorname{tr}_{+}(3423[1+2]1)}{\operatorname{tr}_{-}(3423[1+2]1)}, \, \frac{\operatorname{tr}_{+}(5232[1+2]4)}{\operatorname{tr}_{-}(5232[1+2]4)}, \, \frac{\operatorname{tr}_{+}(5143[1+2]1)}{\operatorname{tr}_{-}(5143[1+2]1)}, \right. \nn \\
& \left. \frac{\operatorname{tr}_{+}(4151[1+2]5)}{\operatorname{tr}_{-}(4151[1+2]5)}\right\}.
\end{align}
The final class, $\mathbf{W}_{SR-2}$, is made by letters in terms of $\tilde{\Omega}$ as defined above in Eq. \eqref{eq:lettertemplate2},
\begin{align}
\mathbf{W}_{SR-2} := & \left\{\tilde{\Omega}\left(d_{12}+d_{23}-d_{45}+m_t^2, \, (\Delta_1)^2, \, (d_{12}+m_t^2)^2 \beta^2 \right), \right. \nn \\
& \left. \tilde{\Omega}\left(d_{12}+d_{15}-d_{34}+m_t^2, \, (\Delta_2)^2, \, (d_{12}+m_t^2)^2 \beta^2 \right), \right. \nn \\
& \left. \tilde{\Omega}\left(d_{23},\, (\Delta_3)^2 (d_{12}+d_{23}+m_t^2)^2, \, (d_{12}+m_t^2)^2 \beta^2 \right), \right. \nn \\
& \left. \tilde{\Omega}\left(d_{12}-d_{45}+m_t^2, \, (\Delta_3)^2 (d_{12}+d_{23}+m_t^2)^2, \, (\Delta_1)^2 \right), \right. \nn \\
& \left. \tilde{\Omega}\left(-((d_{12}+m_t^2) (d_{15}-d_{23}+d_{45})), \, (\Delta_4)^2 (d_{12}+m_t^2)^2 (d_{15}-d_{23})^2,\, \beta^2 \, d_{45}^2 (d_{12}+m_t^2)^2 \right), \right. \nn \\
& \left. \tilde{\Omega}\left(d_{12} d_{15}-d_{12} d_{23}-d_{15} d_{45}+d_{15} m_t^2-d_{23} m_t^2, \, d_{34}^2 \, (\Delta_1)^2 , \, \frac{\operatorname{tr}_5^2}{16}\right), \right. \nn \\
& \left. \tilde{\Omega}\left(d_{12} d_{15}-d_{12} d_{23}+d_{15} m_t^2+d_{23} d_{34}-d_{23} m_t^2, \, d_{45}^2 \, (\Delta_2)^2 , \, \frac{\operatorname{tr}^2_5}{16}\right), \right. \nn \\
& \left. \tilde{\Omega}\left(d_{12}^2+d_{12} (d_{15}+d_{23}-d_{34}-d_{45}+2 m_t^2)-(d_{45}-m_t^2) (d_{15}-d_{34}+m_t^2) \right. \right. \nn \\
& \left.\left.+d_{23} (m_t^2-d_{34}), \, (d_{12}+m_t^2)^2 (d_{12}-d_{34}-d_{45}+m_t^2)^2 \beta^2 , \, \frac{\operatorname{tr}^2_5}{16}\right), \right. \nn \\
& \left. \tilde{\Omega}\left(d_{12} (d_{15}-d_{23}-d_{34})-(d_{15}-d_{34}) (d_{45}-m_t^2)-d_{23} (d_{34}+m_t^2),\, d_{34}^2 (d_{12}+m_t^2)^2 \beta^2 , \, \frac{\operatorname{tr}^2_5}{16}\right), \right. \nn \\
& \left. \tilde{\Omega}\left(d_{12} (d_{15}-d_{23}-d_{34})+d_{15} (m_t^2-d_{45})-m_t^2 (d_{23}+d_{34})+d_{34} d_{45}, \right.\right. \nn \\
& \left.\left. d_{34}^2 (\Delta_3)^2 (d_{12}+d_{23}+m_t^2)^2 , \, \frac{\operatorname{tr}^2_5}{16}\right), \right. \nn \\
& \left. \tilde{\Omega}\left(d_{15} d_{45}+d_{23} d_{34}-d_{34} d_{45}, \, (\Delta_4)^2 (d_{12}+m_t^2)^2 (d_{15}-d_{23})^2, \, \frac{\operatorname{tr}^2_5}{16}\right) \right\}.
\end{align}
We observe encouraging patterns between these letters and those observed in other five-particle kinematic configurations which suggest a general alphabet for all polylogarihmic two-loop integrals with five legs or fewer can be described with similar letters.

\subsection{Symbol level structure}

While a completely analytic solution for the master integrals is beyond the
scope of this article, using the weight zero terms from the boundary values we
are able to construct the symbol of the master integrals~\cite{Goncharov:2010jf,Duhr:2011zq} by iteratively expanding the
canonical form differential equation in $\eps$,
\begin{equation}
\vec{\cI}(\vec{x},\epsilon) = \sum_k \epsilon^k \, \vec{\cI}^{(k)}(\vec{x}).
\end{equation} 
At each order the result is obtained by integrating over the previous one:
\begin{equation} \label{eq:k_order_sol}
\vec{\cI}^{(k)}(\vec{x}) = \int \sum_i c_i \, d \log(w_i(\vec{x})) \, \vec{\cI}^{(k-1)}(\vec{x}),
\end{equation}
where at weight $0$, $\vec{\cI}^{(0)}$ is just the vector of boundary
conditions and it is made of rational numbers. For the system of MIs under
study $\vec{\cI}^{(0)}$ has the following form:
\begin{align}
\vec{\cI}^{(0)} = & \left\{\frac{5}{6},0,0,\frac{5}{24},0,0,\frac{1}{6},\frac{19}{24},\frac{5}{6},0,-1,0,\frac{11}{24},0,\frac{5}{12},\frac{1}{2},0,0,0,0,0,0,\frac{1}{6},\frac{5}{12},0,0,0,0,0,0,0,0,0,-\frac{1}{6}, \right. \nn \\
& \left.-\frac{1}{6},0,\frac{1}{6},0,0,0,-1,0,0,0,0,0,\frac{1}{6},0,0,-\frac{1}{6},\frac{1}{2},\frac{1}{2},0,-\frac{1}{6},0,1,0,0,\frac{1}{6},1,\frac{1}{2},0,0,0,\frac{1}{2},\frac{1}{2},\frac{1}{2},\frac{1}{2},\frac{1}{2}, \right. \nn \\
& \left. 0,0,\frac{1}{4},0,0,0,0,1,1,0,-\frac{1}{2},-\frac{1}{2},-\frac{1}{2},0,-\frac{1}{2},0,-\frac{1}{2},1,1\right\}.
\end{align}
By iterating the expression in Eq. \eqref{eq:k_order_sol} we can write $\vec{\cI}^{(k)}(\vec{x})$ as:
\begin{equation} \label{eq:k_order_sol2}
\vec{\cI}^{(k)}(\vec{x}) = \sum_{i_1, \cdots , i_k} e_{i_1, \cdots , i_k} \int d \log(w_{i_1}(\vec{x})) \cdots d \log(w_{i_k}(\vec{x})),
\end{equation}
where $e_{i_1, \cdots , i_k}$ are given by products of the matrices $c_i$ in
Eq. \eqref{eq:dematrix}. The expression in Eq. \eqref{eq:k_order_sol2} is not
enough to obtain an analytic expression for the MIs, however, it contains
analytic information at the integrand level which is encoded in the symbol
definition \cite{Goncharov:2010jf,Duhr:2011zq}:
\begin{equation} \label{eq:symbol}
\mathcal{S}\left[ \vec{\cI}^{(k)}(\vec{x}) \right] = \sum_{i_1, \cdots , i_k} e_{i_1, \cdots , i_k} \left[ w_{i_1}(\vec{x}), \cdots , w_{i_k}(\vec{x}) \right].
\end{equation}
Using the information provided in the ancillary files together with the
descriptions in the literature~\cite{Goncharov:2010jf,Duhr:2011zq,Duhr:2012fh}
and some help from the \textsc{PolyLogTools} package~\cite{Duhr:2019tlz}, it is
straightforward to construct explicitly the symbol of the master integrals.
This symbol level expression can be used to perform a useful consistency check
on our results since it carries information about the discontinuities of the
Feynman integrals. The so-called \emph{first entry condition}~\cite{Gaiotto:2011dt} states that $e_{i_1, \cdots , i_k} = 0$ if the first
entry, $w_{i_1}(\vec{x})$, in the symbol \eqref{eq:symbol} does not correspond
to a physical channel of the topology. Checking this condition for our integrals can be simply
stated as expanding the symbol level expression to weight one (logarithmic
terms only) and checking that the only discontinuities appear in the invariants,
\begin{equation}
\mathcal{T} = \left\{s_{12}, \, s_{23} - m_t^2, \, s_{34}, \, s_{45}, \, s_{15} - m_t^2\right\},
\end{equation}
which we have confirmed to be true.

\section{Numerical solution of the differential equations} \label{sec:num}

As a proof of concept of our work, we discuss in this section a numerical solution 
for the system of differential equations associated to the master integrals. 
The  system has been integrated semi-analytically exploiting the generalised
power series expansion method \cite{Francesco:2019yqt}, as implemented in the
package \textsc{DiffExp} \cite{Hidding:2020ytt}. Since we are interested in a
numerical evaluation of the master integrals, we integrated the system using
high-precision numerical boundary conditions. This evaluation has been done exploiting the
auxiliary mass flow method  \cite{Liu:2017jxz,Liu:2021wks,Liu:2022tji}, by means of the package
\textsc{AMFlow} \cite{Liu:2022chg}. The boundary values are evaluated at the rational 
point chosen arbitrarily in the Euclidean region:
\begin{equation} \label{eq:bound_point}
\vec{x}_0 := \left\{ -\frac{2}{17}, -\frac{17}{13}, -\frac{19}{7}, -\frac{23}{5}, -\frac{11}{3}, 1\right\},
\end{equation}
with a precision of $O(100)$ digits.
All the relevant material for the numerical evaluation is given in the ancillary files:
\begin{itemize}
\item \texttt{anc/DiffExp/boundary\_value.m}: a set of numerical boundary conditions;
\item \texttt{anc/DiffExp/DEQs/d\_1.m}: the dlog matrix in the \textsc{DiffExp} format;
\item \texttt{anc/DiffExp/analytic\_continuation.m}: the list of polynomials needed for the analytic continuation;
\item \texttt{anc/DiffExp/DIFFEXP\_run.wl}: a \textsc{Mathematica} file for the numerical evaluation of the MIs with \textsc{DiffExp};
\item \texttt{anc/boundary/run.wl}: an \textsc{AMFlow} script to generate high-precision boundary conditions.
\end{itemize}

Our numerical tests with \textsc{DiffExp} have not been optimised for a
realistic phase-space integration required by phenomenological studies. As a
result it is not possible to quote any sensible analysis of the evaluation
times since in our tests, all benchmark points were transported from the same
Euclidean boundary point. While this was useful to establish that the analytic
continuation was performed correctly a different strategy would likely be
beneficial during the evaluation of multiple points. It has been shown for
other processes that iterating in short steps around an initial high precision
grid of evaluations can lead to a highly efficient implementation suitable for
phase-space integration~\cite{Becchetti:2020wof,Abreu:2020jxa,Bonciani:2021zzf,Bonciani:2022jmb,Armadillo:2022bgm,Abreu:2021smk}.
High precision boundary terms valid in a particular phase-space region can also
easily computed using auxiliary mass flow method if required.

\subsection{Benchmark points}

We now give some benchmark points for the pentagon-box MIs $\cI_{1}$ and $\cI_{2}$. 
Interestingly, the third master integral in this sector, $\cI_3$, is
zero up to and including weight 4 for all the points that we studied. 

We consider benchmark points for the physical phase-space region in the scattering channel $45 \rightarrow 123$:
\begin{align}
\mathcal{R} := & \left\{p_1^2 > 0, \, p_2^2 >0, \, d_{12} >0, \, d_{15} < 0, \, d_{23} >0, \, d_{34} < 0, \, d_{45} > 0, \, \operatorname{tr}_5^2 < 0  \right\}.
\end{align}
In particular we consider the following five points:
\begin{align} \label{eq:phys_p}
\vec{x}_1 = & \left\{\frac{13}{80},\frac{19}{200},-\frac{11}{80},\frac{1}{2},-\frac{81}{400},\frac{1}{16}\right\}, \nn \\
\vec{x}_2 = &  \left\{\frac{107}{400},\frac{7}{200},-\frac{17}{200},\frac{1}{2},-\frac{93}{400},\frac{1}{16}\right\}, \nn \\
\vec{x}_3 = & \left\{\frac{91}{400},\frac{23}{200},-\frac{21}{200},\frac{1}{2},-\frac{77}{400},\frac{1}{16}\right\}, \nn \\
\vec{x}_4 = & \left\{\frac{271}{400},\frac{259}{200},-\frac{222}{25},\frac{37}{2},-\frac{3441}{400},\frac{1}{16}\right\}, \nn \\
\vec{x}_5 = & \left\{\frac{271}{400},\frac{1221}{200},-\frac{222}{25},\frac{37}{2},-\frac{2479}{400},\frac{1}{16}\right\}.
\end{align}

\begin{center}
\renewcommand{\arraystretch}{1.5}
\begin{tabular}{| m{1cm} || m{2.3cm} | m{2.3cm} | m{2.3cm} | m{2.3cm} | m{2.3cm} |} 
 \hline
  & $\vec{x}_1$ & $\vec{x}_2$ & $\vec{x}_3$ & $\vec{x}_4$ & $\vec{x}_5$ \\[5pt]
 \hline\hline
 $\cI_1^{(0)}$ & $\frac{5}{6}$ & $\frac{5}{6}$ & $\frac{5}{6}$ & $\frac{5}{6}$ & $\frac{5}{6}$ \\[5pt]
 \hline
 $\cI_1^{(1)}$ & $2.1892384+4.1887902 i$ & $3.7462547+4.1887902 i$ & $2.0349747+4.1887902 i$ & $-3.8483012+4.1887902 i$ & $-5.9157644+4.1887902 i$ \\[5pt]
 \hline
 $\cI_1^{(2)}$ & $-4.0886316+9.4351407 i$ & $0.601470+16.615964 i$ & $-4.9774769+10.3252137 i$ & $-2.102532-29.186022 i$ & $9.928524-35.681149 i$ \\[5pt]
 \hline
 $\cI_1^{(3)}$ & $-6.9367835+6.1424776 i$ & $-11.982563+29.534555 i$ & $-21.690194+10.540708 i$ & $-89.442855+18.056883 i$ & $58.305031+71.732816 i$ \\[5pt]
 \hline
 $\cI_1^{(4)}$ & $-51.557014+40.311095 i$ & $-50.707105+81.832621 i$ & $-141.376078+1.757813 i$ & $-51.44856+237.86399 i$ & $-277.01306+85.51492 i$ \\[5pt]
 \hline
\end{tabular}
\captionof{table}{Benchmark points for the pentagon-box master integrals $\cI_1$. $\cI_1^{(k)}$ indicates the $k$-th order term in the $\epsilon$-expansion of the integral.}
\end{center}

\begin{center}
\renewcommand{\arraystretch}{1.5}
\begin{tabular}{| m{1cm} || m{2.3cm} | m{2.3cm} | m{2.3cm} | m{2.3cm} | m{2.3cm} |} 
 \hline
  & $\vec{x}_1$ & $\vec{x}_2$ & $\vec{x}_3$ & $\vec{x}_4$ & $\vec{x}_5$ \\[5pt]
 \hline\hline
 $\cI_2^{(0)}$ & 0 & 0 & 0 & 0 & 0 \\[5pt]
 \hline
 $\cI_2^{(1)}$ & 0 & 0 & 0 & 0 & 0 \\[5pt]
 \hline
 $\cI_2^{(2)}$ & 0 & 0 & 0 & 0 & 0 \\[5pt]
 \hline
 $\cI_2^{(3)}$ & $0.15787753-0.49701005 i$ & $0.09544126-0.39795332 i$ & $0.23166742-0.52220052 i$ & $0.03401419-0.28601824 i$ & $0.16100404-0.57235050 i$ \\[5pt]
 \hline
 $\cI_2^{(4)}$ & $1.1713578-2.2750822 i$ & $0.8565234-1.9943250 i$ & $1.6259689-2.5557664 i$ & $0.00744603+1.08835475 i$ & $-0.2359265+1.8438365 i$ \\[5pt]
 \hline
\end{tabular}
\captionof{table}{Benchmark points for the pentagon-box master integrals $\cI_2$. $\cI_2^{(k)}$ indicates the $k$-th order term in the $\epsilon$-expansion of the integral.}
\end{center}

\subsection{Numerical checks}

We briefly comment on the numerical checks that we performed in order to
validate our results. The numerical checks have been done by comparing the numerical
results, obtained with \textsc{DiffExp}, with respect to a fully numerical
evaluations performed with \textsc{AMFlow}. We made checks for several values
of the kinematic invariants and we found full agreement between the two
methods.

\subsection{Remark on square roots numerical evaluation}

We finish this section with a comment about the square root implementation within our \textsc{DiffExp} setup. 
In order to be able to run \textsc{DiffExp}, the differential equations file has to
contain only irreducible square roots. As it can be seen from Eq. \eqref{eq:sqrt} the square roots $\Delta_3$
and $\Delta_4$ contain a perfect square at denominator, hence they are not irreducible. Consequently a replacement 
rule has to be applied within the \textsc{DiffExp} setup. Specifically, we made the following replacement in generating the
differential equations file:
\begin{align} \label{eq:roots_rep}
\Delta_3 \rightarrow & \operatorname{sign}(d_{12} + d_{23} + m_t^2) \frac{\sqrt{2 \left(d_{12}+d_{23}-d_{45}\right) m_t^2+\left(d_{12}+d_{23}\right){}^2+m_t^4}}{d_{12} + d_{23} + m_t^2}, \nn \\
\Delta_4 \rightarrow & \operatorname{sign}(d_{15} - d_{23}) \frac{\sqrt{\left(\left(d_{15}-d_{23}\right){}^2+2 d_{34} d_{45}\right) m_t^2+d_{12} \left(d_{15}-d_{23}\right){}^2}}{\sqrt{d_{12} + m_t^2}(d_{15} - d_{23})}.
\end{align}
The sign in Eq. \eqref{eq:roots_rep} depends on the boundary point that is used 
within \textsc{DiffExp}. As an example, for the setup that is given in the ancillary 
\textsc{DiffExp} files the sign is negative both for $\Delta_3$ and $\Delta_4$, because
we are using the boundary point $\vec{x}_0$ in Eq. \eqref{eq:bound_point}. Moreover, the square roots $\Delta_3$
and $\Delta_4$ appear as normalisation factors in the definition of the UT basis for the MIs 18 and 31.
Therefore, in order to have a consistent numerical evaluation of these MIs, the sign prefactors in Eq. \eqref{eq:roots_rep}
have to be kept into account when generating new sets of boundary conditions. 

\section{Conclusions} \label{sec:conclusion}

In this article we have considered a set of master integrals required to
describe $pp\to t\bar{t}j$ at two-loops in QCD in the planar limit. While we
have limited ourselves to a semi-analytic evaluation of the integrals using
the method of generalised series expansions, the identification of a `dlog'
representation of the differential equation is the first step towards a well
defined special function representation as has been achieved in massless
propagator cases ~\cite{Gehrmann:2018yef,Chicherin:2020oor,Chicherin:2021dyp}. We also observe
some simple structure in the choices of UT integrals which we hope will be of
use when treating the other planar topologies.

An analytic computation of $pp\to t\bar{t}j$ at two-loops in QCD remains a
considerable challenge, yet in the planar limit (excluding corrections from
closed heavy fermion loops) the prospects look quite reasonable. Of course, as
soon as elliptic curves (or more complicated geometries) become relevant, the
problem quickly grows in complexity, both for finding a good choice of MIs and reconstructing the differential equation and by the fact that the space of special functions the integrals evaluate to is often unknown. Nevertheless, the successful application of
the generalised series expansion together with the high precision boundary
values obtained through the auxiliary mass flow method, offers hope that
representations suitable for phenomenological applications may be achievable
in the near future. 

Beyond the phenomenological applications of this work, the analytic results obtained for the differential equations
and the alphabet structure could also be of interest in some more theoretical contexts, such as cluster algebras 
 \cite{Chicherin:2020umh,He:2021esx} or recent studies concerning the singularities structure of Feynman
 integrals \cite{Hannesdottir:2021kpd,Bourjaily:2022vti,Flieger:2022xyq}.

\section{Acknowledgements}

We thank Simone Zoia and Heribertus Bayu Hartanto for many helpful discussions. This project received
funding from the European Union's Horizon 2020 research and innovation programmes \textit{High precision
multi-jet dynamics at the LHC} (consolidator grant agreement No 772099), \textit{EWMassHiggs} (Marie Sk{\l}odowska Curie Grant agreement ID: 101027658), European Research Council starting grant BOSON 101041109 as well as from the Villum Fonden research grant 00025445.

\appendix

\section{UT integrals for sectors with few than five external legs} \label{app:utbasis}

In this section we give explicitly the expressions for the UT basis of the non five-point MIs.

\subsection*{Sector: $\cI_{11}$, $\cI_{12}$}

\begin{minipage}{0.4\textwidth}
\hspace{\parindent}
\begin{figure}[H]
\centering
\includegraphics[width=3.8 cm]{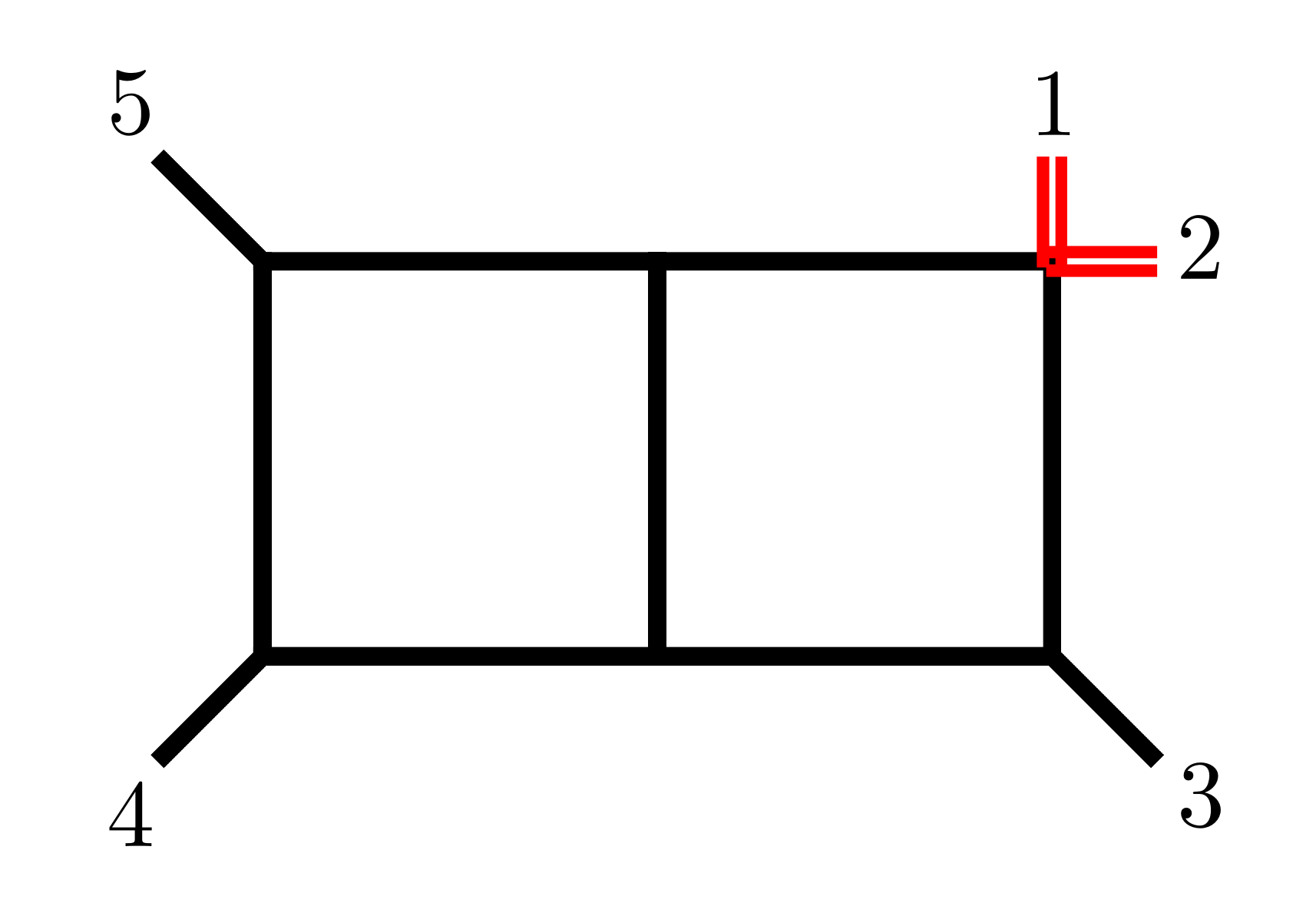}  
\end{figure}
\end{minipage}
\begin{minipage}{0.54\textwidth}
\begin{align}\begin{split}
\cI_{11} = & -8 d_{34} \, d_{45}^2 \, \epsilon ^4 \, I_{1,0,1,1,1,1,1,1}^{0,0,0}   \\
\cI_{12} = & -4 d_{45} \, (-d_{12}+d_{45}-m_t^2) \, \epsilon ^4 \, I_{1,0,1,1,1,1,1,1}^{1,0,0}
\end{split}\end{align}
\end{minipage}%

\subsection*{Sector: $\cI_{13}$, $\cI_{14}$, $\cI_{15}$}

\hspace{\parindent}
\begin{figure}[H]
\includegraphics[width=3.8cm]{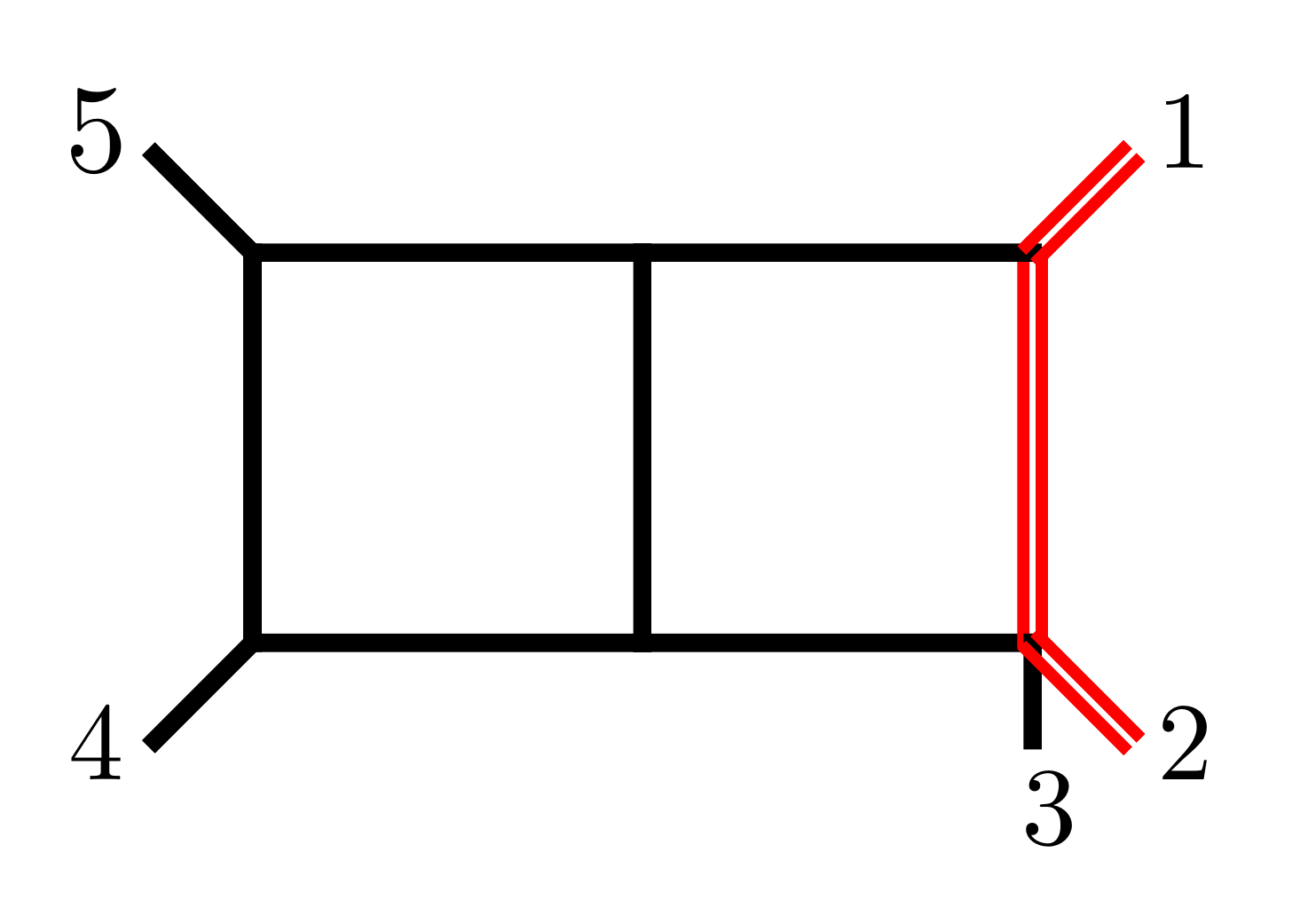}  
\end{figure}

\begin{align}
\cI_{13} = \, & 8 d_{15} \, d_{45}^2 \, \epsilon ^4 \, I_{1,1,0,1,1,1,1,1}^{0,0,0}  \nn \\
\cI_{14} = \, &2 d_{45} \, \Delta _1 \, \epsilon ^4 \, I_{1,1,0,1,1,1,1,1}^{1,0,0} \nn \\
\cI_{15} = \, & 4 d_{45}^2 \epsilon ^4 \, I_{1,1,0,1,1,1,1,1}^{0,1,0} + \frac{3 d_{23} \, \epsilon \,  (2 \epsilon -1) (3 \epsilon -2) (3 \epsilon -1)}{4 d_{15} \, d_{45}} \, I_{0,0,0,1,1,0,0,1}^{0,0,0}  \nn \\
& + \frac{4 d_{23} \, d_{45} \, m_t^2 \, \epsilon ^2 (2 \epsilon -1)}{d_{15}} \, I_{1,2,0,1,0,1,0,1}^{0,0,0} - \frac{4 d_{23} \, d_{45} \, m_t^2 \, \epsilon ^3}{d_{15}} \, I_{1,2,0,0,0,1,1,1}^{0,0,0} \nn \\
& -\frac{\epsilon  (2 \epsilon -1) (3 \epsilon -2) \left(10 d_{23} \, \epsilon -2 d_{23}+4 \epsilon \, m_t^2-m_t^2\right)}{8 d_{15} \, d_{23}} \, I_{0,1,0,0,0,0,1,1}^{0,0,0} \nn \\
& -\frac{3 \epsilon ^2 (2 \epsilon -1) (3 \epsilon -2) m_t^2}{8 d_{15}^2} \, I_{0,1,0,0,0,1,0,1}^{0,0,0} -\frac{3 \epsilon ^2 (2 \epsilon -1) (3 \epsilon -1) \left(2 d_{23}+m_t^2\right)}{4 d_{15}} \, I_{0,1,0,1,0,1,0,1}^{0,0,0} \nn \\
& -\frac{\epsilon  (2 \epsilon -1) \left(11 d_{23} \, \epsilon \, m_t^2-2 d_{23} \, m_t^2+4 d_{23}^2 \, \epsilon +4 \epsilon \, m_t^4-m_t^4\right)}{4 d_{15} \, d_{23}} \, I_{0,2,0,0,0,0,1,1}^{0,0,0} \nn \\
& -\frac{3 \epsilon ^2 (2 \epsilon -1) m_t^2 \left(d_{15}+m_t^2\right)}{4 d_{15}^2} \, I_{0,2,0,0,0,1,0,1}^{0,0,0} +  \frac{3 d_{23} \, \epsilon ^2 (2 \epsilon -1) (3 \epsilon -1)}{2 d_{15}} \, I_{1,0,0,1,0,1,0,1}^{0,0,0} \nn \\
&-\frac{6 d_{23} \left(d_{15}-d_{23}+d_{45}\right) \epsilon ^4}{d_{15}} \, I_{1,1,0,0,0,1,1,1}^{0,0,0} -\frac{6 d_{23} \left(d_{23}-d_{45}\right) \epsilon ^3 (2 \epsilon -1)}{d_{15}} \, I_{1,1,0,1,0,1,0,1}^{0,0,0}
\end{align}

\subsection*{Sector: $\cI_{27}$}

\begin{minipage}{0.4\textwidth}
\hspace{\parindent}
\begin{figure}[H]
\centering
\includegraphics[width=3.5cm]{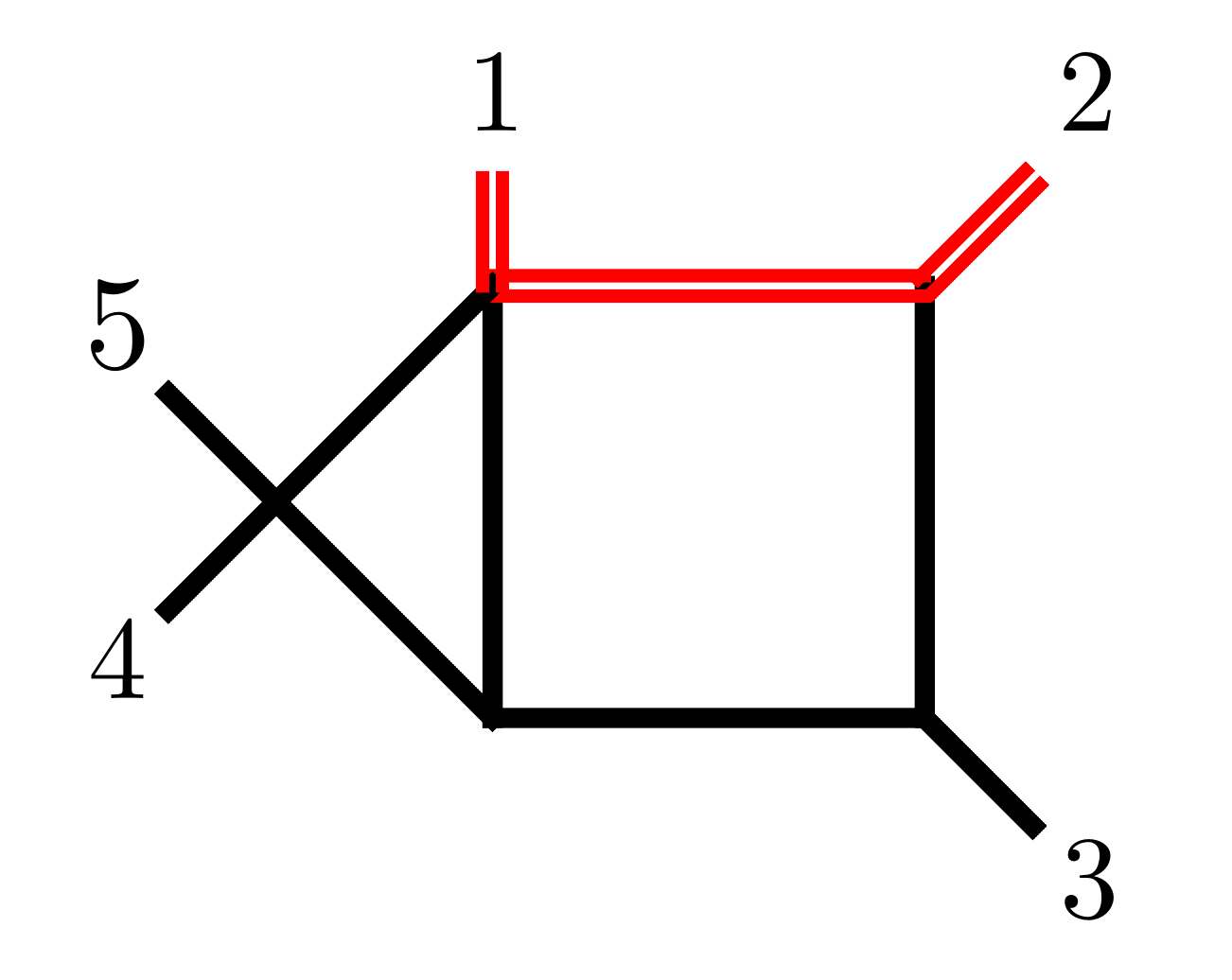}  
\end{figure}
\end{minipage}
\begin{minipage}{0.54\textwidth}
\begin{align}\begin{split}
\cI_{27} = & 4 d_{23} \, \left(d_{12}-d_{45}+m_t^2\right) \,\epsilon ^4 \, I_{0,1,1,1,1,0,1,1}^{0,0,0} 
\end{split}\end{align}
\end{minipage}%

\subsection*{Sector: $\cI_{28}$}

\begin{minipage}{0.4\textwidth}
\hspace{\parindent}
\begin{figure}[H]
\centering
\includegraphics[width=3.5cm]{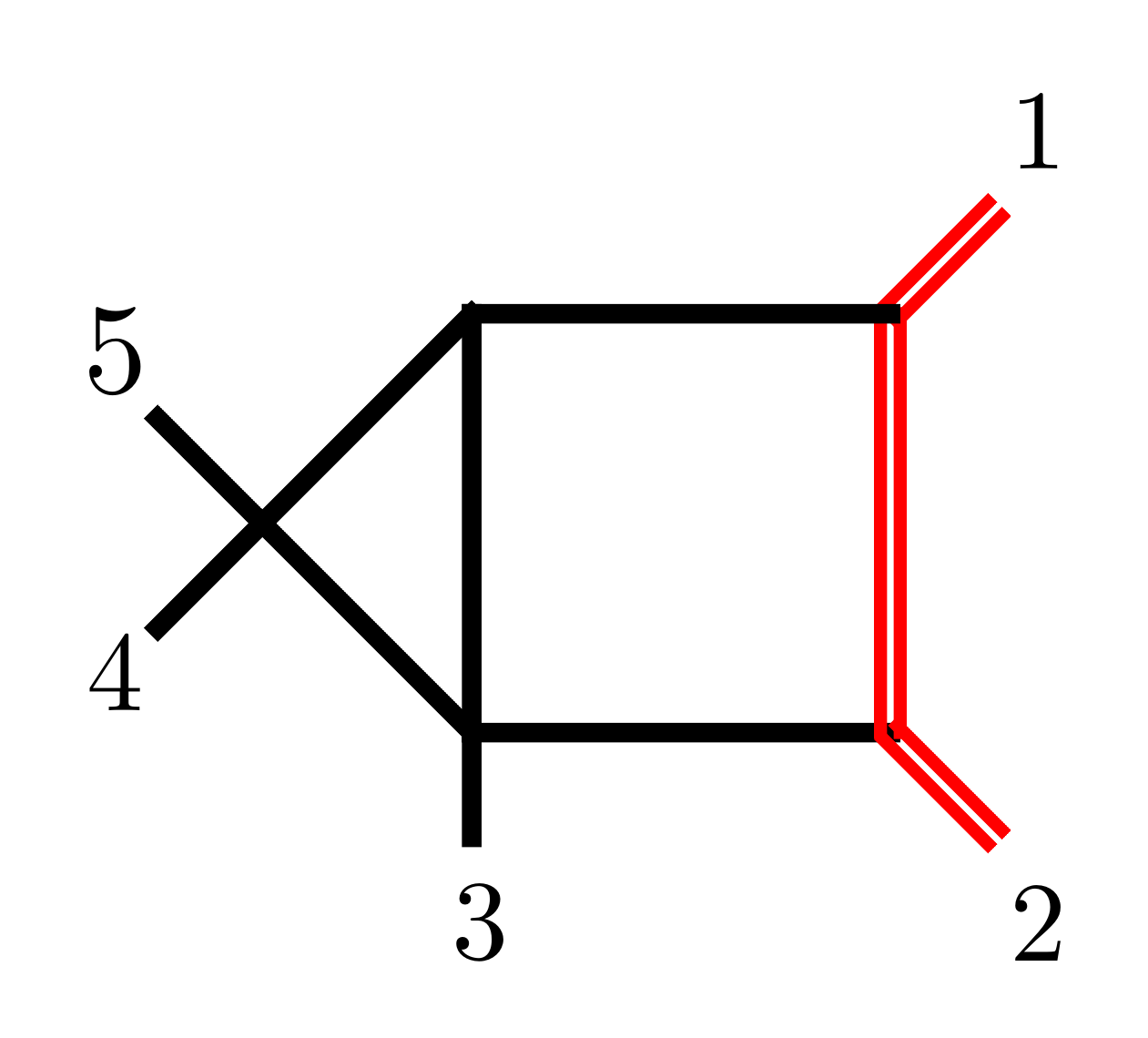}  
\end{figure}
\end{minipage}
\begin{minipage}{0.54\textwidth}
\begin{align}\begin{split}
\cI_{28} = & 2 \Delta _1 \, \left(d_{12}+m_t^2\right) \,  \epsilon ^4 \, I_{1,1,1,0,1,0,1,1}^{0,0,0}
\end{split}\end{align}
\end{minipage}%

\subsection*{Sector: $\cI_{29}$}

\begin{minipage}{0.4\textwidth}
\hspace{\parindent}
\begin{figure}[H]
\centering
\includegraphics[width=3.5cm]{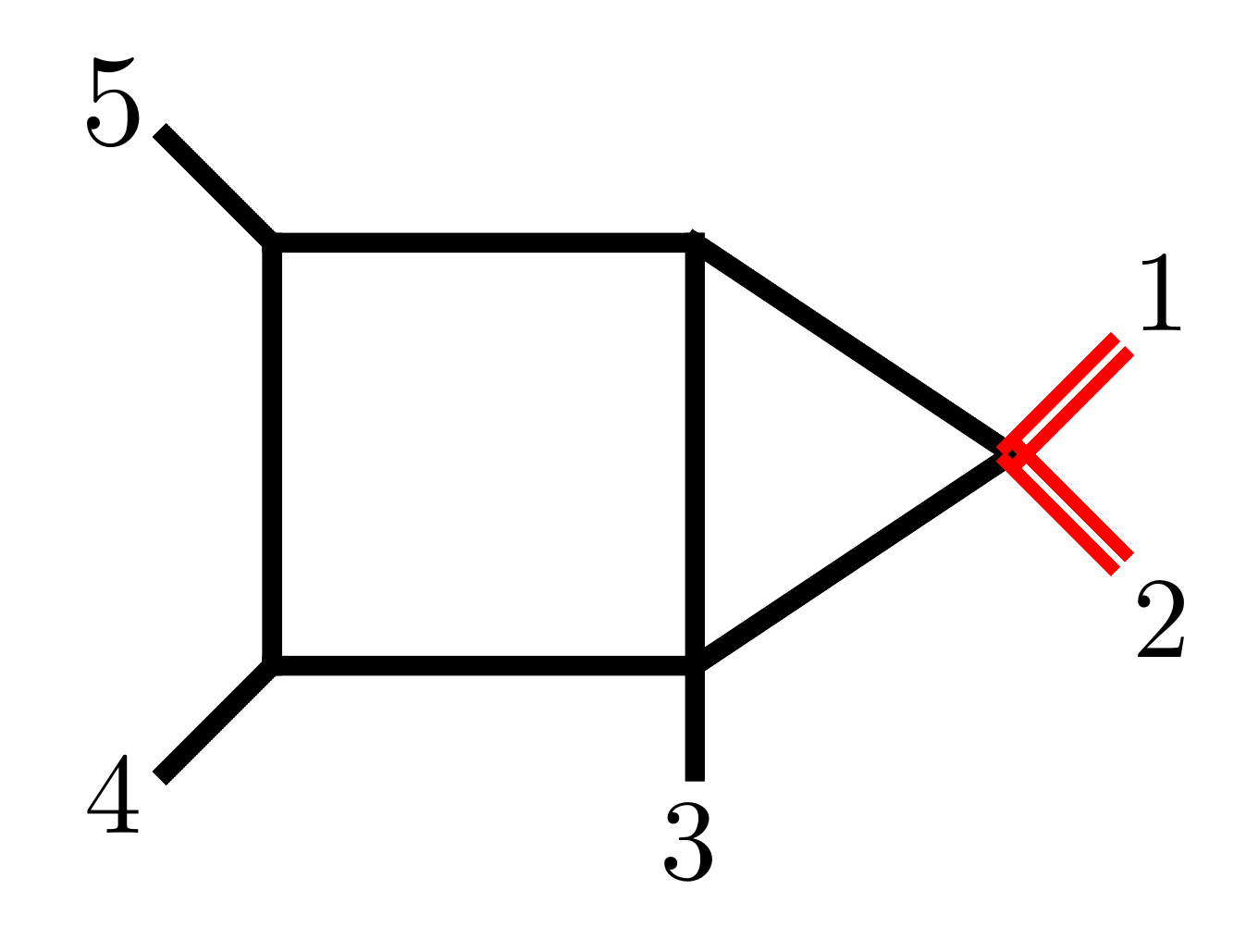}  
\end{figure}
\end{minipage}
\begin{minipage}{0.54\textwidth}
\begin{align}\begin{split}
\cI_{29} = & 4 d_{45} \, \left(d_{12}-d_{34}+m_t^2\right) \, \epsilon ^4 \, I_{1,0,1,0,1,1,1,1}^{0,0,0}
\end{split}\end{align}
\end{minipage}%

\subsection*{Sector: $\cI_{30}$}

\begin{minipage}{0.4\textwidth}
\hspace{\parindent}
\begin{figure}[H]
\centering
\includegraphics[width=3.5cm]{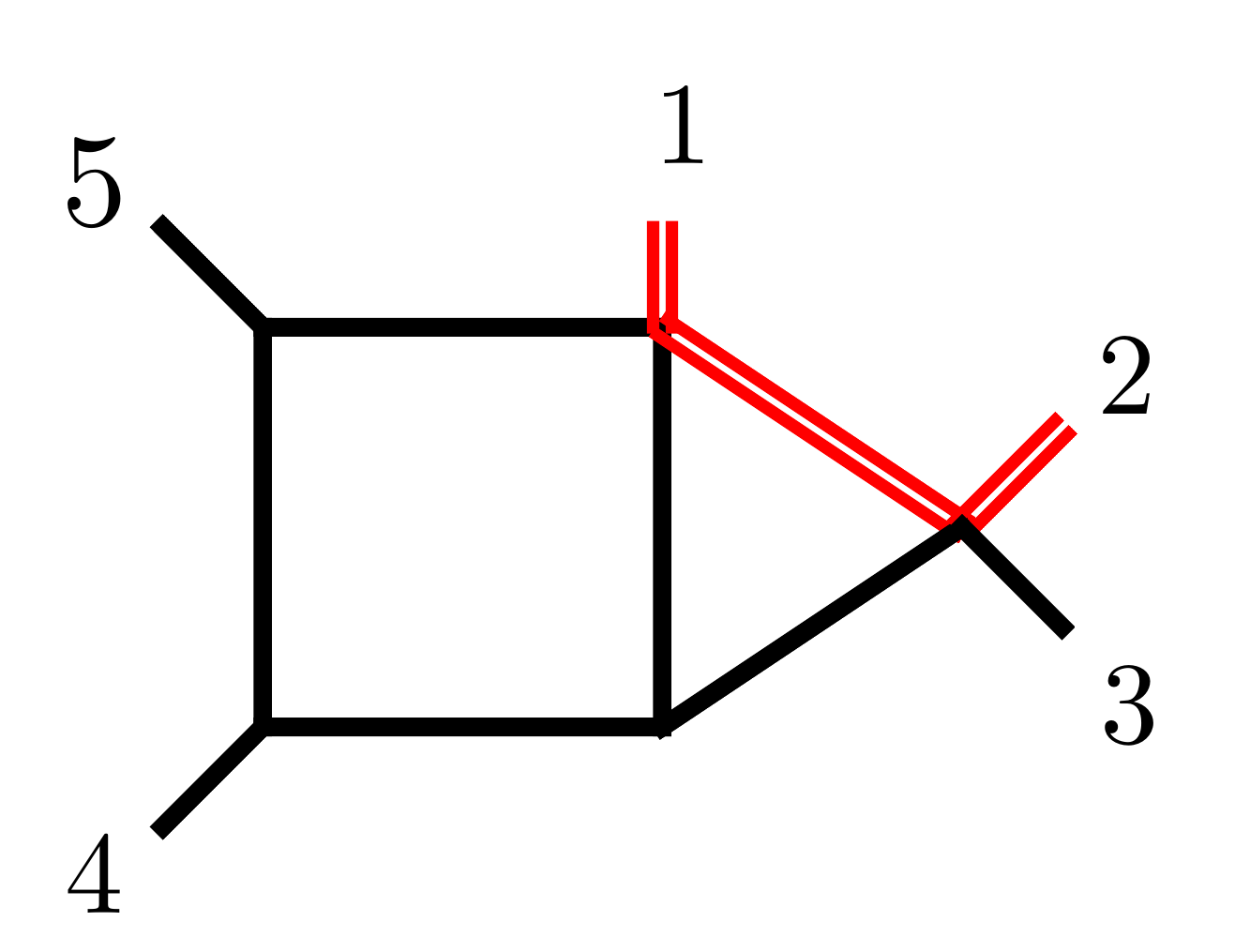}  
\end{figure}
\end{minipage}
\begin{minipage}{0.54\textwidth}
\begin{align}\begin{split}
\cI_{30} = & -4 \left(d_{15}-d_{23}\right) \, d_{45} \, \epsilon ^4 \, I_{0,1,0,1,1,1,1,1}^{0,0,0} 
\end{split}\end{align}
\end{minipage}%

\subsection*{Sector: $\cI_{31}, \, \cI_{32}, \, \cI_{33}, \, \cI_{34}, \, \cI_{35}$}

\begin{minipage}{0.4\textwidth}
\hspace{\parindent}
\begin{figure}[H]
\centering
\includegraphics[width=3.5cm]{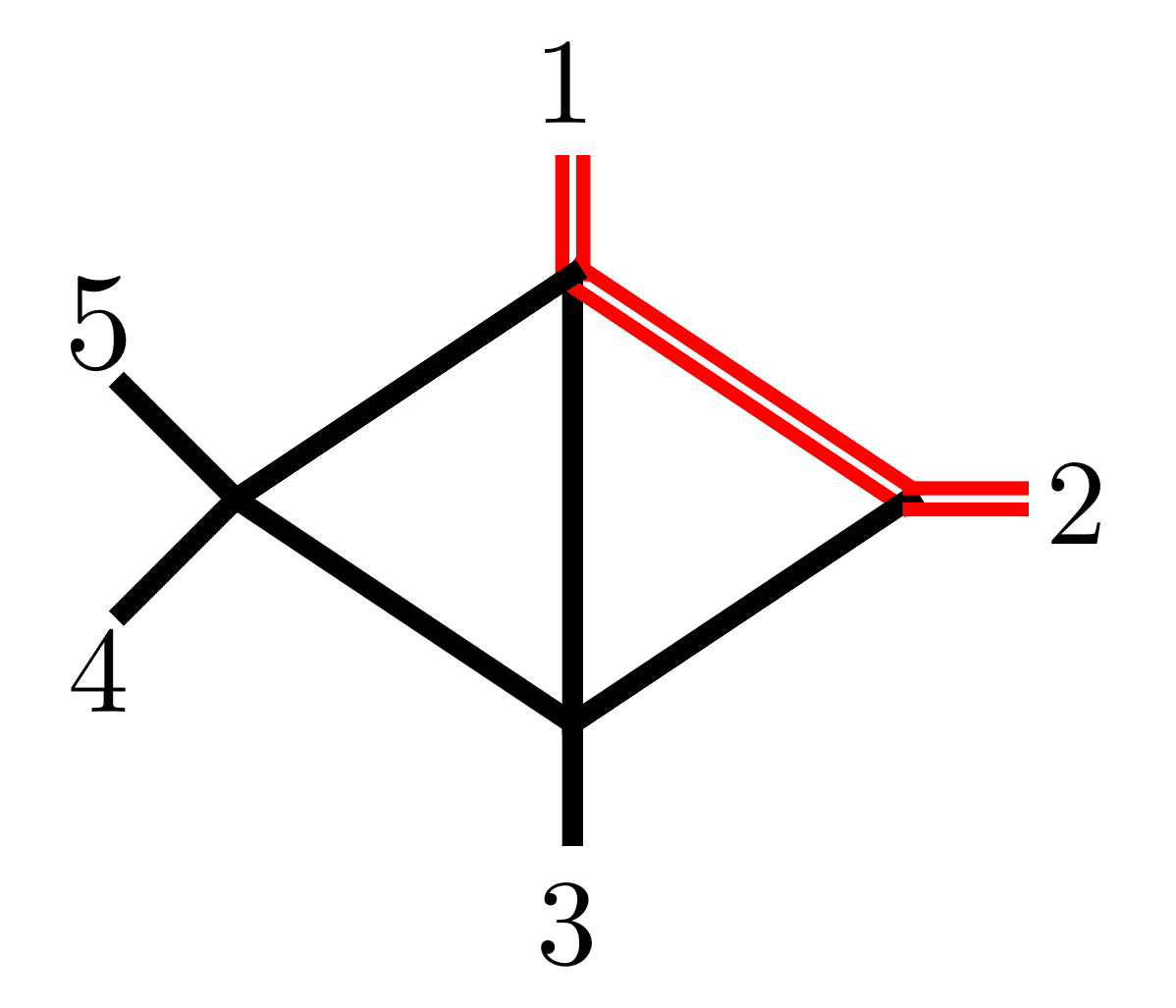}  
\end{figure}
\end{minipage}
\begin{minipage}{0.54\textwidth}
\begin{align}\begin{split}
\cI_{31} = & 2 \Delta _3 \, \left(d_{12}+d_{23}+m_t^2\right) \, \epsilon ^4 \, I_{0,1,1,0,1,0,1,1}^{0,0,0}  \\
\cI_{32} = & 2  m_t^2 \, \left(d_{12}-d_{45}+m_t^2\right) \, \epsilon ^3 \, I_{0,2,1,0,1,0,1,1}^{0,0,0}  \\
\cI_{33} = & 4 \beta \, d_{45} \, \left(d_{12}+m_t^2\right) \, \epsilon ^3 \, I_{0,1,1,0,2,0,1,1}^{0,0,0}   \\
\cI_{34} = & 4 d_{23} \, d_{45} \, \epsilon ^3 \, I_{0,1,1,0,1,0,2,1}^{0,0,0}  \\
\cI_{35} = & 4 d_{23} \, \left(d_{12}+m_t^2\right) \, \epsilon ^3 \, I_{0,1,1,0,1,0,1,2}^{0,0,0}
\end{split}\end{align}
\end{minipage}%

\subsection*{Sectors: $\cI_{36}, \, \cI_{37}$ and $\cI_{46}, \, \cI_{47}$}

\begin{minipage}{0.4\textwidth}
\hspace{\parindent}
\begin{figure}[H]
\centering
\includegraphics[width=3.5cm]{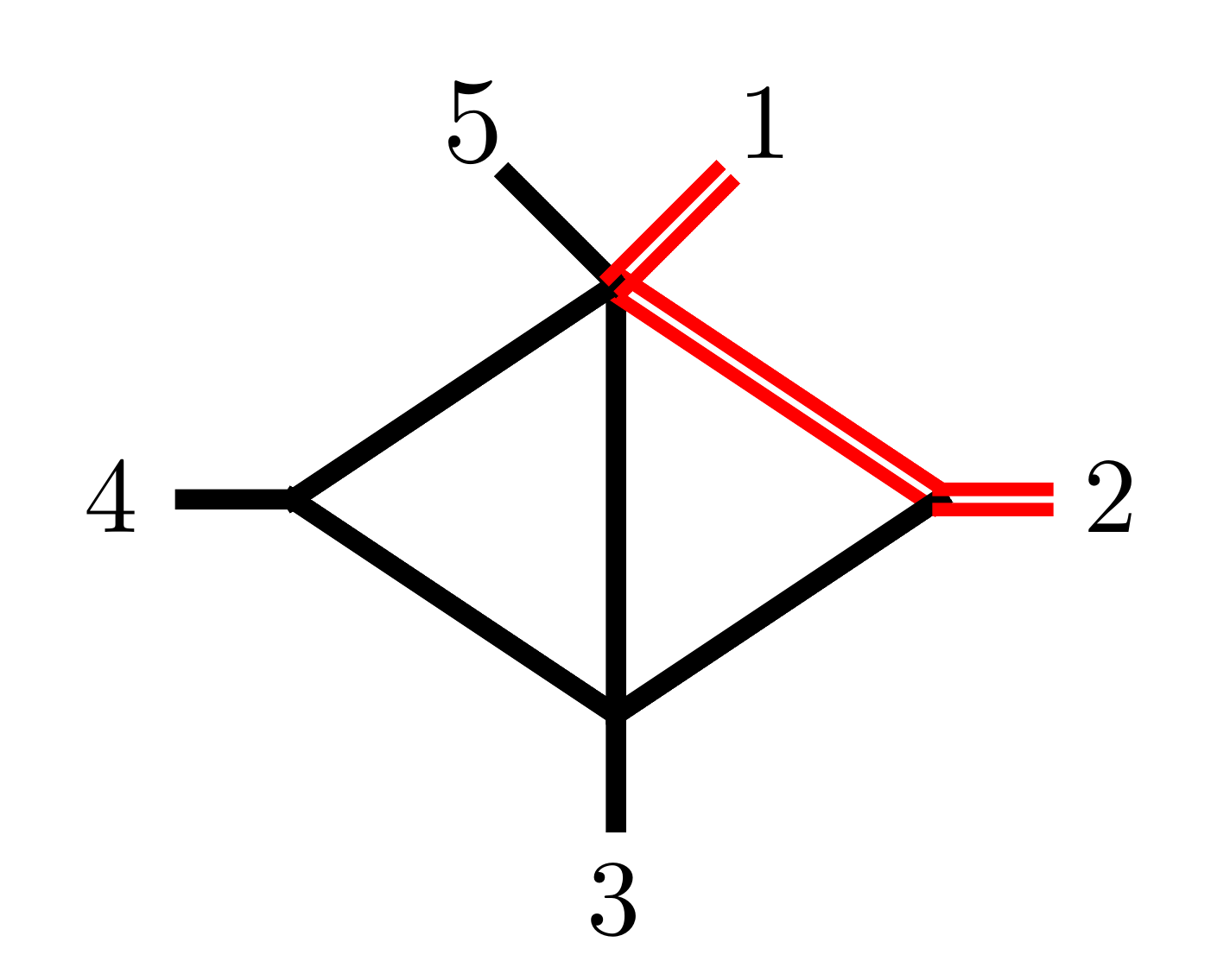} 
\includegraphics[width=3.5cm]{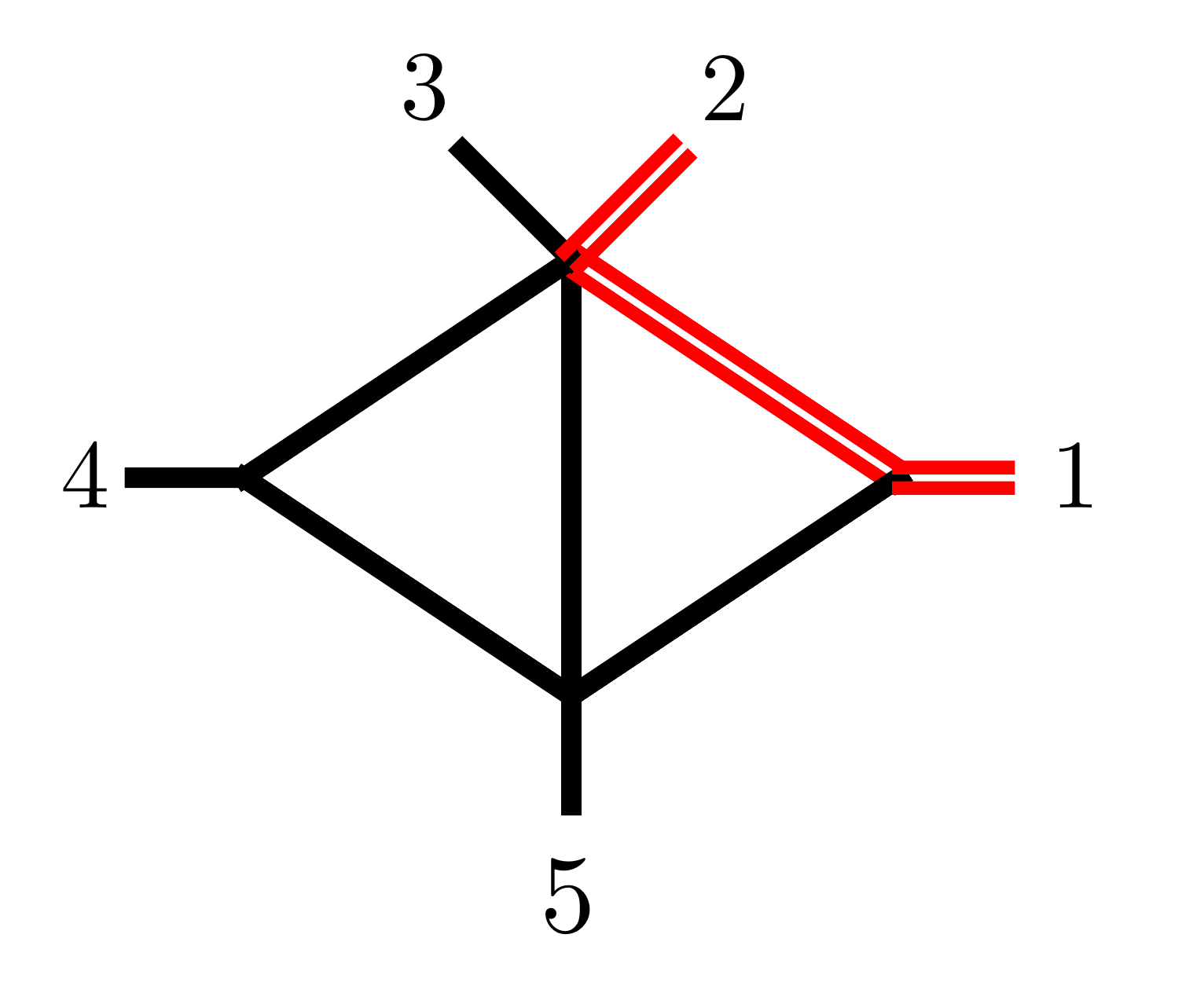} 
\end{figure}
\end{minipage}
\begin{minipage}{0.54\textwidth}
\begin{align}\begin{split}
\cI_{36} = & -2 \left(d_{15}-d_{23}-d_{34}\right) \, \epsilon ^4 \, I_{0,1,1,0,0,1,1,1}^{0,0,0} \\
\cI_{37} = & 2 d_{34} \, m_t^2 \, \epsilon ^3 \, I_{0,2,1,0,0,1,1,1}^{0,0,0} \\
\cI_{46} = & 2 \left(d_{15}-d_{23}+d_{45}\right) \, \epsilon ^4 \, I_{1,1,0,0,0,1,1,1}^{0,0,0} \\
\cI_{47} = & 2 d_{45} \, m_t^2 \, \epsilon ^3 \, I_{1,2,0,0,0,1,1,1}^{0,0,0}
\end{split}\end{align}
\end{minipage}%

\subsection*{Sector: $\cI_{38}, \, \cI_{39}$}

\begin{minipage}{0.4\textwidth}
\hspace{\parindent}
\begin{figure}[H]
\centering
\includegraphics[width=3.5cm]{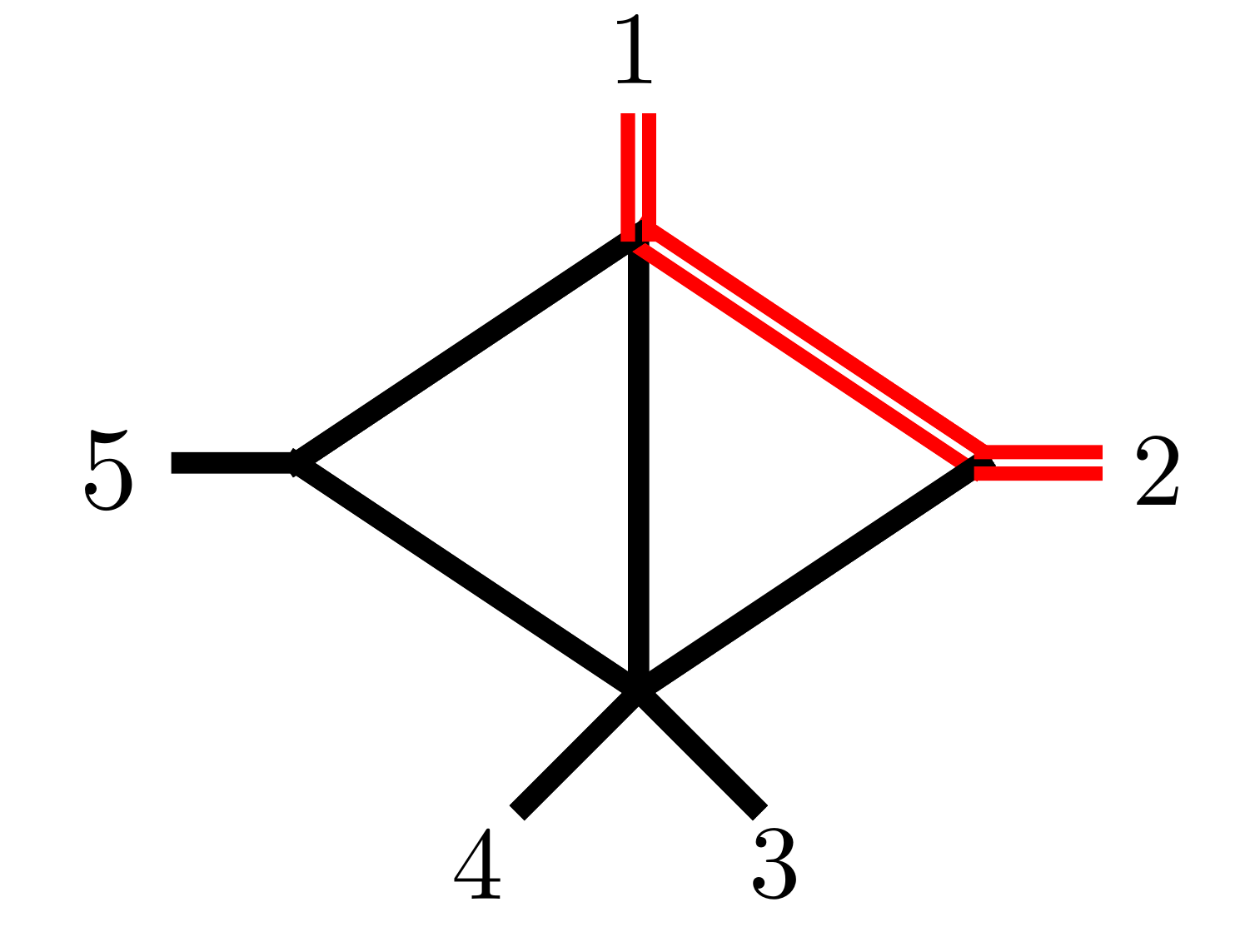}  
\end{figure}
\end{minipage}
\begin{minipage}{0.54\textwidth}
\begin{align}\begin{split}
\cI_{38} = & 2 \left(d_{12}+d_{15}-d_{34}+m_t^2\right) \, \epsilon ^4 \, I_{0,1,1,0,1,1,0,1}^{0,0,0}  \\
\cI_{39} = & 2 m_t^2 \, \left(d_{12}-d_{34}+m_t^2\right) \, \epsilon ^3 \, I_{0,2,1,0,1,1,0,1}^{0,0,0}
\end{split}\end{align}
\end{minipage}%

\subsection*{Sector: $\cI_{40}, \, \cI_{41}$}

\begin{minipage}{0.4\textwidth}
\hspace{\parindent}
\begin{figure}[H]
\centering
\includegraphics[width=3.5cm]{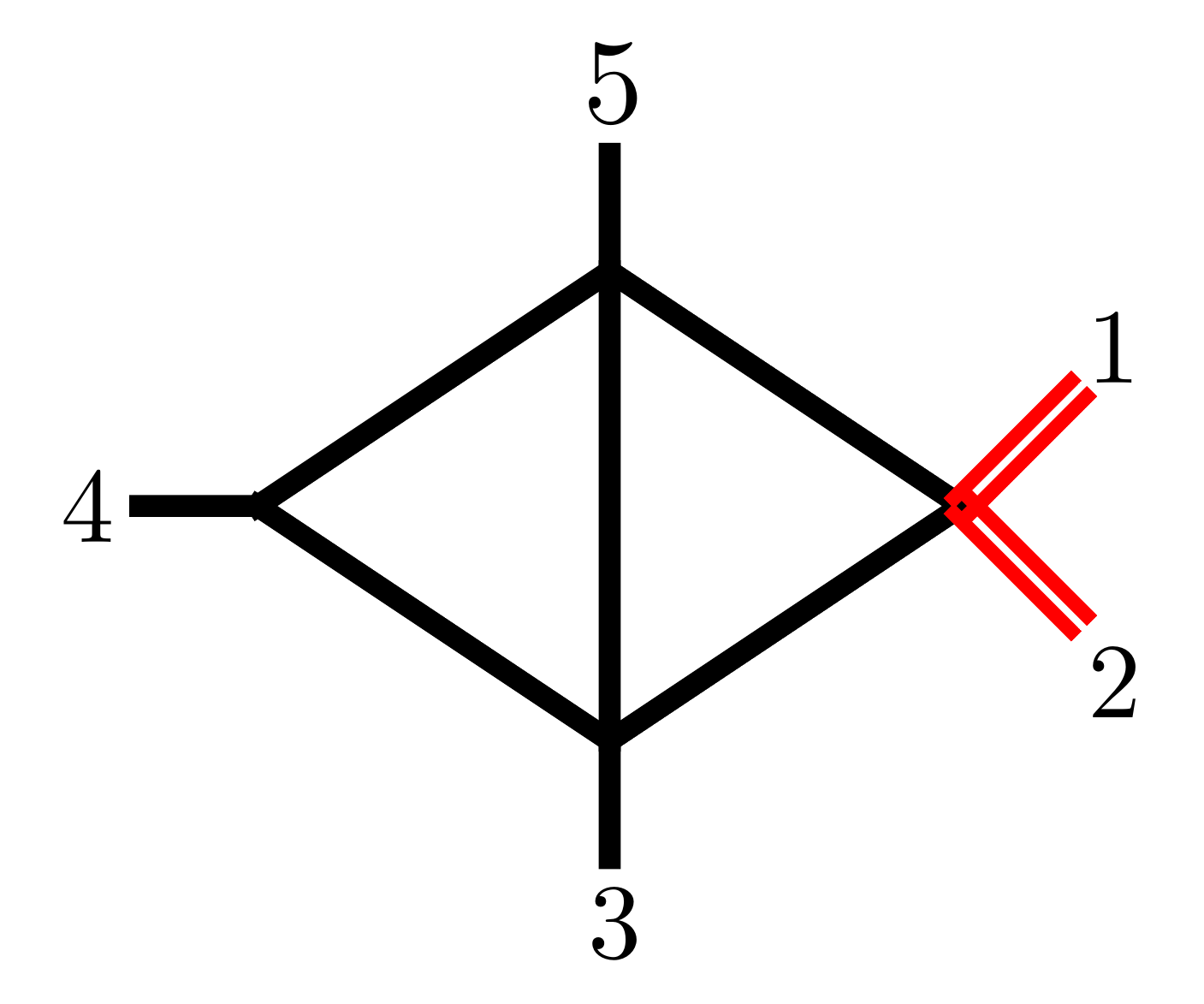}  
\end{figure}
\end{minipage}
\begin{minipage}{0.54\textwidth}
\begin{align}\begin{split}
\cI_{40} = & 4 \left(d_{34}+d_{45}\right) \, \epsilon ^4 \, I_{1,0,1,0,0,1,1,1}^{0,0,0} \\
\cI_{41} = & 4 d_{34} \, d_{45} \, \epsilon ^3 \, I_{1,0,1,0,0,1,1,2}^{0,0,0}
\end{split}\end{align}
\end{minipage}%

\subsection*{Sector: $\cI_{42}$}

\begin{minipage}{0.4\textwidth}
\hspace{\parindent}
\begin{figure}[H]
\centering
\includegraphics[width=3.5cm]{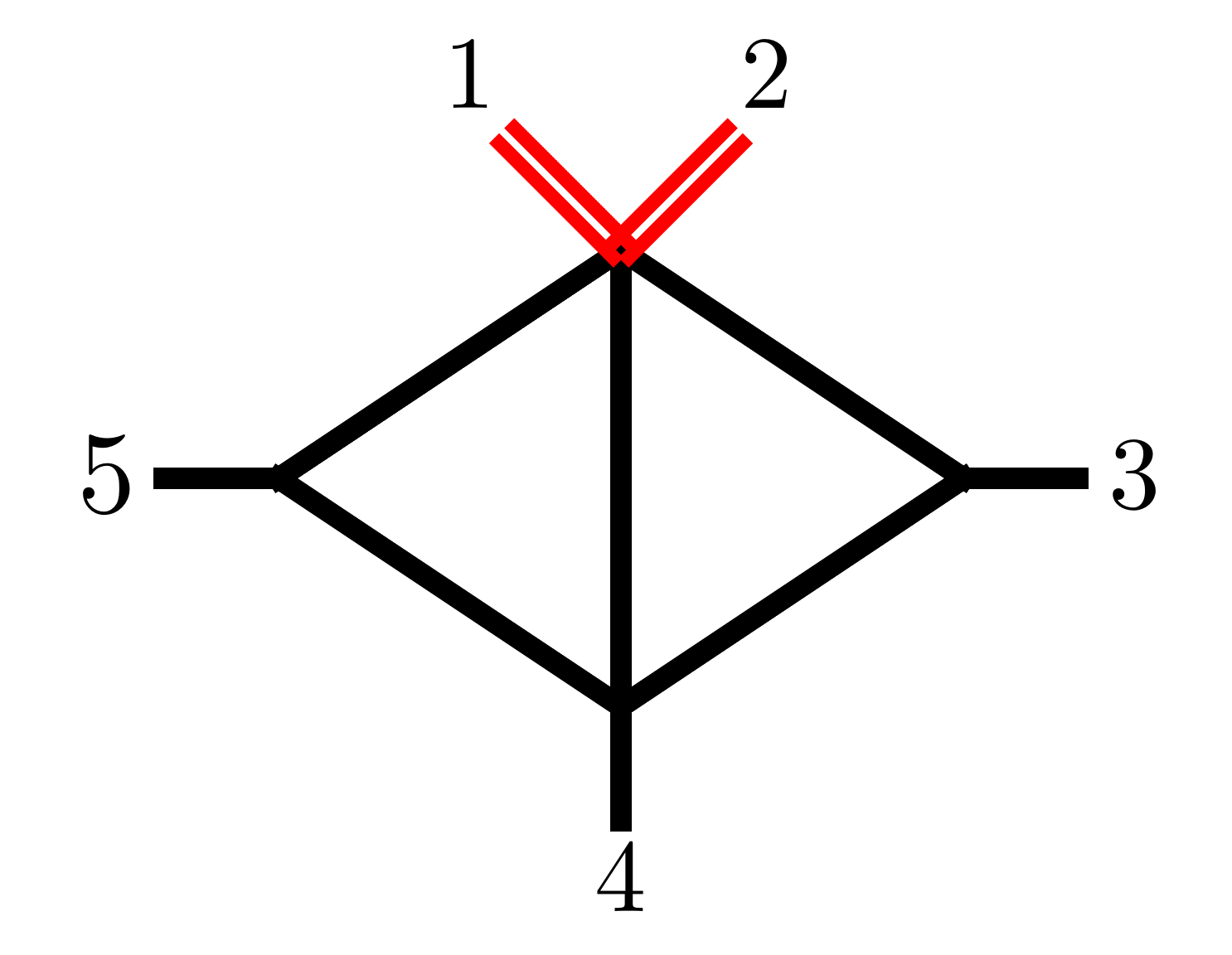}  
\end{figure}
\end{minipage}
\begin{minipage}{0.54\textwidth}
\begin{align}\begin{split}
\cI_{42} = & 2 \left(d_{12}-d_{34}-d_{45}+m_t^2\right) \,  \epsilon ^4 \, I_{0,0,1,1,1,1,0,1}^{0,0,0}
\end{split}\end{align}
\end{minipage}%

\subsection*{Sector: $\cI_{43}$}

\begin{minipage}{0.4\textwidth}
\hspace{\parindent}
\begin{figure}[H]
\centering
\includegraphics[width=3.5cm]{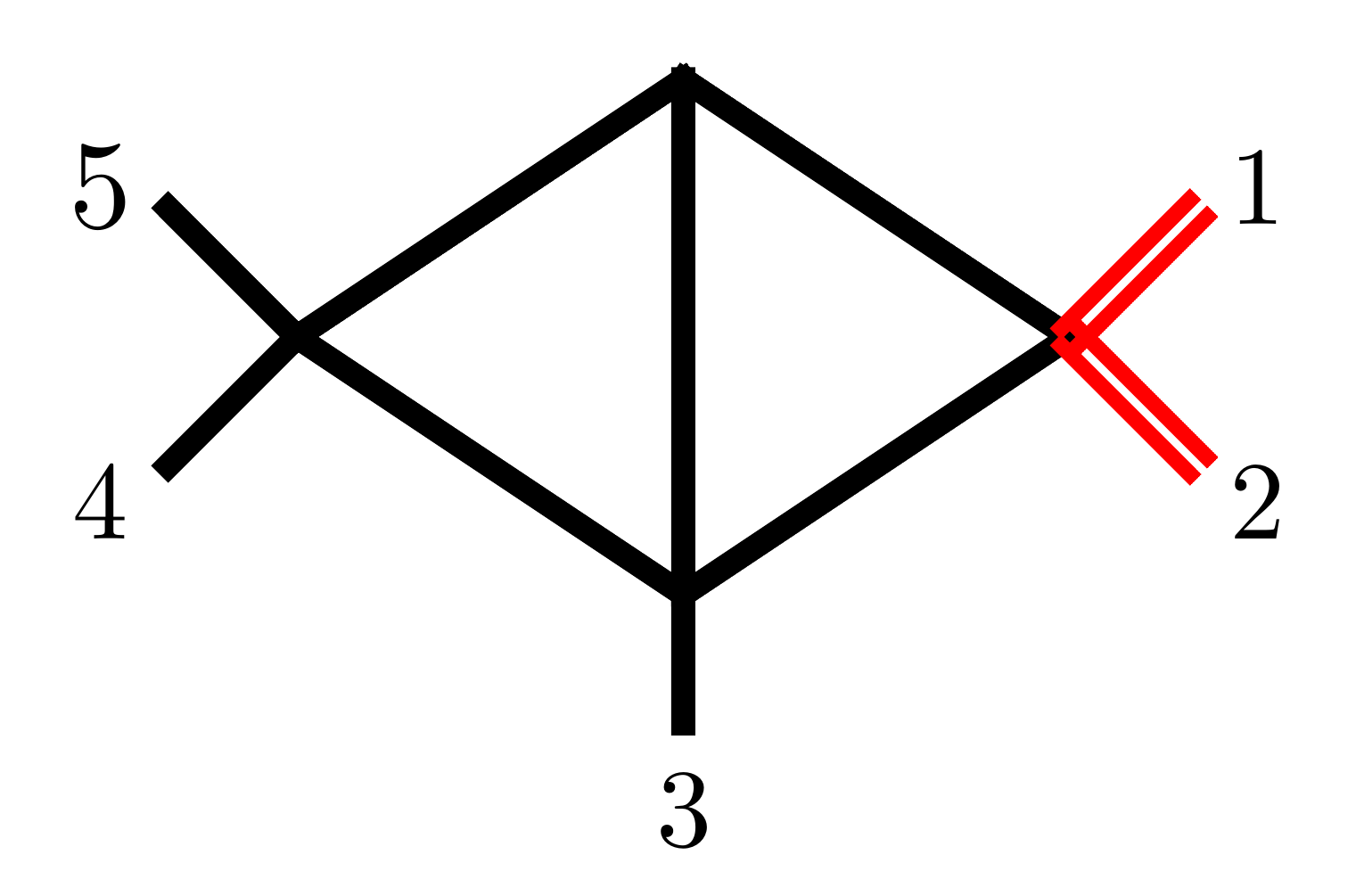}  
\end{figure}
\end{minipage}
\begin{minipage}{0.54\textwidth}
\begin{align}\begin{split}
\cI_{43} = & 2  \left(d_{12}-d_{45}+m_t^2\right) \, \epsilon ^4 \, I_{1,0,1,0,1,0,1,1}^{0,0,0}
\end{split}\end{align}
\end{minipage}%

\subsection*{Sector: $\cI_{44}, \, \cI_{45}$}

\begin{minipage}{0.4\textwidth}
\hspace{\parindent}
\begin{figure}[H]
\centering
\includegraphics[width=3.5cm]{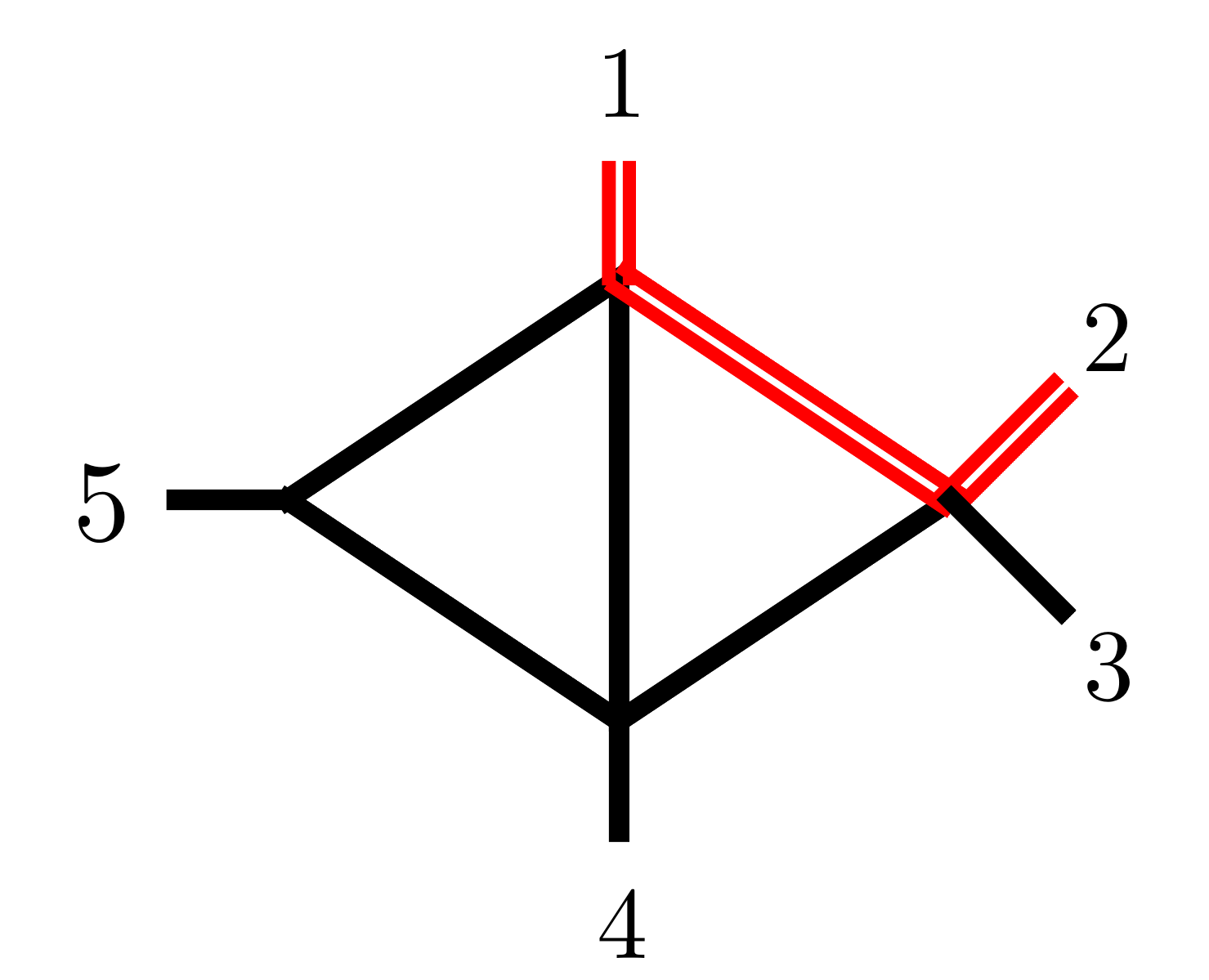}  
\end{figure}
\end{minipage}
\begin{minipage}{0.54\textwidth}
\begin{align}\begin{split}
\cI_{44} = & 2 \left(d_{15}+d_{45}\right) \, \epsilon ^4 \, I_{0,1,0,1,1,1,0,1}^{0,0,0} \\
\cI_{45} = & 2  \left(2 d_{15} d_{23}-d_{45} m_t^2\right) \, \epsilon ^3 \, I_{0,2,0,1,1,1,0,1}^{0,0,0}
\end{split}\end{align}
\end{minipage}%

\subsection*{Sector: $\cI_{48}$}

\begin{minipage}{0.4\textwidth}
\hspace{\parindent}
\begin{figure}[H]
\centering
\includegraphics[width=3.5cm]{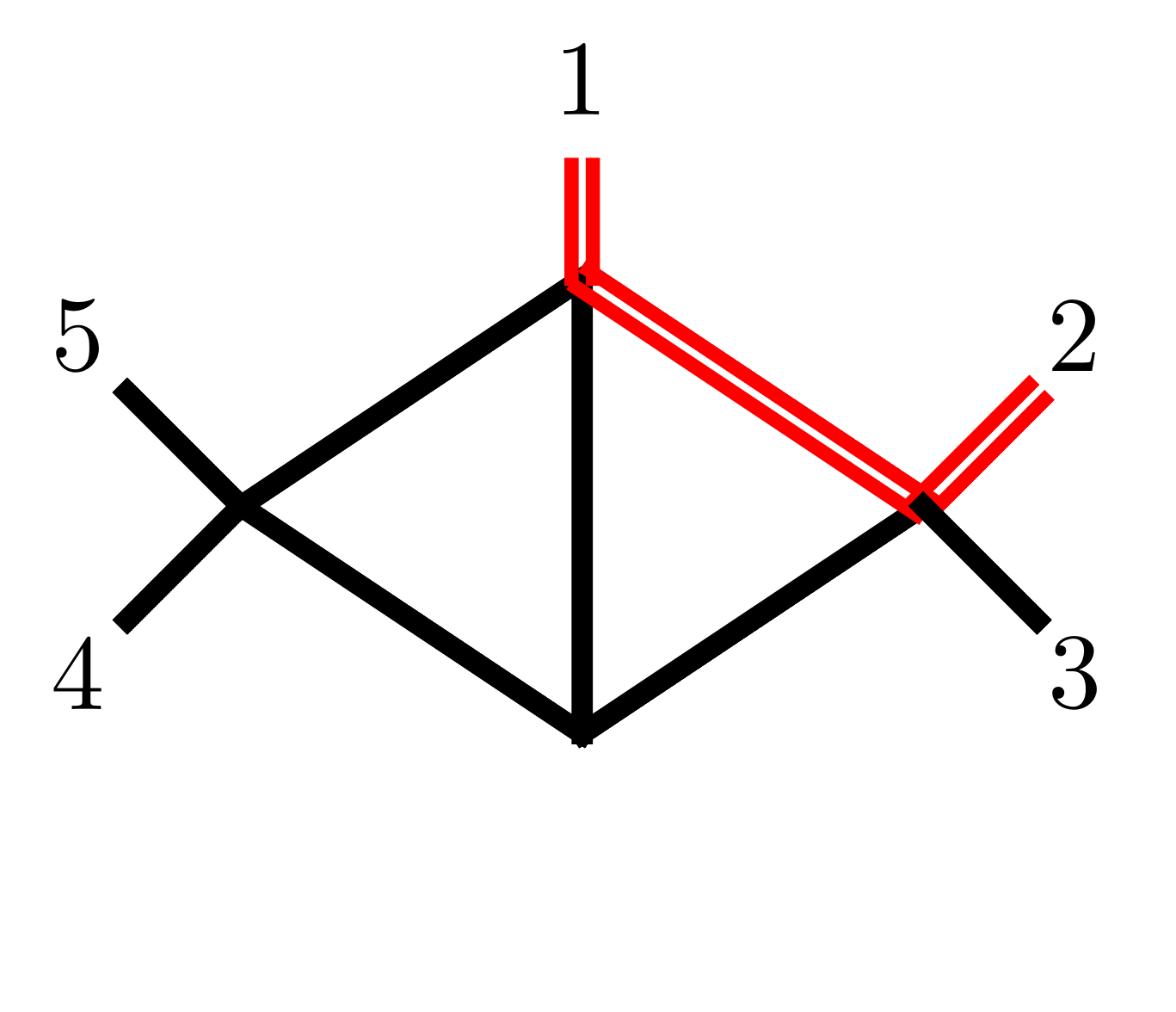}  
\end{figure}
\end{minipage}
\begin{minipage}{0.54\textwidth}
\begin{align}\begin{split}
\cI_{48} = & \Delta_1 \, \epsilon ^4 \, I_{0,1,0,1,1,0,1,1}^{0,0,0}
\end{split}\end{align}
\end{minipage}%

\subsection*{Sectors: $\cI_{49}, \, \cI_{50}$ and $\cI_{53}, \, \cI_{54}$}

\begin{minipage}{0.4\textwidth}
\hspace{\parindent}
\begin{figure}[H]
\centering
\includegraphics[width=3.cm]{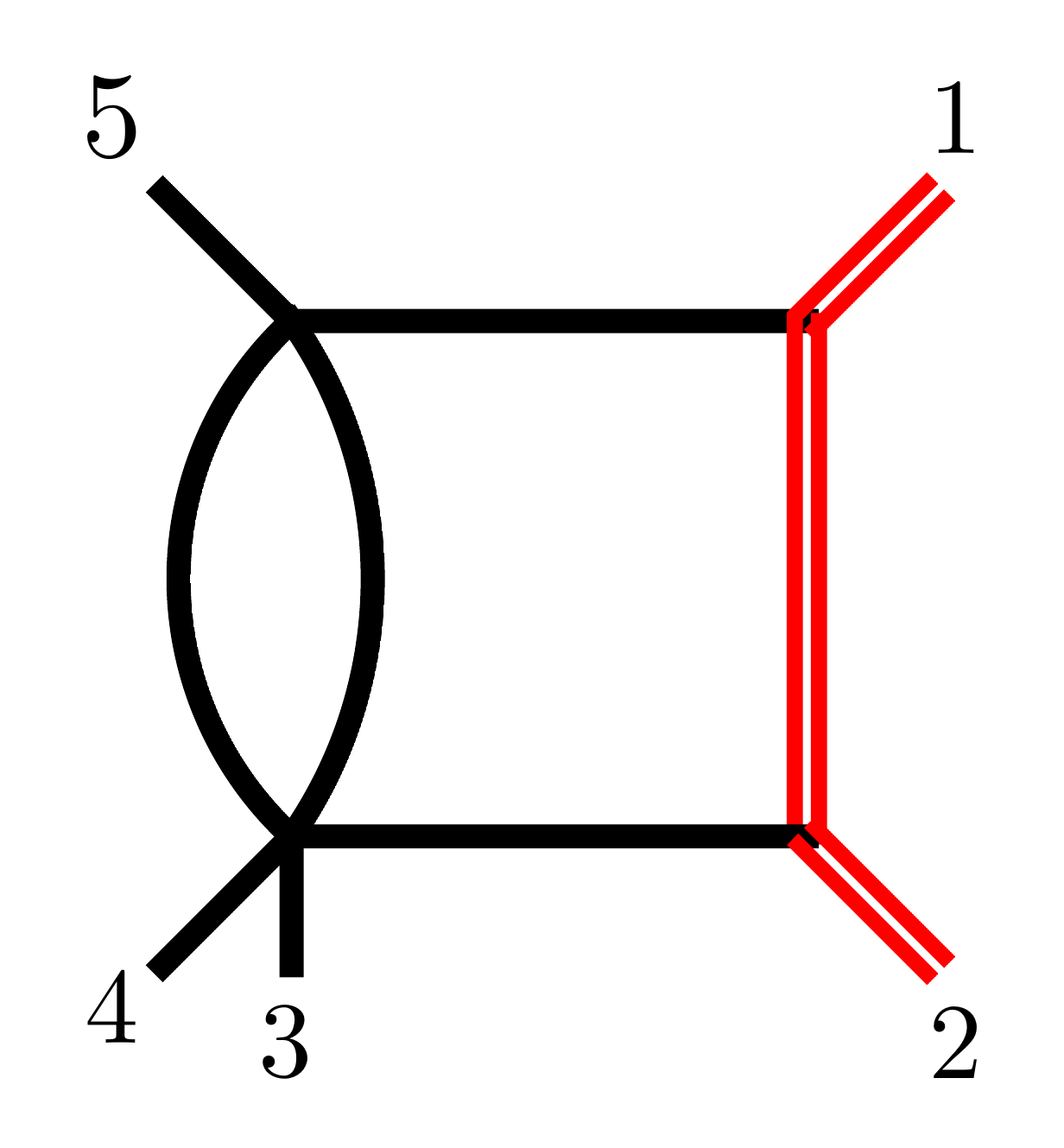}  
\includegraphics[width=3.cm]{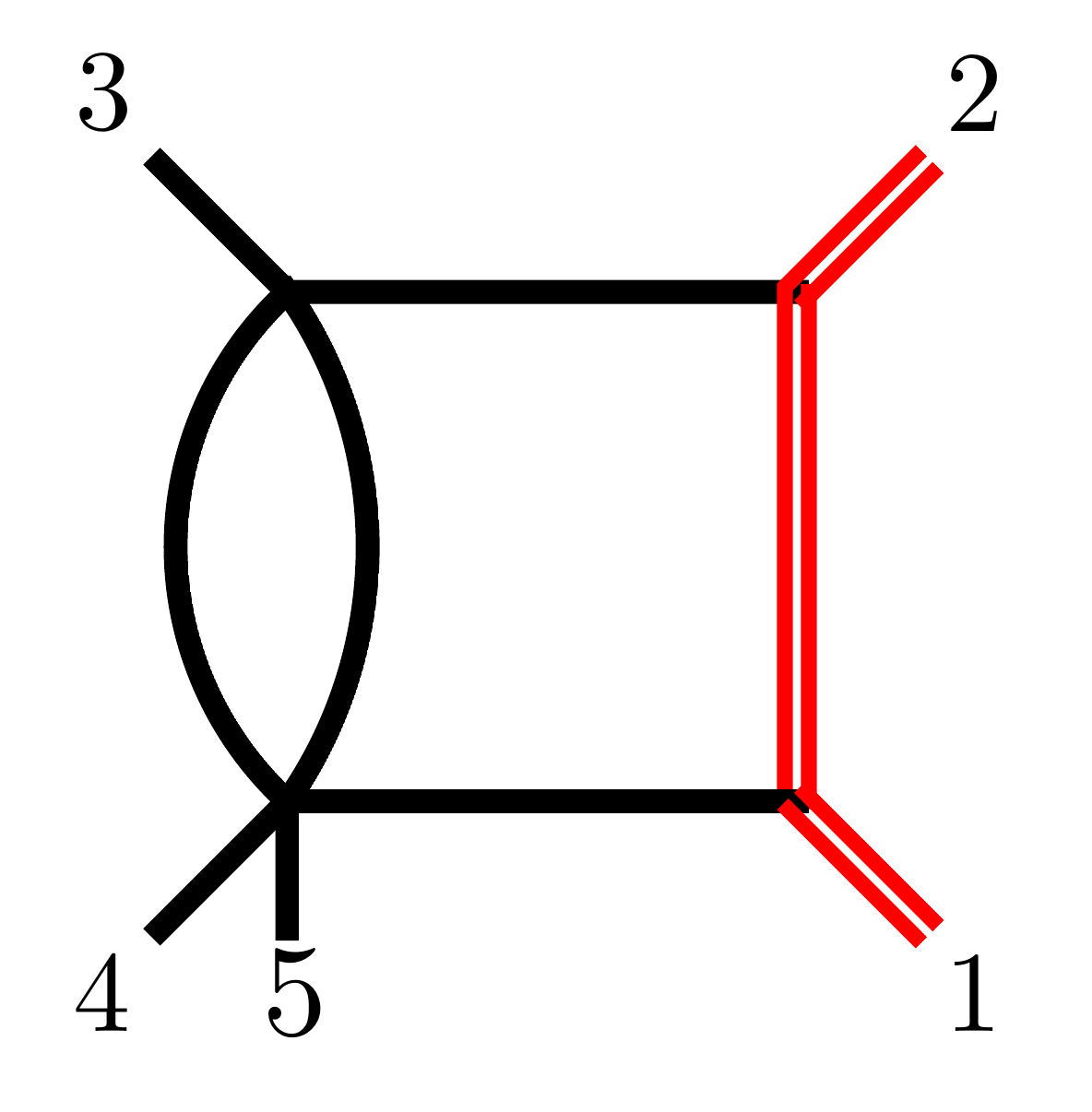}  
\end{figure}
\end{minipage}
\begin{minipage}{0.54\textwidth}
\begin{align}\begin{split}
\cI_{49} = & -2 \beta \, (2 \epsilon -1) \left(d_{12}+m_t^2\right) \, \epsilon ^3 \, I_{1,1,1,0,0,1,0,1}^{0,0,0}  \\
\cI_{50} = & 4 d_{15} \, \left(d_{12}+m_t^2\right) \, \epsilon ^3 \, I_{1,1,1,0,0,1,0,2}^{0,0,0} \\
\cI_{53} = &-2 \beta \, (2 \epsilon -1) \left(d_{12}+m_t^2\right) \, \epsilon ^3 \, I_{1,1,1,0,0,0,1,1}^{0,0,0}  \\
\cI_{54} = & 4 d_{23} \, \left(d_{12}+m_t^2\right) \, \epsilon ^3 \, I_{1,1,1,0,0,0,1,2}^{0,0,0}
\end{split}\end{align}
\end{minipage}%

\subsection*{Sector: $\cI_{51}$}

\begin{minipage}{0.4\textwidth}
\hspace{\parindent}
\begin{figure}[H]
\centering
\includegraphics[width=3cm]{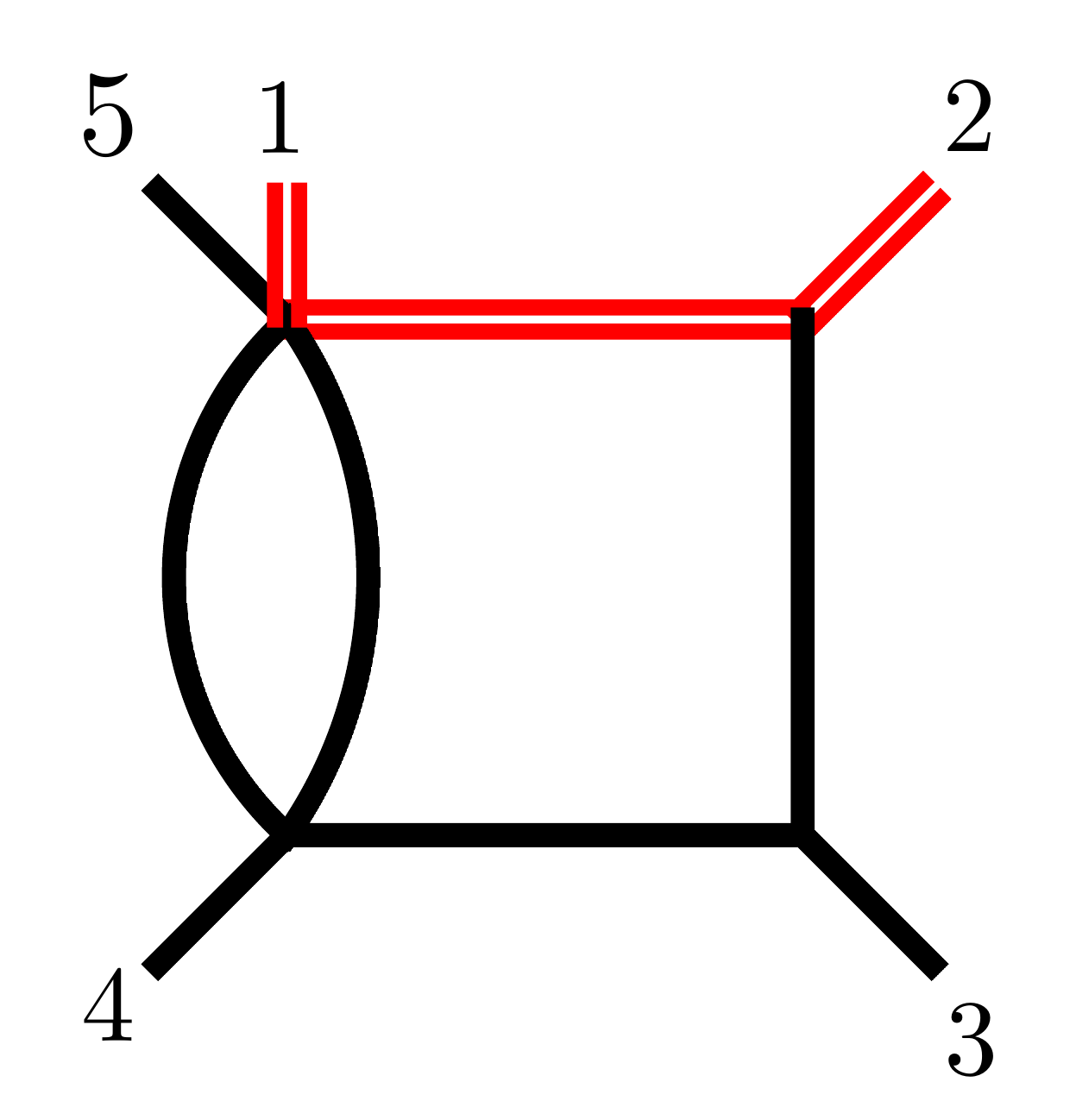}  
\end{figure}
\end{minipage}
\begin{minipage}{0.54\textwidth}
\begin{align}\begin{split}
\cI_{51} = & -2 d_{23} \, (2 \epsilon -1) \, \epsilon ^3 \, I_{0,1,1,1,0,1,0,1}^{0,0,0}
\end{split}\end{align}
\end{minipage}%

\subsection*{Sector: $\cI_{52}$}

\begin{minipage}{0.4\textwidth}
\hspace{\parindent}
\begin{figure}[H]
\centering
\includegraphics[width=3cm]{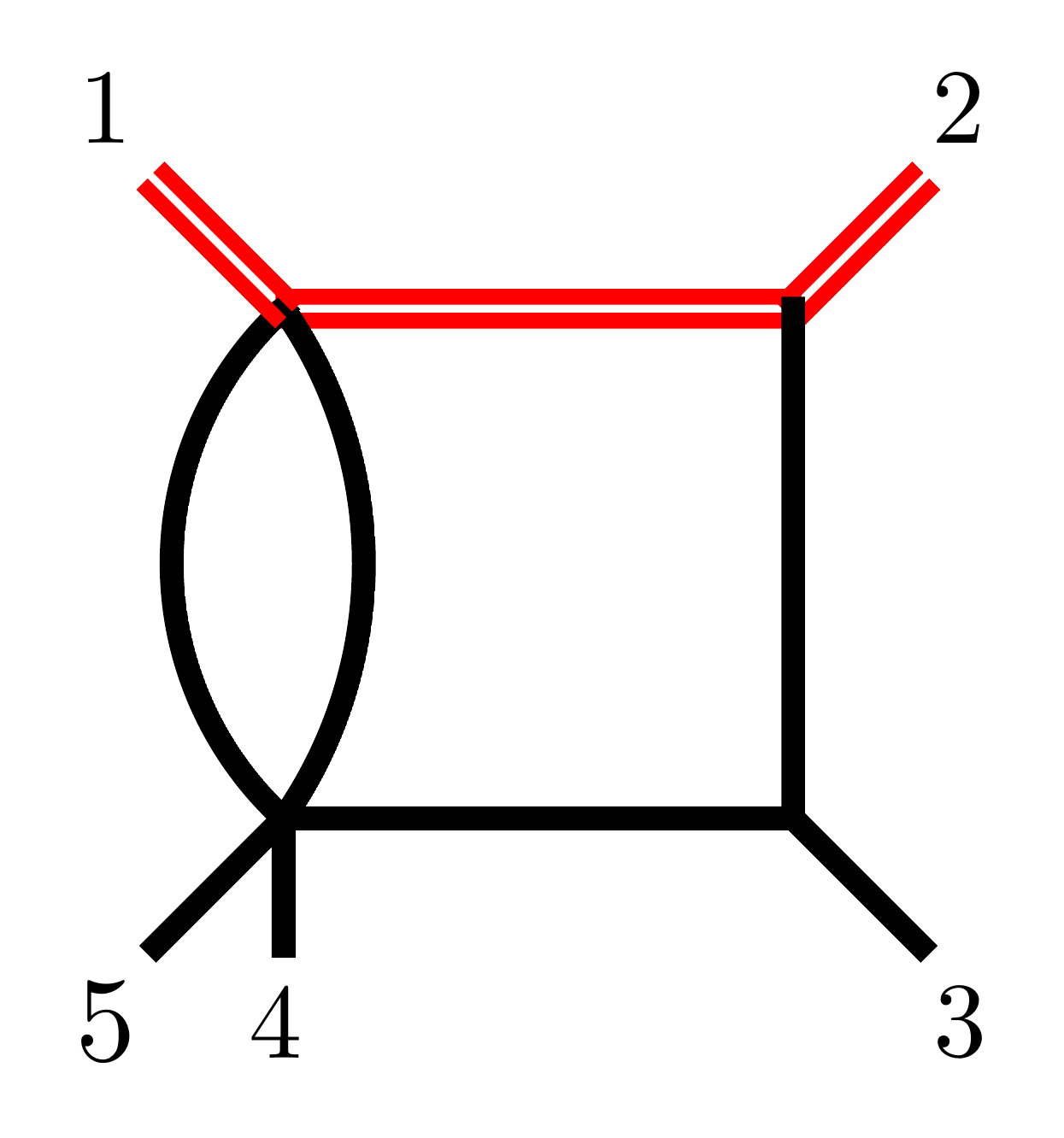}  
\end{figure}
\end{minipage}
\begin{minipage}{0.54\textwidth}
\begin{align}\begin{split}
\cI_{52} = & -2 d_{23} \, (2 \epsilon -1) \, \epsilon ^3 \, I_{0,1,1,1,1,0,0,1}^{0,0,0}
\end{split}\end{align}
\end{minipage}%

\subsection*{Sector: $\cI_{55}$}

\begin{minipage}{0.4\textwidth}
\hspace{\parindent}
\begin{figure}[H]
\centering
\includegraphics[width=3cm]{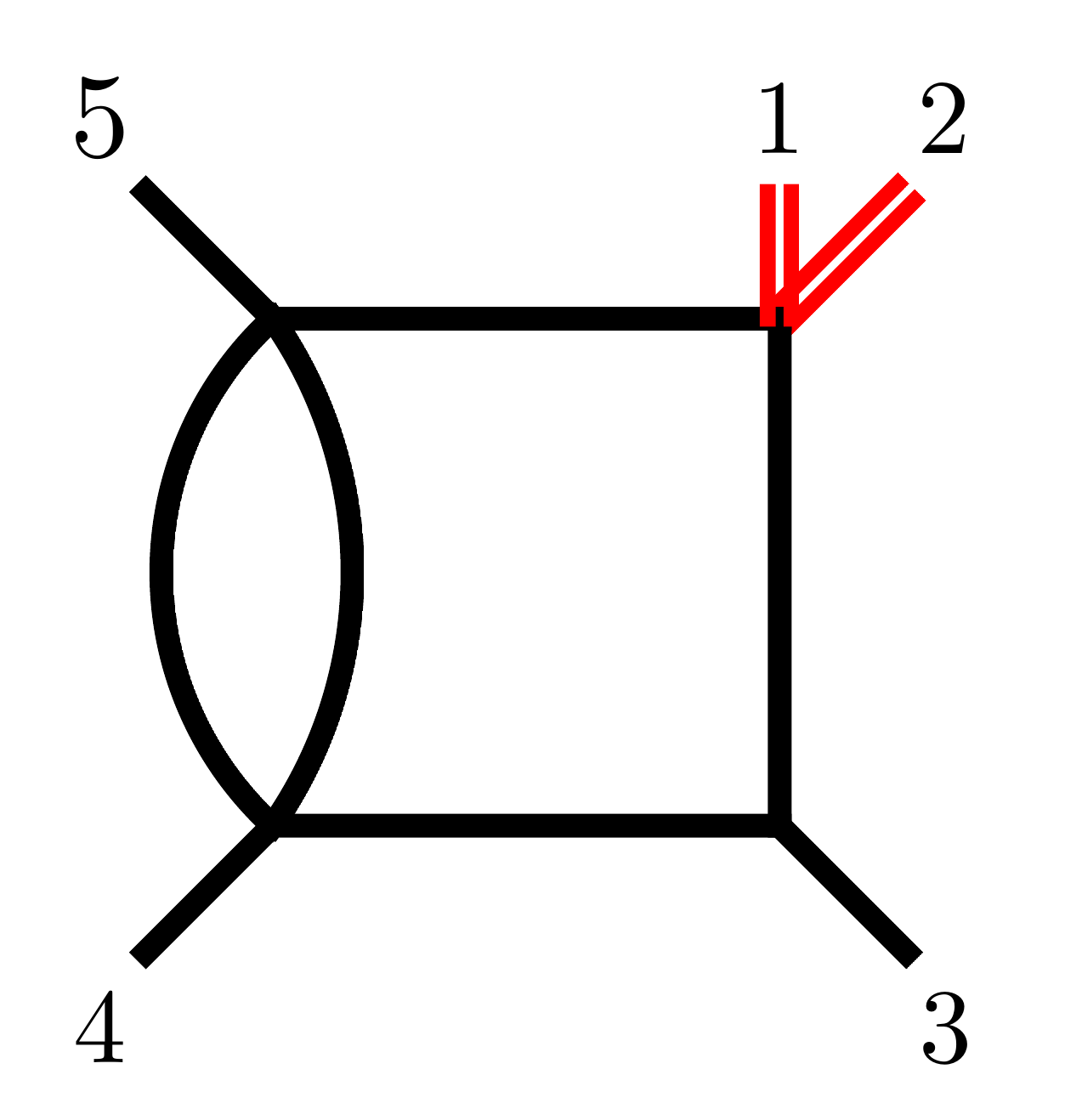}  
\end{figure}
\end{minipage}
\begin{minipage}{0.54\textwidth}
\begin{align}\begin{split}
\cI_{55} = & -2 (2 \epsilon -1) \, \epsilon ^3 \, \left(d_{12}-d_{45}+m_t^2\right) \, I_{1,0,1,1,0,1,0,1}^{0,0,0}
\end{split}\end{align}
\end{minipage}%

\subsection*{Sector: $\cI_{56}, \, \cI_{57}$}

\begin{minipage}{0.4\textwidth}
\hspace{\parindent}
\begin{figure}[H]
\centering
\includegraphics[width=3cm]{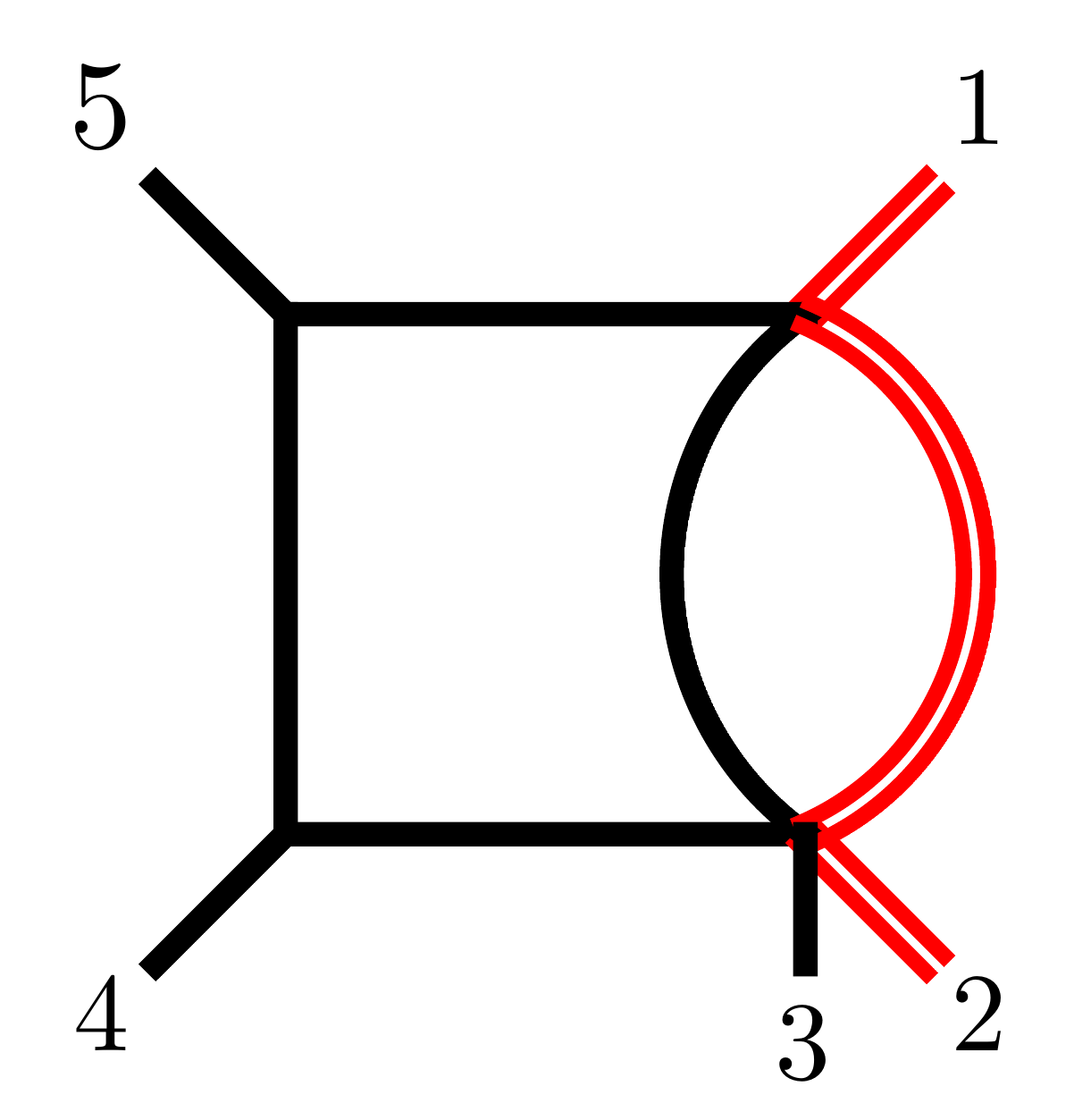}  
\end{figure}
\end{minipage}
\begin{minipage}{0.54\textwidth}
\begin{align}\begin{split}
\cI_{56} = & -2 d_{45} \, \epsilon ^3 \, (2 \epsilon -1) \, I_{0,1,0,0,1,1,1,1}^{0,0,0} \\
& + 4 d_{15} \, d_{45} \, \epsilon ^3 \, I_{0,2,0,0,1,1,1,1}^{0,0,0} \\
\cI_{57} = & 2 d_{45} \, \left(2 d_{15}+m_t^2\right) \, \epsilon ^3 \, I_{0,2,0,0,1,1,1,1}^{0,0,0}
\end{split}\end{align}
\end{minipage}%

\subsection*{Sector: $\cI_{58}, \, \cI_{59}$}

\begin{minipage}{0.4\textwidth}
\hspace{\parindent}
\begin{figure}[H]
\centering
\includegraphics[width=3cm]{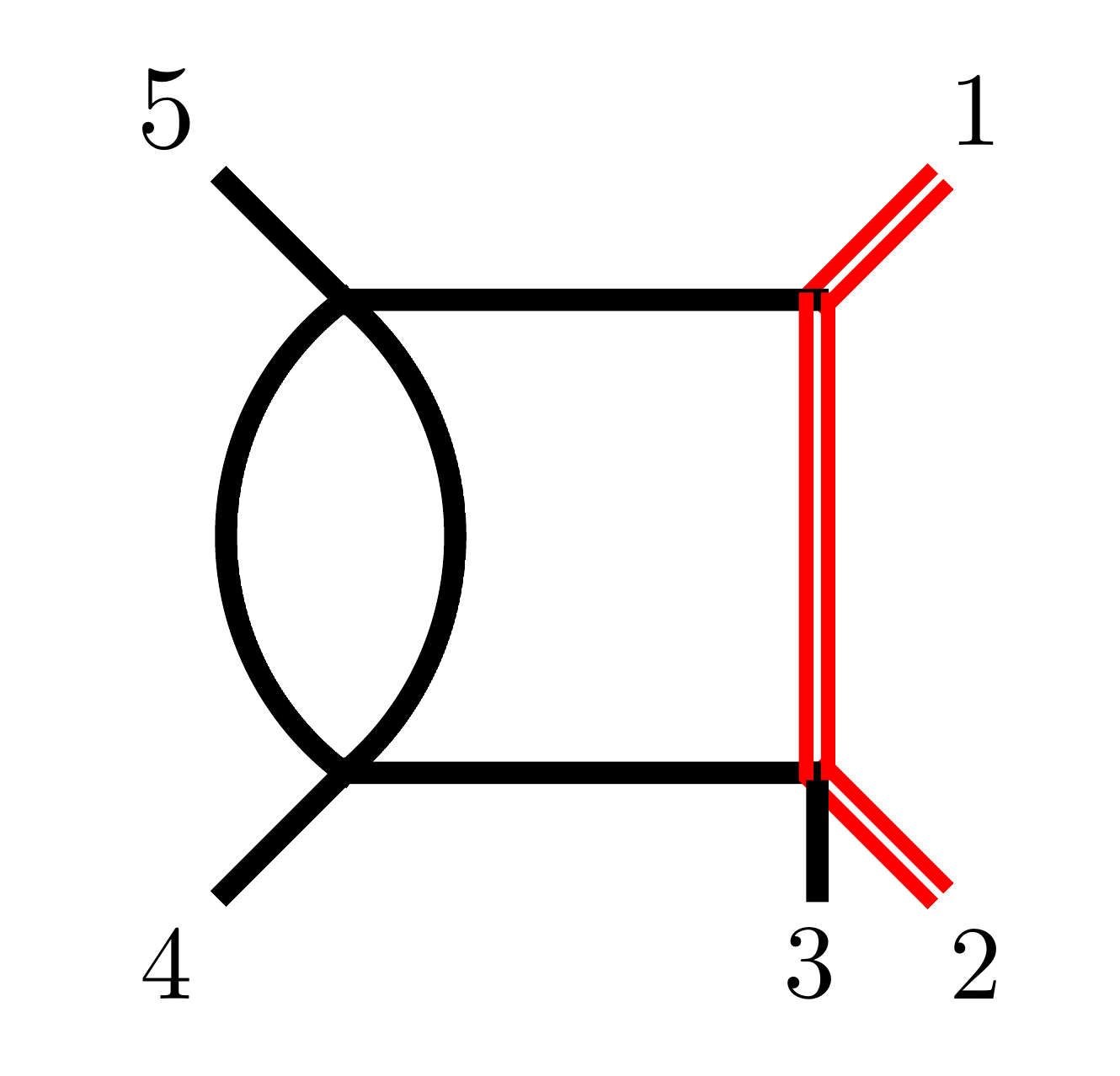}  
\end{figure}
\end{minipage}
\begin{minipage}{0.54\textwidth}
\begin{align}\begin{split}
\cI_{58} = & - \Delta_1 \, \epsilon ^3 \, (2 \epsilon -1) \, I_{1,1,0,1,0,1,0,1}^{0,0,0} \\
\cI_{59} = & \frac{3}{2} \left(d_{23}-d_{45}\right) \, \epsilon ^3 \, (2 \epsilon -1) \, I_{1,1,0,1,0,1,0,1}^{0,0,0} \\
& - d_{45} \, m_t^2 \, \epsilon ^2 \, (2 \epsilon -1) \, I_{1,2,0,1,0,1,0,1}^{0,0,0}
\end{split}\end{align}
\end{minipage}%

\subsection*{Sector: $\cI_{60}$}

\begin{minipage}{0.4\textwidth}
\hspace{\parindent}
\begin{figure}[H]
\centering
\includegraphics[width=3cm]{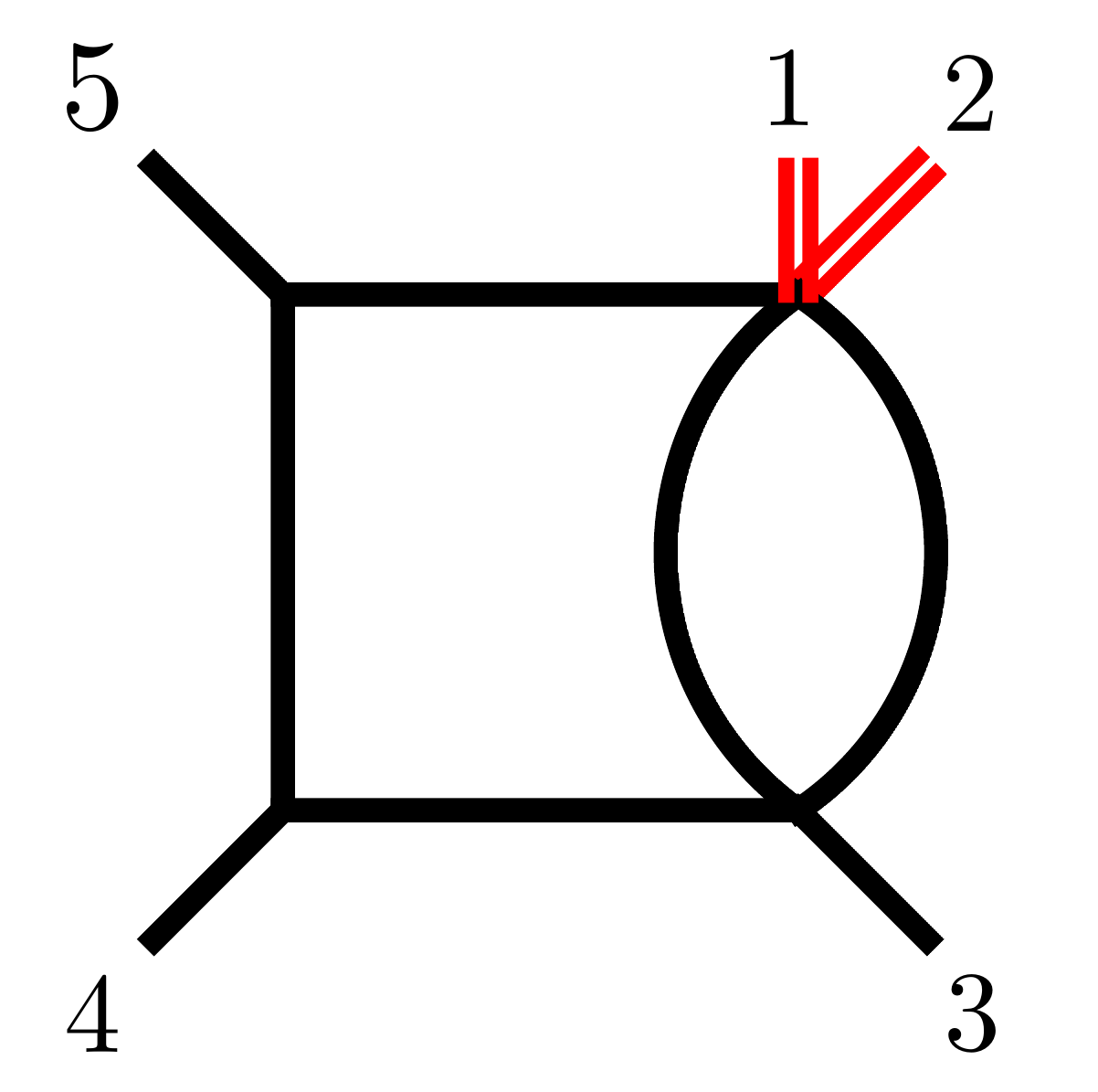}  
\end{figure}
\end{minipage}
\begin{minipage}{0.54\textwidth}
\begin{align}\begin{split}
\cI_{60} = & -2 d_{45} \, \epsilon ^3 \, (2 \epsilon -1) \, I_{0,0,1,0,1,1,1,1}^{0,0,0}
\end{split}\end{align}
\end{minipage}%

\subsection*{Sector: $\cI_{61}$}

\begin{minipage}{0.4\textwidth}
\hspace{\parindent}
\begin{figure}[H]
\centering
\includegraphics[width=3.5cm]{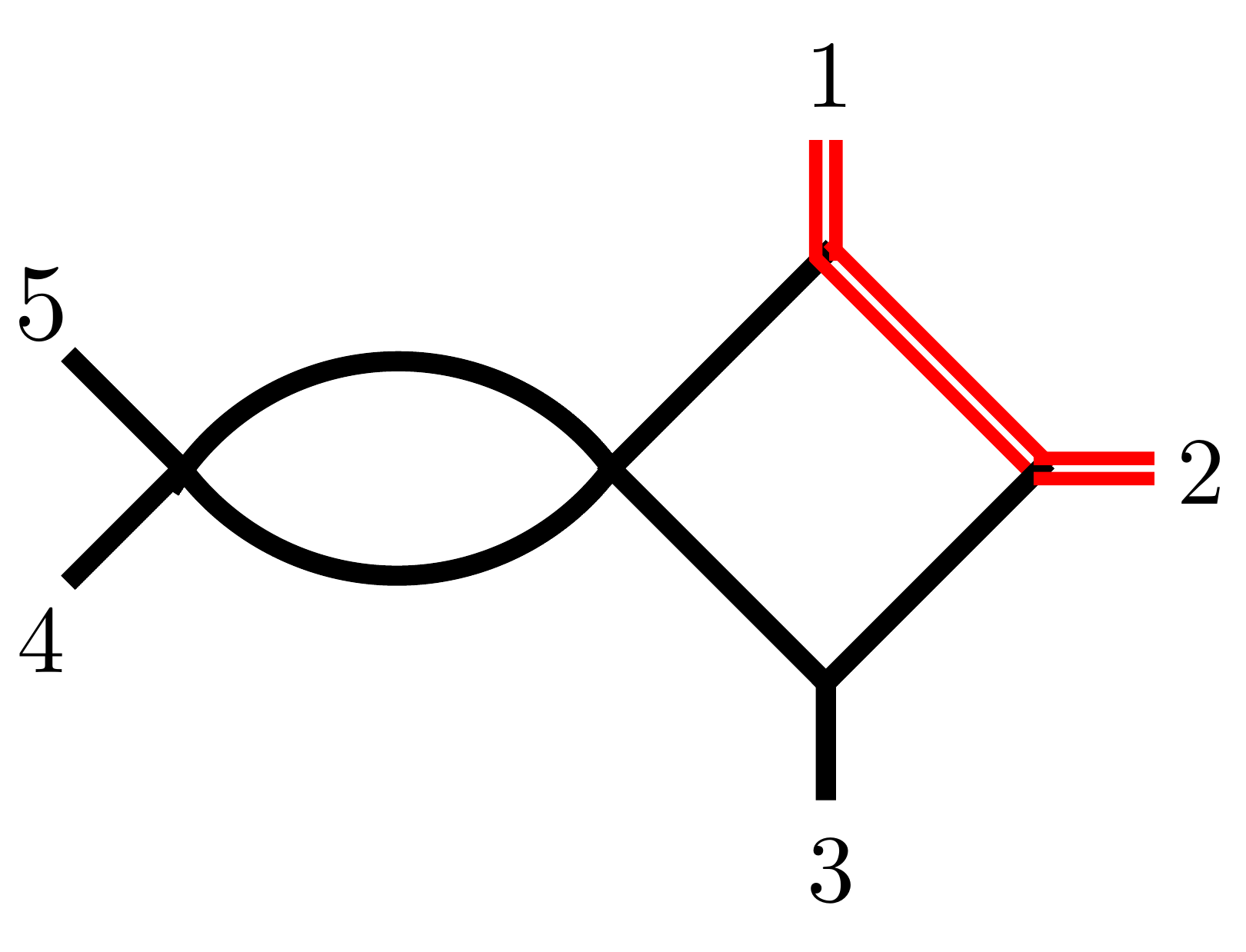}  
\end{figure}
\end{minipage}
\begin{minipage}{0.54\textwidth}
\begin{align}\begin{split}
\cI_{61} = & -4 d_{23} \, \left(d_{12}+m_t^2\right) \, \epsilon ^3 \, (2 \epsilon -1) \, I_{1,1,1,1,1,0,1,0}^{0,0,0}
\end{split}\end{align}
\end{minipage}%

\subsection*{Sector: $\cI_{62}$}

\begin{minipage}{0.4\textwidth}
\hspace{\parindent}
\begin{figure}[H]
\centering
\includegraphics[width=3.5cm]{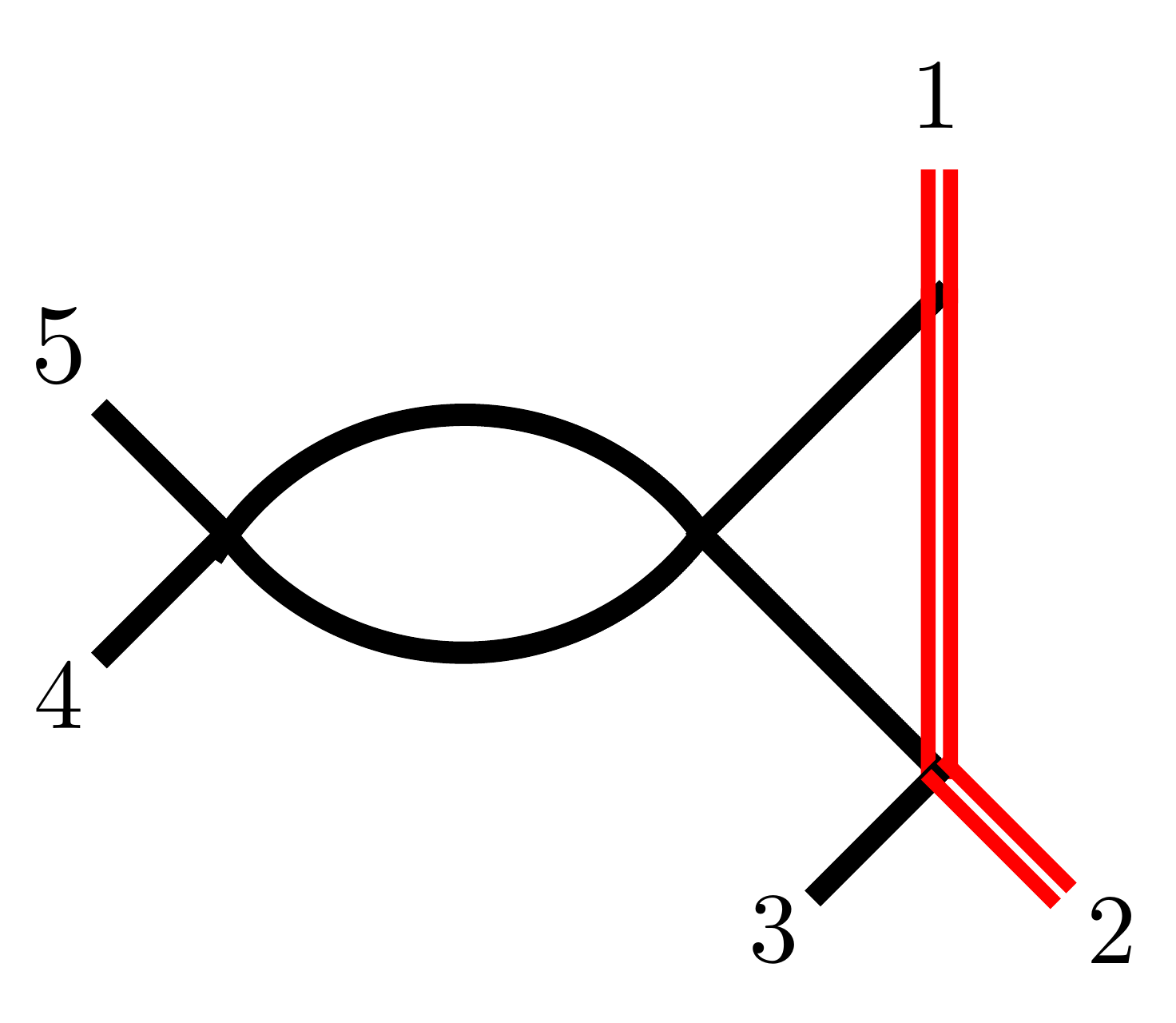}  
\end{figure}
\end{minipage}
\begin{minipage}{0.54\textwidth}
\begin{align}\begin{split}
\cI_{62} = & - \Delta_1 \, \epsilon ^3 \, (2 \epsilon -1) \, I_{1,1,0,1,1,0,1,0}^{0,0,0}
\end{split}\end{align}
\end{minipage}%

\subsection*{Sector: $\cI_{63}$}

\begin{minipage}{0.4\textwidth}
\hspace{\parindent}
\begin{figure}[H]
\centering
\includegraphics[width=3cm]{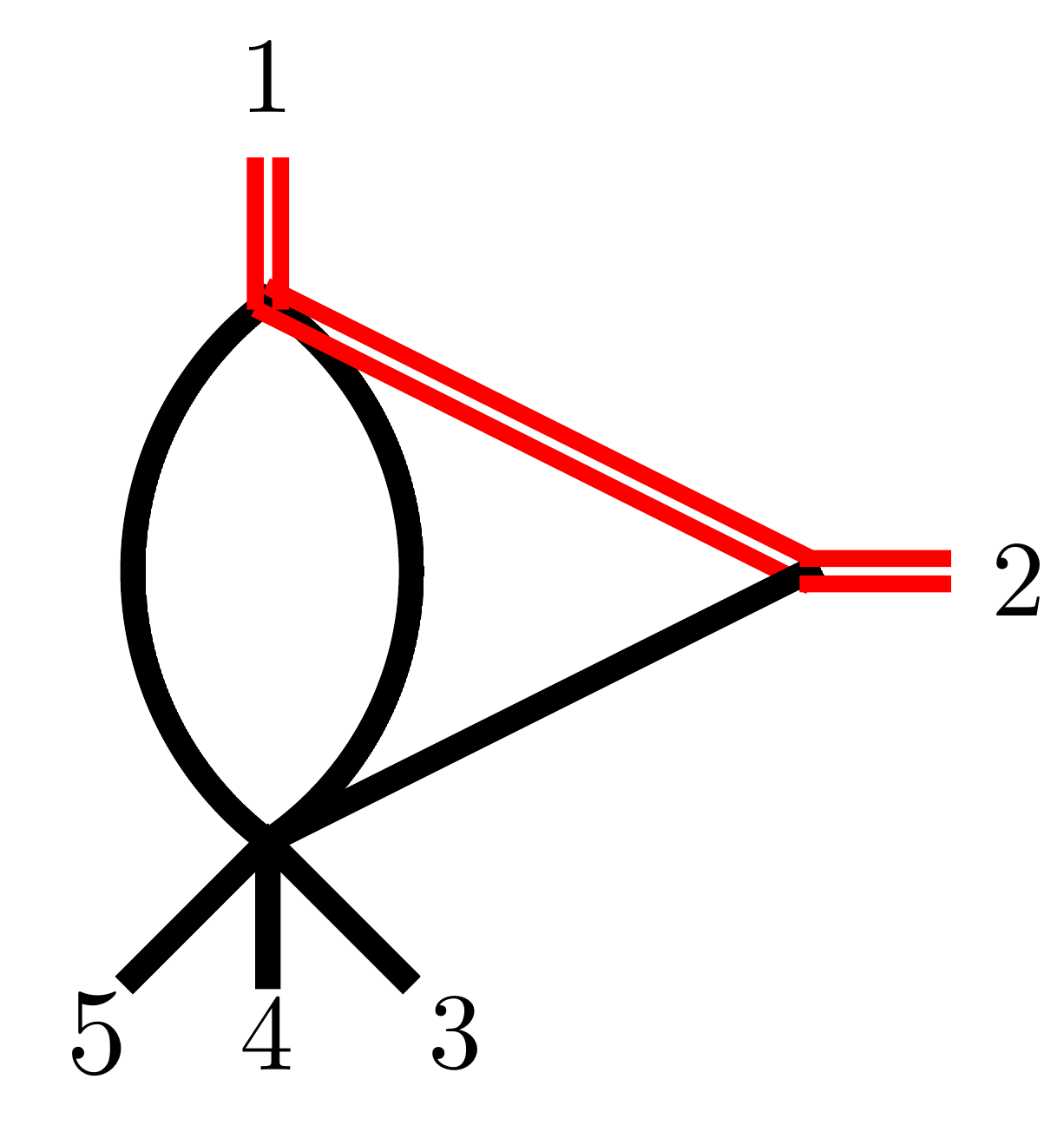}  
\end{figure}
\end{minipage}
\begin{minipage}{0.54\textwidth}
\begin{align}\begin{split}
\cI_{63} = & \beta \, \epsilon ^2 \, (2 \epsilon -1) (3 \epsilon -1) \, I_{0,1,1,0,1,0,0,1}^{0,0,0} \\
& - \frac{\beta \,  \epsilon ^2 \, (2 \epsilon -1) (3 \epsilon -2) (3 \epsilon -1)}{2 (4 \epsilon -1) \, m_t^2} \, I_{0,1,0,0,1,0,0,1}^{0,0,0}
\end{split}\end{align}
\end{minipage}%

\subsection*{Sectors: $\cI_{64}$ and  $\cI_{75}$}

\hspace{\parindent}
\begin{figure}[H]
\includegraphics[width=3cm]{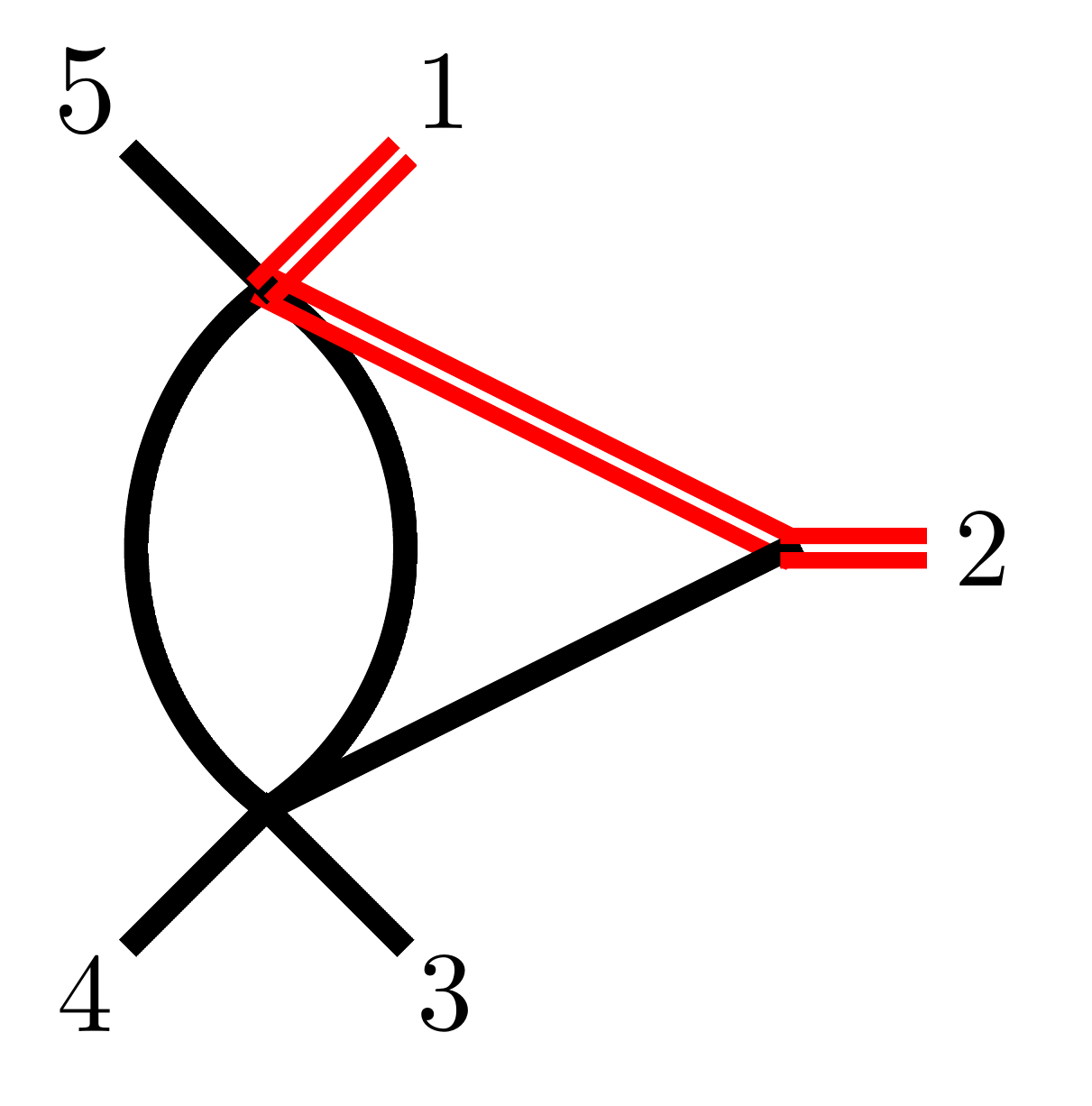}
\includegraphics[width=3cm]{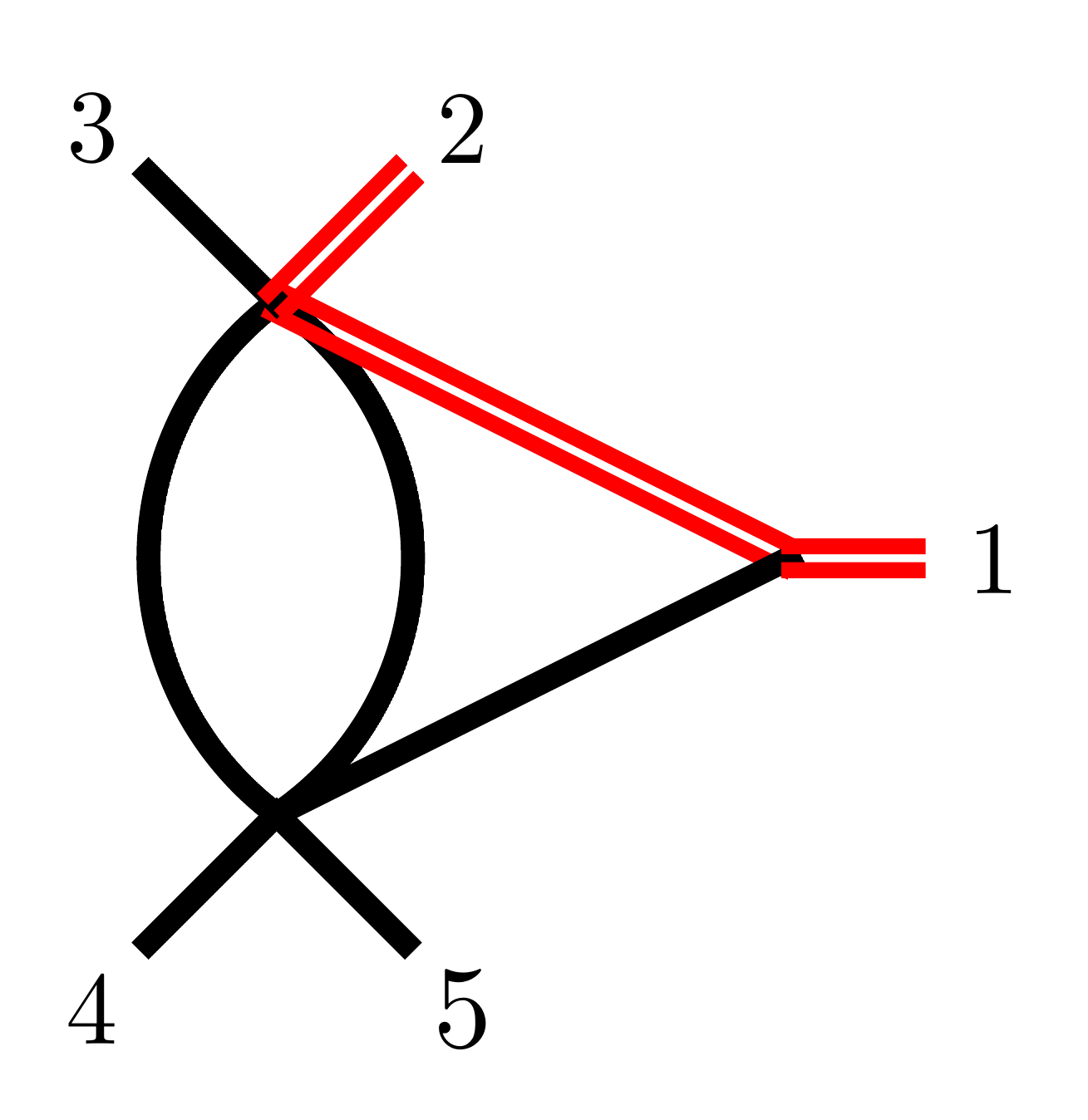}    
\end{figure}

\begin{flalign}
\cI_{64} = & \frac{\Delta _2 \, \epsilon ^2 \, (2 \epsilon -1) (3 \epsilon -1)}{2 d_{34}} \, I_{0,1,1,0,0,1,0,1}^{0,0,0} + \frac{3 \Delta _2 \, \epsilon ^2 \, (\epsilon -1) (2 \epsilon -1) \left(d_{15}+m_t^2\right)}{4 d_{15}^2 \, d_{34} \, m_t^2} \, I_{0,1,0,0,0,1,0,1}^{0,1,0} \nn \\
& -\frac{\Delta _2 \, \epsilon ^2\, (2 \epsilon -1) \left(-5 d_{15} \epsilon \, m_t^2+3 d_{15} \, m_t^2+2 d_{15}^2 \, \epsilon -2 d_{15}^2-4 \epsilon \, m_t^4+3 m_t^4\right)}{4 d_{15}^2 \, d_{34}  \, m_t^2} \, I_{0,1,0,0,0,1,0,1}^{0,0,0} && \nn \\
\cI_{75} = &  \frac{\Delta _1 \, \epsilon ^2 \, (2 \epsilon -1) (3 \epsilon -1)}{2 d_{45}} \, I_{1,1,0,0,0,0,1,1}^{0,0,0} + \frac{\Delta _1 \, \epsilon ^2 \, (2 \epsilon -1) (3 \epsilon -2)}{4 d_{23} \, d_{45}} \, I_{0,1,0,0,0,0,1,1}^{0,0,0}  \nn \\
& + \frac{\Delta _1 \, \epsilon ^2 \, (2 \epsilon -1) \left(d_{23}+m_t^2\right)}{2 d_{23} \, d_{45}} \, I_{0,2,0,0,0,0,1,1}^{0,0,0} &&
\end{flalign}

\subsection*{Sector: $\cI_{65}$}

\begin{minipage}{0.4\textwidth}
\hspace{\parindent}
\begin{figure}[H]
\centering
\includegraphics[width=3cm]{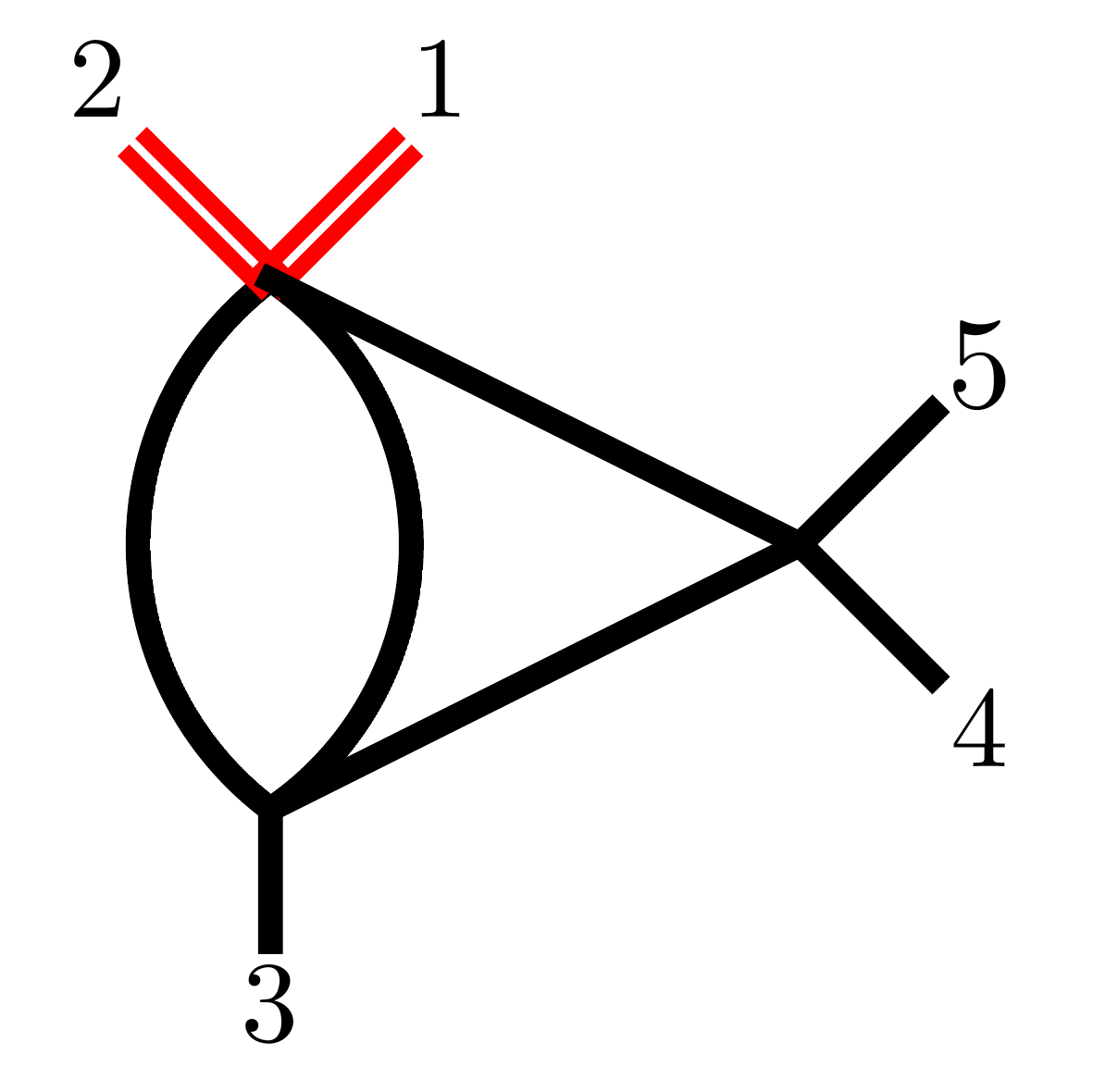}  
\end{figure}
\end{minipage}
\begin{minipage}{0.54\textwidth}
\begin{align}\begin{split}
\cI_{65} = & \epsilon ^2 \, (2 \epsilon -1) (3 \epsilon -1) \, I_{0,0,1,0,1,0,1,1}^{0,0,0}
\end{split}\end{align}
\end{minipage}%

\subsection*{Sector: $\cI_{66}$}

\begin{minipage}{0.4\textwidth}
\hspace{\parindent}
\begin{figure}[H]
\centering
\includegraphics[width=3cm]{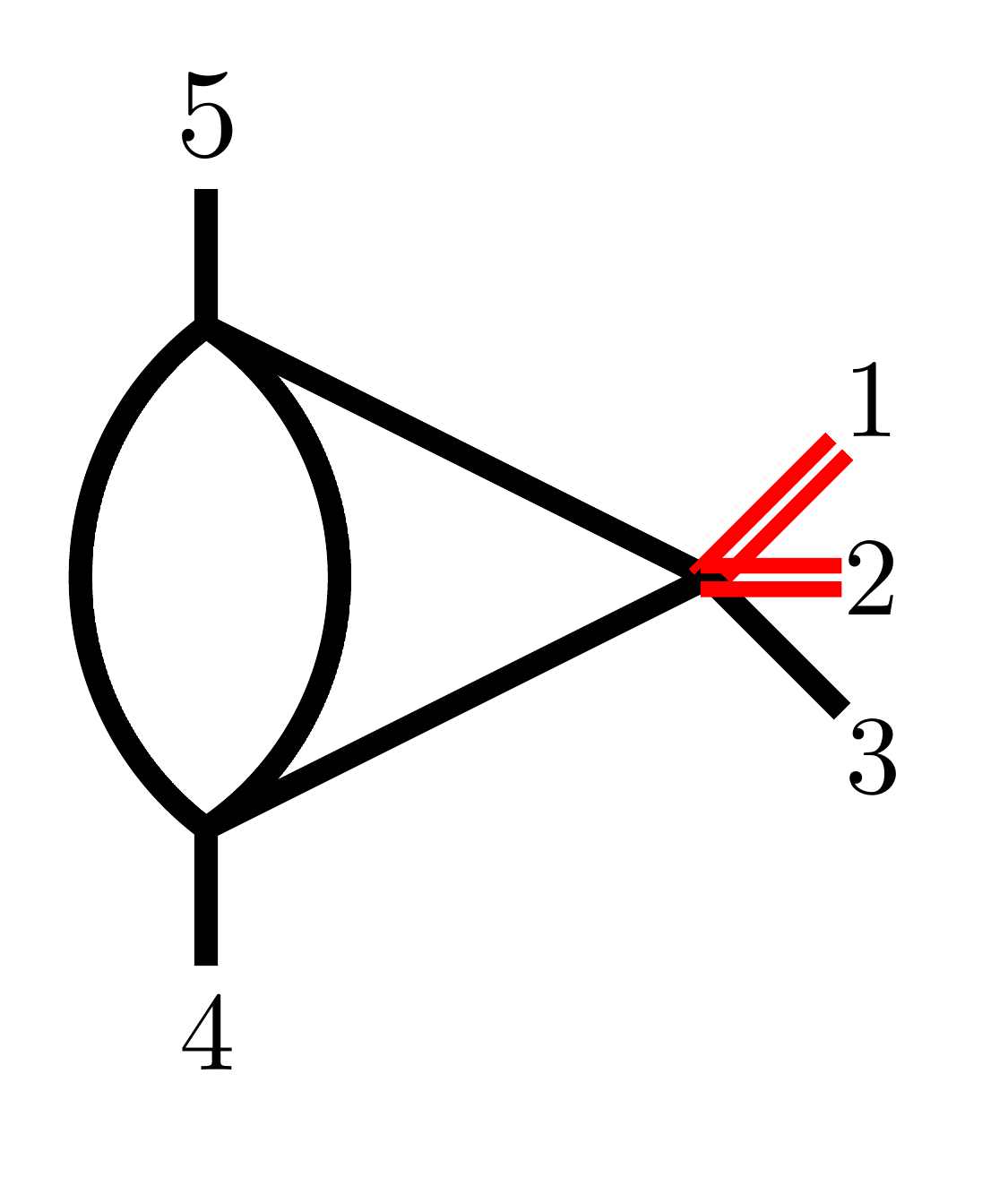}  
\end{figure}
\end{minipage}
\begin{minipage}{0.54\textwidth}
\begin{align}\begin{split}
\cI_{66} = & \epsilon ^2 \, (2 \epsilon -1) (3 \epsilon -1) \,I_{1,0,0,1,0,1,0,1}^{0,0,0}
\end{split}\end{align}
\end{minipage}%

\subsection*{Sectors: $\cI_{67}$ and $\cI_{68}$}

\begin{minipage}{0.4\textwidth}
\begin{figure}[H]
\includegraphics[width=3cm]{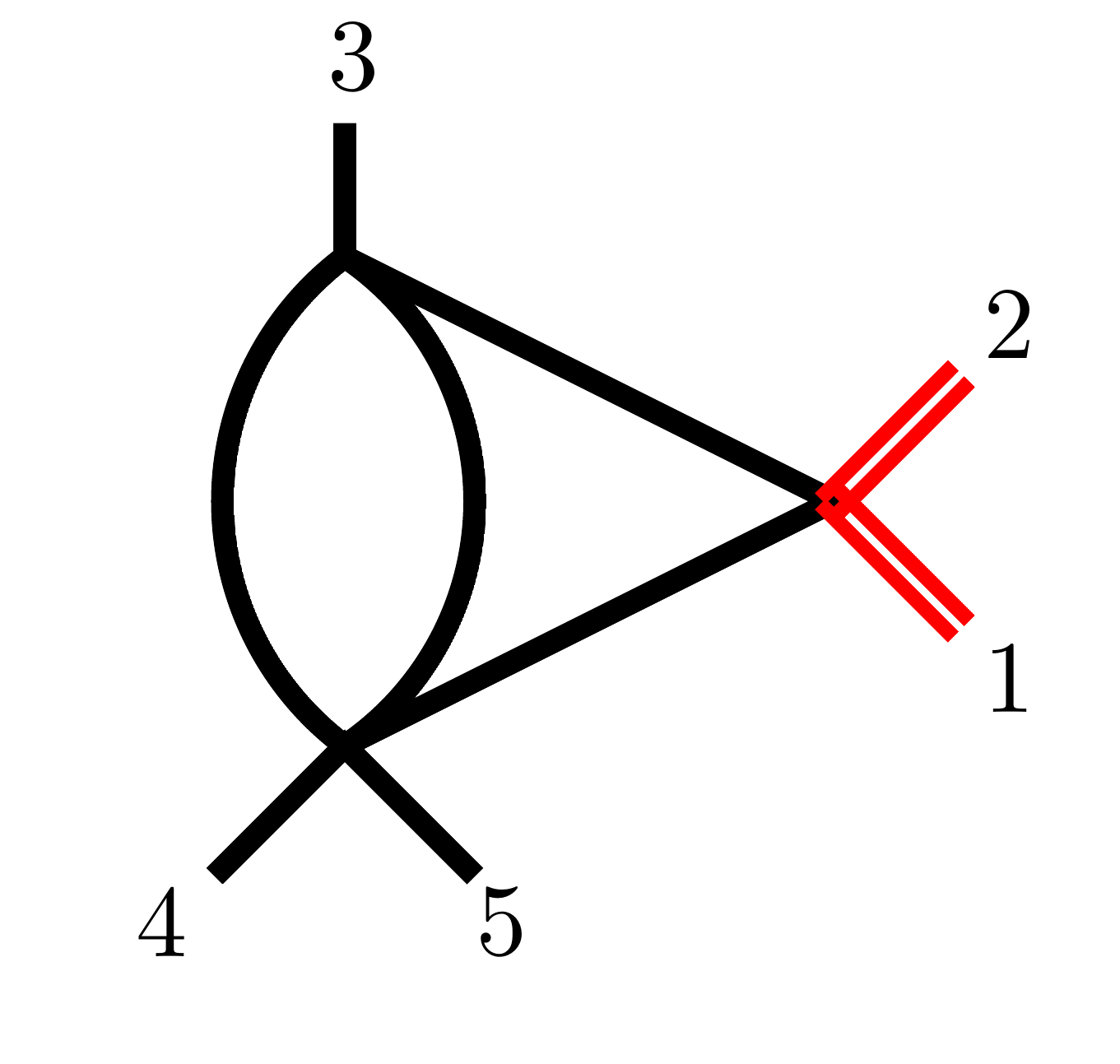}
\includegraphics[width=3cm]{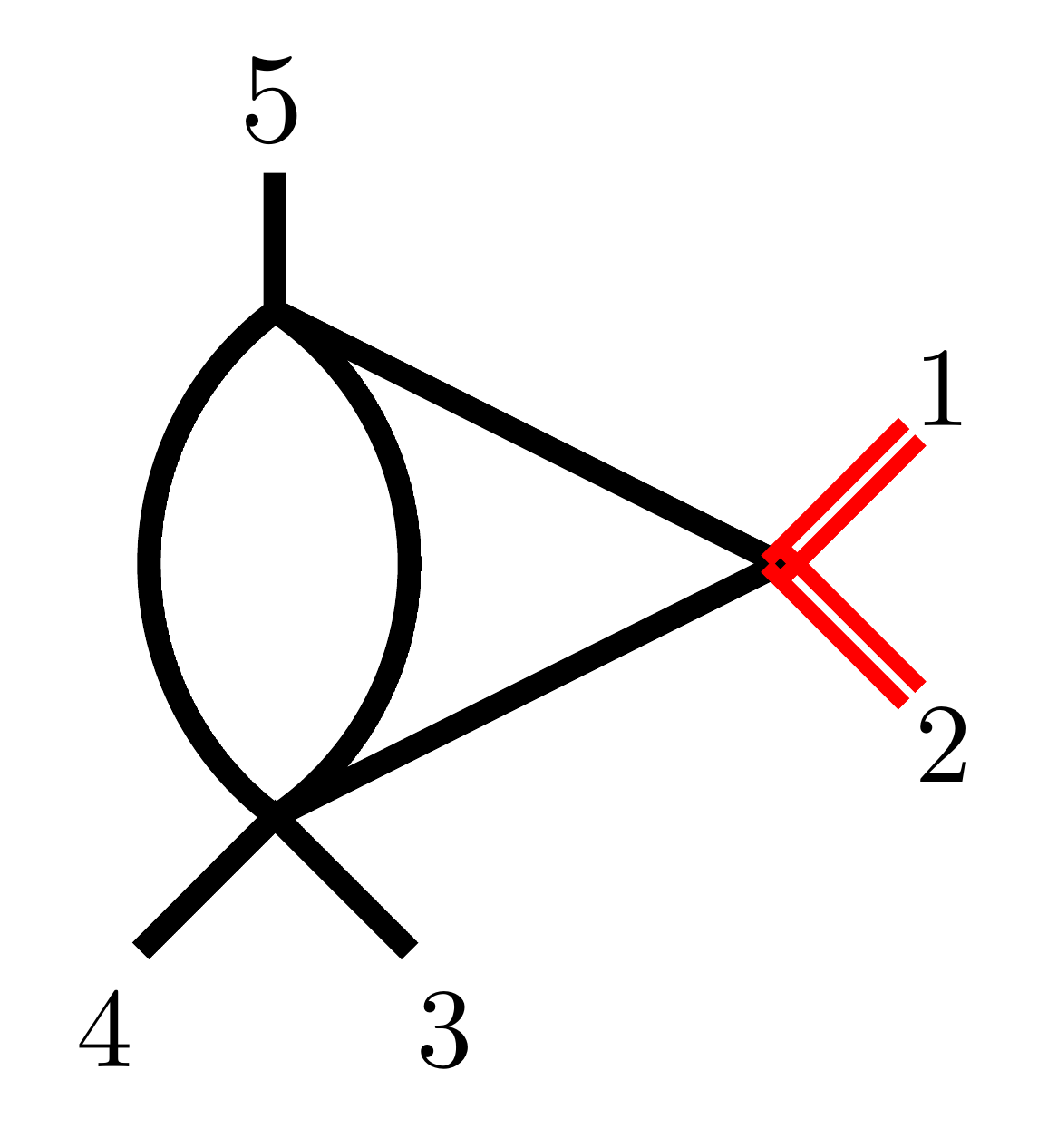}   
\end{figure}
\end{minipage}
\begin{minipage}{0.54\textwidth}
\begin{align}\begin{split}
\cI_{67} = & \epsilon ^2 \, (2 \epsilon -1) (3 \epsilon -1) \, I_{1,0,1,0,0,0,1,1}^{0,0,0} \\
\cI_{68} = & \epsilon ^2 \, (2 \epsilon -1) (3 \epsilon -1) \,  I_{1,0,1,0,0,1,0,1}^{0,0,0}
\end{split}\end{align}
\end{minipage}%

\subsection*{Sector: $\cI_{69}, \, \cI_{70}, \, \cI_{71}$}

\begin{minipage}{0.4\textwidth}
\hspace{\parindent}
\begin{figure}[H]
\centering
\includegraphics[width=3cm]{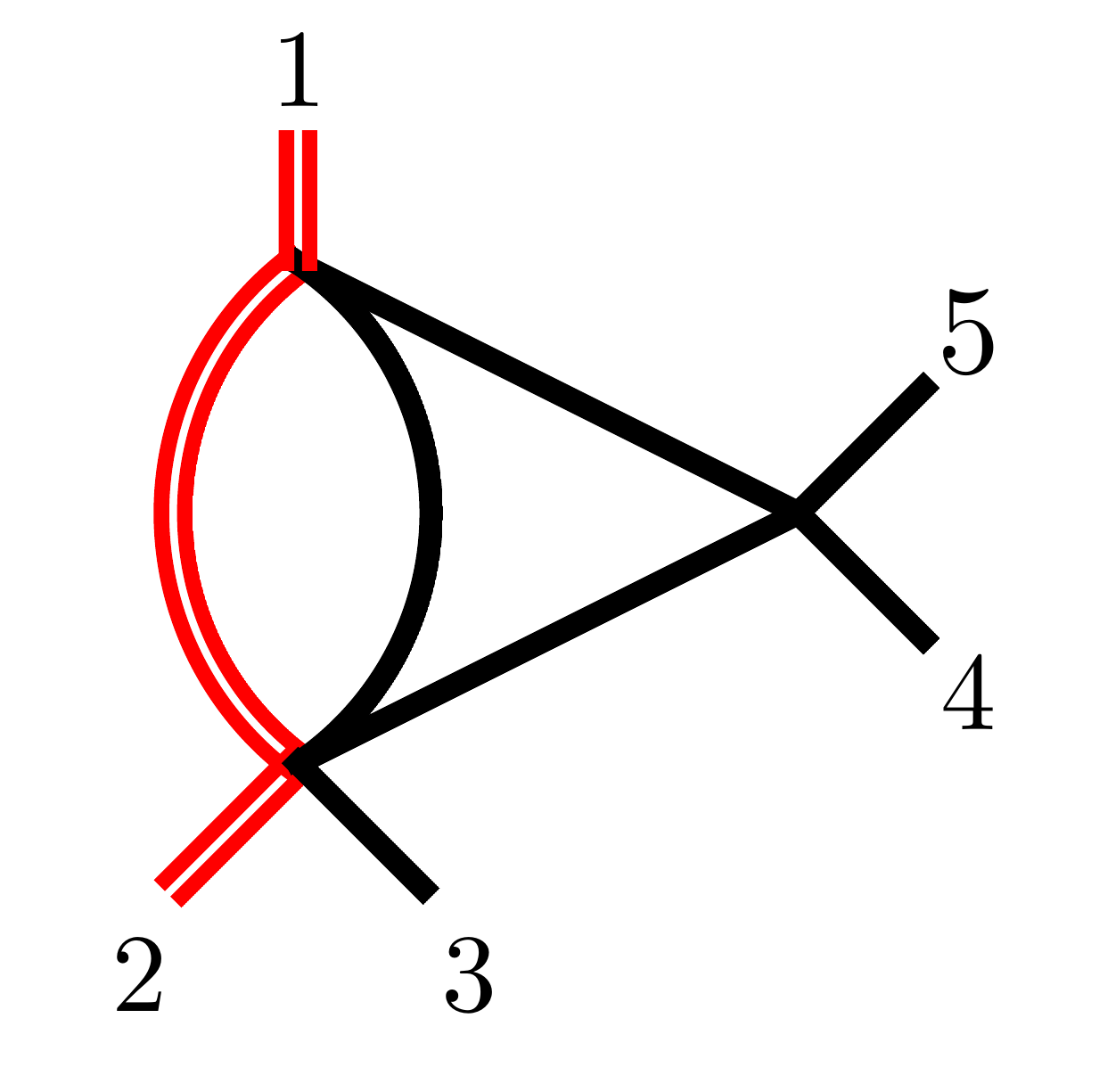}  
\end{figure}
\end{minipage}
\begin{minipage}{0.54\textwidth}
\begin{align}\begin{split}
\cI_{69} = & \epsilon ^2 \, (2 \epsilon -1) (3 \epsilon -1) \, I_{0,1,0,0,1,0,1,1}^{0,0,0} \\
& - \epsilon ^3 \, \left(d_{23}-d_{45}\right) \, I_{0,1,0,0,1,0,1,2}^{0,0,0}  \\
& + \epsilon ^2 (4 \epsilon -1) \, m_t^2 \, I_{0,2,0,0,1,0,1,1}^{0,0,0}  \\
\cI_{70} = & \Delta _1 \, \epsilon ^3 \, I_{0,2,0,0,1,0,1,1}^{0,0,0}  \\
\cI_{71} = & \Delta _1 \, \epsilon ^3 \, I_{0,1,0,0,1,0,1,2}^{0,0,0}
\end{split}\end{align}
\end{minipage}%

\subsection*{Sector: $\cI_{72}, \, \cI_{73}$}

\begin{minipage}{0.4\textwidth}
\hspace{\parindent}
\begin{figure}[H]
\centering
\includegraphics[width=3cm]{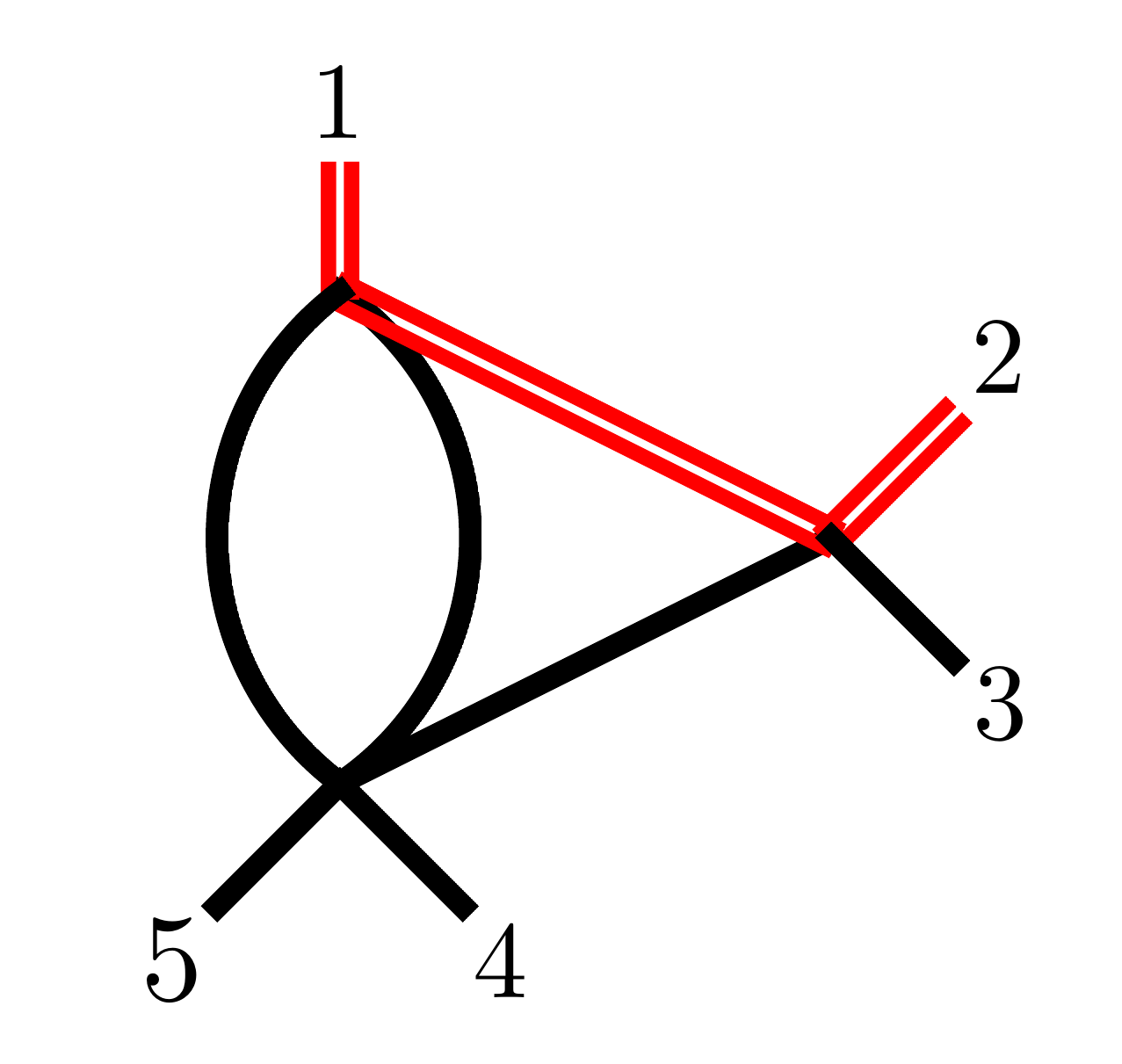}  
\end{figure}
\end{minipage}
\begin{minipage}{0.54\textwidth}
\begin{align}\begin{split}
\cI_{72} = & \frac{\epsilon ^2 \, (2 \epsilon -1) (3 \epsilon -1) \left(2 d_{23}+m_t^2\right)}{2 d_{23}} \, I_{0,1,0,1,1,0,0,1}^{0,0,0} \\
& + \frac{\epsilon ^3 \, \left(-d_{45} m_t^2+d_{23}^2-d_{45} d_{23}\right)}{d_{23}} \, I_{0,1,0,1,1,0,0,2}^{0,0,0} \\
& -\frac{\epsilon ^2 \, (2 \epsilon -1) (3 \epsilon -2) (3 \epsilon -1) \left(d_{23}+m_t^2\right)}{4 d_{23} \, m_t^2 \, (4 \epsilon -1)} \, I_{0,1,0,0,1,0,0,1}^{0,0,0}  \\
\cI_{73} = & \Delta _1 \, \epsilon ^3 \, I_{0,1,0,1,1,0,0,2}^{0,0,0}
\end{split}\end{align}
\end{minipage}%

\subsection*{Sector: $\cI_{74}$}

\begin{minipage}{0.4\textwidth}
\hspace{\parindent}
\begin{figure}[H]
\centering
\includegraphics[width=3cm]{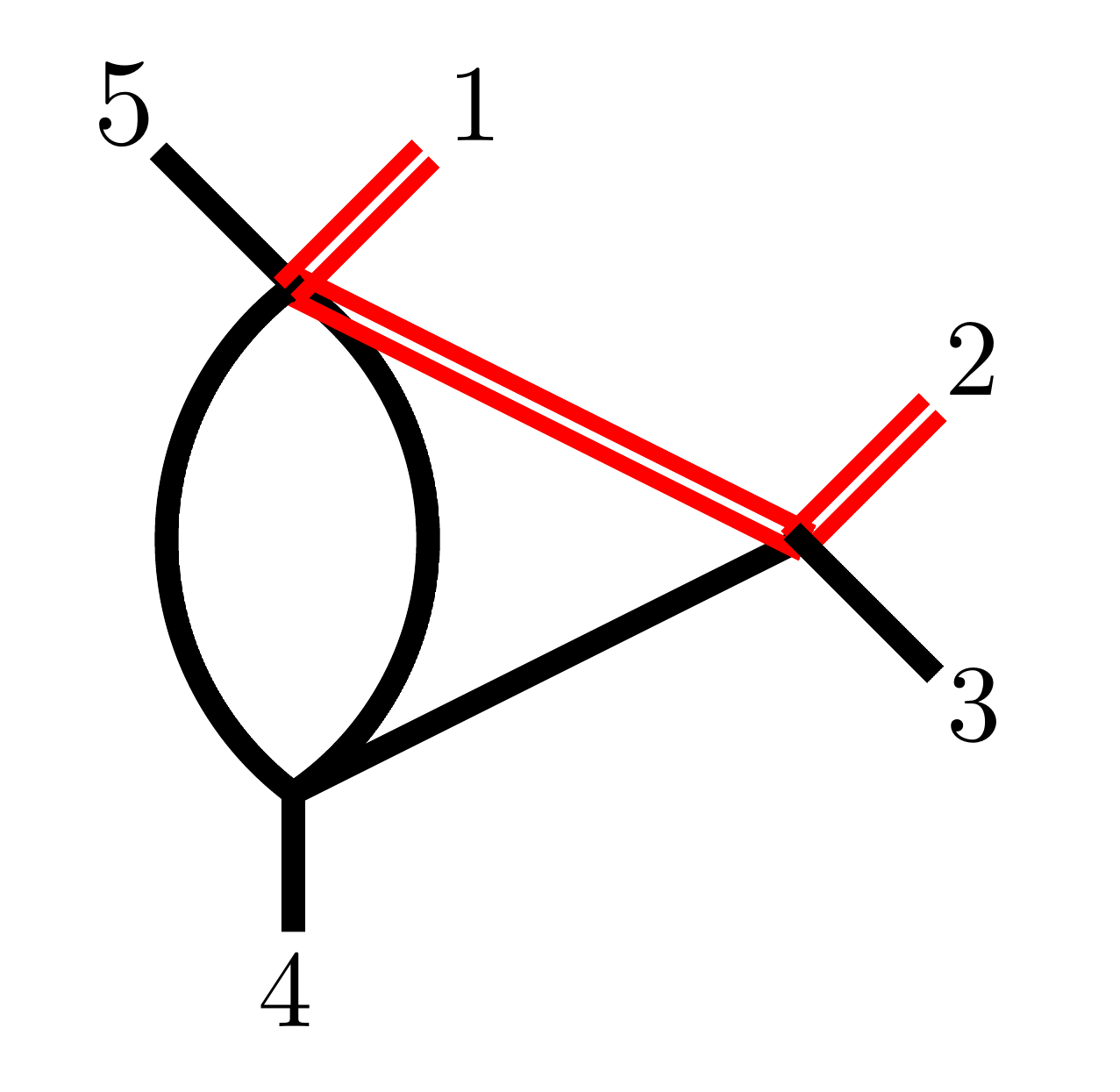}  
\end{figure}
\end{minipage}
\begin{minipage}{0.54\textwidth}
\begin{align}\begin{split}
\cI_{74} = & \frac{\epsilon ^2 \, (2 \epsilon -1) (3 \epsilon -1) \left(2 d_{23}+m_t^2\right)}{2 d_{23}} \, I_{0,1,0,1,0,1,0,1}^{0,0,0} \\
& + \frac{\epsilon ^2 \, (2 \epsilon -1) \left(d_{15}+m_t^2\right) \left(2 d_{23}+m_t^2\right)}{2 d_{15} \, d_{23}} \, I_{0,2,0,0,0,1,0,1}^{0,0,0} \\
& + \frac{\epsilon ^2 \, (2 \epsilon -1) (3 \epsilon -2) \left(2 d_{23}+m_t^2\right)}{4 d_{15} \, d_{23}} \, I_{0,1,0,0,0,1,0,1}^{0,0,0} 
\end{split}\end{align}
\end{minipage}%

\subsection*{Sector: $\cI_{76}$}

\begin{minipage}{0.4\textwidth}
\hspace{\parindent}
\begin{figure}[H]
\centering
\includegraphics[width=3.5cm]{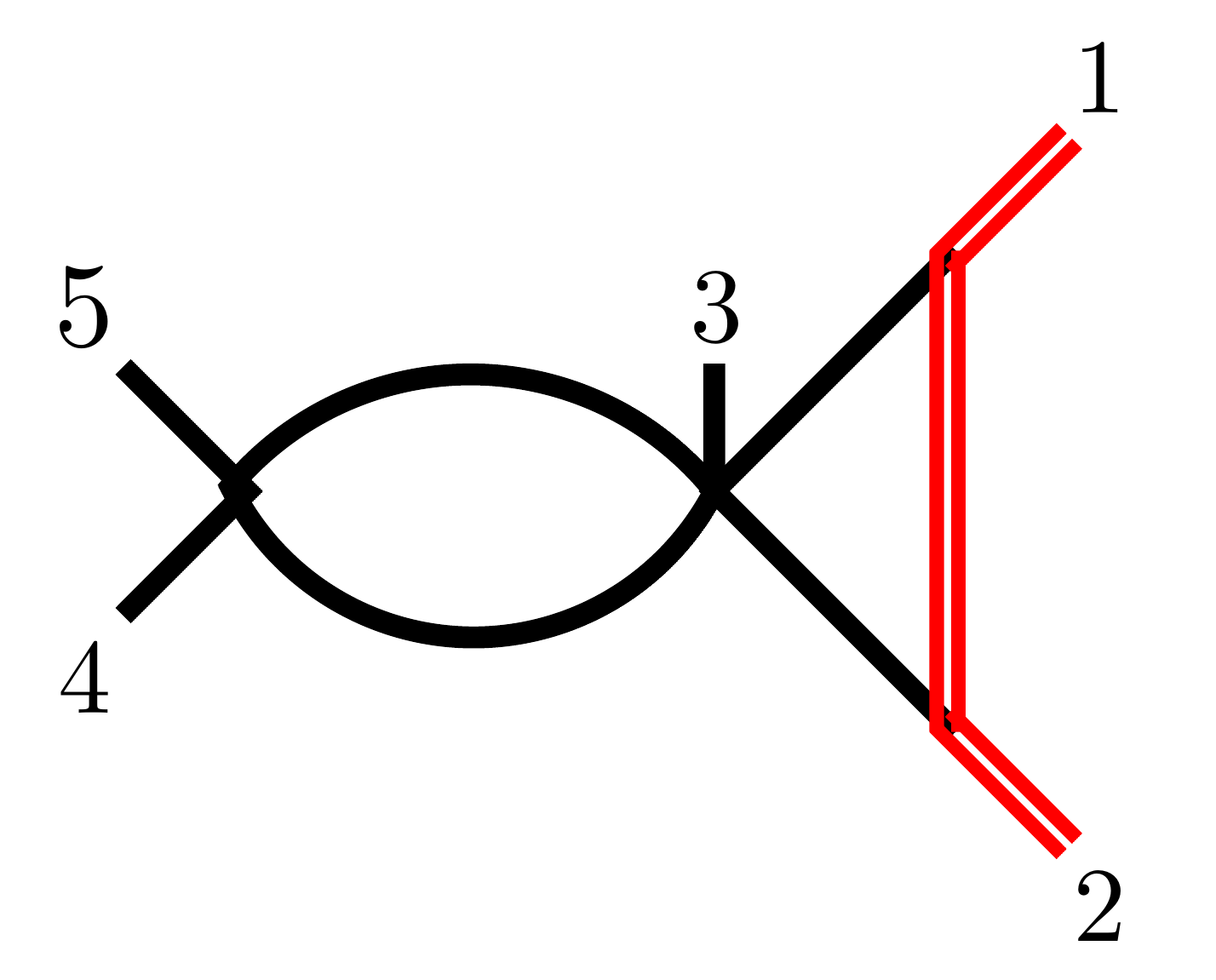}  
\end{figure}
\end{minipage}
\begin{minipage}{0.54\textwidth}
\begin{align}\begin{split}
\cI_{76} = & -2 \beta \,  \epsilon ^3 \, (2 \epsilon -1) \left(d_{12}+m_t^2\right) \, I_{1,1,1,0,1,0,1,0}^{0,0,0}
\end{split}\end{align}
\end{minipage}%

\subsection*{Sector: $\cI_{77}$}

\begin{minipage}{0.4\textwidth}
\hspace{\parindent}
\begin{figure}[H]
\centering
\includegraphics[width=3.5cm]{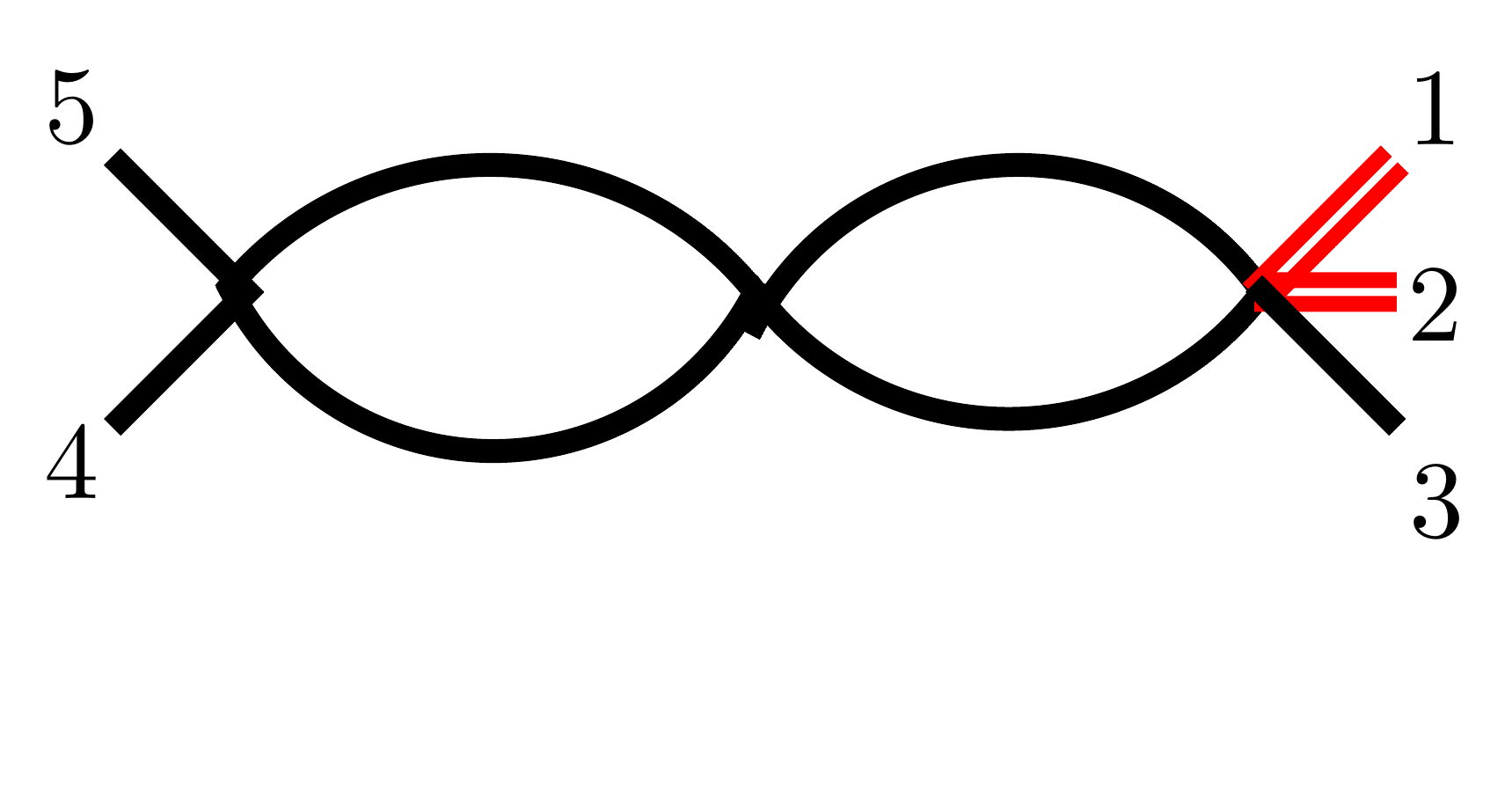}  
\end{figure}
\end{minipage}
\begin{minipage}{0.54\textwidth}
\begin{align}\begin{split}
\cI_{77} = & \epsilon ^2 \, (2 \epsilon -1)^2 \, I_{1,0,0,1,1,0,1,0}^{0,0,0}
\end{split}\end{align}
\end{minipage}%

\subsection*{Sector: $\cI_{78}$}

\begin{minipage}{0.4\textwidth}
\hspace{\parindent}
\begin{figure}[H]
\centering
\includegraphics[width=3.5cm]{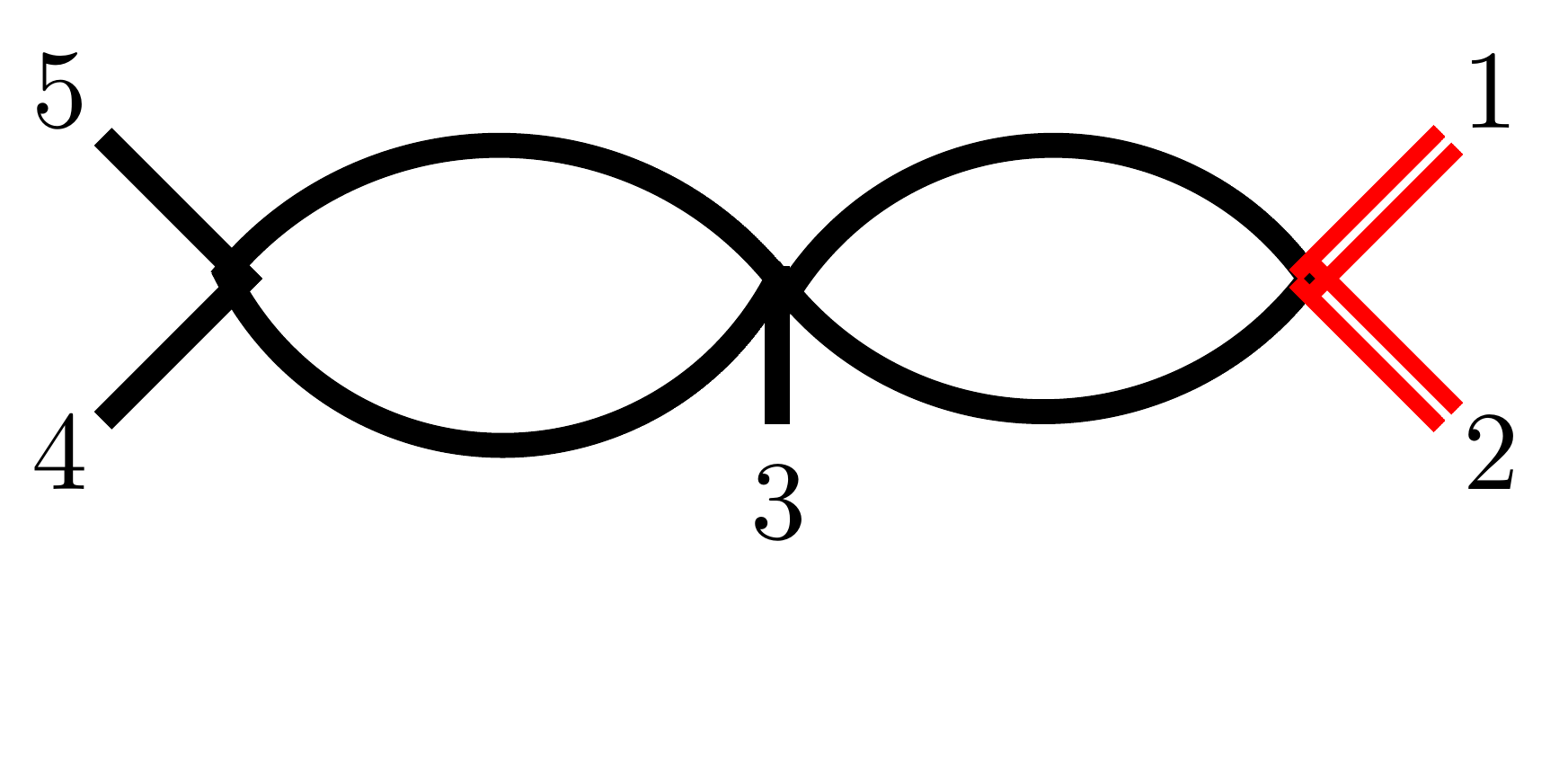}  
\end{figure}
\end{minipage}
\begin{minipage}{0.54\textwidth}
\begin{align}\begin{split}
\cI_{78} = & \epsilon ^2 \, (2 \epsilon -1)^2 \, I_{1,0,1,0,1,0,1,0}^{0,0,0}
\end{split}\end{align}
\end{minipage}%

\subsection*{Sector: $\cI_{79}$}

\begin{minipage}{0.4\textwidth}
\hspace{\parindent}
\begin{figure}[H]
\centering
\includegraphics[width=3.5cm]{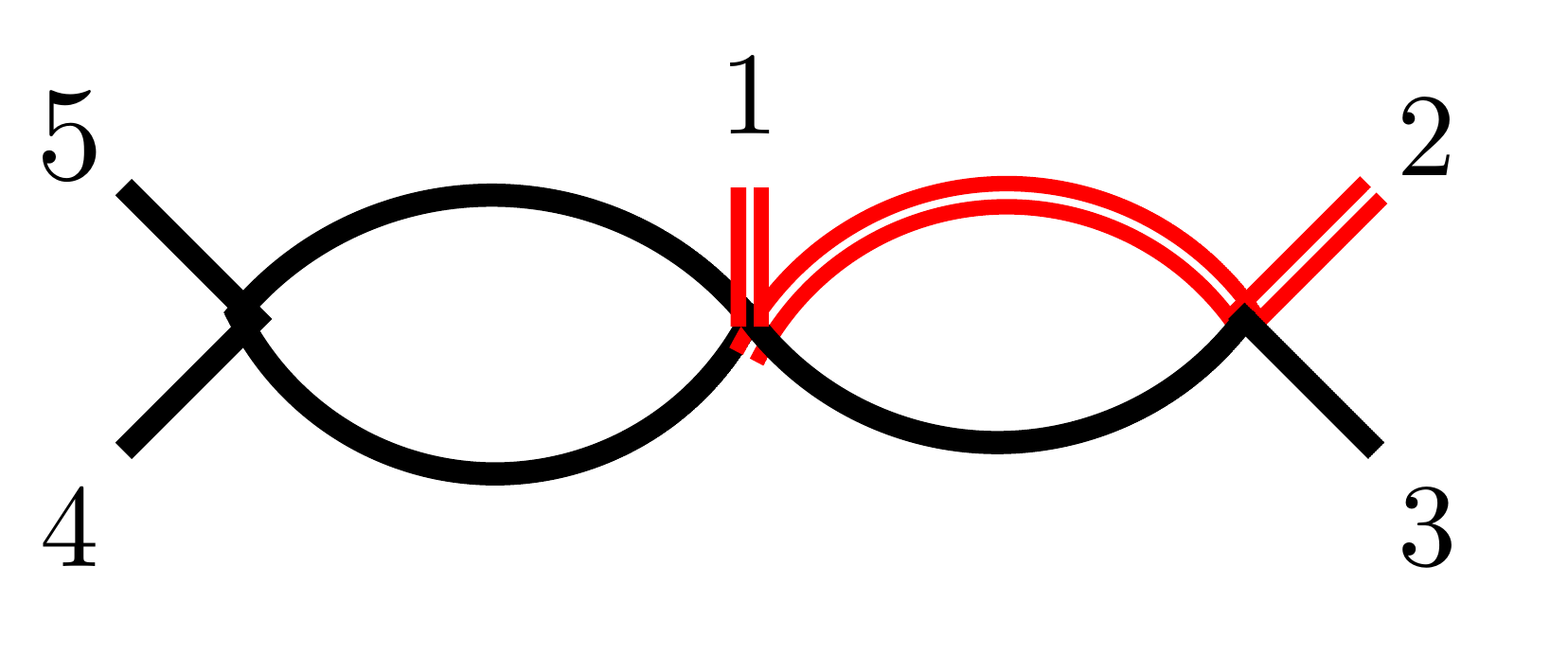}  
\end{figure}
\end{minipage}
\begin{minipage}{0.54\textwidth}
\begin{align}\begin{split}
\cI_{79} = & \frac{\epsilon ^2 \, (2 \epsilon -1)^2 \left(2 d_{23}+m_t^2\right)}{2 d_{23}} \, I_{0,1,0,1,1,0,1,0}^{0,0,0} \\
& - \frac{\epsilon ^2 \, (\epsilon -1) (2 \epsilon -1) \left(2 d_{23}+m_t^2\right)}{2 d_{23} \, m_t^2} \, I_{0,1,0,0,1,0,1,0}^{0,0,0}  
\end{split}\end{align}
\end{minipage}%

\subsection*{Sectors: $\cI_{80}$ and  $\cI_{81}$}

\begin{minipage}{0.4\textwidth}
\hspace{\parindent}
\begin{figure}[H]
\centering
\includegraphics[width=2.8cm]{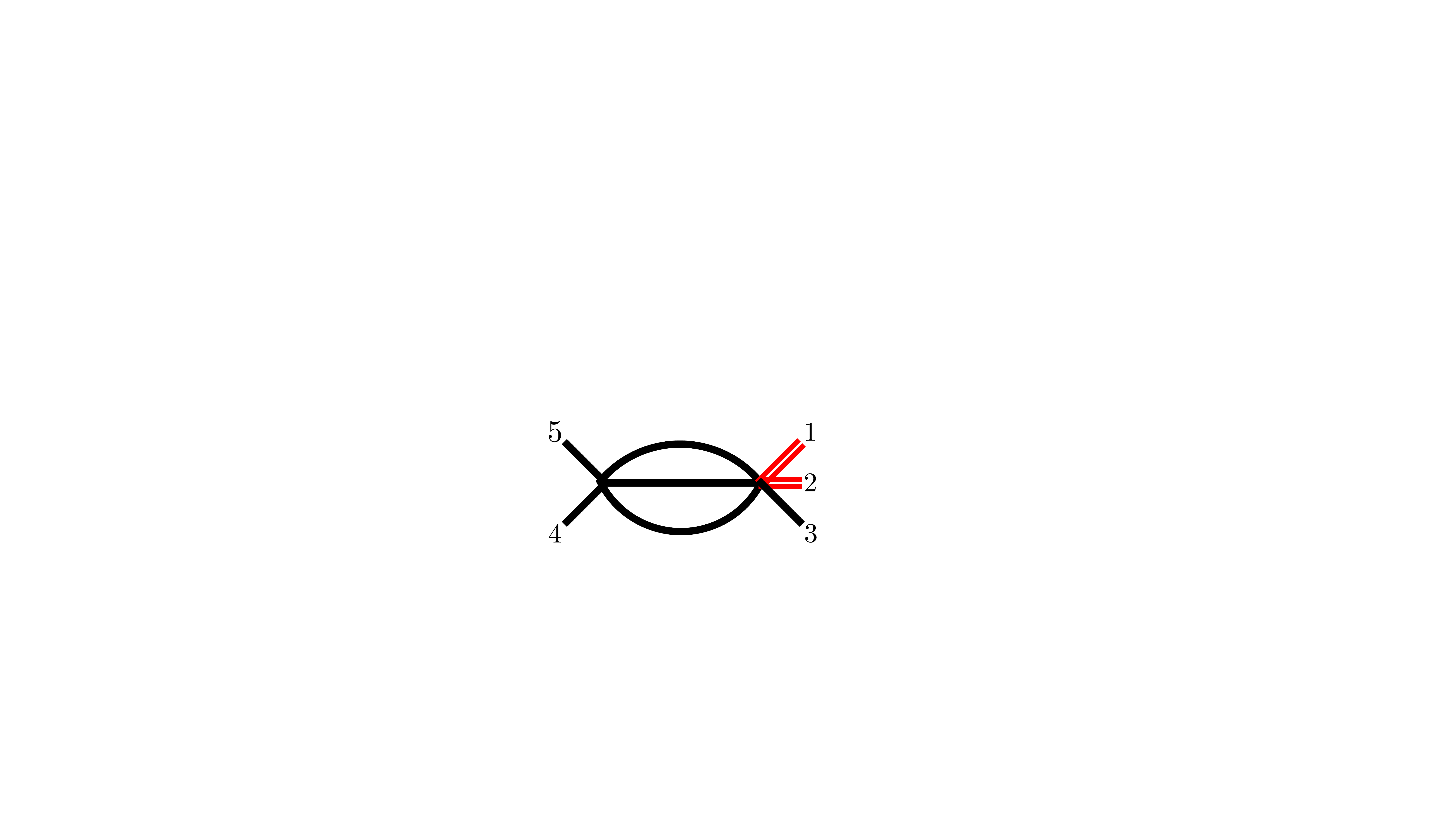}  
\includegraphics[width=2.8cm]{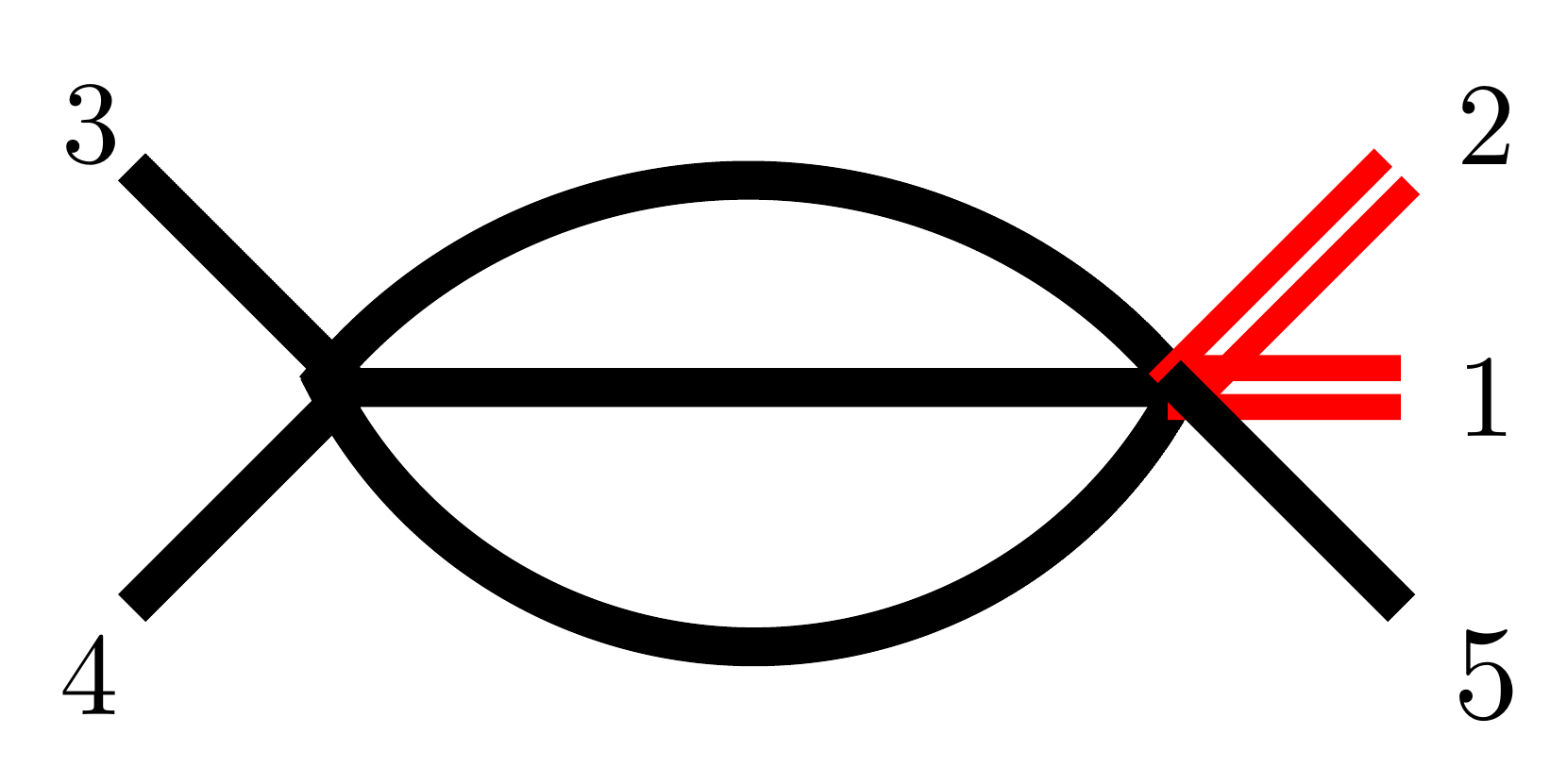}  
\end{figure}
\end{minipage}
\begin{minipage}{0.54\textwidth}
\begin{align}\begin{split}
\cI_{80} = & -\frac{\epsilon \, (2 \epsilon -1) (3 \epsilon -2) (3 \epsilon -1)}{2 d_{45}} \, I_{0,0,0,1,1,0,0,1}^{0,0,0} \\
\cI_{81} = & -\frac{\epsilon \, (2 \epsilon -1) (3 \epsilon -2) (3 \epsilon -1)}{2 d_{34}} \, I_{0,0,1,0,0,1,0,1}^{0,0,0}
\end{split}\end{align}
\end{minipage}%

\subsection*{Sector: $\cI_{82}$}

\begin{minipage}{0.4\textwidth}
\hspace{\parindent}
\begin{figure}[H]
\centering
\includegraphics[width=2.8cm]{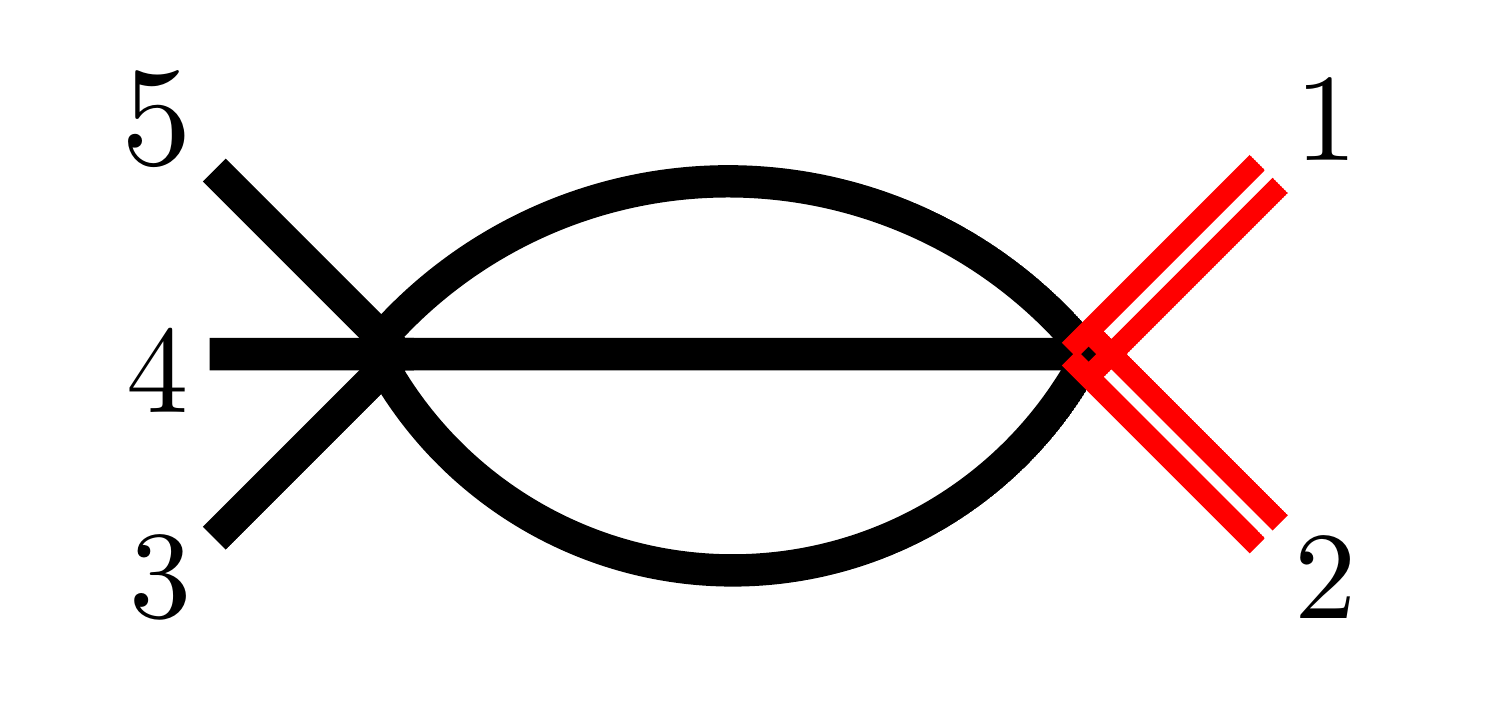}  
\end{figure}
\end{minipage}
\begin{minipage}{0.54\textwidth}
\begin{align}\begin{split}
\cI_{82} = &-\frac{\epsilon \, (2 \epsilon -1) (3 \epsilon -2) (3 \epsilon -1)}{2 \left(d_{12}+m_t^2\right)} \, I_{0,0,1,0,1,0,0,1}^{0,0,0}
\end{split}\end{align}
\end{minipage}%

\subsection*{Sectors: $\cI_{83}, \, \cI_{84}$ and $\cI_{85}, \, \cI_{86}$}

\hspace{\parindent}
\begin{figure}[H]
\raisebox{0.1cm}{\includegraphics[width=2.8cm]{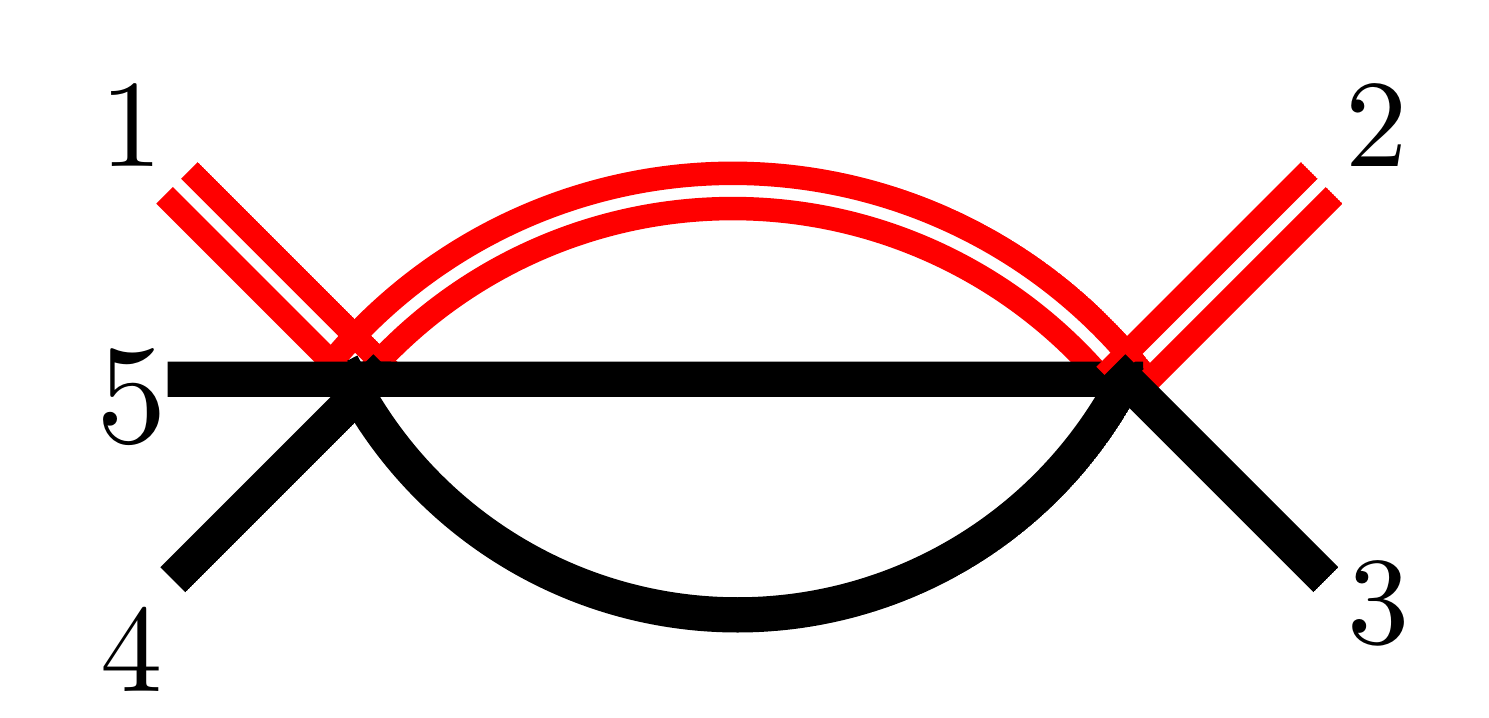}}  
\includegraphics[width=2.8cm]{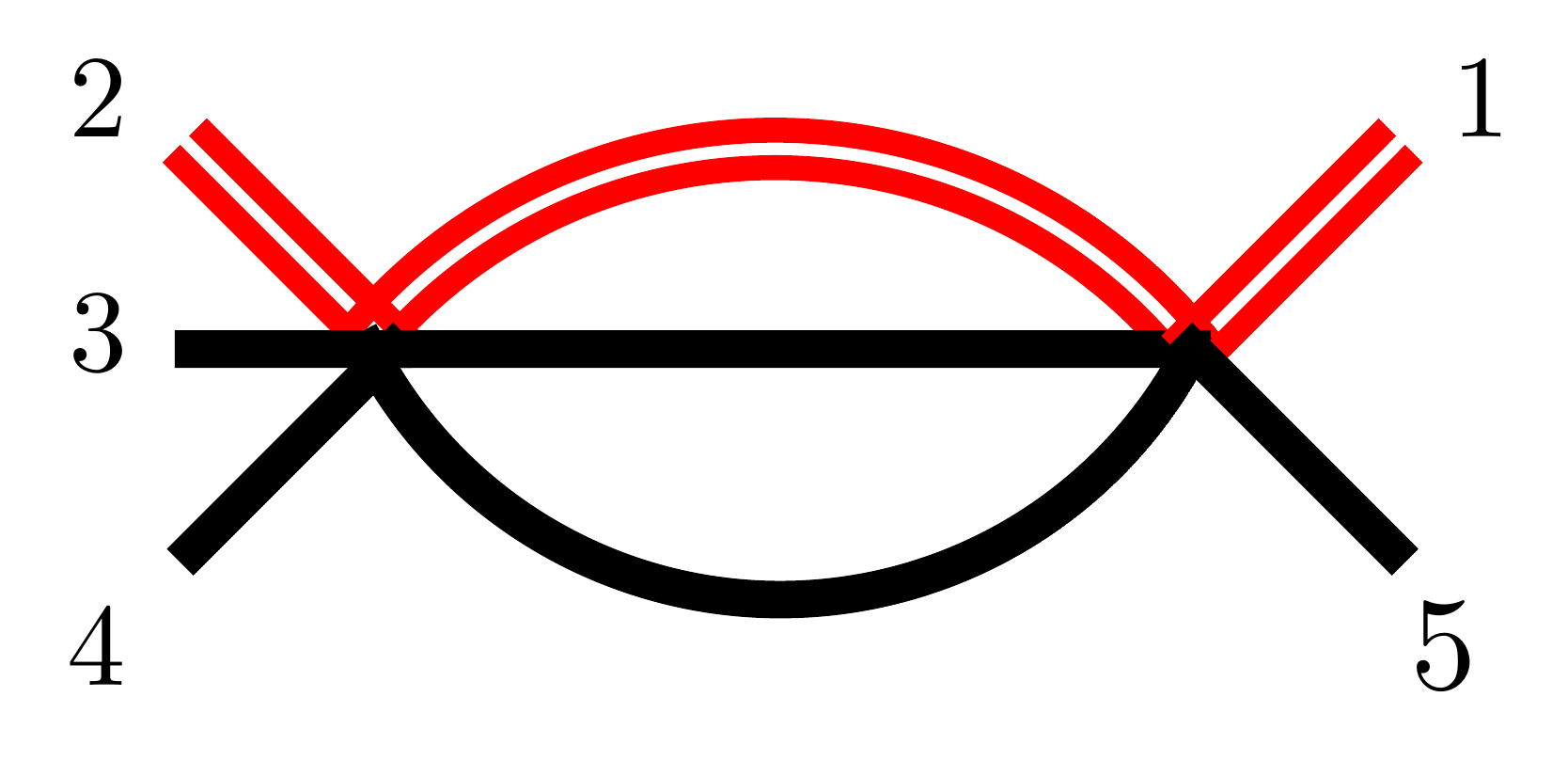}  
\end{figure}

\begin{flalign}
\cI_{83} = & -\frac{\epsilon \, (2 \epsilon -1) (3 \epsilon -2) (4 \epsilon -1) \left(2 d_{23}+m_t^2\right)}{4 d_{23}^2} \, I_{0,1,0,0,0,0,1,1}^{0,0,0} \nn \\
& -\frac{\epsilon \, (2 \epsilon -1) \left(2 d_{23}+m_t^2\right) \left(\epsilon \, d_{23} +4 \epsilon \, m_t^2-m_t^2\right)}{2 d_{23}^2} \, I_{0,2,0,0,0,0,1,1}^{0,0,0}&& \nn \\
\cI_{84} = & \frac{\epsilon ^2 \, (2 \epsilon -1) (3 \epsilon -2)}{2 d_{23}} \, I_{0,1,0,0,0,0,1,1}^{0,0,0} + \frac{\epsilon ^2 \, (2 \epsilon -1) \left(d_{23}+m_t^2\right)}{d_{23}} \, I_{0,2,0,0,0,0,1,1}^{0,0,0} && \nn \\
\cI_{85}  = & -\frac{\epsilon \, (2 \epsilon -1) (3 \epsilon -2) (4 \epsilon -1) \left(2 d_{15}+m_t^2\right)}{4 d_{15}^2} \, I_{0,1,0,0,0,1,0,1}^{0,0,0} \nn \\
&  -\frac{\epsilon \, (2 \epsilon -1) \left(2 d_{15}+m_t^2\right) \left(\epsilon \, d_{15} +4 \epsilon \, m_t^2-m_t^2\right)}{2 d_{15}^2} \, I_{0,2,0,0,0,1,0,1}^{0,0,0} && \nn \\
\cI_{86} = & \frac{\epsilon ^2 \, (2 \epsilon -1) (3 \epsilon -2)}{2 d_{15}} \, I_{0,1,0,0,0,1,0,1}^{0,0,0} + \frac{\epsilon ^2 \, (2 \epsilon -1) \left(d_{15}+m_t^2\right)}{d_{15}} \, I_{0,2,0,0,0,1,0,1}^{0,0,0} &&
\end{flalign}

\subsection*{Sector: $\cI_{87}$}

\begin{minipage}{0.4\textwidth}
\hspace{\parindent}
\begin{figure}[H]
\centering
\includegraphics[width=2.8cm]{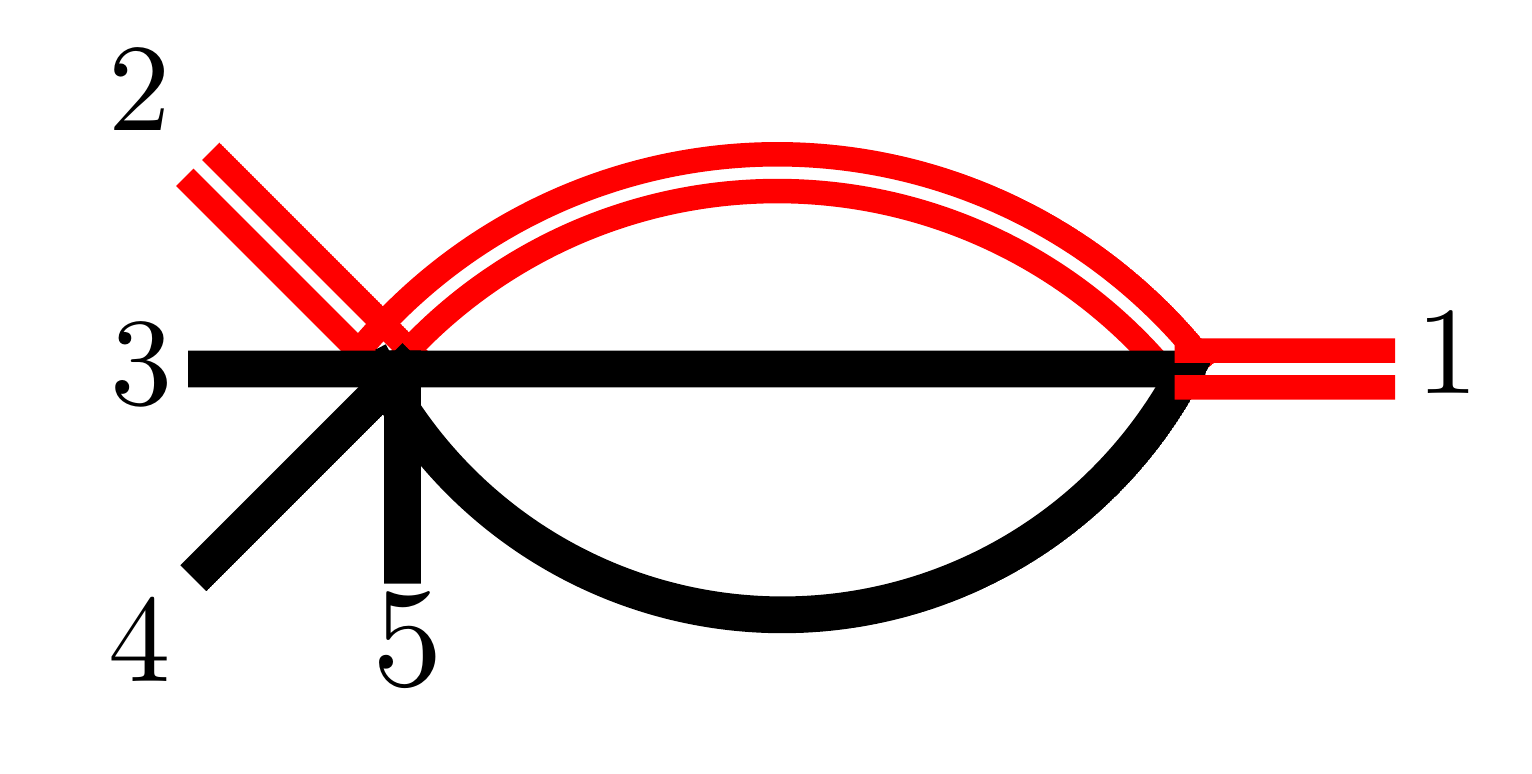}  
\end{figure}
\end{minipage}
\begin{minipage}{0.54\textwidth}
\begin{align}\begin{split}
\cI_{87} = & \frac{\epsilon ^2 \, (2 \epsilon -1) (3 \epsilon -2) (3 \epsilon -1)}{(4 \epsilon -1) m_t^2} \, I_{0,1,0,0,1,0,0,1}^{0,0,0}
\end{split}\end{align}
\end{minipage}%

\subsection*{Sector: $\cI_{88}$}

\begin{minipage}{0.4\textwidth}
\hspace{\parindent}
\begin{figure}[H]
\centering
\includegraphics[width=2.8cm]{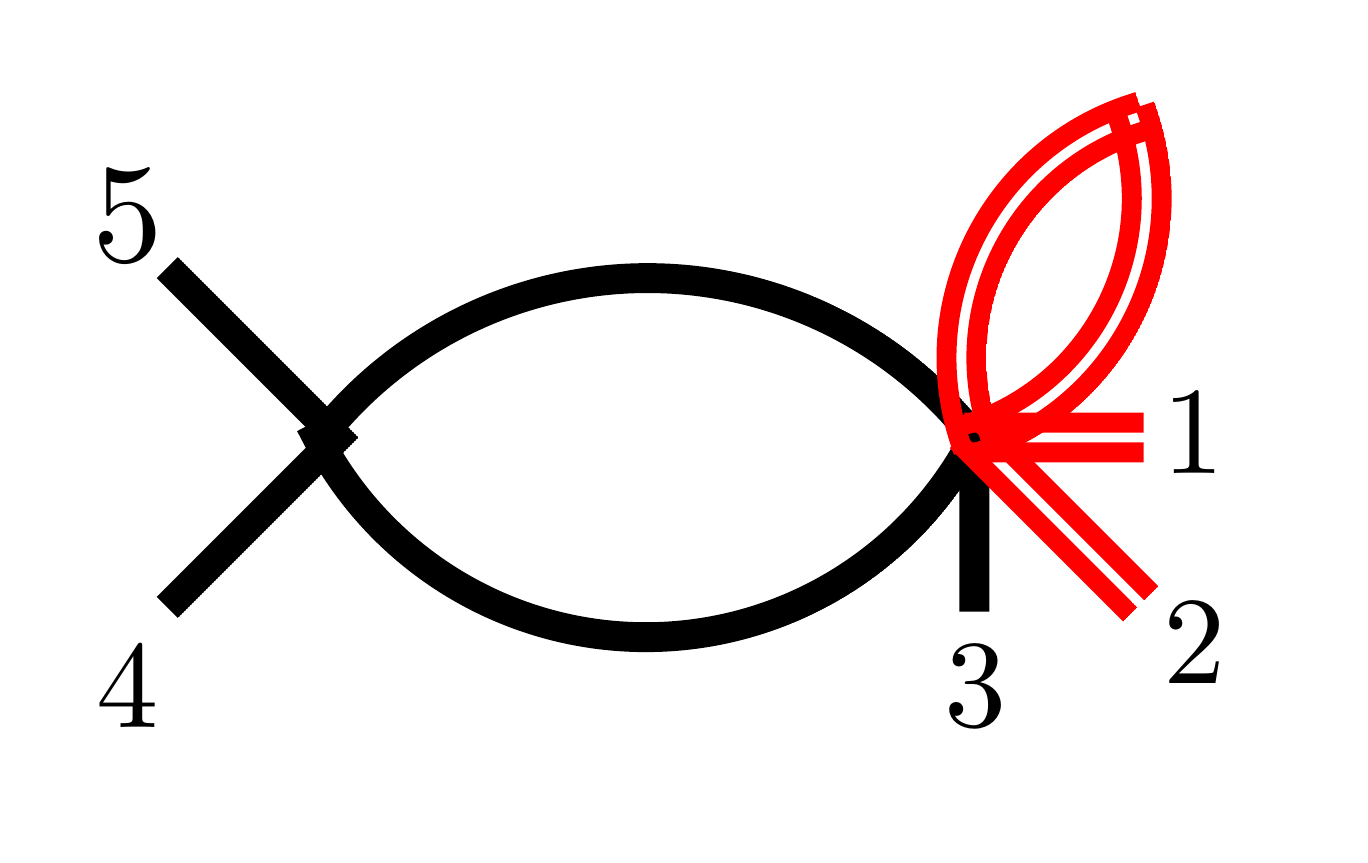}  
\end{figure}
\end{minipage}
\begin{minipage}{0.54\textwidth}
\begin{align}\begin{split}
\cI_{88} = & \frac{\epsilon ^2 \, (\epsilon -1) (2 \epsilon -1)}{m_t^2} \, I_{0,1,0,0,1,0,1,0}^{0,0,0}
\end{split}\end{align}
\end{minipage}%

\bibliographystyle{JHEP}
\bibliography{ppttj_431}

\providecommand{\href}[2]{#2}\begingroup\raggedright\begin{thebibliography}{100}

\bibitem{Czakon:2019yrx}
M.~Czakon, S.~Dulat, T.-J. Hou, J.~Huston, A.~Mitov, A.~S. Papanastasiou
  et~al., \emph{{An exploratory study of the impact of CMS double-differential
  top distributions on the gluon parton distribution function}},
  \href{http://dx.doi.org/10.1088/1361-6471/abb1b6}{\emph{J. Phys. G} {\bf 48}
  (2020) 015003}, [\href{http://arxiv.org/abs/1912.08801}{{\tt 1912.08801}}].

\bibitem{Cooper-Sarkar:2020twv}
A.~M. Cooper-Sarkar, M.~Czakon, M.~A. Lim, A.~Mitov and A.~S. Papanastasiou,
  \emph{{Simultaneous extraction of $\alpha_s$ and $m_t$ from LHC $t\bar{t}$
  differential distributions}},  \href{http://arxiv.org/abs/2010.04171}{{\tt
  2010.04171}}.

\bibitem{Alioli:2013mxa}
S.~Alioli, P.~Fernandez, J.~Fuster, A.~Irles, S.-O. Moch, P.~Uwer et~al.,
  \emph{{A new observable to measure the top-quark mass at hadron colliders}},
  \href{http://dx.doi.org/10.1140/epjc/s10052-013-2438-2}{\emph{Eur. Phys. J.
  C} {\bf 73} (2013) 2438}, [\href{http://arxiv.org/abs/1303.6415}{{\tt
  1303.6415}}].

\bibitem{Bevilacqua:2017ipv}
G.~Bevilacqua, H.~B. Hartanto, M.~Kraus, M.~Schulze and M.~Worek, \emph{{Top
  quark mass studies with $ t\overline{t}j $ at the LHC}},
  \href{http://dx.doi.org/10.1007/JHEP03(2018)169}{\emph{JHEP} {\bf 03} (2018)
  169}, [\href{http://arxiv.org/abs/1710.07515}{{\tt 1710.07515}}].

\bibitem{Alioli:2022lqo}
S.~Alioli, J.~Fuster, M.~V. Garzelli, A.~Gavardi, A.~Irles, D.~Melini et~al.,
  \emph{{Phenomenology of $ t\overline{t}j $ + X production at the LHC}},
  \href{http://dx.doi.org/10.1007/JHEP05(2022)146}{\emph{JHEP} {\bf 05} (2022)
  146}, [\href{http://arxiv.org/abs/2202.07975}{{\tt 2202.07975}}].

\bibitem{Dittmaier:2007wz}
S.~Dittmaier, P.~Uwer and S.~Weinzierl, \emph{{NLO QCD corrections to t anti-t
  + jet production at hadron colliders}},
  \href{http://dx.doi.org/10.1103/PhysRevLett.98.262002}{\emph{Phys. Rev.
  Lett.} {\bf 98} (2007) 262002},
  [\href{http://arxiv.org/abs/hep-ph/0703120}{{\tt hep-ph/0703120}}].

\bibitem{Dittmaier:2008uj}
S.~Dittmaier, P.~Uwer and S.~Weinzierl, \emph{{Hadronic top-quark pair
  production in association with a hard jet at next-to-leading order QCD:
  Phenomenological studies for the Tevatron and the LHC}},
  \href{http://dx.doi.org/10.1140/epjc/s10052-008-0816-y}{\emph{Eur. Phys. J.
  C} {\bf 59} (2009) 625--646}, [\href{http://arxiv.org/abs/0810.0452}{{\tt
  0810.0452}}].

\bibitem{Melnikov:2010iu}
K.~Melnikov and M.~Schulze, \emph{{NLO QCD corrections to top quark pair
  production in association with one hard jet at hadron colliders}},
  \href{http://dx.doi.org/10.1016/j.nuclphysb.2010.07.003}{\emph{Nucl. Phys. B}
  {\bf 840} (2010) 129--159}, [\href{http://arxiv.org/abs/1004.3284}{{\tt
  1004.3284}}].

\bibitem{Alioli:2011as}
S.~Alioli, S.-O. Moch and P.~Uwer, \emph{{Hadronic top-quark pair-production
  with one jet and parton showering}},
  \href{http://dx.doi.org/10.1007/JHEP01(2012)137}{\emph{JHEP} {\bf 01} (2012)
  137}, [\href{http://arxiv.org/abs/1110.5251}{{\tt 1110.5251}}].

\bibitem{Czakon:2015cla}
M.~Czakon, H.~B. Hartanto, M.~Kraus and M.~Worek, \emph{{Matching the
  Nagy-Soper parton shower at next-to-leading order}},
  \href{http://dx.doi.org/10.1007/JHEP06(2015)033}{\emph{JHEP} {\bf 06} (2015)
  033}, [\href{http://arxiv.org/abs/1502.00925}{{\tt 1502.00925}}].

\bibitem{Bevilacqua:2015qha}
G.~Bevilacqua, H.~B. Hartanto, M.~Kraus and M.~Worek, \emph{{Top Quark Pair
  Production in Association with a Jet with Next-to-Leading-Order QCD Off-Shell
  Effects at the Large Hadron Collider}},
  \href{http://dx.doi.org/10.1103/PhysRevLett.116.052003}{\emph{Phys. Rev.
  Lett.} {\bf 116} (2016) 052003}, [\href{http://arxiv.org/abs/1509.09242}{{\tt
  1509.09242}}].

\bibitem{Bevilacqua:2016jfk}
G.~Bevilacqua, H.~B. Hartanto, M.~Kraus and M.~Worek, \emph{{Off-shell Top
  Quarks with One Jet at the LHC: A comprehensive analysis at NLO QCD}},
  \href{http://dx.doi.org/10.1007/JHEP11(2016)098}{\emph{JHEP} {\bf 11} (2016)
  098}, [\href{http://arxiv.org/abs/1609.01659}{{\tt 1609.01659}}].

\bibitem{Gutschow:2018tuk}
C.~G\"utschow, J.~M. Lindert and M.~Sch\"onherr, \emph{{Multi-jet merged
  top-pair production including electroweak corrections}},
  \href{http://dx.doi.org/10.1140/epjc/s10052-018-5804-2}{\emph{Eur. Phys. J.
  C} {\bf 78} (2018) 317}, [\href{http://arxiv.org/abs/1803.00950}{{\tt
  1803.00950}}].

\bibitem{ATLAS:2019guf}
{\scshape ATLAS} collaboration, G.~Aad et~al., \emph{{Measurement of the
  top-quark mass in $t\bar{t}+1$-jet events collected with the ATLAS detector
  in $pp$ collisions at $\sqrt{s}=8$ TeV}},
  \href{http://dx.doi.org/10.1007/JHEP11(2019)150}{\emph{JHEP} {\bf 11} (2019)
  150}, [\href{http://arxiv.org/abs/1905.02302}{{\tt 1905.02302}}].

\bibitem{CMS:2020grm}
{\scshape CMS} collaboration, A.~M. Sirunyan et~al., \emph{{Measurement of the
  cross section for $\text{t}\bar{\text{t}}$ production with additional jets
  and b jets in pp collisions at $\sqrt{s}=$ 13 TeV}},
  \href{http://dx.doi.org/10.1007/JHEP07(2020)125}{\emph{JHEP} {\bf 07} (2020)
  125}, [\href{http://arxiv.org/abs/2003.06467}{{\tt 2003.06467}}].

\bibitem{Gehrmann:2015bfy}
T.~Gehrmann, J.~Henn and N.~Lo~Presti, \emph{{Analytic form of the two-loop
  planar five-gluon all-plus-helicity amplitude in QCD}},
  \href{http://dx.doi.org/10.1103/PhysRevLett.116.062001}{\emph{Phys. Rev.
  Lett.} {\bf 116} (2016) 062001}, [\href{http://arxiv.org/abs/1511.05409}{{\tt
  1511.05409}}].

\bibitem{Papadopoulos:2015jft}
C.~G. Papadopoulos, D.~Tommasini and C.~Wever, \emph{{The Pentabox Master
  Integrals with the Simplified Differential Equations approach}},
  \href{http://dx.doi.org/10.1007/JHEP04(2016)078}{\emph{JHEP} {\bf 04} (2016)
  078}, [\href{http://arxiv.org/abs/1511.09404}{{\tt 1511.09404}}].

\bibitem{Gehrmann:2018yef}
T.~Gehrmann, J.~M. Henn and N.~A. Lo~Presti, \emph{{Pentagon functions for
  massless planar scattering amplitudes}},
  \href{http://dx.doi.org/10.1007/JHEP10(2018)103}{\emph{JHEP} {\bf 10} (2018)
  103}, [\href{http://arxiv.org/abs/1807.09812}{{\tt 1807.09812}}].

\bibitem{Abreu:2018aqd}
S.~Abreu, L.~J. Dixon, E.~Herrmann, B.~Page and M.~Zeng, \emph{{The two-loop
  five-point amplitude in $\mathcal{N} =4$ super-Yang-Mills theory}},
  \href{http://dx.doi.org/10.1103/PhysRevLett.122.121603}{\emph{Phys. Rev.
  Lett.} {\bf 122} (2019) 121603}, [\href{http://arxiv.org/abs/1812.08941}{{\tt
  1812.08941}}].

\bibitem{Chicherin:2018old}
D.~Chicherin, T.~Gehrmann, J.~M. Henn, P.~Wasser, Y.~Zhang and S.~Zoia,
  \emph{{All Master Integrals for Three-Jet Production at
  Next-to-Next-to-Leading Order}},
  \href{http://dx.doi.org/10.1103/PhysRevLett.123.041603}{\emph{Phys. Rev.
  Lett.} {\bf 123} (2019) 041603}, [\href{http://arxiv.org/abs/1812.11160}{{\tt
  1812.11160}}].

\bibitem{Chicherin:2020oor}
D.~Chicherin and V.~Sotnikov, \emph{{Pentagon Functions for Scattering of Five
  Massless Particles}},
  \href{http://dx.doi.org/10.1007/JHEP12(2020)167}{\emph{JHEP} {\bf 20} (2020)
  167}, [\href{http://arxiv.org/abs/2009.07803}{{\tt 2009.07803}}].

\bibitem{Abreu:2020jxa}
S.~Abreu, H.~Ita, F.~Moriello, B.~Page, W.~Tschernow and M.~Zeng,
  \emph{{Two-Loop Integrals for Planar Five-Point One-Mass Processes}},
  \href{http://dx.doi.org/10.1007/JHEP11(2020)117}{\emph{JHEP} {\bf 11} (2020)
  117}, [\href{http://arxiv.org/abs/2005.04195}{{\tt 2005.04195}}].

\bibitem{Canko:2020ylt}
D.~D. Canko, C.~G. Papadopoulos and N.~Syrrakos, \emph{{Analytic representation
  of all planar two-loop five-point Master Integrals with one off-shell leg}},
  \href{http://dx.doi.org/10.1007/JHEP01(2021)199}{\emph{JHEP} {\bf 01} (2021)
  199}, [\href{http://arxiv.org/abs/2009.13917}{{\tt 2009.13917}}].

\bibitem{Chicherin:2021dyp}
D.~Chicherin, V.~Sotnikov and S.~Zoia, \emph{{Pentagon functions for one-mass
  planar scattering amplitudes}},
  \href{http://dx.doi.org/10.1007/JHEP01(2022)096}{\emph{JHEP} {\bf 01} (2022)
  096}, [\href{http://arxiv.org/abs/2110.10111}{{\tt 2110.10111}}].

\bibitem{Abreu:2021smk}
S.~Abreu, H.~Ita, B.~Page and W.~Tschernow, \emph{{Two-loop hexa-box integrals
  for non-planar five-point one-mass processes}},
  \href{http://dx.doi.org/10.1007/JHEP03(2022)182}{\emph{JHEP} {\bf 03} (2022)
  182}, [\href{http://arxiv.org/abs/2107.14180}{{\tt 2107.14180}}].

\bibitem{Badger:2018enw}
S.~Badger, C.~Br\o{}nnum-Hansen, H.~B. Hartanto and T.~Peraro, \emph{{Analytic
  helicity amplitudes for two-loop five-gluon scattering: the single-minus
  case}}, \href{http://dx.doi.org/10.1007/JHEP01(2019)186}{\emph{JHEP} {\bf 01}
  (2019) 186}, [\href{http://arxiv.org/abs/1811.11699}{{\tt 1811.11699}}].

\bibitem{Chicherin:2018yne}
D.~Chicherin, T.~Gehrmann, J.~Henn, P.~Wasser, Y.~Zhang and S.~Zoia,
  \emph{{Analytic result for a two-loop five-particle amplitude}},
  \href{http://dx.doi.org/10.1103/PhysRevLett.122.121602}{\emph{Phys. Rev.
  Lett.} {\bf 122} (2019) 121602}, [\href{http://arxiv.org/abs/1812.11057}{{\tt
  1812.11057}}].

\bibitem{Abreu:2018zmy}
S.~Abreu, J.~Dormans, F.~Febres~Cordero, H.~Ita and B.~Page, \emph{{Analytic
  Form of Planar Two-Loop Five-Gluon Scattering Amplitudes in QCD}},
  \href{http://dx.doi.org/10.1103/PhysRevLett.122.082002}{\emph{Phys. Rev.
  Lett.} {\bf 122} (2019) 082002}, [\href{http://arxiv.org/abs/1812.04586}{{\tt
  1812.04586}}].

\bibitem{Abreu:2019odu}
S.~Abreu, J.~Dormans, F.~Febres~Cordero, H.~Ita, B.~Page and V.~Sotnikov,
  \emph{{Analytic Form of the Planar Two-Loop Five-Parton Scattering Amplitudes
  in QCD}}, \href{http://dx.doi.org/10.1007/JHEP05(2019)084}{\emph{JHEP} {\bf
  05} (2019) 084}, [\href{http://arxiv.org/abs/1904.00945}{{\tt 1904.00945}}].

\bibitem{Abreu:2018jgq}
S.~Abreu, F.~Febres~Cordero, H.~Ita, B.~Page and V.~Sotnikov, \emph{{Planar
  Two-Loop Five-Parton Amplitudes from Numerical Unitarity}},
  \href{http://dx.doi.org/10.1007/JHEP11(2018)116}{\emph{JHEP} {\bf 11} (2018)
  116}, [\href{http://arxiv.org/abs/1809.09067}{{\tt 1809.09067}}].

\bibitem{Badger:2019djh}
S.~Badger, D.~Chicherin, T.~Gehrmann, G.~Heinrich, J.~Henn, T.~Peraro et~al.,
  \emph{{Analytic form of the full two-loop five-gluon all-plus helicity
  amplitude}},
  \href{http://dx.doi.org/10.1103/PhysRevLett.123.071601}{\emph{Phys. Rev.
  Lett.} {\bf 123} (2019) 071601}, [\href{http://arxiv.org/abs/1905.03733}{{\tt
  1905.03733}}].

\bibitem{Abreu:2020cwb}
S.~Abreu, B.~Page, E.~Pascual and V.~Sotnikov, \emph{{Leading-Color Two-Loop
  QCD Corrections for Three-Photon Production at Hadron Colliders}},
  \href{http://dx.doi.org/10.1007/JHEP01(2021)078}{\emph{JHEP} {\bf 01} (2021)
  078}, [\href{http://arxiv.org/abs/2010.15834}{{\tt 2010.15834}}].

\bibitem{Abreu:2021oya}
S.~Abreu, F.~F. Cordero, H.~Ita, B.~Page and V.~Sotnikov, \emph{{Leading-color
  two-loop QCD corrections for three-jet production at hadron colliders}},
  \href{http://dx.doi.org/10.1007/JHEP07(2021)095}{\emph{JHEP} {\bf 07} (2021)
  095}, [\href{http://arxiv.org/abs/2102.13609}{{\tt 2102.13609}}].

\bibitem{Chawdhry:2020for}
H.~A. Chawdhry, M.~Czakon, A.~Mitov and R.~Poncelet, \emph{{Two-loop
  leading-color helicity amplitudes for three-photon production at the LHC}},
  \href{http://dx.doi.org/10.1007/JHEP06(2021)150}{\emph{JHEP} {\bf 06} (2021)
  150}, [\href{http://arxiv.org/abs/2012.13553}{{\tt 2012.13553}}].

\bibitem{Hartanto:2019uvl}
H.~B. Hartanto, S.~Badger, C.~Br\o{}nnum-Hansen and T.~Peraro, \emph{{A
  numerical evaluation of planar two-loop helicity amplitudes for a W-boson
  plus four partons}},
  \href{http://dx.doi.org/10.1007/JHEP09(2019)119}{\emph{JHEP} {\bf 09} (2019)
  119}, [\href{http://arxiv.org/abs/1906.11862}{{\tt 1906.11862}}].

\bibitem{Agarwal:2021grm}
B.~Agarwal, F.~Buccioni, A.~von Manteuffel and L.~Tancredi, \emph{{Two-loop
  leading colour QCD corrections to $q \bar{q} \to \gamma \gamma g$ and $q g
  \to \gamma \gamma q$}},
  \href{http://dx.doi.org/10.1007/JHEP04(2021)201}{\emph{JHEP} {\bf 04} (2021)
  201}, [\href{http://arxiv.org/abs/2102.01820}{{\tt 2102.01820}}].

\bibitem{Chawdhry:2021mkw}
H.~A. Chawdhry, M.~Czakon, A.~Mitov and R.~Poncelet, \emph{{Two-loop
  leading-colour QCD helicity amplitudes for two-photon plus jet production at
  the LHC}}, \href{http://dx.doi.org/10.1007/JHEP07(2021)164}{\emph{JHEP} {\bf
  07} (2021) 164}, [\href{http://arxiv.org/abs/2103.04319}{{\tt 2103.04319}}].

\bibitem{Agarwal:2021vdh}
B.~Agarwal, F.~Buccioni, A.~von Manteuffel and L.~Tancredi, \emph{{Two-Loop
  Helicity Amplitudes for Diphoton Plus Jet Production in Full Color}},
  \href{http://dx.doi.org/10.1103/PhysRevLett.127.262001}{\emph{Phys. Rev.
  Lett.} {\bf 127} (2021) 262001}, [\href{http://arxiv.org/abs/2105.04585}{{\tt
  2105.04585}}].

\bibitem{Badger:2021imn}
S.~Badger, C.~Br\o{}nnum-Hansen, D.~Chicherin, T.~Gehrmann, H.~B. Hartanto,
  J.~Henn et~al., \emph{{Virtual QCD corrections to gluon-initiated diphoton
  plus jet production at hadron colliders}},
  \href{http://dx.doi.org/10.1007/JHEP11(2021)083}{\emph{JHEP} {\bf 11} (2021)
  083}, [\href{http://arxiv.org/abs/2106.08664}{{\tt 2106.08664}}].

\bibitem{Badger:2021ega}
S.~Badger, H.~B. Hartanto, J.~Kry\'s and S.~Zoia, \emph{{Two-loop
  leading-colour QCD helicity amplitudes for Higgs boson production in
  association with a bottom-quark pair at the LHC}},
  \href{http://dx.doi.org/10.1007/JHEP11(2021)012}{\emph{JHEP} {\bf 11} (2021)
  012}, [\href{http://arxiv.org/abs/2107.14733}{{\tt 2107.14733}}].

\bibitem{Badger:2022ncb}
S.~Badger, H.~B. Hartanto, J.~Kry\'s and S.~Zoia, \emph{{Two-loop leading
  colour helicity amplitudes for W$^{±}$\ensuremath{\gamma} + j production at
  the LHC}}, \href{http://dx.doi.org/10.1007/JHEP05(2022)035}{\emph{JHEP} {\bf
  05} (2022) 035}, [\href{http://arxiv.org/abs/2201.04075}{{\tt 2201.04075}}].

\bibitem{Chawdhry:2019bji}
H.~A. Chawdhry, M.~L. Czakon, A.~Mitov and R.~Poncelet, \emph{{NNLO QCD
  corrections to three-photon production at the LHC}},
  \href{http://dx.doi.org/10.1007/JHEP02(2020)057}{\emph{JHEP} {\bf 02} (2020)
  057}, [\href{http://arxiv.org/abs/1911.00479}{{\tt 1911.00479}}].

\bibitem{Kallweit:2020gcp}
S.~Kallweit, V.~Sotnikov and M.~Wiesemann, \emph{{Triphoton production at
  hadron colliders in NNLO QCD}},
  \href{http://dx.doi.org/10.1016/j.physletb.2020.136013}{\emph{Phys. Lett. B}
  {\bf 812} (2021) 136013}, [\href{http://arxiv.org/abs/2010.04681}{{\tt
  2010.04681}}].

\bibitem{Chawdhry:2021hkp}
H.~A. Chawdhry, M.~Czakon, A.~Mitov and R.~Poncelet, \emph{{NNLO QCD
  corrections to diphoton production with an additional jet at the LHC}},
  \href{http://dx.doi.org/10.1007/JHEP09(2021)093}{\emph{JHEP} {\bf 09} (2021)
  093}, [\href{http://arxiv.org/abs/2105.06940}{{\tt 2105.06940}}].

\bibitem{Badger:2021ohm}
S.~Badger, T.~Gehrmann, M.~Marcoli and R.~Moodie, \emph{{Next-to-leading order
  QCD corrections to diphoton-plus-jet production through gluon fusion at the
  LHC}}, \href{http://dx.doi.org/10.1016/j.physletb.2021.136802}{\emph{Phys.
  Lett. B} {\bf 824} (2022) 136802},
  [\href{http://arxiv.org/abs/2109.12003}{{\tt 2109.12003}}].

\bibitem{Hartanto:2022qhh}
H.~B. Hartanto, R.~Poncelet, A.~Popescu and S.~Zoia, \emph{{NNLO QCD
  corrections to $Wb\bar{b}$ production at the LHC}},
  \href{http://arxiv.org/abs/2205.01687}{{\tt 2205.01687}}.

\bibitem{Kotikov:1990kg}
A.~V. Kotikov, \emph{{Differential equations method: New technique for massive
  Feynman diagrams calculation}},
  \href{http://dx.doi.org/10.1016/0370-2693(91)90413-K}{\emph{Phys. Lett. B}
  {\bf 254} (1991) 158--164}.

\bibitem{Remiddi:1997ny}
E.~Remiddi, \emph{{Differential equations for Feynman graph amplitudes}},
  {\emph{Nuovo Cim. A} {\bf 110} (1997) 1435--1452},
  [\href{http://arxiv.org/abs/hep-th/9711188}{{\tt hep-th/9711188}}].

\bibitem{Henn:2013pwa}
J.~M. Henn, \emph{{Multiloop integrals in dimensional regularization made
  simple}}, \href{http://dx.doi.org/10.1103/PhysRevLett.110.251601}{\emph{Phys.
  Rev. Lett.} {\bf 110} (2013) 251601},
  [\href{http://arxiv.org/abs/1304.1806}{{\tt 1304.1806}}].

\bibitem{Tkachov:1981wb}
F.~V. Tkachov, \emph{{A Theorem on Analytical Calculability of Four Loop
  Renormalization Group Functions}},
  \href{http://dx.doi.org/10.1016/0370-2693(81)90288-4}{\emph{Phys. Lett.} {\bf
  100B} (1981) 65--68}.

\bibitem{Chetyrkin:1981qh}
K.~G. Chetyrkin and F.~V. Tkachov, \emph{{Integration by Parts: The Algorithm
  to Calculate beta Functions in 4 Loops}},
  \href{http://dx.doi.org/10.1016/0550-3213(81)90199-1}{\emph{Nucl. Phys. B}
  {\bf 192} (1981) 159--204}.

\bibitem{Laporta:2001dd}
S.~Laporta, \emph{{High precision calculation of multiloop Feynman integrals by
  difference equations}},
  \href{http://dx.doi.org/10.1016/S0217-751X(00)00215-7,
  10.1142/S0217751X00002157}{\emph{Int. J. Mod. Phys.} {\bf A15} (2000)
  5087--5159}, [\href{http://arxiv.org/abs/hep-ph/0102033}{{\tt
  hep-ph/0102033}}].

\bibitem{vonManteuffel:2014ixa}
A.~von Manteuffel and R.~M. Schabinger, \emph{{A novel approach to integration
  by parts reduction}},
  \href{http://dx.doi.org/10.1016/j.physletb.2015.03.029}{\emph{Phys. Lett. B}
  {\bf 744} (2015) 101--104}, [\href{http://arxiv.org/abs/1406.4513}{{\tt
  1406.4513}}].

\bibitem{Peraro:2016wsq}
T.~Peraro, \emph{{Scattering amplitudes over finite fields and multivariate
  functional reconstruction}},
  \href{http://dx.doi.org/10.1007/JHEP12(2016)030}{\emph{JHEP} {\bf 12} (2016)
  030}, [\href{http://arxiv.org/abs/1608.01902}{{\tt 1608.01902}}].

\bibitem{Peraro:2019svx}
T.~Peraro, \emph{{FiniteFlow: multivariate functional reconstruction using
  finite fields and dataflow graphs}},
  \href{http://dx.doi.org/10.1007/JHEP07(2019)031}{\emph{JHEP} {\bf 07} (2019)
  031}, [\href{http://arxiv.org/abs/1905.08019}{{\tt 1905.08019}}].

\bibitem{Lee:2017qql}
R.~N. Lee, A.~V. Smirnov and V.~A. Smirnov, \emph{{Solving differential
  equations for Feynman integrals by expansions near singular points}},
  \href{http://dx.doi.org/10.1007/JHEP03(2018)008}{\emph{JHEP} {\bf 03} (2018)
  008}, [\href{http://arxiv.org/abs/1709.07525}{{\tt 1709.07525}}].

\bibitem{Mandal:2018cdj}
M.~K. Mandal and X.~Zhao, \emph{{Evaluating multi-loop Feynman integrals
  numerically through differential equations}},
  \href{http://dx.doi.org/10.1007/JHEP03(2019)190}{\emph{JHEP} {\bf 03} (2019)
  190}, [\href{http://arxiv.org/abs/1812.03060}{{\tt 1812.03060}}].

\bibitem{Francesco:2019yqt}
F.~Moriello, \emph{{Generalised power series expansions for the elliptic planar
  families of Higgs + jet production at two loops}},
  \href{http://dx.doi.org/10.1007/JHEP01(2020)150}{\emph{JHEP} {\bf 01} (2020)
  150}, [\href{http://arxiv.org/abs/1907.13234}{{\tt 1907.13234}}].

\bibitem{Hidding:2020ytt}
M.~Hidding, \emph{{DiffExp, a Mathematica package for computing Feynman
  integrals in terms of one-dimensional series expansions}},
  \href{http://arxiv.org/abs/2006.05510}{{\tt 2006.05510}}.

\bibitem{Liu:2017jxz}
X.~Liu, Y.-Q. Ma and C.-Y. Wang, \emph{{A Systematic and Efficient Method to
  Compute Multi-loop Master Integrals}},
  \href{http://dx.doi.org/10.1016/j.physletb.2018.02.026}{\emph{Phys. Lett. B}
  {\bf 779} (2018) 353--357}, [\href{http://arxiv.org/abs/1711.09572}{{\tt
  1711.09572}}].

\bibitem{Liu:2021wks}
X.~Liu and Y.-Q. Ma, \emph{{Multiloop corrections for collider processes using
  auxiliary mass flow}},  \href{http://arxiv.org/abs/2107.01864}{{\tt
  2107.01864}}.

\bibitem{Liu:2022tji}
Z.-F. Liu and Y.-Q. Ma, \emph{{Automatic computation of Feynman integrals
  containing linear propagators via auxiliary mass flow}},
  \href{http://arxiv.org/abs/2201.11636}{{\tt 2201.11636}}.

\bibitem{Liu:2022chg}
X.~Liu and Y.-Q. Ma, \emph{{AMFlow: a Mathematica Package for Feynman integrals
  computation via Auxiliary Mass Flow}},
  \href{http://arxiv.org/abs/2201.11669}{{\tt 2201.11669}}.

\bibitem{Bonciani:2016qxi}
R.~Bonciani, V.~Del~Duca, H.~Frellesvig, J.~M. Henn, F.~Moriello and V.~A.
  Smirnov, \emph{{Two-loop planar master integrals for Higgs$\to 3$ partons
  with full heavy-quark mass dependence}},
  \href{http://dx.doi.org/10.1007/JHEP12(2016)096}{\emph{JHEP} {\bf 12} (2016)
  096}, [\href{http://arxiv.org/abs/1609.06685}{{\tt 1609.06685}}].

\bibitem{Bonciani:2019jyb}
R.~Bonciani, V.~Del~Duca, H.~Frellesvig, J.~M. Henn, M.~Hidding, L.~Maestri
  et~al., \emph{{Evaluating a family of two-loop non-planar master integrals
  for Higgs + jet production with full heavy-quark mass dependence}},
  \href{http://dx.doi.org/10.1007/JHEP01(2020)132}{\emph{JHEP} {\bf 01} (2020)
  132}, [\href{http://arxiv.org/abs/1907.13156}{{\tt 1907.13156}}].

\bibitem{Frellesvig:2019byn}
H.~Frellesvig, M.~Hidding, L.~Maestri, F.~Moriello and G.~Salvatori, \emph{{The
  complete set of two-loop master integrals for Higgs + jet production in
  QCD}}, \href{http://dx.doi.org/10.1007/JHEP06(2020)093}{\emph{JHEP} {\bf 06}
  (2020) 093}, [\href{http://arxiv.org/abs/1911.06308}{{\tt 1911.06308}}].

\bibitem{Becchetti:2020wof}
M.~Becchetti, R.~Bonciani, V.~Del~Duca, V.~Hirschi, F.~Moriello and
  A.~Schweitzer, \emph{{Next-to-leading order corrections to light-quark mixed
  QCD-EW contributions to Higgs boson production}},
  \href{http://dx.doi.org/10.1103/PhysRevD.103.054037}{\emph{Phys. Rev. D} {\bf
  103} (2021) 054037}, [\href{http://arxiv.org/abs/2010.09451}{{\tt
  2010.09451}}].

\bibitem{Armadillo:2022bgm}
T.~Armadillo, R.~Bonciani, S.~Devoto, N.~Rana and A.~Vicini, \emph{{Two-loop
  mixed QCD-EW corrections to neutral current Drell-Yan}},
  \href{http://arxiv.org/abs/2201.01754}{{\tt 2201.01754}}.

\bibitem{Bonciani:2021zzf}
R.~Bonciani, L.~Buonocore, M.~Grazzini, S.~Kallweit, N.~Rana, F.~Tramontano
  et~al., \emph{{Mixed Strong-Electroweak Corrections to the Drell-Yan
  Process}},
  \href{http://dx.doi.org/10.1103/PhysRevLett.128.012002}{\emph{Phys. Rev.
  Lett.} {\bf 128} (2022) 012002}, [\href{http://arxiv.org/abs/2106.11953}{{\tt
  2106.11953}}].

\bibitem{Badger:2022mrb}
S.~Badger, M.~Becchetti, E.~Chaubey, R.~Marzucca and F.~Sarandrea,
  \emph{{One-loop QCD helicity amplitudes for pp \textrightarrow{} $
  t\overline{t}j $ to O(\ensuremath{\varepsilon}$^{2}$)}},
  \href{http://dx.doi.org/10.1007/JHEP06(2022)066}{\emph{JHEP} {\bf 06} (2022)
  066}, [\href{http://arxiv.org/abs/2201.12188}{{\tt 2201.12188}}].

\bibitem{Chetyrkin:1979bj}
K.~G. Chetyrkin, A.~L. Kataev and F.~V. Tkachov, \emph{{Higher Order
  Corrections to Sigma-t (e+ e- ---\ensuremath{>} Hadrons) in Quantum
  Chromodynamics}},
  \href{http://dx.doi.org/10.1016/0370-2693(79)90596-3}{\emph{Phys. Lett. B}
  {\bf 85} (1979) 277--279}.

\bibitem{Lee:2012cn}
R.~N. Lee, \emph{{Presenting LiteRed: a tool for the Loop InTEgrals
  REDuction}},  \href{http://arxiv.org/abs/1212.2685}{{\tt 1212.2685}}.

\bibitem{Lee:2013mka}
R.~N. Lee, \emph{{LiteRed 1.4: a powerful tool for reduction of multiloop
  integrals}}, \href{http://dx.doi.org/10.1088/1742-6596/523/1/012059}{\emph{J.
  Phys. Conf. Ser.} {\bf 523} (2014) 012059},
  [\href{http://arxiv.org/abs/1310.1145}{{\tt 1310.1145}}].

\bibitem{Arkani-Hamed:2010zjl}
N.~Arkani-Hamed, J.~L. Bourjaily, F.~Cachazo, S.~Caron-Huot and J.~Trnka,
  \emph{{The All-Loop Integrand For Scattering Amplitudes in Planar N=4 SYM}},
  \href{http://dx.doi.org/10.1007/JHEP01(2011)041}{\emph{JHEP} {\bf 01} (2011)
  041}, [\href{http://arxiv.org/abs/1008.2958}{{\tt 1008.2958}}].

\bibitem{Arkani-Hamed:2010pyv}
N.~Arkani-Hamed, J.~L. Bourjaily, F.~Cachazo and J.~Trnka, \emph{{Local
  Integrals for Planar Scattering Amplitudes}},
  \href{http://dx.doi.org/10.1007/JHEP06(2012)125}{\emph{JHEP} {\bf 06} (2012)
  125}, [\href{http://arxiv.org/abs/1012.6032}{{\tt 1012.6032}}].

\bibitem{Badger:2016ozq}
S.~Badger, G.~Mogull and T.~Peraro, \emph{{Local integrands for two-loop
  all-plus Yang-Mills amplitudes}},
  \href{http://dx.doi.org/10.1007/JHEP08(2016)063}{\emph{JHEP} {\bf 08} (2016)
  063}, [\href{http://arxiv.org/abs/1606.02244}{{\tt 1606.02244}}].

\bibitem{Gehrmann:2014bfa}
T.~Gehrmann, A.~von Manteuffel, L.~Tancredi and E.~Weihs, \emph{{The two-loop
  master integrals for $q\overline{q} \to VV$}},
  \href{http://dx.doi.org/10.1007/JHEP06(2014)032}{\emph{JHEP} {\bf 06} (2014)
  032}, [\href{http://arxiv.org/abs/1404.4853}{{\tt 1404.4853}}].

\bibitem{Argeri:2014qva}
M.~Argeri, S.~Di~Vita, P.~Mastrolia, E.~Mirabella, J.~Schlenk, U.~Schubert
  et~al., \emph{{Magnus and Dyson Series for Master Integrals}},
  \href{http://dx.doi.org/10.1007/JHEP03(2014)082}{\emph{JHEP} {\bf 03} (2014)
  082}, [\href{http://arxiv.org/abs/1401.2979}{{\tt 1401.2979}}].

\bibitem{Lee:2014ioa}
R.~N. Lee, \emph{{Reducing differential equations for multiloop master
  integrals}}, \href{http://dx.doi.org/10.1007/JHEP04(2015)108}{\emph{JHEP}
  {\bf 04} (2015) 108}, [\href{http://arxiv.org/abs/1411.0911}{{\tt
  1411.0911}}].

\bibitem{Lee:2020zfb}
R.~N. Lee, \emph{{Libra: A package for transformation of differential systems
  for multiloop integrals}},
  \href{http://dx.doi.org/10.1016/j.cpc.2021.108058}{\emph{Comput. Phys.
  Commun.} {\bf 267} (2021) 108058},
  [\href{http://arxiv.org/abs/2012.00279}{{\tt 2012.00279}}].

\bibitem{Gituliar:2017vzm}
O.~Gituliar and V.~Magerya, \emph{{Fuchsia: a tool for reducing differential
  equations for Feynman master integrals to epsilon form}},
  \href{http://dx.doi.org/10.1016/j.cpc.2017.05.004}{\emph{Comput. Phys.
  Commun.} {\bf 219} (2017) 329--338},
  [\href{http://arxiv.org/abs/1701.04269}{{\tt 1701.04269}}].

\bibitem{Prausa:2017ltv}
M.~Prausa, \emph{{epsilon: A tool to find a canonical basis of master
  integrals}}, \href{http://dx.doi.org/10.1016/j.cpc.2017.05.026}{\emph{Comput.
  Phys. Commun.} {\bf 219} (2017) 361--376},
  [\href{http://arxiv.org/abs/1701.00725}{{\tt 1701.00725}}].

\bibitem{Dlapa:2020cwj}
C.~Dlapa, J.~Henn and K.~Yan, \emph{{Deriving canonical differential equations
  for Feynman integrals from a single uniform weight integral}},
  \href{http://dx.doi.org/10.1007/JHEP05(2020)025}{\emph{JHEP} {\bf 05} (2020)
  025}, [\href{http://arxiv.org/abs/2002.02340}{{\tt 2002.02340}}].

\bibitem{Dlapa:2021qsl}
C.~Dlapa, X.~Li and Y.~Zhang, \emph{{Leading singularities in Baikov
  representation and Feynman integrals with uniform transcendental weight}},
  \href{http://dx.doi.org/10.1007/JHEP07(2021)227}{\emph{JHEP} {\bf 07} (2021)
  227}, [\href{http://arxiv.org/abs/2103.04638}{{\tt 2103.04638}}].

\bibitem{Abreu:2018rcw}
S.~Abreu, B.~Page and M.~Zeng, \emph{{Differential equations from unitarity
  cuts: nonplanar hexa-box integrals}},
  \href{http://dx.doi.org/10.1007/JHEP01(2019)006}{\emph{JHEP} {\bf 01} (2019)
  006}, [\href{http://arxiv.org/abs/1807.11522}{{\tt 1807.11522}}].

\bibitem{Chicherin:2019xeg}
D.~Chicherin, T.~Gehrmann, J.~M. Henn, P.~Wasser, Y.~Zhang and S.~Zoia,
  \emph{{The two-loop five-particle amplitude in $ \mathcal{N} $ = 8
  supergravity}}, \href{http://dx.doi.org/10.1007/JHEP03(2019)115}{\emph{JHEP}
  {\bf 03} (2019) 115}, [\href{http://arxiv.org/abs/1901.05932}{{\tt
  1901.05932}}].

\bibitem{Abreu:2019rpt}
S.~Abreu, L.~J. Dixon, E.~Herrmann, B.~Page and M.~Zeng, \emph{{The two-loop
  five-point amplitude in $ \mathcal{N} $ = 8 supergravity}},
  \href{http://dx.doi.org/10.1007/JHEP03(2019)123}{\emph{JHEP} {\bf 03} (2019)
  123}, [\href{http://arxiv.org/abs/1901.08563}{{\tt 1901.08563}}].

\bibitem{Chen:2021gjv}
L.-B. Chen and J.~Wang, \emph{{Analytic two-loop master integrals for tW
  production at hadron colliders: I *}},
  \href{http://dx.doi.org/10.1088/1674-1137/ac2a1e}{\emph{Chin. Phys. C} {\bf
  45} (2021) 123106}, [\href{http://arxiv.org/abs/2106.12093}{{\tt
  2106.12093}}].

\bibitem{Heller:2019gkq}
M.~Heller, A.~von Manteuffel and R.~M. Schabinger, \emph{{Multiple
  polylogarithms with algebraic arguments and the two-loop EW-QCD Drell-Yan
  master integrals}},
  \href{http://dx.doi.org/10.1103/PhysRevD.102.016025}{\emph{Phys. Rev. D} {\bf
  102} (2020) 016025}, [\href{http://arxiv.org/abs/1907.00491}{{\tt
  1907.00491}}].

\bibitem{Zoia:2021zmb}
S.~Zoia, \emph{{Modern Analytic Methods for Computing Scattering Amplitudes:
  With Application to Two-Loop Five-Particle Processes}}.
\newblock PhD thesis, Aff1= Department of Physics, University of Turin, Turin,
  Italy, GRID:grid.7605.4, Munich U., Munich U., 2022.
\newblock 10.1007/978-3-031-01945-6.

\bibitem{Chaubey:2022hlr}
E.~Chaubey, M.~Kaur and A.~Shivaji, \emph{{Master integrals for $ \mathcal{O}
  $(\ensuremath{\alpha}\ensuremath{\alpha}$_{s}$) corrections to H
  \textrightarrow{} ZZ$^{*}$}},
  \href{http://dx.doi.org/10.1007/JHEP10(2022)056}{\emph{JHEP} {\bf 10} (2022)
  056}, [\href{http://arxiv.org/abs/2205.06339}{{\tt 2205.06339}}].

\bibitem{Chicherin:2017dob}
D.~Chicherin, J.~Henn and V.~Mitev, \emph{{Bootstrapping pentagon functions}},
  \href{http://dx.doi.org/10.1007/JHEP05(2018)164}{\emph{JHEP} {\bf 05} (2018)
  164}, [\href{http://arxiv.org/abs/1712.09610}{{\tt 1712.09610}}].

\bibitem{Goncharov:2010jf}
A.~B. Goncharov, M.~Spradlin, C.~Vergu and A.~Volovich, \emph{{Classical
  Polylogarithms for Amplitudes and Wilson Loops}},
  \href{http://dx.doi.org/10.1103/PhysRevLett.105.151605}{\emph{Phys. Rev.
  Lett.} {\bf 105} (2010) 151605}, [\href{http://arxiv.org/abs/1006.5703}{{\tt
  1006.5703}}].

\bibitem{Duhr:2011zq}
C.~Duhr, H.~Gangl and J.~R. Rhodes, \emph{{From polygons and symbols to
  polylogarithmic functions}},
  \href{http://dx.doi.org/10.1007/JHEP10(2012)075}{\emph{JHEP} {\bf 10} (2012)
  075}, [\href{http://arxiv.org/abs/1110.0458}{{\tt 1110.0458}}].

\bibitem{Duhr:2012fh}
C.~Duhr, \emph{{Hopf algebras, coproducts and symbols: an application to Higgs
  boson amplitudes}},
  \href{http://dx.doi.org/10.1007/JHEP08(2012)043}{\emph{JHEP} {\bf 08} (2012)
  043}, [\href{http://arxiv.org/abs/1203.0454}{{\tt 1203.0454}}].

\bibitem{Duhr:2019tlz}
C.~Duhr and F.~Dulat, \emph{{PolyLogTools \textemdash{} polylogs for the
  masses}}, \href{http://dx.doi.org/10.1007/JHEP08(2019)135}{\emph{JHEP} {\bf
  08} (2019) 135}, [\href{http://arxiv.org/abs/1904.07279}{{\tt 1904.07279}}].

\bibitem{Gaiotto:2011dt}
D.~Gaiotto, J.~Maldacena, A.~Sever and P.~Vieira, \emph{{Pulling the straps of
  polygons}}, \href{http://dx.doi.org/10.1007/JHEP12(2011)011}{\emph{JHEP} {\bf
  12} (2011) 011}, [\href{http://arxiv.org/abs/1102.0062}{{\tt 1102.0062}}].

\bibitem{Bonciani:2022jmb}
R.~Bonciani, V.~Del~Duca, H.~Frellesvig, M.~Hidding, V.~Hirschi, F.~Moriello
  et~al., \emph{{Next-to-leading-order QCD Corrections to Higgs Production in
  association with a Jet}},  \href{http://arxiv.org/abs/2206.10490}{{\tt
  2206.10490}}.

\bibitem{Chicherin:2020umh}
D.~Chicherin, J.~M. Henn and G.~Papathanasiou, \emph{{Cluster algebras for
  Feynman integrals}},
  \href{http://dx.doi.org/10.1103/PhysRevLett.126.091603}{\emph{Phys. Rev.
  Lett.} {\bf 126} (2021) 091603}, [\href{http://arxiv.org/abs/2012.12285}{{\tt
  2012.12285}}].

\bibitem{He:2021esx}
S.~He, Z.~Li and Q.~Yang, \emph{{Notes on cluster algebras and some all-loop
  Feynman integrals}},
  \href{http://dx.doi.org/10.1007/JHEP06(2021)119}{\emph{JHEP} {\bf 06} (2021)
  119}, [\href{http://arxiv.org/abs/2103.02796}{{\tt 2103.02796}}].

\bibitem{Hannesdottir:2021kpd}
H.~S. Hannesdottir, A.~J. McLeod, M.~D. Schwartz and C.~Vergu,
  \emph{{Implications of the Landau equations for iterated integrals}},
  \href{http://dx.doi.org/10.1103/PhysRevD.105.L061701}{\emph{Phys. Rev. D}
  {\bf 105} (2022) L061701}, [\href{http://arxiv.org/abs/2109.09744}{{\tt
  2109.09744}}].

\bibitem{Bourjaily:2022vti}
J.~L. Bourjaily, C.~Vergu and M.~von Hippel, \emph{{Landau Singularities and
  Higher-Order Roots}},  \href{http://arxiv.org/abs/2208.12765}{{\tt
  2208.12765}}.

\bibitem{Flieger:2022xyq}
W.~Flieger and W.~J. Torres~Bobadilla, \emph{{Landau and leading singularities
  in arbitrary space-time dimensions}},
  \href{http://arxiv.org/abs/2210.09872}{{\tt 2210.09872}}.

\end{thebibliography}\endgroup

\end{document}